\documentclass{emulateapj}

\newcommand{\Te}{$T_{\rm eff}$}
\newcommand{\gl}{$\log g$}
\newcommand{\Fe}{[Fe/H]}
\newcommand{\fem}{$<Fe>$}
\def\etal{et~al.\ }

\newbox\grsign \setbox\grsign=\hbox{$>$} \newdimen\grdimen \grdimen=\ht\grsign
\newbox\simlessbox \newbox\simgreatbox
\setbox\simgreatbox=\hbox{\raise.5ex\hbox{$>$}\llap
     {\lower.5ex\hbox{$\sim$}}}\ht1=\grdimen\dp1=0pt
\setbox\simlessbox=\hbox{\raise.5ex\hbox{$<$}\llap
     {\lower.5ex\hbox{$\sim$}}}\ht2=\grdimen\dp2=0pt
\def\simgreat{\mathrel{\copy\simgreatbox}}
\def\simless{\mathrel{\copy\simlessbox}}
\newbox\simppropto
\setbox\simppropto=\hbox{\raise.5ex\hbox{$\sim$}\llap
     {\lower.5ex\hbox{$\propto$}}}\ht2=\grdimen\dp2=0pt

\slugcomment{To Appear in The Astrophysical Journal Supplement Series}

\shorttitle{Population Synthesis in the Blue IV}
\shortauthors{Ricardo P. Schiavon}

\begin{document}


\title{Population Synthesis in the Blue IV. Accurate Model Predictions
for Lick Indices and UBV Colors in Single Stellar Populations}


\author{Ricardo P. Schiavon\altaffilmark{1}}
\affil{Department of Astronomy, University of Virginia, P.O. Box 400325,
Charlottesville, VA 22904-4325}
\email{ripisc@virginia.edu}




\begin{abstract}
We present a new set of model predictions for 16 Lick absorption line
indices from $H\delta$ through Fe5335, and UBV colors for single stellar
populations with ages ranging between 1 and 15 Gyr, and [Fe/H] ranging
from --1.3 to +0.3, and variable abundance ratios.  The models are based
on accurate stellar parameters for the Jones library stars and a new
set of fitting functions describing the behavior of line indices as a
function of effective temperature, surface gravity, and iron abundance.
The abundances of several key elements in the library stars have been
obtained from the literature in order to characterize the abundance
pattern of the stellar library, thus allowing us to produce model
predictions for any set of abundance ratios desired.  We develop a
method to estimate mean ages and abundances of iron, carbon, nitrogen,
magnesium and calcium that explores the sensitivity of the various
indices modeled to those parameters.  The models are compared to high
S/N data for Galactic clusters spanning the range of ages, metallicities
and abundance pattern of interest. Essentially all line indices are
matched when the known cluster parameters are adopted as input. Cluster
spectroscopic ages determined from different Balmer line indices are
consistent to within $\sim$ 1~Gyr.  The models can predict confidently
the above elemental abundances to within $\pm$ 0.1 dex and ages to
within $\pm$ 1 (0.5) Gyr for old (intermediate-age) stellar populations.
Comparing the models to high-quality data for galaxies in the nearby
universe, we reproduce previous results regarding the enhancement of
light elements and the spread in the mean luminosity-weighted ages of
early-type galaxies. When the results from the analysis of blue and red
indices are contrasted, we find good consistency in the [Fe/H] that is
inferred from different Fe indices. Applying our method to 
stacked SDSS spectra of early-type
galaxies brighter than $L^\star$, we find mean luminosity-weighted
ages of the order of $\sim$ 8 Gyr and iron abundances slightly below
solar. Abundance ratios, [X/Fe], tend to be higher than solar, and
are positively correlated with galaxy luminosity.  Of all elements,
nitrogen is the more strongly correlated with galaxy luminosity, which
seems to indicate secondary nitrogen enrichment.  If that interpretation
is correct, this result may impose a lower limit of 50-200 Myr to the
timescale of star formation in early-type galaxies. Unlike clusters,
galaxies show a systematic effect whereby higher-order, bluer, Balmer
lines yield younger ages than $H\beta$. This age discrepancy is stronger
for lower luminosity galaxies. We examine four possible scenarios to
explain this trend. Contamination of the bluer indices by a metal-poor
stellar population with a blue horizontal branch cannot account for the
data. Blue stragglers and abundance-ratio effects cannot be ruled out,
as they can potentially satisfy the data, even though this can only be
achieved by resorting to extreme conditions, such as extremely high [O/Fe]
or specific blue-straggler frequencies. The most likely explanation is the
presence of small amounts of a young/intermediate-age stellar population
component.  We simulate this effect by producing two-component models and
show that they provide a reasonably good match to the data when the mass
fraction of the young component is typically a few \%.  If confirmed, this
result implies star formation has been extended in early-type galaxies,
and more so in less massive galaxies, which seems to lend support to the
``downsizing'' scenario.  Moreover, it implies that stellar population
synthesis models are capable of constraining not only the mean ages of
stellar populations in galaxies, but also their age spread.

\end{abstract}




\section{Introduction}

With the consolidation of the cold dark matter scenario for structure
formation (e.g., Blumenthal \etal 1984), the study of galaxy evolution
is entering an era of high precision, such that crucial questions can
only be answered on the basis of accurate data and models. For instance,
roughly half of all stellar mass in today's universe inhabits early-type
galaxies (Fukugita, Hogan \& Peebles 1998) yet a definitive picture of
the history of star formation in these systems is still lacking. That
is one of the chief motivations for the construction of stellar population
synthesis models.  High accuracy is required for such models because the
spectrophotometric evolution of stellar populations proceeds at a very
slow pace after the first Gyr or so, which makes it very hard to extract
reliable age information from the integrated light of galaxies when most
of the stars are old.

Stellar population synthesis aims at discerning the stellar mix in
galaxies from their integrated spectral energy distributions. With that
intent, models are computed which predict the evolution of magnitudes,
colors, and absorption line indices of stellar populations. Comparisons
of these models with the observations should constrain the age and metal
abundance distribution of stars in galaxies, thus yielding constraints
on their histories of star formation and chemical enrichment. The problem
is complicated, however, as the spectral energy distributions of stellar
populations respond to variations of different parameters in degenerate
ways. One popular example is the age-metallicity degeneracy (e.g., Faber
1972, 1973, O'Connell 1980, Rose 1985, Renzini 1986, Worthey 1994),
whereby stellar population colors and most absorption line strengths
respond similarly to variations of age and metallicity. Major improvement
was brought by the introduction of the Lick/IDS system of equivalent
widths (Burstein \etal 1984, Gorgas \etal 1993, Worthey \etal 1994),
which systematized the measurements of absorption line strengths in the
spectra of stars and galaxies. Later on, with the development of models to
predict the strengths of these indices as a function of stellar population
parameters (e.g., Worthey \etal 1994, Worthey 1994, Bressan, Chiosi \&
Fagotto 1994, Weiss, Peletier \& Matteucci 1995, Borges \etal 1995), key
aspects of the evolution of early-type galaxies were unveiled. Worthey,
Faber \& Gonz\'alez (1992) showed that giant ellipticals are characterized
by enhancement of the abundances of light elements (see also Peterson
1976, O'Connell 1980, Peletier 1989), which possibly indicates that the
bulk of their stars were formed in a rapid ($\simless$ 1~Gyr) star formation 
event. Later on, Worthey (1994) showed that the $H\beta$ index is more 
sensitive to age than to metallicity, thus allowing to break the age-metallicity
degeneracy. Further extension of the models towards higher-order Balmer
lines was accomplished by Worthey \& Ottaviani (1997). Despite some
controversy as to how clean an age indicator a given Balmer line is,
this sparked a world-wide industry to estimate mean-ages and metallicities
of stellar populations in galaxies. Until very recently, however, most
of the work has been focussed on the ``green'' Lick indices: $H\beta$,
Fe5270, Fe5335, Mg$_2$, and Mg $b$. The blue indices ($\lambda \simless
4500 {\rm\AA}$) were for a while relegated to a relative ostracism due
to problems in the calibration of original Lick/IDS data in the blue and
to intrinsic modeling difficulties related to the higher crowding of
lines in that spectral region, which renders a clean absorption line strength
measurement extremely difficult.

But the integrated spectra of early-type galaxies in the blue contain
a wealth of information for those who take the challenge, as
demonstrated by the pioneering work of Rose (1985, 1994). Moreover,
combining accurate models in the blue to those currently available for
red indices adds the benefit of a wider baseline, which proves to be
extremely advantageous for stellar population studies (O'Connell 1976).
Another important benefit of constructing consistent models within
a large baseline that includes the blue spectral region lies in the
need to interpret the integrated spectra of remote galaxies. Ongoing
surveys based on 8-10 m class telescopes are obtaining large amounts
of high-quality spectroscopic data for galaxies at z $\sim$ 1 (e.g.,
DEEP survey, Davis \etal 2003; VIRMOS-VLT Deep Survey, Le F\'evre \etal
2001, K20, Cimatti \etal 2002, Gemini Deep Deep Survey, Abraham \etal
2004). Because of strong telluric emission lines in the far red/near
infrared, only the blue spectral region is accessible from the ground
for galaxies at the involved redshifts, using current instrumental
capabilities. Therefore, models which are consistent from the blue to
the red are crucial, so that the mean ages and metallicities of remote
systems, which are necessarily based on blue spectra, can be safely tied
to those measured in nearby galaxies.

In this paper, we present a new set of model predictions for line
indices and UBV colors of single stellar populations. Our goal is to
produce models that are accurate and consistent throughout the spectral
range going from $\lambda\lambda$ 4000 to 5400 ${\rm\AA}$. This is the 
fourth paper of a series dedicated to the study of stellar populations in the 
optical, with emphasis in the blue spectral region. In Schiavon \etal (2002a,b,
hereafter Papers I and II), we studied the integrated spectrum of the
moderately metal-rich Galactic globular cluster 47~Tuc and in Schiavon,
Caldwell \& Rose (2004, hereafter Paper III) we constructed and analyzed
the integrated spectrum of the metal-rich, intermediate-age Galactic
open cluster, M~67.

Our models are based on a new set of fitting functions for indices in
the Lick system. We omit on purpose the ``IDS'' part of the usual 
nomenclature of this system of equivalent widths because, as it will be 
seen in Section~\ref{abslines}, our models are not in the Lick/IDS
system, as they are {\it not} based on index measurements taken in 
the standard Lick/IDS stellar library (Burstein \etal 1984, Gorgas 
\etal 1993, Worthey \etal 1994). Instead, the index measurements 
that form the backbone of our models were taken in a much more
recent spectral library, by Jones (1999). The spectra from that library
are flux-calibrated, thus being unaffected by the response curve of the 
old Lick Image Dissector Scanner. However, for reasons that will 
become clear in Section~\ref{abslines}, our line indices are measured
at the relatively low resolution of the original Lick/IDS system. We are 
aware of the ongoing work on the construction of better spectral libraries 
with higher resolution than that of the Lick system, and spanning a wide 
range of stellar parameters (e.g., Le Borgne \etal 2003, Valdes \etal 2004). 
However, when this project started, these libraries were not available publicly. 
Moreover, our main goal is to apply these models to study distant giant early-type 
galaxies, whose spectra are irretrievably smoothed by their high velocity 
dispersions to resolutions that are comparable to that of the Lick system 
($\sim$ 8 ${\rm\AA}$). 

It is also important to justify here our reasons to adopt fitting functions,
even though in Papers I and II we produced synthetic spectra of
single stellar populations. The main reason is that the degree of 
accuracy needed for such model predictions cannot be achieved
on the basis of the Jones (1999) library, because of its relatively limited
coverage of stellar parameter space. Moreover, fitting functions are very 
convenient for a number of reasons. For instance, they can be easily implemented
in any evolutionary synthesis code. In addition, models based on fitting
functions can also be corrected to yield line index predictions for
varying abundance patterns.

A key feature of our models is related to the stellar parameters
adopted for the library stars. They have to be homogeneous, internally
accurate, and free of important systematic effects. Accuracy is very
important, for it allows a precise assessment of the behavior of stellar
observables as a function of fundamental parameters. This aspect of our
models was very carefully crafted.

One of the chief applications of stellar population synthesis is the
estimation of mean elemental abundances of stars in remote systems from
absorption line indices measured in their integrated spectra. Ratios of elemental
abundances, such as that between magnesium and iron, hold important clues
for the history of star formation and chemical evolution of galaxies
(e.g., Matteucci \& Tornamb\`e 1987; Wheeler, Sneden \& Truran 1989;
Peletier 1989, Worthey, Faber \& Gonz\'alez 1992; Edvardsson \etal 1993;
McWilliam 1997; Worthey 1998). Therefore, models that are able to convert,
for instance, Mg $b$ and Fe4383 measurements into a mean [Mg/Fe] for a
given galaxy are highly desirable.

This is unfortunately not very easy to achieve because of two
reasons. First, the integrated spectra of galaxies are smoothed due to
the intrinsic dispersion of the velocities of their member stars along
the line of sight. As a result, all the absorption lines are blended,
making it impossible to isolate absorption features that cleanly indicate
the abundance of a given chemical species. Second, detailed abundance
patterns for the majority of the stars used in the construction of the
models---hence the abundance pattern of the models themselves---are 
unknown.

A method to address these difficulties was devised by Trager \etal (2000)
and further developed by Proctor \& Sansom (2002), Thomas, Maraston \&
Bender (2003a), and Thomas, Maraston \& Korn (2004), and Korn,
Maraston \& Thomas (2005). The core of this
method resides in the use of sensitivities of Lick indices to
variations in the elemental abundances of all the relevant chemical
species with absorption lines included in each index passband and
pseudo-continuum windows. Trager \etal used the Tripicco \& Bell (1995)
tabulations of index sensitivities computed from spectrum synthesis
adopting model atmospheres of stars with representative stellar
parameters. The sensitivities were used to integrate the effect of
abundance ratio variations of the main Lick indices in the green region.
That allowed them to estimate by how much the mean [Mg/Fe] of the stellar
populations of elliptical galaxies in their sample depart from that of
the spectral library used as input in their models (which they assumed
to be equal to solar). The method was extended by Thomas \etal (2003a)
to include all the Lick/IDS indices in Worthey \etal (1994). Later on,
Korn \etal (2005) computed new sensitivities that include also the higher
order Balmer lines defined by Worthey \& Ottaviani (1997).

It is vital for the success of this method that the abundance pattern
of the models be well-known. As emphasized by Thomas \etal (2003a),
models that are based on empirical stellar libraries are characterized
by an abundance pattern that is equal to that of the stars that make
up the adopted spectral library. Therefore, we decided to survey the
literature for elemental abundance determinations of the stars in the
spectral library adopted in our models. We provide mean abundance ratios
as a function of [Fe/H] for several important chemical elements. This
information is used in Section~\ref{aenhance}, in combination with
our fitting functions and the Korn \etal (2005) sensitivity tables,
to produce model predictions for varying abundance patterns. We present
a detailed study of the response of line indices to age and elemental
abundance variations, in order to explore ways in which our models
can be used to constrain those parameters for intermediate-age and old
stellar populations. On the basis of the insights gained in this study,
we develop a new method to estimate the mean luminosity-weighted age
of stellar populations, as well as their abundances of iron, magnesium,
calcium, carbon, and nitrogen.

In a degenerate problem like that of stellar population synthesis, knowing
the answers that are to be expected for a given set of input parameters
is priceless. Therefore, a detailed comparison of our model predictions
for single stellar populations with accurate data for well-known Galactic
clusters is performed. Our goal is to match all 16 indices for a small
sample of clusters which spans the entire range of stellar population
parameters of interest. We hope to convince the reader that significant
advance has been made towards meeting this initial goal. At the end of
this exercise, we show that the ages and metal abundances derived from
application of our models to the integrated light of stellar populations
is meaningful in an absolute sense, i.e., they are consistent with
metallicity and age scales defined by known local systems, such as
Galactic clusters and stars in the solar neighborhood.

Once we are convinced that the comparison of the models with cluster data
gives satisfactory results, we turn our attention to galaxies. In this
paper, we refrain from pursuing a detailed analysis of the galaxy data
and instead make more qualitative comparisons between data and models
and discuss what can be learned therefrom.  We first compare our models
to the data from Trager \etal (2000a,b, henceforth simply Trager \etal
2000) in order to make sure that we reproduce some well established
results, such as the $\alpha$-enhancement characteristic of massive
early-type galaxies and the spread in the mean ages of their stellar
populations. Next, we take advantage of the wide baseline covered by
our models to compare them to measurements taken on stacked spectra of
early-type galaxies from the Sloan Digital Sky Survey (Eisenstein \etal
2003). We determine the mean abundances of iron, magnesium, calcium,
and, for the first time, those of carbon and nitrogen for the stars in
early-type galaxies. The behavior of these abundances as a function
of galaxy luminosity is studied.  Of all abundance ratios studied,
[N/Fe] is the one that seems to be the most strongly correlated with
galaxy luminosity, perhaps indicating an important secondary source of
nitrogen enrichment in these galaxies. If confirmed, this result may
be telling us that there is a minimum duration for the star formation
in early-type galaxies, which is set by the lifetimes of the stars
contributing secondary nitrogen. If the characteristic masses of these
stars, as proposed by Chiappini, Matteucci \& Ballero (2005), range
between 4 and 8 $M_\odot$, these timescales are of the order of 50-200
Myr. A more accurate prediction can be obtained on the basis of detailed
chemical evolution modeling.

Regarding stellar ages, unlike what we found in the case of clusters,
models do not match the data consistently throughout the spectral region
considered.  Specifically, bluer Balmer lines tend to indicate younger
mean ages than $H\beta$.  This might be one of the many instances in
the field of stellar population synthesis where, when the models do
not match the data, there might be something interesting to be learned.
We argue that we are detecting an age spread in the stellar content of
early-type galaxies.

This paper is organized as follows. In Section 2 we describe the stellar 
library used in the models and the determination of the stellar parameters
and abundance pattern of its constituent stars. The fitting functions 
are presented in Section 3. Model construction is presented in Section 4.
In Sections 5 and 6 the models are compared with cluster and galaxy data,
respectively. Our conclusions are summarized in Section 7. The reader who
is not interested in model construction details should go directly to 
Section 5.

\section{Stellar Library}

We adopted the library of stellar spectra by Jones (1999). It consists
of spectra of 684 stars collected with the Coude-feed telescope at Kitt
Peak National Observatory. The spectra cover the region that goes from
3820 to 5410 ${\rm\AA}$ with a gap between 4500 and 4780 ${\rm\AA}$,
with a resolution of 1.8 ${\rm\AA}$. More detailed information about
the spectral library can be found in Jones \& Worthey (1995, hereafter
JW95), Jones (1999), Vazdekis (1999), and Paper I). The Jones spectral
library is very comprehensive, encompassing all spectral types and
luminosity classes that contribute relevantly to the integrated light
of early-type galaxies in the optical. 

\subsection{Photometric Data}

UBV photometry for all the stars was taken from the SIMBAD database.
Str\"omgren photometry for the dwarfs, which was used in the stellar
parameter determinations, was taken from the compilation of Hauck 
\& Mermilliod (1998). 

\begin{figure*}
\plotone{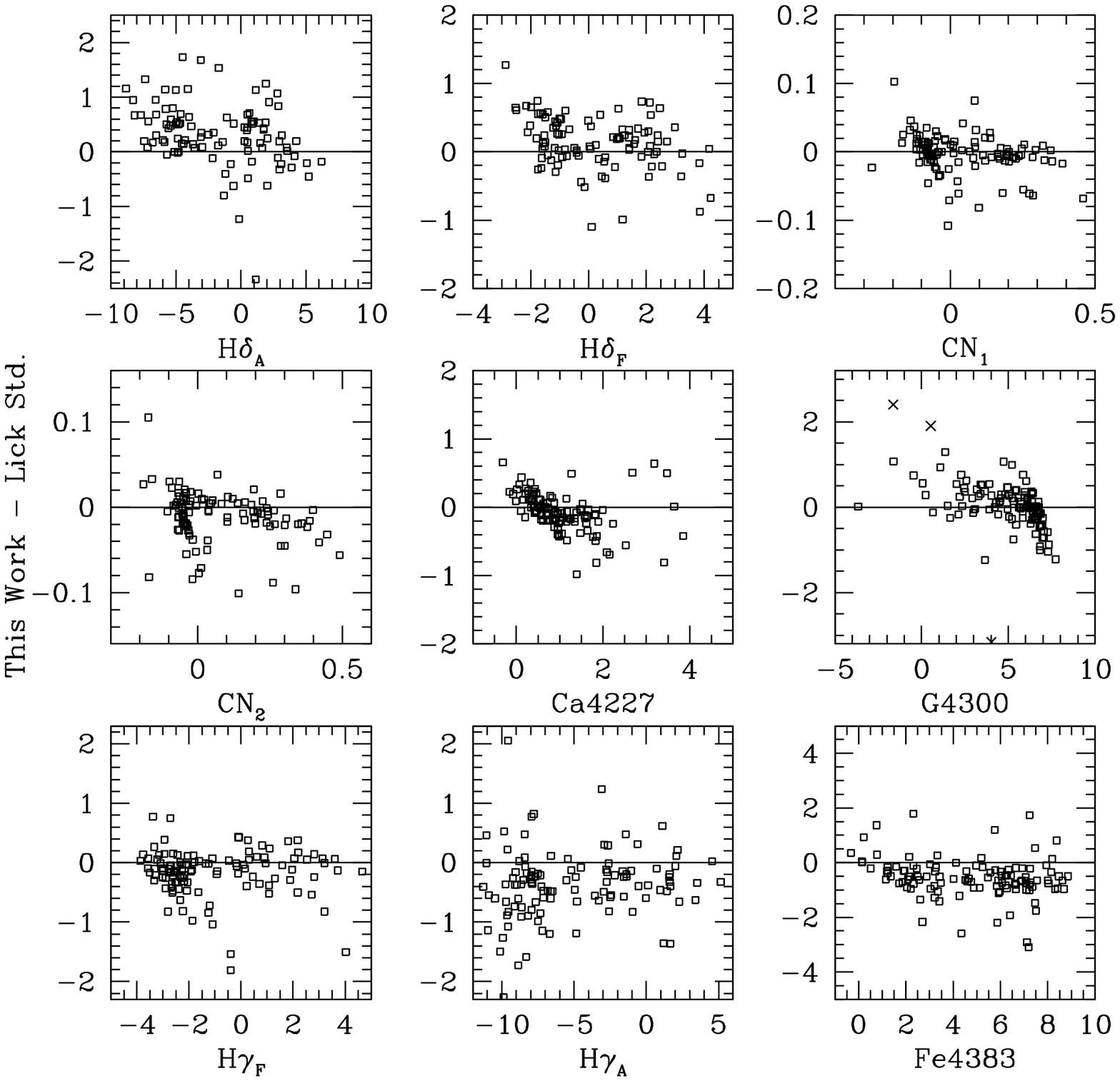}
\caption{a. Comparison of the line indices measured in our spectra for
Lick/IDS standards with the values tabulated by Worthey \etal (1994)
and Worthey \& Ottaviani (1997). For all indices only minor zero-point
corrections are needed to convert our data to the Lick/IDS system. Crosses
mark stars removed from the zero-point estimate (see text for details).
}
\label{calib}
\end{figure*}

\setcounter{figure}{0}
\begin{figure*}
\plotone{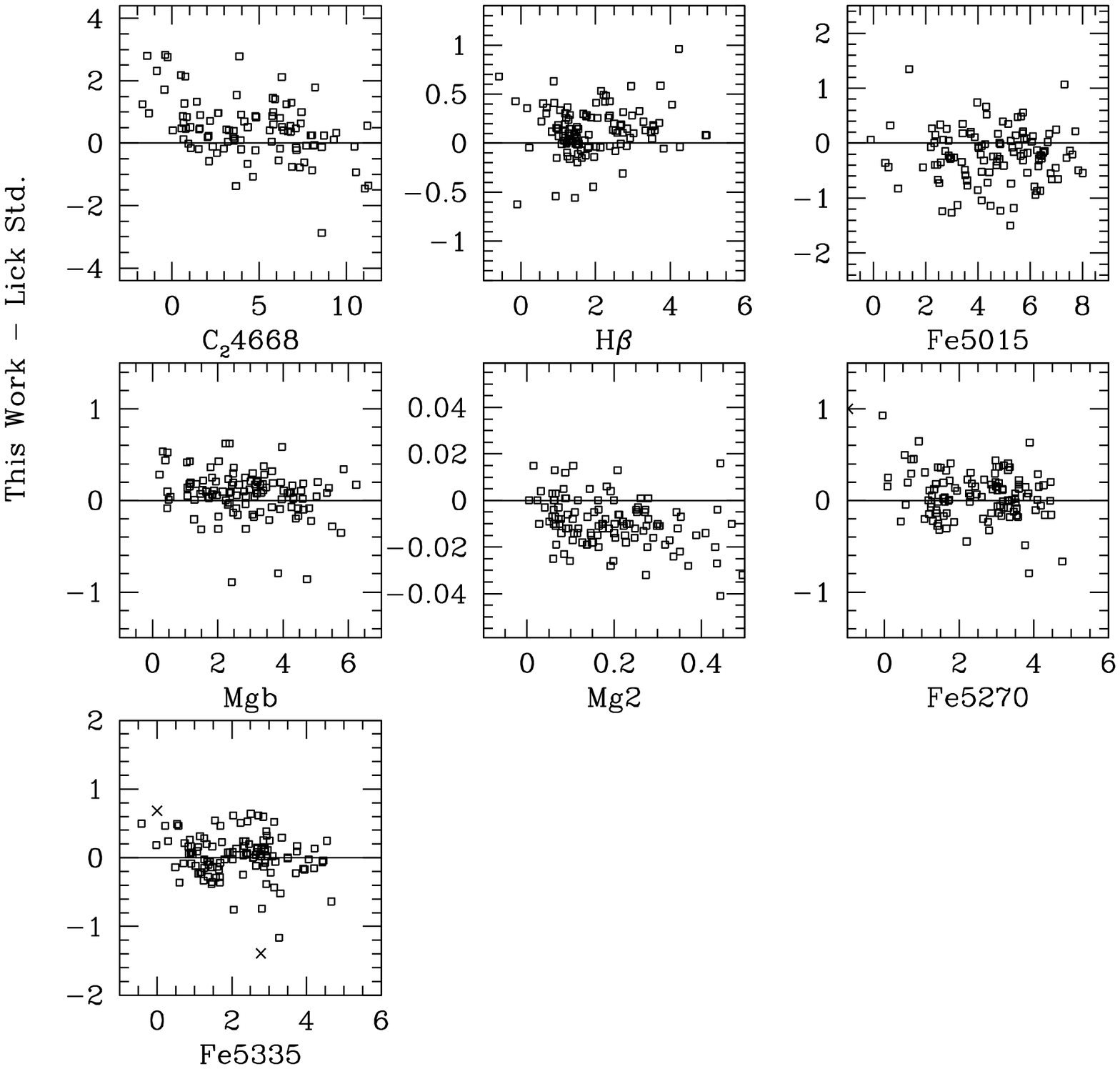}
\caption{b.}
\label{calib}
\end{figure*}

\subsection{Absorption Line Indices} \label{abslines}

We measured the equivalent widths (EWs) of a number of absorption lines,
following the definitions given by Worthey \etal (1994) and Worthey \&
Ottaviani (1997). The somewhat limited spectral coverage of the Jones
spectral library prevented us from measuring several interesting line
indices, such as Ca4455, Fe4531, C$_2$4668, and all indices redder than
$\sim$ 5400 ${\rm\AA}$ (but see Section~\ref{cdois}). Nevertheless, the
remaining Lick/IDS indices that can be modeldeficiencyed on the basis of the Jones
spectral library still provide us with a rich set of spectral indicators
which are sensitive to the ages of old stellar populations, as well as to
the abundances of key elements for the understanding of galaxy chemical
evolution, such as iron, magnesium, calcium, carbon and nitrogen.

Another limitation of the Jones library refers to its coverage of
stellar parameters, whereby some important loci of stellar parameter
space are not represented with sufficient density. In order to address
this deficiency, and enhance the robustness of our fitting functions in
those stellar parameter regions, we decided to supplement our data with
index measurements from Worthey \etal (1994) for stars hotter than 7000
K, M giants and K-M dwarfs. For that purpose, we need to determine the
conversion between our EWs and the Lick/IDS system.  A detailed recipe
to perform this determination has been given by Worthey \& Ottaviani
(1997), and is followed here.  The most important part of the conversion
involves degrading the resolution of the Jones spectra (1.8 ${\rm\AA}$)
to match the lower, variable resolution of Lick/IDS spectra (8.5--11
${\rm\AA}$). This was achieved by gaussian-convolving the Jones spectra
in order to match the Lick/IDS resolution. The resolution of the original
Lick/IDS spectra at the central wavelength of each index was obtained
from graphical interpolation in Figure 7 of Worthey \& Ottaviani (1997).
In Table~\ref{zeropoints} we provide the resolution FWHM assumed for
each index.

Equivalent widths are somewhat dependent on the software used to perform
the actual measurements.  In the initial stages of this project, all
index measurements were performed using a script based on the IRAF
{\tt bplot} routine (i.e., {\tt splot} in batch mode). Unfortunately,
however, we later realized that {\tt bplot} did not consider fractionary
pixels. That means that the wavelengths of the pseudo-continuum and
passband definitions actually employed in the measurements were not the
input numbers, but were instead the wavelengths of the pixels that were
nearest to those of the original definitions. That error introduced in
our EWs systematic effects that were a function of the actual grid of
wavelengths defined by the dispersion solution for each spectrum and
which, of course, were more severe for lower resolution spectra. As a
result, the index measurements had to be retaken, this time using the
LECTOR\footnote{See http://www.iac.es/galeria/vazdekis/index.html}
program, by A. Vazdekis, and {\it all} the numbers 
in this paper (in particular, the index fitting functions,
see Section~\ref{ffs}) had to be re-derived.

\subsubsection{The C$_2$4668 Index} \label{cdois}

At a later stage in this project, we realized that carbon abundances
affect the blue spectral region importantly enough that one would want to
nail them down as tightly as possible. In order to improve our confidence
in our carbon abundance determinations, we decided to add to our models
the $C_2$4668 index, which is extremely sensitive to carbon abundances
(Trippico \& Bell 1995). This became possible when the Indo-US spectral
library, by Valdes \etal (2004) became publicly available. This new
spectral library covers the entire spectral range between $\sim$ 3500
and 9500 ${\rm\AA}$ (with a resolution of $\sim$ 1 ${\rm\AA}$, FWHM),
without gaps such ast that in the Jones (1999) library. Most importantly,
the set of 1273 stars in the Indo-US library contains almost all the Jones
(1999) stars, for which we obtained accurate stellar parameters (see
Section~\ref{stelpar}), so that incorporating the C$_2$4668 index in
our models depended only on getting accurate measurements from Indo-US
spectra.  Such measurements were performed after convolving the Indo-US
spectra into the Lick resolution (Table~\ref{zeropoints}). Equivalent
widths, fitting functions, and model predictions for other Lick indices
not covered by the Jones (1999) library, such as Ca4455, Fe4531, and all
indices redder than $\sim$ 5400 ${\rm\AA}$ will be presented elsewhere.
Models for all the other indices presented in this paper are based on 
measurements taken on spectra from the Jones (1999) library.

\subsubsection{Zero-points}

In this Section we compare the index measurements taken in the
smoothed Jones spectra with the standard Lick/IDS measurements
from Worthey \etal (1994) and Worthey \& Ottaviani (1997) for stars
common to the two spectral libraries in order to derive zero-point
transformations between the two index systems. Such comparisons are
shown in Figures~\ref{calib}a-b, where the residual differences between
measurements in the two sets of spectra are plotted as a function
of index strength. These Figures are worthy of some contemplation
and a few thoughts. First, we call attention to the large scatter
found for all the indices (the standard deviations are listed in
Table~\ref{zeropoints}). This is a striking result, given the fact that
the EWs were measured in very high S/N spectra of bright stars. In fact,
such a large scatter should not come as a surprise, as any spectroscopist
is acquainted with the fact that even EW measurements taken in repeat
spectra of the same star, taken with the same instrumental setup,
are also characterized by a sizable scatter, which is probably due
to a combination of wavelength-calibration, background-subtraction,
and flat-fielding errors, low resolution, poor determination of the
latter, cosmic-ray residuals, bad pixels, variations in spectrograph
focus along an observing night, and a myriad of other possible factors
that can spoil the measurement of an equivalent width. Second, zero-point
differences are found for some indices, most notably the wider-baseline
molecular-band indices such as Mg$_2$, CN$_2$ and G4300. Zero-point
differences are also found for some narrower indices, such as $H\delta_A$,
$H\gamma_A$, and Mg $b$.  Such differences, especially in the case of the
wide-baseline indices, are mostly (but not only, see below) due to the
fact that, contrary to case of the Jones spectra, those of the Lick/IDS
standards are not flux-calibrated, so that wide-band indices measured in
the latter are liable to be affected by the response curve of the Lick
Image Dissector Scanner. Third, there is a hint of a systematic trend of
the residuals as a function of index strength for some of the indices,
like CN$_2$, Ca4227, G4300, and Mg$_2$. Such trends are not uncommon
(see, for instance, Paper III), and Figures \ref{calib}a-b highlight the
importance, in any observational work dealing with Lick indices, to secure
large amounts of standard star spectra, covering a wide range of index
values, in order to guarantee a safe conversion into the Lick system.

\begin{deluxetable*}{crrrrrrrr} 
\tabletypesize{\scriptsize}
\tablecaption{Zero-point conversions between the equivalent width
systems of this work and the original Lick/IDS (Worthey \etal 1994)}
\tablewidth{0pt}
\tablehead{\colhead{Index} & $H\delta_A$ & $H\delta_F$ &
           CN$_1$ & CN$_2$ & Ca4227 & 
           G4300 & $H\gamma_A$ & $H\gamma_F$ }
\startdata
Resolution (${\rm\AA}$)  & 10.9 & 10.9 & 10.3 & 10.3 & 10.3 & 
               9.7 & 9.5 & 9.5 \\
Zero point$^2$ (${\rm\AA}$) &  +0.38 &  +0.16 & --0.0027 & --0.011 & --0.11 &  +0.04 & --0.45 & --0.16 \\
r.m.s.  (${\rm\AA}$)  &  0.67 &  0.43 &  0.029  &  0.029 &  0.35  &  0.46 &  0.54 &  0.39  \\
 & & & & & & & & \\
\noalign{\hrulefill}
 & & & & & & & & \\
 & Fe4383 &  C$_2$4668 &  $H\beta$  & Fe5015   &   Mg$_2$ &  Mg $b$  & Fe5270  & Fe5335     \\
 & & & & & & & & \\
\noalign{\hrulefill}
 & & & & & & & & \\
Resolution & 9.3 & 8.7 & 8.4 & 8.4 & 8.4 & 8.4 &    8.4 & 8.4 \\
 Zero point$^2$ & --0.55 &  +0.45 & +0.13   & --0.20   & --0.0095 & +0.06 & +0.059  &  +0.030 \\
 r.m.s. &  0.96 &  0.93 & 0.24   &  0.73  &  0.010 &  0.29 &  0.25 & 0.30  \\
\enddata
\tablenotetext{1}{FWHM}
\tablenotetext{2}{Zero point = $I_{Schiavon} - I_{WFBG}$} 
\label{zeropoints}
\end{deluxetable*}

It is very difficult, from the comparisons shown in Figures~\ref{calib}a-b
alone, to have an idea of the true quality of our EW measurements,
given that they are compared with lower quality measurements taken with
the Lick/IDS instrument. In Figures~\ref{calib2}a-b, our measurements
are compared to those taken in high quality, flux-calibrated, spectra
presented in Paper III for stars in common with this program. These were
taken with the FAST spectrograph (Fabricant \etal 1998), attached to the
1.5 m telescope at the Fred Whipple Observatory.  One can see that the
residuals are much smaller than those between our index measurements and
the original Lick/IDS standard values (Figures~\ref{calib}a-b). Comparison
between Figures~\ref{calib}a-b and \ref{calib2}a-b should serve as
an eloquent statement of the vast improvement in the quality of the
equivalent widths upon which our models are based. The line indices
of standard stars are of course in the very root of our models and,
without such high quality measurements, the task of making accurate
model predictions would be hopeless.

\begin{figure*}
\plotone{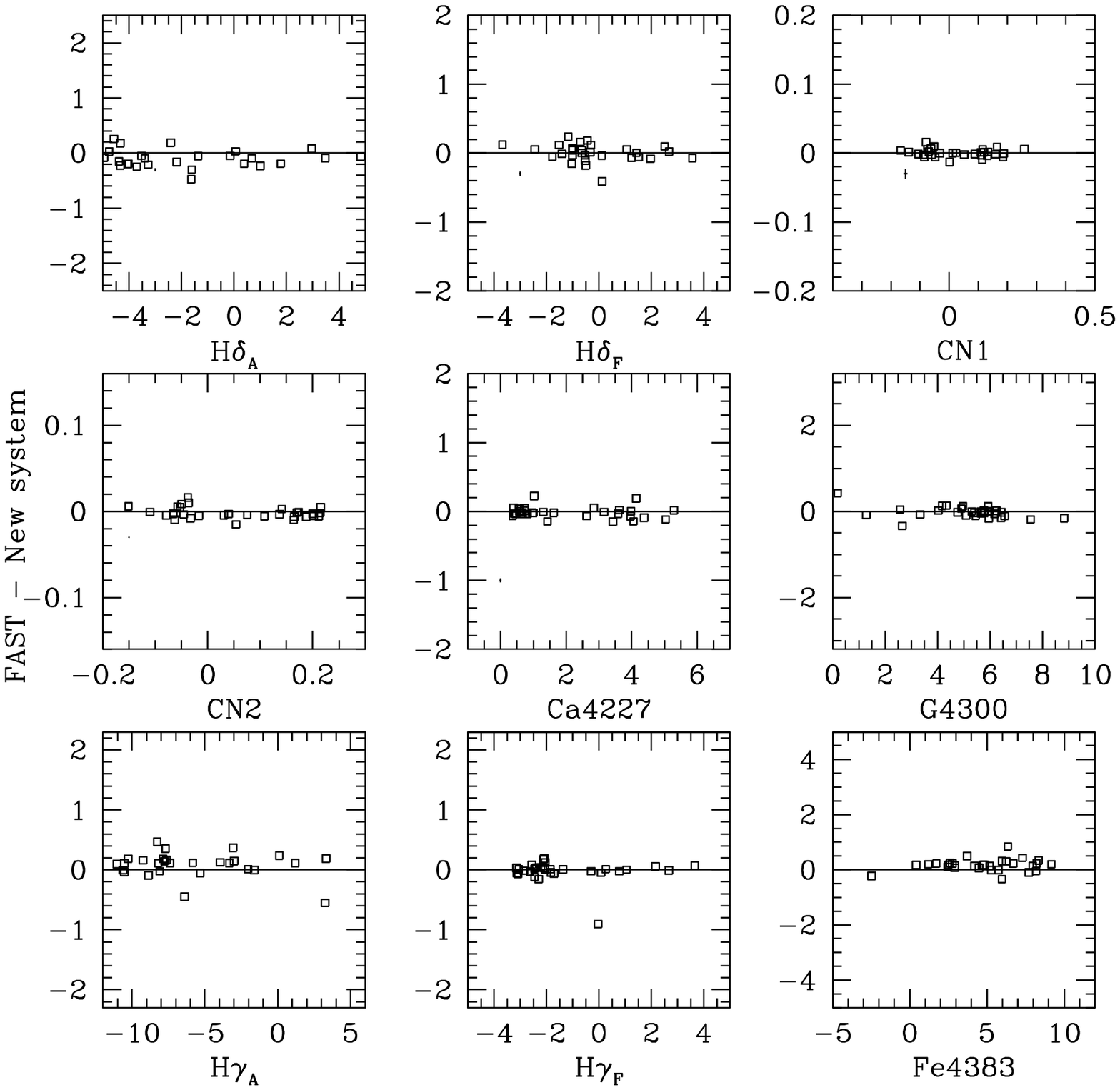}
\caption{a. Comparison of the line indices measured in the Jones
(1999) spectra for Lick standards with those obtained in Paper III
(see text). Compare the scatter of the residuals with that seen in
Figure~\ref{calib}. The much lower scatter seen here provides an
assessment of the much better quality of the index measurements upon
which our models are based.
}
\label{calib2}
\end{figure*}

\setcounter{figure}{1}
\begin{figure*}
\plotone{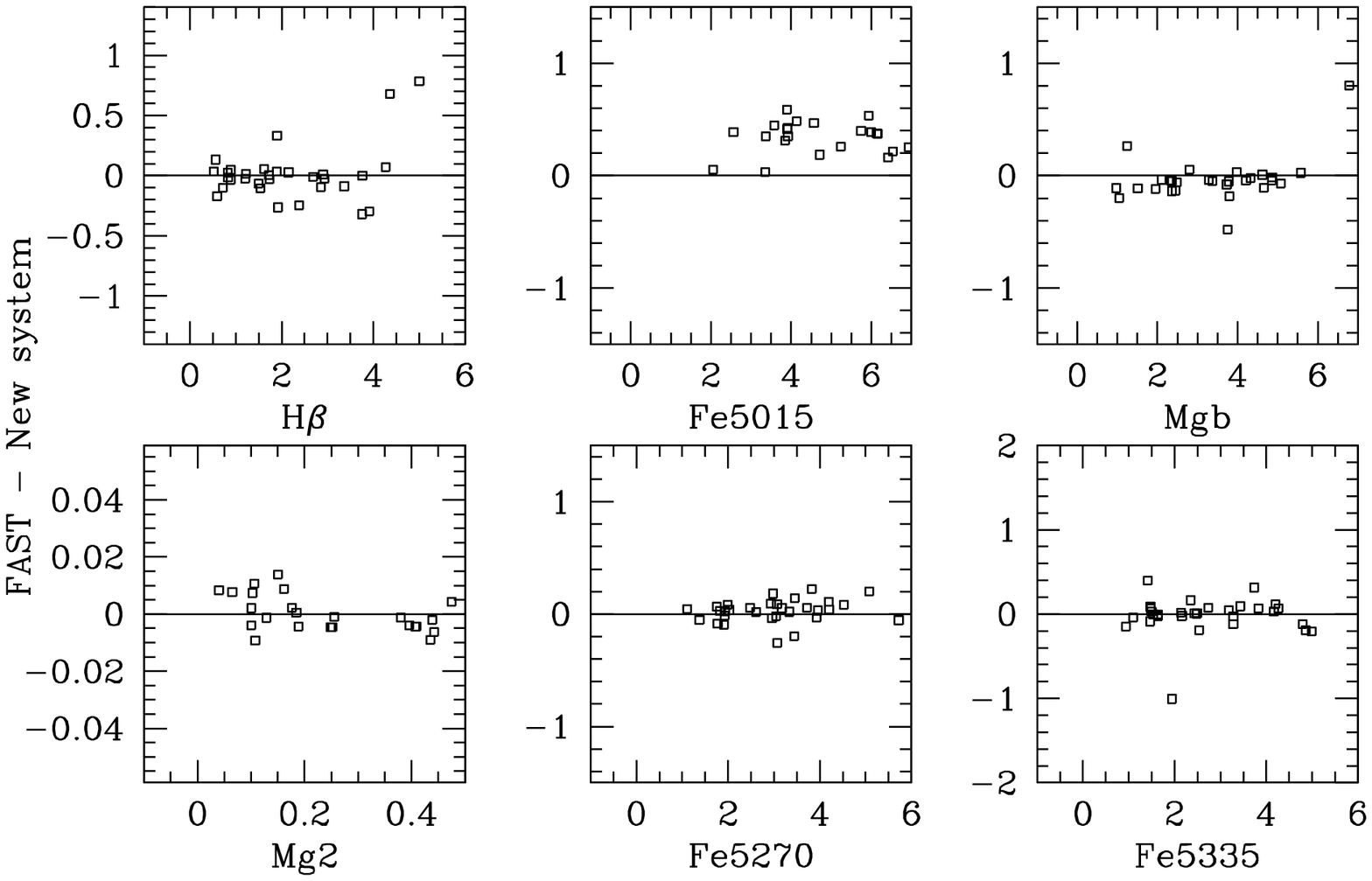}
\caption{b.} 
\label{calib2}
\end{figure*}

It is interesting to note, however, that even when flux-calibrated
spectra are employed, there are zero-point differences between this
work's and FAST measurements, as clearly visible in the case of Fe5015
(Figure~\ref{calib2}b) and, to a lesser extent, $H\delta_A$, $H\gamma_A$,
and Fe4383. This serves as a demonstration that flux calibration alone
cannot eliminate the need for zero-point determinations, based on extensive
measurements taken on high quality standard star spectra. In other words,
any equivalent width measurement necessarily depends on the instrumental
set up and reduction techniques employed in obtaining the spectra, so
that conversion into the equivalent system defined by a given set of
standard values will always be necessary. While it is true that most
zero-point differences in Figures~\ref{calib2}a-b are very small, the
increasing quality of both models and data will certainly push the need
towards higher and higher accuracy measurements, thus requiring precise
zero-point determinations.

In view of the above considerations we decided to maintain measurements
taken on flux-calibrated spectra, like those of the Jones library,
because they are more easily reproducible by observers using different
instrumental setups. Therefore, we decided {\it not} to convert our EWs
into the Lick/IDS system, but rather redefine it on the basis of the
measurements performed on the Jones spectral library. To that effect, the
EWs of all line indices are given in Table A1 in the Appendix. Observers
wishing to compare their data to the models presented in this paper
should seek to reproduce the EWs of the Jones standards provided in
Table A1 with measurements performed in spectra obtained with their
own instrumental setups.

Finally, in order to conform with the vast amount of previous work based
on the Lick/IDS system, we estimate zero-point conversions between
our EWs and the Lick/IDS standards. These conversions are listed in
Table~\ref{zeropoints}. Those wishing to compare the models presented
here with measurements taken in the Lick/IDS system should first take those
conversions into account.

\subsection{Stellar Parameters} \label{stelpar}

A fundamental aspect of the construction of stellar population models is
the set of stellar parameters adopted for the library stars.  Here we
provide a brief summary of our method to determine the effective
temperatures (\Te), iron abundances (\Fe) and surface gravities (\gl)
of F and G dwarfs, and G and K giants. A more detailed description can
be found in Paper I. Since the main focus of that paper was reproducing
the observations of a mildly metal-rich Galactic globular cluster
(47~Tuc), we did not provide a detailed account of our procedure to
determine the stellar parameters of stars outside the above range of
spectra types. Therefore, we briefly summarize here the determination
of stellar parameters for FG dwarfs and GK giants, and in the following
subsections we describe the determinations for cooler and hotter stars.

For dwarf stars, \Te\ and \Fe\ were estimated from Str\"omgren photometry
adopting the calibrations by Alonso \etal (1996) and Schuster \&
Nissen (1989, as revised by Clementini \etal 1999). For giants those
parameters are based on the \Te\ and \Fe\ scales of Soubiran, Katz \&
Cayrel (1998), through the construction of a relation between absorption
line features and stellar parameters for a sub-sample of the library
stars in common with that study (see details in Paper I). For all the
stars, \gl\ was determined using {\it Hipparcos} parallaxes, bolometric
corrections inferred from the calibrations by Alonso \etal (1995, 1999)
and an assumed mass. A discussion of the uncertainties in these stellar
parameter determinations can be found in Paper I.

\subsubsection{M Giants}

M giants dominate the integrated light of old stellar populations at
wavelengths redder than $\sim$ 6500 ${\rm\AA}$ (Schiavon \& Barbuy 1999,
Schiavon, Barbuy \& Bruzual 2000). In the optical, even though their
total contribution to the integrated light is not dominant, they affect
the equivalent widths of key absorption features, such as Mg $b$ and
$H\beta$ (Paper III). Therefore, it is important to correctly estimate
their stellar parameters in order to account for their contribution
to the integrated light. However, this is not an easy task, because
despite the tremendous progress made in the last decade or so towards
the understanding of the atmospheres of these stars, fundamental stellar
parameters are known for very few M giants (metallicities for virtually
none of those in the field). Likewise, the representation of M giants
in current stellar libraries is scarce.

Our procedure to determine the stellar parameters of M giants was
the following. We searched the literature for determinations of \Te\
of library stars from a fundamental method, such as angular diameter
measurements (Ridgway \etal 1980, Dyck, van Belle \& Thompson 1998,
Perrin \etal 1998), or from the infrared flux method (Blackwell,
Lynas-Gray \& Petford 1991,  Alonso \etal 1999), which is known to be
fairly model (thus, metallicity) independent. This sample was further
supplemented by stars for which \Te\ determinations from any of those
two methods were not available, but for which they could be determined
from their $(V-K)$ colors, adopting the relation by Perrin \etal (1998).

The above stars were used to define a standard relation between \Te\ and
the equivalent width of a TiO band measured in our spectra. The latter
was used to infer the \Te\ of stars for which fundamental determinations
are currently lacking.

The stars used to define the standard relation between \Te\ and EW
of TiO are listed in Table~\ref{tiostands}. In Table~\ref{tioindex}
we show the pseudo-continua and passbands adopted in the measurement of
the EWs of the TiO bands in the library spectra. We fitted a 4th order
polynomial to the relation between EW(TiO) and \Te, whose coefficients
are given in Table~\ref{polytio}. The latter was applied to infer the \Te\
of the remaining library M giants.

TiO bands are known to be strongly sensitive to metallicity, especially
when they are not saturated. Since the metallicities of all M giants in
the library are unknown, we were forced to ignore this effect. Another
caveat concerns the strong photometric and spectral variability
characteristic of M giants, some of them possibly being variables of Mira
type. As a consequence, a given \Te\ determined either from photometry
or from the strength of a TiO band is strongly dependent on the epoch
of the observation.  Therefore, the \Te\ determined by our method should
be taken with caution.

In order to improve the reliability of our fitting functions in the M
giant regime, we enlarged our sample by inclusion of M giants from Worthey
\etal (1994) but only in cases for which independent \Te\ determinations
from one of the methods above were available in the literature. These
stars are listed in Table~\ref{wortheygiants}.

\subsubsection{K and M Dwarfs}

The stellar parameters for low mass dwarfs were determined as follows.
Effective temperatures came from interpolation in the Baraffe \etal
(1998) isochrones for given $(V-I)_0$ or $(V-K)_0$ colors, adopting
metallicites from the Cayrel de Strobel \etal (1997) catalogue. Gravities
were estimated from Hipparcos parallaxes adopting masses interpolated
in the Baraffe \etal isochrones and the metallicities were taken from
the Cayrel de Strobel \etal catalogue.

\subsubsection{O, B, and A Stars}

As in the case of F and G field dwarfs, stellar parameters of hotter
stars were inferred from Str\"omgren photometry. We used a combination
of the $\beta$, $a_0$, and $c_1$ indices. The $\beta$ index (Crawford
1958; Crawford \& Perry 1966) provides a photometric measurement of the
strength of $H\beta$ and therefore it is a good \Te\ indicator for A and
F stars.  The calibration adopted in Paper I for F and G stars cannot
be used here, though, as Alonso \etal did not extend to stars as hot
as spectral type A. Therefore, for stars with $\beta > 2.7$ we
adopted a relation obtained from a polynomial fit to the data of Smalley
\& Dworetsky (1993). Metallicities for these stars were computed from
the $m_1$ index (Str\"omgren 1966), adopting the relation by Smalley
(1993) and the standard calibration from Perry, Olsen \& Crawford (1987).

The $\beta$ index cannot be used along the entire BA sequence to determine
uniquely \Te, because Balmer lines get weaker for stars hotter than
A0. Therefore, the \Te s of stars with $-0.034 \leq (b-y) \leq 0.066$
were determined from the $a_0$ parameter (Str\"omgren 1966), and for
bluer stars, from the $c_1$ index (Str\"omgren 1966), in both cases
adopting the calibrations by Ribas \etal (1997).

\subsubsection{Comparison with Other Determinations}

The final stellar parameters are listed in Table A1 in the Appendix.
We compared these values with those obtained by JW95, who constructed
fitting functions and SPS models based on the same spectral library
we adopt in this work, so that the number of stars in common is
maximum. Overall, there is no major systematic difference between the
two sets of stellar parameters. The average differences (Ours -- JW95)
are as follows: $\Delta T_{\rm eff}$
= 19 $\pm$ 260 K, $\Delta \log g$ = --0.06 $\pm$ 0.4, and $\Delta$
[Fe/H] = --0.02 $\pm$ 0.2.  Moreover, the 1$-\sigma$ error bars in \Te\
and \gl\ drop to $\sim$ 130 K and $\sim$ 0.2 dex when roughly 30 stars
hotter than 7000 K, for which stellar parameters are more uncertain,
are excluded from the statistics.

However, further scrutiny reveals the presence of systematic differences
worthy of mention, for instance when we split the comparisons between
dwarfs and giants. It makes sense to look into comparisons within these
sub-samples, because different procedures are followed to determine
stellar parameters for dwarfs and giants both in this work and by JW95.

\bigskip
\centerline{\it Dwarfs}
\smallskip

We first focus on dwarf stars. In Figure \ref{pardwarfs} differences
between the two sets of stellar parameters (this work -- JW95) are
compared as a function of JW95 values. The most outstanding differences
revealed by the comparisons in Figure \ref{pardwarfs} are those between
the two sets of \Fe s. Our values are on average 0.15--0.2 dex higher
than those of JW95. While JW95 adopted \Fe s from Edvardsson \etal
(1993), ours are based on Str\"omgren photometry using the calibration
from Schuster \& Nissen (1989), as revised by Clementini \etal (1999).
The latter explains the discrepancy, as Clementini \etal added an extra
0.15 dex to Schuster \& Nissen's \Fe\ values.

There are also systematic differences, albeit more subtle, between the
two sets of \Te 's. Our \Te s are hotter by up to 250 K (average $\sim$
100 K) for stars hotter than 6200 K. JW95's \Te s for dwarf stars are
based on broadband color--\Te\ calibrations from the literature, while
ours come from Str\"omgren photometry, thus being consistent with the
values estimated by Edvardsson \etal (1993). In fact, JW95 note that their
\Te s were cooler than those of Edvardsson \etal by a similar amount,
and they decided to use stars in common with Edvardsson \etal to convert
the latter set of \Te\ into their own. Since our \Te -scale is already
consistent with that of Edvardsson \etal (1993), the difference found
here is not surprising.

\begin{deluxetable*}{lcrcccc} 
\tablecaption{Standard stars defining the \Te\ scale of the library
M Giants}
\tablewidth{0pt}
\tablehead{\colhead{Star}  &  \colhead{\Te}  &  \colhead{$M_V$}  &  
       \colhead{$(B-V)_0$}  &  \colhead{$(V-I)_0$}  &  \colhead{$(V-K)_0$} &
       \colhead{EW$_{TiO}$} }
\startdata
HD~~29139    & 3947 & --0.68 & 1.53 & 2.150 & 3.630 & 19.25   \\
HD~~39853    & 3881 & --1.33 & 1.51 &  ...  &  ...  & 17.75   \\
HD~~44033    & 3870 & --0.61 & 1.55 &  ...  &  ...  & 21.85   \\
HD~~44478    & 3610 & --1.42 & 1.60 & 2.950 & 4.740 & 40.92   \\
HD~~44537    & 3055 & --5.53 & 1.91 & 2.620 & 4.340 & 25.07   \\
HD~~63302    & 4500 & --3.03 & 1.80 &  ...  &  ...  & 19.02   \\
HD~~78712    & 3110 &   0.64 & 1.37 & 5.340 & 7.720 & 78.09   \\
HD~~99167    & 3890 & --0.50 & 1.55 &  ...  &  ...  & 20.87   \\
HD~~99998    & 3891 & --1.57 & 1.56 &  ...  & 1.900 & 13.32   \\
HD~102212    & 3844 & --0.94 & 1.51 & 2.240 & 3.900 & 23.45   \\
HD~110281    & 3950 &   0.55 & 1.70 &  ...  & 8.700 & 8.885   \\
HD~112300    & 3673 & --0.62 & 1.57 & 2.840 & 4.560 & 41.46   \\
HD~114961    & 2921 &   1.23 & 1.26 &  ...  &  ...  & 78.77   \\
HD~120933    & 3681 & --1.68 & 1.62 & 2.780 & 2.170 & 36.52   \\
HD~123657    & 3506 & --0.65 & 1.46 &  ...  &  ...  & 58.93   \\
HD~126327    & 2786 &   1.91 & 1.24 &  ...  &  ...  & 77.16   \\
HD~131918    & 3956 & --0.67 & 1.50 &  ...  &  ...  & 11.74   \\
HD~138481    & 3919 & --2.22 & 1.58 & 2.160 &  ...  & 18.41   \\
HD~139669    & 3919 & --2.04 & 1.58 &  ...  &  ...  & 18.34   \\
HD~148783    & 3449 & --0.41 & 1.25 & 4.580 & 6.79  & 69.23   \\
HD~149161    & 3952 & --0.17 & 1.47 & 2.010 & 3.57  & 16.55   \\
HD~180928    & 4008 & --0.13 & 1.34 & 1.860 & 3.38  & 12.39   \\
\enddata
\label{tiostands}
\end{deluxetable*}

\begin{deluxetable}{ccc} 
\tablecaption{TiO index used to estimate \Te\ for library M giants not included
in Table~\ref{tiostands}}
\tablewidth{0pt}
\tablehead{\colhead{Blue ``continuum'' (${\rm\AA}$)}  &  \colhead{Passband (${\rm\AA}$)}  &
           \colhead{Red ``continuum'' (${\rm\AA}$)}  }
\startdata
4947.56 -- 4952.65    &    4952.65 -- 5046.85   &   5157.19 -- 5163.97 \\
\enddata
\label{tioindex}
\end{deluxetable}

\begin{deluxetable}{ccccc} 
\tablecaption{Power Series Coefficients of the Relation between \Te\ and
EW$_{TiO}^1$
}
\tablewidth{0pt}
\tablehead{\colhead{$a_0$} & \colhead{$a_1$} & \colhead{$a_2$} &
           \colhead{$a_3$} & \colhead{$a_4$} }
\startdata
3722.06 & 45.5016 & --2.74143 & 5.23202$\times 10^{-2}$ & --3.34287$\times
10^{-4}$ \\
\enddata
\tablenotetext{1}{Where $$ T_{eff} = \sum_{i=0}^{4} a_i \times EW^i  $$ }
\label{polytio}
\end{deluxetable}

\begin{deluxetable}{lccl} 
\tablecaption{M Giants from Worthey \etal (1994)}
\tablewidth{0pt}
\tablehead{\colhead{Star}  &  \colhead{$(V-K)_0$} &\colhead{\Te}  &  
       \colhead{\Te\ Source} }
\startdata
HD~~~4656    &  3.53 & 4075  &  Richichi \etal (1999)  \\
HD~~17709   &  3.67 & 3921  &  (V--K)  \\
HD~~18191   &  6.80 & 3442  &  Dyck \etal (1998)  \\
HD~~47914   &  ...  & 3975  &  Alonso \etal (1999)  \\
HD~~60522   &  3.78 & 3883  &  (V--K)  \\
HD~~62721   &  3.55 & 3961  &  Alonso \etal (1999)  \\
HD~~70272   &  3.61 & 3943  &  (V--K)  \\
HD~~94705   &  6.61 & 3299  &  (V--K)  \\
HD~175865  &  6.23 & 3749  &  Dyck \etal (1998)  \\
HD~218329  &  3.79 & 3879  &  (V--K)  \\
HD~219734  &  4.18 & 3761  &  (V--K)  \\
\enddata
\label{wortheygiants}
\end{deluxetable}

No substantial systematic effect is seen for \gl, but the scatter
is higher for this parameter.  This is not surprising. Uncertainties in
\gl\ are usually large\ because they are affected by uncertainties
in \Te, adopted mass, distance and bolometric correction.

\bigskip
\centerline{Giants}
\smallskip

Figure~\ref{pargiants} repeats Figure \ref{pardwarfs} restricting the
plot to giant stars. While no systematic effect is found for
\gl, our \Fe s tend to be lower than those of JW95, especially in the
high-\Fe\ end, where the average residual reaches $\sim$ --0.25 dex. At
\Fe\ $\sim$ --0.5, the two scales are essentially the same. There is
also a small systematic effect in the \Te\ values in that ours are
slightly lower (on average $\simless$ 100 K) than those of JW95. It
is natural to suppose that the two effects might be correlated, given
the degenerate effects of \Te\ and \Fe\ on colors and absorption line
features. However, there is no correlation between $\Delta T_{\rm eff}$
and $\Delta$ \Fe .  Our atmospheric parameters for giant stars are rooted
in the Soubiran \etal (1998) scale (see Paper I for details), while the
JW95 scale is based on that of Dickow \etal (1970), so we believe that
our parameters, being based on updated stellar parameter determinations,
are more reliable.

In summary, the differences found here are not unexpected, and we stress
that they not only are not substantial but in fact are commensurate
with the uncertainties associated with \Te\ and \Fe\ 
determinations from broadband colors.

\subsubsection{Final Results}

\begin{figure}
\plotone{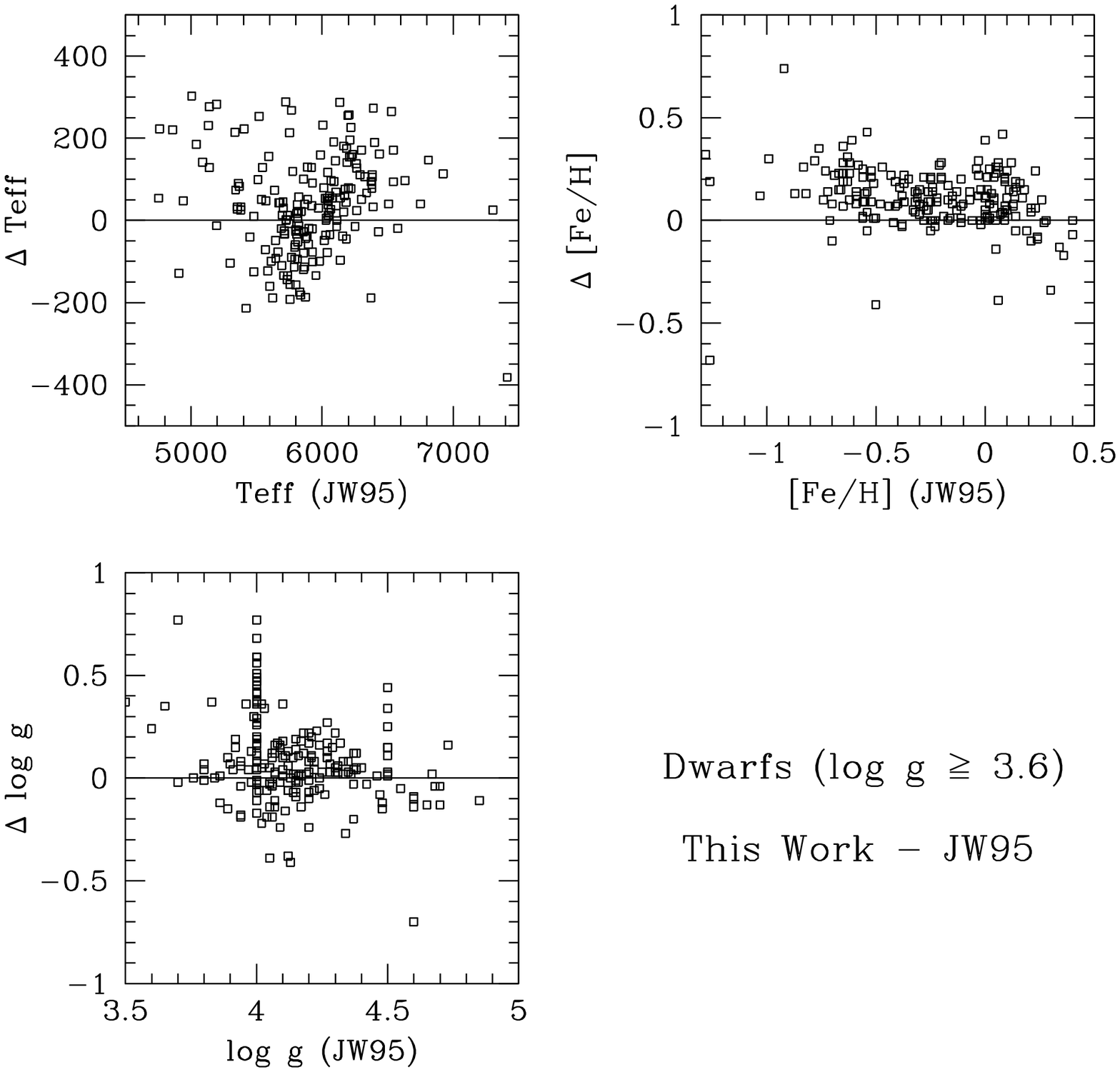}
\caption{Comparison of our stellar parameters with those determined by
Jones \& Worthey (1995) for dwarf stars. The systematic shift in [Fe/H]
is due to a revision of the Schuster \& Nissen (1989) metallicity scale
by Clementini \etal (1999).
} 
\label{pardwarfs}
\end{figure}

\begin{figure}
\plotone{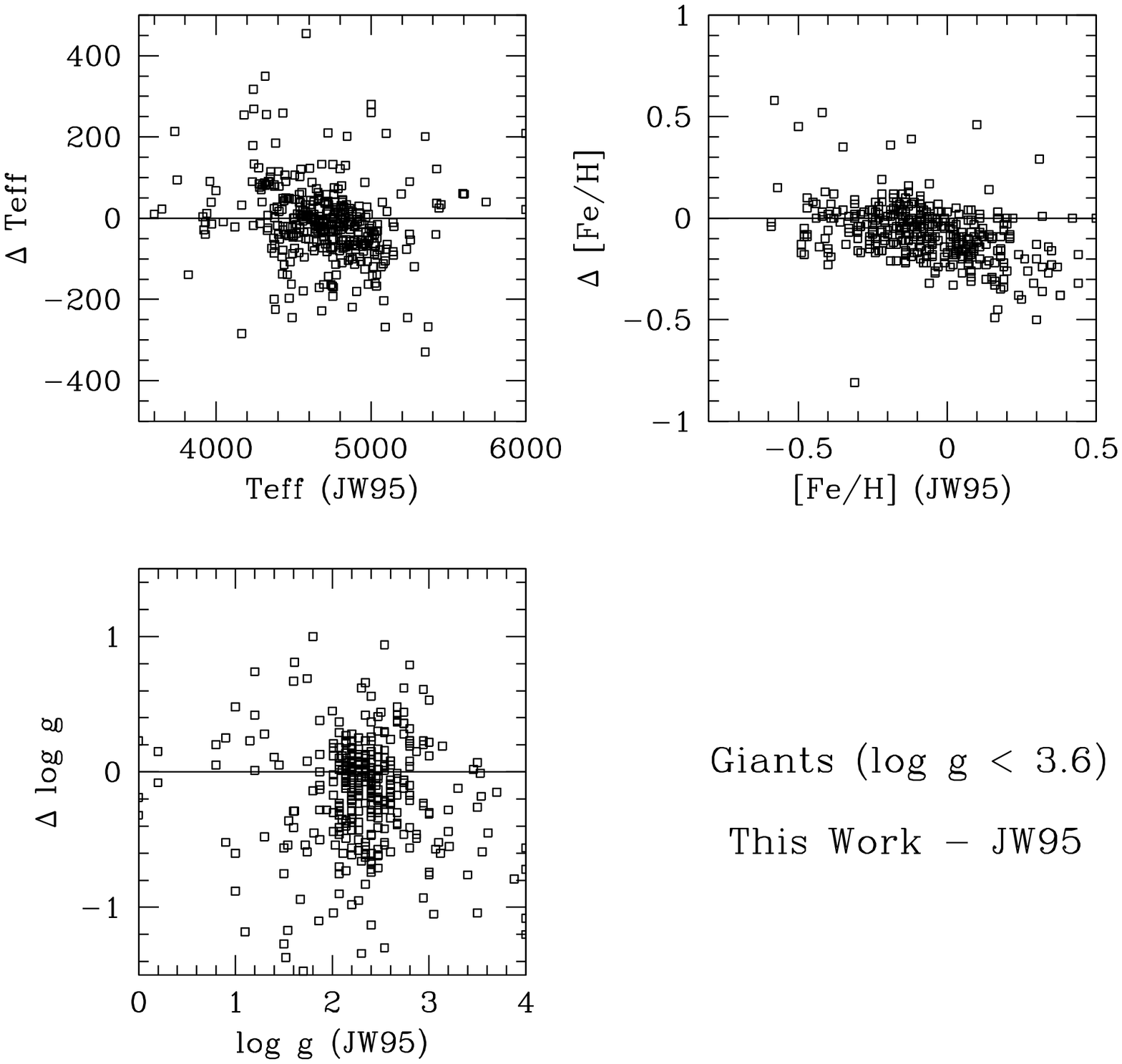}
\caption{Same as Figure~\ref{pardwarfs}, for giant stars. The systematic
effects on [Fe/H] is due to updated stellar parameters by Soubiran \etal 
(1998)
} 
\label{pargiants}
\end{figure}

In spite of the overall agreement in Figures~\ref{pardwarfs} and
\ref{pargiants} and the numbers above, there is in all panels a
significant number of stars that deviate significantly from the identity
relations. This is not negligible, because the spectral library is
somewhat sparse in some areas of stellar parameter space, where a few
stars with badly wrong stellar parameters may have an important weight
on the resulting fitting functions.

We tried to improve the quality of our determinations by inspecting
significantly deviant stars on a case-by-case basis. In \Te\ space, dwarf
(giant) stars cooler than 7000 K were deemed significantly deviant, and
thus worthy of further scrutiny, when our determinations differed from
those of JW95 by more than $\sim$ 200 K (300 K). In \gl\ space, and in the
same \Te\ range, dwarf (giant) stars were considered significantly deviant
when differences were higher than $\sim$ 0.5 ($\sim$ 1.0) dex. In \Fe\
space, we decided to further investigate the cases of all stars cooler
than 7000 K, and for which discrepancies were larger than $\sim$ 0.4 dex.
For approximately 1/3 of the hotter stars we needed to double-check our
determinations, because there the uncertainties are significantly higher,
and therefore agreement with JW95 is poorer.

Determining the best values of \Fe\ was quite laborious, because iron
abundance estimates from any method are subject to larger uncertainties
than the other parameters. For the same reason, the scatter in the
values found in the literature is likewise higher.  Following the above
criteria, we found 59 stars (almost 10\% of the spectral library) for
which our \Fe\ estimates were significantly different from those of JW95,
according to the criteria defined above. For each star, we performed a
critical, non-exhaustive, revision of the available literature, in order
to select those values which we regarded as more robust. Determinations
based on classical abundance analyses of high-resolution spectra were
given precedence, and amongst the latter, higher weight was given to
those involving recent, high S/N CCD observations, and updated model
atmospheres.

Deciding for the best values of \Te\ and \gl\ was relatively
simple, as these determinations tend by themselves to be fairly robust,
and besides it is possible to compare our values with estimates made
using independent methods. According to the above criteria, 27 of our
\Te\ determinations were found to significantly disagree with
those of JW95. In order to decide for the best value, we compared the
two sets of \Te\ with those inferred from photometry from the
literature. Consistency with the observed spectra, and in particular
with the measured EWs of key absorption features was also required in
order to help choosing that which seemed the most reliable \Te\
determination. For stars hotter than $\sim$ 8000 K, the lack
of robust calibrations of \Te\ against photometric indices
other than the ones employed in our own estimates made us resort to
spectroscopic determinations from the literature, based on the analysis of
intermediate-to-high resolution spectra on the basis of model atmospheres.

Deciding for the best choice of \gl\ is very important, as this is
the parameter that, for a given \Te, discriminates between the giant
or dwarf nature of a given star, thus deciding for its allocation as
input for different sets of fitting functions (see Section \ref{ffs}).
Luckily, our \gl\ estimates were found to disagree strongly with those
of JW95 for only 16 stars. In order to decide for the best value, we
looked in the literature for spectroscopic \gl\ determinations.

In most cases, our stellar parameter determinations, being based on recent,
more robust calibrations and high S/N spectra, were found to be in better 
agreement with those from the literature and/or other methods, than those
of JW95. In some cases we gave preference to the latter values, and for
very few stars we chose to adopt values from the literature.

\subsection{The Abundance Pattern of the Library Stars}

As stressed in the Introduction, one can only use a stellar population
synthesis model to estimate the abundance patterns of galaxies if the
abundance pattern of the input model is known. The latter is dictated
by the abundance pattern of the stars used in the construction the
models. The latter probably mirrors the abundance pattern of the solar
neighborhood and as such it should vary as a function of [Fe/H] (e.g.,
Edvardsson \etal 1993). This so-called ``bias'' of stellar population
synthesis models was pointed out, and accounted for, by Thomas \etal
(2003a). In this section we try to characterize the abundance pattern of
our models. This information will make it possible to use these models
to infer accurate abundance ratios from integrated spectra of galaxies
through the method developed by Trager \etal (2000) and Thomas \etal
(2003a).

\begin{figure}
\plotone{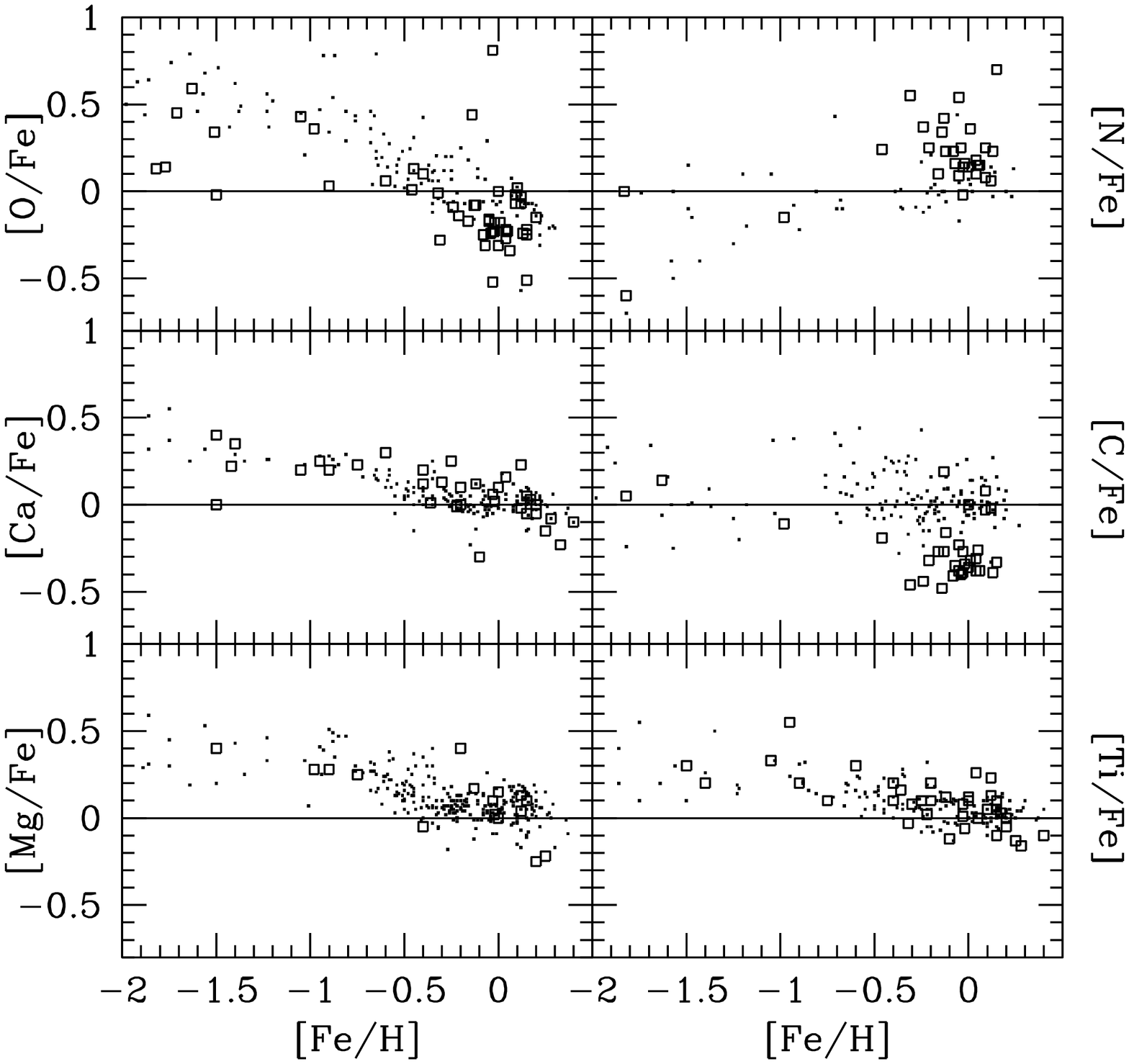}
\caption{Abundance pattern of the input stellar library. Dwarfs and giants
are marked by small dots and open squares, respectively. Nitrogen and
carbon abundances are affected by stellar evolution, therefore they differ
between dwarfs and giants. Overall, dwarf abundances are more homogeneous
and present less scatter than values for giants, so we decide to discard 
the latter.
} 
\label{abrat}
\end{figure}

We searched the literature for abundance determinations of our library
stars. We found data for roughly one third of the entire library
and assume that these stars are representative of the whole sample.
The results are plotted in Figure~\ref{abrat}, where abundance ratios
of some key elements are plotted against [Fe/H]. The sources of the
abundances plotted are as follows: Calcium abundances come from Th\'evenin
(1998), Gratton \etal (2003), and Reddy \etal (2003). Magnesium abundances
come from the latter works and also from Carretta, Gratton \& Sneden
(2000). Oxygen comes from Luck \& Challener (1995), Th\'evenin (1998),
Reddy \etal (2003), Gratton \etal (2003), Carretta \etal (2000), and
Israelian \etal (2004). Titanium abundances were taken from Th\'evenin
(1998) and Gratton \etal (2003). Most of the carbon abundances come from
Carretta \etal (2000), but we also include data from Shi, Zhao \& Chen
(2002), Carbon \etal (1987), and Reddy \etal (2003). Nitrogen abundances
were drawn from Shi \etal (2002), Israelian \etal (2004), Ecuvillon \etal
(2004), Reddy \etal (2003), Carretta \etal (2000), and Carbon \etal
(1987).

In Figure~\ref{abrat}, giant stars are plotted as open squares and dwarfs
as small dots. As expected, the abundance ratios of some elements do
present a significant variation as a function of [Fe/H]. From this figure
it is also clear that there are two groups of elements in terms of the
behavior of their abundances as a function of evolutionary stage. For
magnesium, calcium, titanium, and oxygen, the abundances in giants and
dwarfs seem to be similar. The same is not true for carbon and nitrogen,
though. The abundances of carbon are much lower in giants than in
dwarfs.  Nitrogen, on the other hand, is more abundant in giants than in
dwarfs. These trends are not unexpected. They result from contamination,
during the first dredge-up, of the atmospheres of giant stars by fresh
material processed by the CNO-cycle (e.g., Iben 1964, Brown 1987, Carretta
\etal 2000, Thor\'en, Edvardsson \& Gustafsson 2004). As a consequence,
the giant abundances for these elements do not reflect their original
values, so that they will not be considered here.  For the other elements,
the data on giant stars are consistent with, but more scattered than,
those of dwarfs, so that we decided to eliminate the giant abundances
in the following derivation.

In order to estimate mean values for the abundance ratios of the various
elements as a function of [Fe/H], we fitted low order polynomials
to the relations [X/Fe] vs. [Fe/H]. The results are presented
in Table~\ref{tabxfe} for a number of reference values of [Fe/H]. The
1-$\sigma$ error bars come from the {\it r.m.s.} of the polynomial fits at
different [Fe/H] bins and probably reflect a combination of measurement
errors and intrinsic spread. We chose to present these data in fine [Fe/H]
bins, in spite of the relatively large error bars in the abundance ratios,
in order to facilitate interpolation in the table values.

\begin{deluxetable*}{rcccccc}

\tablecaption{The abundance pattern of the library stars}
\tablewidth{0pt}
\tablehead{\colhead{[Fe/H]} & \colhead{[O/Fe]} & \colhead{[N/Fe]} &
           \colhead{[C/Fe]} & \colhead{[Mg/Fe]} & \colhead{[Ca/Fe]} &
           \colhead{[Ti/Fe]}   }
\startdata
--1.6 & +0.6 $\pm$ 0.1 & 0.0 $\pm$ 0.2 & 0.0 $\pm$ 0.1 & 0.4  $\pm$ 0.1  &  0.32 $\pm$ 0.05 & 0.3 $\pm$ 0.1 \\
--1.4 & +0.5 $\pm$ 0.1 & 0.0 $\pm$ 0.2 & 0.0 $\pm$ 0.1 & 0.4  $\pm$ 0.1  &  0.30 $\pm$ 0.05 & 0.3 $\pm$ 0.1 \\
--1.2 & +0.5 $\pm$ 0.1 & 0.0 $\pm$ 0.2 & 0.0 $\pm$ 0.1 & 0.4  $\pm$ 0.1  &  0.28 $\pm$ 0.05 & 0.3 $\pm$ 0.1 \\
--1.0 & +0.4 $\pm$ 0.1 & 0.0 $\pm$ 0.2 & 0.0 $\pm$ 0.1 & 0.4  $\pm$ 0.1  &  0.26 $\pm$ 0.05 & 0.21 $\pm$ 0.07 \\
--0.8 & +0.3 $\pm$ 0.1 & 0.0 $\pm$ 0.2 & 0.0 $\pm$ 0.1 & 0.29 $\pm$ 0.08 &  0.20 $\pm$ 0.05 & 0.18 $\pm$ 0.07 \\
--0.6 & +0.2 $\pm$ 0.1 & 0.0 $\pm$ 0.2 & 0.0 $\pm$ 0.1 & 0.20 $\pm$ 0.08 &  0.12 $\pm$ 0.05 & 0.14 $\pm$ 0.07 \\
--0.4 & +0.2 $\pm$ 0.1 & 0.0 $\pm$ 0.2 & 0.0 $\pm$ 0.1 & 0.13 $\pm$ 0.08 &  0.06 $\pm$ 0.05 & 0.11 $\pm$ 0.07 \\
--0.2 & +0.1 $\pm$ 0.1 & 0.0 $\pm$ 0.2 & 0.0 $\pm$ 0.1 & 0.08 $\pm$ 0.08 &  0.02 $\pm$ 0.05 & 0.08 $\pm$ 0.07 \\
 0.0  &  0.0 $\pm$ 0.1 & 0.0 $\pm$ 0.2 & 0.0 $\pm$ 0.1 & 0.05 $\pm$ 0.08 &  0.00 $\pm$ 0.05 & 0.04 $\pm$ 0.07 \\
+0.2  &--0.1 $\pm$ 0.1 & 0.0 $\pm$ 0.2 & 0.0 $\pm$ 0.1 & 0.04 $\pm$ 0.08 &--0.01 $\pm$ 0.05 & 0.01 $\pm$ 0.07 \\
\enddata
\label{tabxfe}
\end{deluxetable*}
A few caveats need to be kept in mind when using these numbers.
The first one concerns the oxygen abundances of metal-poor stars, which
are still very controversial (see the review by Kraft 2003). Different
abundance analysis methods, relying on the forbidden lines at $\sim$
6300 ${\rm\AA}$, the triplet at $\sim$ 7770 ${\rm\AA}$ or synthesis of
OH bands in the near-UV and near-IR, yield abundances differing by up
to 0.5 dex at [Fe/H] $\sim$ --1.5.  Probably because our abundances were
compiled from works employing different methods, our mean [O/Fe] values
for [Fe/H] $\simless$ --1.0 fall right in the middle of the range of
current determinations (see Figure 1 of Fulbright \& Johnson 2003). While
that may leave us in a relatively safe position, we caution the reader
that these values might need to be revised once oxygen abundances from
different groups reach agreement.

There also is disagreement in the literature in determinations of carbon
abundances of field stars. On one side, Shi \etal (2002) and Reddy \etal
(2003) find carbon to be overabundant relative to iron in metal-poor
stars and increasingly so with decreasing [Fe/H]. On the other hand,
Carbon \etal (1987) and Carretta \etal (2000) found [C/Fe] $\sim$ 0 and
essentially invariant as a function of [Fe/H].  Finally, Shi \etal (2002)
agree with Carretta \etal for [Fe/H] $\simgreat$ --0.7, but find carbon
overabundances for more metal-poor stars. The [C/Fe] values displayed in
Figure~\ref{abrat} and Table~\ref{tabxfe} are solar and constant with
[Fe/H] because most of the carbon abundances come from Carretta \etal
(2000). As for nitrogen, Shi \etal (2002), Ecuvillon \etal (2004),
and Israelian \etal (2004) all find [N/Fe] $\sim$ 0 and constant within
a very large [Fe/H] range. On the other hand, Reddy \etal (2003) find
[N/Fe] $\sim$ +0.2, in a much smaller range of [Fe/H].

There are three separate issues that should be highlighted here. The
first is related to the uncertainties mentioned above. While we are
not in a position to choose among the various abundance determinations,
we alert the reader for the obvious fact that the numbers provided in
Table~\ref{tabxfe} might need to be revised when future improvements in
abundance determinations come about. The second regards the degree to
which the spectral library in use here, and any other spectral library
for that matter, can be safely assumed to replicate the abundance
pattern of the solar neighborhood in detail. The selection of targets
involved in the production of such spectral libraries is dictated by
criteria that are very different from those involved in standard surveys
of the abundance pattern of Galactic field stars. Therefore, it is not
unlikely that the abundance pattern of the stars in the spectral library
might be biased in different ways. An obvious example of a way in which
this can happen is the inclusion of cluster stars (Worthey \etal 1994),
whose detailed abundance patterns often differ from those found in the
field. Last, but not least, there is the issue of whole regions in the
stellar parameter space where detailed abundances (and sometimes even
just [Fe/H]!) are unknown. That is the case in both ends of the \Te\
spectrum, and is especially worrisome in the case of bright stars such
as M giants and hot stars in general. Fortunately, we are working in
a spectral region where the former contribute little light and are mostly
concerned with an age/metallicity regime where the latter are not very
important. But that should be a reason for concern for work on stellar
populations younger than $\sim$ 1 Gyr, and for any attempt at studying
stellar populations of any age longward of $\sim$ 6000 ${\rm\AA}$.

\section{Fitting Functions} \label{ffs}

\subsection{Procedure}

The polynomial functions describing the relations between the various
spectral indices and stellar parameters were computed through a
general linear least squares method. The spectral library spans a vast
range of stellar types, with \Te\ varying from $\sim$ 3000 to $\sim$
15000 K, and \gl\ and \Fe\ varying by 6 and 3 orders of magnitude,
respectively. Photospheric structure, and with it the dependence of
absorption line indices on stellar parameters, varies greatly within this
large region of stellar parameter space. As a consequence, it is very hard
to devise a single simple mathematical expression capable of accounting
for line index behavior in the whole range of stellar parameters spanned
by the spectral library. For that reason, we decided to split the library
in five major stellar classes and perform the fits separately for each
class. The five sub-regions of stellar parameter space we consider are
roughly: G-K giants, F-G dwarfs, B-A dwarfs, M giants, and K-M dwarfs. The
strict boundaries defining each sub-region vary from index to index,
and are given in Table~\ref{hdeltaa}. Considerable
inter-region overlap was adopted when performing the fits, in order to
ensure a smooth transition between adjacent sub-regions.

The goal when determining index fitting functions is to find the simplest
mathematical representation of the dependence of a given index on stellar
parameters that yet is reasonably accurate. Very simple statistical tools
come in very handy, but cannot be fully trusted, given the specific
limitations of the spectral library in use. It is worth to describe
two illustrative examples. The approach chosen by Worthey \etal (1994)
was that of considering relevant the terms whose inclusion reduces the
overall {\it r.m.s.} of the fit by a given fractional amount. The danger
of this approach in our case resides in the fact that, for instance,
for the giants, the majority of the stars have \Fe\ $\simgreat$ --0.7,
so that the {\it r.m.s.} is not very sensitive to the quality of the
fit for lower metallicity stars. Another approach is that followed by
Cenarro \etal (2002), where an automatic routine searches, among a large
collection of terms, those whose coefficients depart (according to a
t-test) significantly from zero. The problem with that approach is that,
again due to the low density with which the spectral library occupies
certain regions of parameter space, it may happen that a given coefficient
is statistically significant, but unphysical, which may introduce
unrealistic high frequency features in the final fitting function.

\begin{deluxetable*}{rccccccccc} 
\tabletypesize{\scriptsize}
\tablecaption{Coefficients of the fitting function for $H\delta_A$} 
\tablewidth{0pt}
\tablehead{
\colhead{Const.} & \colhead{$\theta_{\rm eff}$} & \colhead{$\theta_{\rm eff}^2$} & 
\colhead{$\theta_{\rm eff}^3$} & \colhead{$\theta_{\rm eff}^4$} & 
\colhead{[Fe/H]} & \colhead{[Fe/H]$^2$}  & \colhead{[Fe/H]$^3$}  &
\colhead{$\theta_{\rm eff}$ [Fe/H]} & 
\colhead{$\theta_{\rm eff}^3$ [Fe/H]} 
}
\startdata
\cutinhead{G-K Giants, 3600 $\leq T_{\rm eff} \leq$ 8000 K, $\log g \leq
3.6$, No.=390, r.m.s.=0.39}

 -166.6082 &
  788.0336 &
-1219.6762 &
  758.3718 &
 -163.8285 &
   28.2545 &
   -1.3291 &
   -0.1187 &
  -45.0030 &
   12.6246 \\

 & & & & & & & & & \cr

\cutinhead{F-G Dwarfs, 4500 $\leq T_{\rm eff} \leq$ 9000 K, $\log g \geq
3.0$, No.=259, r.m.s.=0.68}

  330.5350 &
-1431.6421 &
 2460.9693 &
-1932.4266 &
  568.7570 &
    2.5741 &
   -1.3796 &
   -0.3409 &
   -5.5392 &
      ... \\

 & & & & & & & & & \cr

\cutinhead{B-A Dwarfs, 7000 $\leq T_{\rm eff} \leq$ 20000 K, No.=48,
r.m.s.=1.29}

   11.0180 &
 -119.9733 &
  437.8068 &
 -390.1295 &
      ... &
      ... &
      ... &
      ... &
      ... &
      ... \\

 & & & & & & & & & \cr

\cutinhead{M Giants, 2000 $\leq T_{\rm eff} \leq$ 4100 K, $\log g \leq
3.6$, No.=33, r.m.s.=0.86}

    -94.1659 &
     95.3567 &
     -9.4888 &
     -8.3640 &
      ... &
      ... &
      ... &
      ... &
      ... &
      ... \\

 & & & & & & & & & \cr

\cutinhead{K-M Dwarfs, 2000 $\leq T_{\rm eff} \leq$ 4800 K, $\log g \geq$
3.6, No.=21, r.m.s.=1.59}

 & & & & & & & & & \cr

    -61.4588 &
     71.8074 &
    -19.7836 &
      ... &
      ... &
      ... &
      ... &
      ... &
      ... &
      ... \\

\enddata
\label{hdeltaa}
\tablenotetext{1}{Fitting function coefficients for the other Lick indices
are available in electronic form.}
\end{deluxetable*}

We addressed this problem by trying to combine the best from each of
the above approaches. We started by following the procedure of Cenarro
\etal (2002) where a first fit was attempted adopting a polynomial
with 25 terms involving products of different powers of \Te, \gl\ and
\Fe. A t-test was then applied to verify and remove terms which were
not statistically significant. Then a new fit based on the reduced set
of terms was performed and the procedure iterated until only terms with
t $\simless$ 0.01 survived. This was all performed automatically. The
next step was to examine the quality of the fits interactively, removing
terms that seem unphysical or otherwise unnecessary, while monitoring
how their removal affects the final {\it r.m.s.} of the fit. We also
adopted a $\sigma$-clipping procedure, whereby stars departing by more
than (typically) 2-3 $\sigma$ from the solution were removed from the
sample and the fit redone. We adopted at most one $\sigma$-clipping
iteration for each fit and typically more than 97\% of the input
stars were preserved at each fitting set. Automatic $\sigma$-clipping
was turned off in regions of parameter space where poor statistics,
due to the scarcity of input stars, prevented a robust estimate of
$\sigma$. That was the case for the fits for dwarfs cooler than $\sim$
5000 K, giants cooler than $\sim$ 4000 K, all stars hotter than $\sim$
8000 K, and giants more metal-poor than \Fe\ $\sim$ --1.0.

\subsection{Results}

\begin{figure}
\plotone{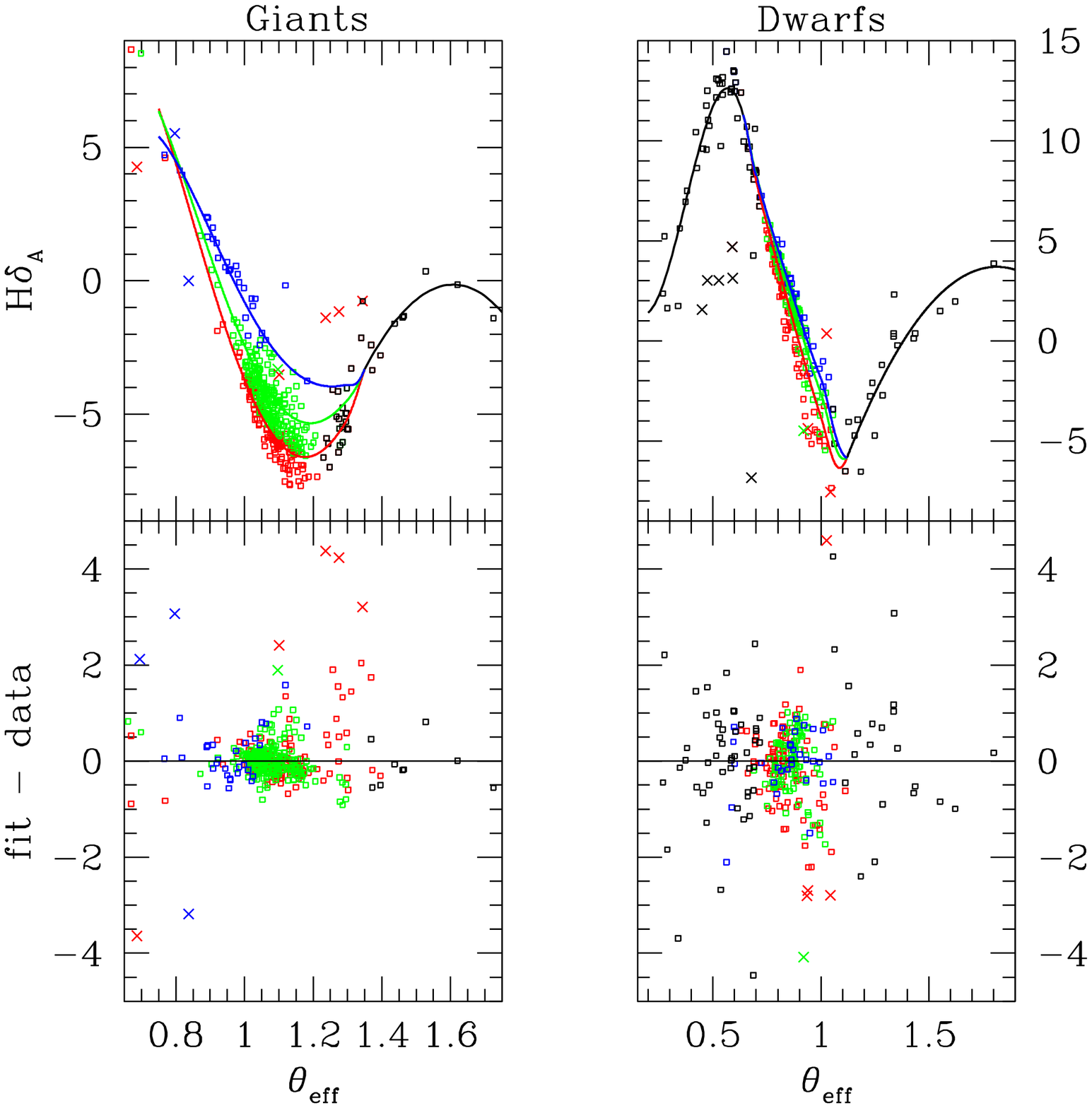}
\caption{a. {\it Upper panels}: fitting functions over-plotted on
$H\delta_A$ measurements for dwarfs and giants, as indicated on top of
each panel. The data are color-coded according to metallicity. Red dots
correspond to [Fe/H] $\geq$ --0.1, green dots correspond to --0.6 $\leq$
[Fe/H] $<$ --0.1, and blue dots correspond to [Fe/H] $<$ --0.6. The
red curves correspond to fitting functions computed for [Fe/H] = +0.1,
the green to [Fe/H] = --0.35, and the blue to [Fe/H]=--1.0. A \Te\ vs
\gl\ relation from Girardi \etal (2000)'s 5 Gyr old isochrone for solar
metallicity is assumed. Black dots and lines are adopted for stars in
regions of parameter space where the fits are [Fe/H]-independent. {\it
Bottom panels}: Residuals of the fits. $\theta_{\rm eff}$ = 5040/$T_{\rm
eff}$. In all plots crosses indicate stars that were rejected by the
fitting routine. {\it Similar figures for all the other indices are 
available electronically and will appear in the final version of the
paper in the ApJS.}
}
\label{fhda}
\end{figure}

The fitting functions obtained according to the procedure delineated
above are presented in Table~\ref{hdeltaa} and
displayed in Figure~\ref{fhda}. In the Figures we
adopt a cosine-weighted interpolation to represent the plots in the
boundaries between the different plotting regions, following Cenarro
\etal (2002). The reader should keep in mind that the plots shown in
Figure~\ref{fhda} are limited representations of the
fitting functions presented in Table~\ref{hdeltaa}. Most
indices depend on three variables, \Te\ , \Fe, and \gl\ through most of
the parameter space. Yet, the plots only allow us to display the index
variations as a function of the two most important parameters, \Te\
and \Fe. Therefore, we must assume a \gl\ value for the indices that
do depend on this parameter, which can vary by as much as 5 orders of
magnitudes in the sample considered here. We did so by adopting a \Te\
$\times$ \gl\ relation interpolated in the isochrones from Girardi \etal
(2000) for 5 Gyr and solar metallicity.

From the Figures and Tables it can be seen that the fits look fairly
robust for G-K giants, and F-G dwarfs, which are the stellar types
that are best represented in the spectral library. For these stars,
reliable estimates of the behavior of the indices as a function
of effective temperature, metallicity and surface gravity could be
achieved. Outside these regions of parameter space, the low density
of the spectral library (especially in the case of the very cool stars)
and uncertainties in the stellar parameters made it very difficult to
obtain estimates of the response of spectral indices to metallicity
and surface gravity. Therefore, our fitting functions for all indices,
except Fe5270 and Fe5335, are solely dependent on effective temperature
for K-M dwarfs, M giants and B-A dwarfs. For those two indices we could
not obtain fitting functions that would extend into the K-M dwarf regime
without a moderately strong dependence on [Fe/H].

The boundaries listed in Tables\ref{hdeltaa} are the
ones adopted to produce the fits. Those attempting to reproduce our
polynomial fits in Table~\ref{hdeltaa} should adopt
those boundaries as input in their programs.  The latter boundaries
should not be confused with those provided in Table \ref{boundaries}
which specify the regions of parameter space within which the various
fitting functions should be applied. Those are meant to be used by
stellar population synthesis modelers wishing to adopt our fitting
functions for the various Lick indices.  The reader will notice that
the boundaries in Table~\ref{boundaries} are in general contained
within those of Table~\ref{hdeltaa}, for a given index
and stellar family. This is to ensure that application of our fitting
functions be restricted to regions of parameter space where they are
well constrained by the input stellar data. We {\it strongly caution}
the reader against trusting extrapolations of the fitting functions away
from the boundaries given in Table~\ref{boundaries}, as in many cases
the functions behave in a strongly non-physical way outside the fitting
region. In the case of the CN indices, we could not find polynomial
functions capable of describing index behavior as a function of \Te and
[Fe/H] in a satisfactory fashion in the metal-poor regime.  Therefore we
caution readers against trusting either the fitting functions or the
single stellar population models for those indices below [Fe/H]=--1.0.

Blue indices tend to display a marked sensitivity to $\log g$ in stars
hotter than $\sim$ 8000 K and for very low gravities ($\log g \simless
2$). All Balmer lines tend to be considerably weaker in the spectra of
B-A super-giants than in that of dwarfs and giants of the same \Te. At
such high \Te s Balmer lines are very strong, and their wings tend to be
stronger for higher gravities. In the spectra of B-A dwarfs, the wings
of the Balmer lines are so strong that they dominate the absorption at
$\lambda \simless 4500$ K and have an impact on all other absorption
line indices in that spectral region. As a result, indices like CN$_1$,
CN$_2$, Ca4227, and G4300 present a dependence on $\log g$ that is similar
in strength to that of the Balmer lines, but with opposite sign. Because
the spectral library has just a handful of B-A super-giants, this effect
could not be modelled in a reliable fashion, and we decided to exclude
these very low surface gravity stars from our fits.  Therefore, the
fitting functions for hot stars should only be applied to stars with
$\log g \simgreat 2$, for which no dependence of Balmer lines (and the
other spectral indices) on $\log g$ could be perceived in our data.

\begin{deluxetable*}{lcccccc} 
\tabletypesize{\scriptsize}
\tablecaption{Intervals of Applicability of the 
fitting functions}
\tablehead{
\colhead{Index} &
\colhead{$T_{\rm eff\,\, min}$ (K)} &
\colhead{$T_{\rm eff\,\, max}$ (K)} &
\colhead{$\log g_{\rm min}$} &
\colhead{$\log g_{\rm max}$} &
\colhead{[Fe/H]$_{\rm min}$} &
\colhead{[Fe/H]$_{\rm max}$}
}
\startdata
$H\delta_A$ -- G-K Giants & 3790 &  6000 & ... & 3.6 & --1.3 & +0.3 \\
$H\delta_A$ -- F-G Dwarfs & 5100 &  7500 & 3.0 & ... & --2.0 & +0.3 \\
$H\delta_A$ -- B-A Dwarfs & 7500 & 18000 & 3.0 & ... &  ...  &  ... \\
$H\delta_A$ -- M Giants   & 2800 &  3790 & ... & 3.6 &  ...  &  ... \\
$H\delta_A$ -- K-M Dwarfs & 3200 &  5100 & 4.0 & ... &  ...  &  ... \\
\enddata
\label{boundaries}
\tablenotetext{1}{Intervals for other indices are available in electronic
form.}
\end{deluxetable*}

\section{Model Predictions for Single Stellar Populations}

\subsection{The Base Models} \label{basemod}

The fitting functions presented in Section~\ref{ffs} were combined with
theoretical isochrones in order to produce predictions of integrated
indices of single stellar populations. The isochrones employed were those
from the Padova group for both the solar-scaled (Girardi \etal 2000)
and $\alpha$-enhanced cases (Salasnich \etal 2000). There are several
other groups producing state-of-the-art stellar evolutionary tracks
and theoretical isochrones (e.g., Charbonnel \etal 1999, Yi \etal 2001,
Kim \etal 2002, Pietrinferni \etal 2004, Jimenez \etal 2004) and it is
very important to study the dependence of the results on the stellar
evolution prescriptions. This will be discussed in a
future paper. Absorption line indices and UBV absolute magnitudes were
computed for the parameters listed in Table~\ref{ssppar}. In column
(1) of Table~\ref{ssppar} we list a model reference number, in columns
(2) and (3) we list the mass fraction of elements heavier than He (Z)
and that of He (Y). The iron abundance, overall metallicity, and mean
$\alpha$-enhancement for the mixture adopted by the Padova group, given by
[Fe/H], [Z/H] and [$\alpha$/Fe], are listed in columns (4) through (6).
The mean $\alpha$-enhancement of the spectral library is listed in Column
(7). Finally, column (8) contains the range of ages encompassed by each
model set.

Throughout this paper we refer to the models summarized in
Table~\ref{ssppar} as our {\it base models}, which result from the mere
combination of our fitting functions derived in Section~\ref{ffs} and
the Padova isochrones.  We note that, except for models 3 through 5,
the $\alpha$-enhancement of the spectral library is inconsistent with
that of the theoretical isochrones adopted (we assume here that a $\sim$
0.1 dex mismatch is negligible). This condition is not unique to our
base models. In fact, other well-known stellar population synthesis
models in the literature (e.g., Worthey 1994, Vazdekis 1999, Bruzual \&
Charlot 2003, Le Borgne \etal 2004, Lee \& Worthey 2005), which are based
on similar combinations of theoretical isochrones and empirical stellar
libraries, are afflicted by the same inconsistency.  In principle, this
issue can and has been addressed via adoption of the response functions of
Tripicco \& Bell (1995), Houdashelt \etal (2002), or Korn \etal (2005),
as discussed above (e.g., Trager \etal 2000, Thomas \etal 2003a, Thomas
\etal 2004, Lee \& Worthey 2005). These models, however, are corrected
for an {\it assumed} abundance pattern of the stars that make up the
stellar library and therefore are also lacking in consistency. To
our knowledge, the only attempts so far at full consistency between
theoretical isochrones and stellar library are those of Coelho (2004)
and this work. The former is based on synthetic spectra and the latter
are discussed in Section~\ref{aratios}.

\begin{deluxetable*}{cccrrrrc} 
\tablecaption{Stellar Population Parameters of the Base Models}
\tablewidth{0pt}
\tablehead{\colhead{Model No.} &
   \colhead{Z} & \colhead{Y} &  \colhead{[Fe/H]}  & \colhead{[Z/H]}
   &\colhead{[$\alpha$/Fe]$_{iso}$} &
           \colhead{[$\alpha$/Fe]$_{lib}$} & \colhead{Age Range (Gyr)} }
\startdata 1 & 0.001 & 0.23 & --1.31 & --1.31 & 0.0   &  +0.38  &
3.5--15.8  \\ 2 & 0.004 & 0.24 & --0.71 & --0.71 & 0.0   &  +0.20  &
3.5--15.8  \\ 3 & 0.008 & 0.25 & --0.40 & --0.40 & 0.0   &  +0.13  &
1.5--15.8  \\ 4 & 0.019 & 0.273 &  0.00 &   0.00 & 0.0   &   0.02  &
0.8--15.8  \\ 5 & 0.030 & 0.300 &  +0.22 &  +0.22 & 0.0   & --0.02  &
0.8--15.8  \\ 6 & 0.008 & 0.25 & --0.75 & --0.40 & +0.42 &  +0.20  &
3.5--15.8  \\ 7 & 0.019 & 0.273 & --0.36 &   0.00 & +0.42 &  +0.13  &
1.5--15.8  \\ 8 & 0.040 & 0.32 & +0.01 &  +0.37 & +0.42 &   0.02  &
0.8--15.8  \\ 9 & 0.070 & 0.39 & +0.33 &  +0.68 & +0.42 & --0.02  &
0.8--15.8  \\

\enddata
\label{ssppar}
\end{deluxetable*}

Our computations were performed as follows.  If $\Im$ is the line index
in the integrated spectrum of a model single stellar population, it is
given by

\begin{equation}
\Im = \sum_i^N \phi_i \,\, f_i \,\, I_i  \label{eq1}
\end{equation}

\noindent when $\Im$ and $I_i$ are defined in terms of an equivalent
width, and

\begin{equation}
\Im = -2.5 \,\, \log \left(\sum_i^N \phi_i \,\, f_i \,\, 10^{-0.4 \,\,
I_i} \right) \label{eq2}
\end{equation}

\noindent when $\Im$ and $I_i$ are defined in terms of a magnitude. In
equations (\ref{eq1}) and (\ref{eq2}), $I_i$ is the index computed from
our fitting functions for the stellar parameters corresponding to the
$i$-th evolutionary stage. $N$ is the number of evolutionary stages in
the theoretical isochrone adopted. $\phi_i$ is the relative number of
stars at the $i$-th position in the isochrone, which is given by the
initial mass function (IMF) of the stellar population. For simplicity,
we adopt a power-law mass function, given by

\begin{equation}
\phi_i = \int A \,\, m_i^{1-x} dm \label{eq3}
\end{equation}

\noindent where $m_i$ is the mass at the $i$-th evolutionary stage in
the isochrone and $A$ is a normalization constant which is chosen so
that the entire stellar population has 1 $M_\odot$. The integration is
performed within a narrow interval centered on $m_i$. For a Salpeter IMF,
$x = 1.35$.

The term $f_i$ in equations (\ref{eq1}) and (\ref{eq2}) gives the
weight in flux for stars at the $i$th position of the isochrone. It
is computed from interpolation between broad band absolute magnitudes
to the index central wavelength. Absolute magnitudes in the U, B,
and V bands were computed using the calibrations described in Paper I.
Integrated magnitudes for single stellar populations in these bands were
computed according to equation~\ref{eq2}, with $f_i = 1$ and making 
$I_i$ equal to the absolute magnitude of the $i$-th position in the
isochrone. The results are provided in tables in the Appendix. Tables A2 
and A3 provide Lick index predictions, and Tables A4 and A5
list predictions for UBV magnitudes/colors. In the following section we
compare our predictions to those obtained when the fitting functions of
Worthey \etal (1994) are employed.

\subsection{New vs. Old Fitting Functions} \label{comparisons}

\begin{figure*}
\plotone{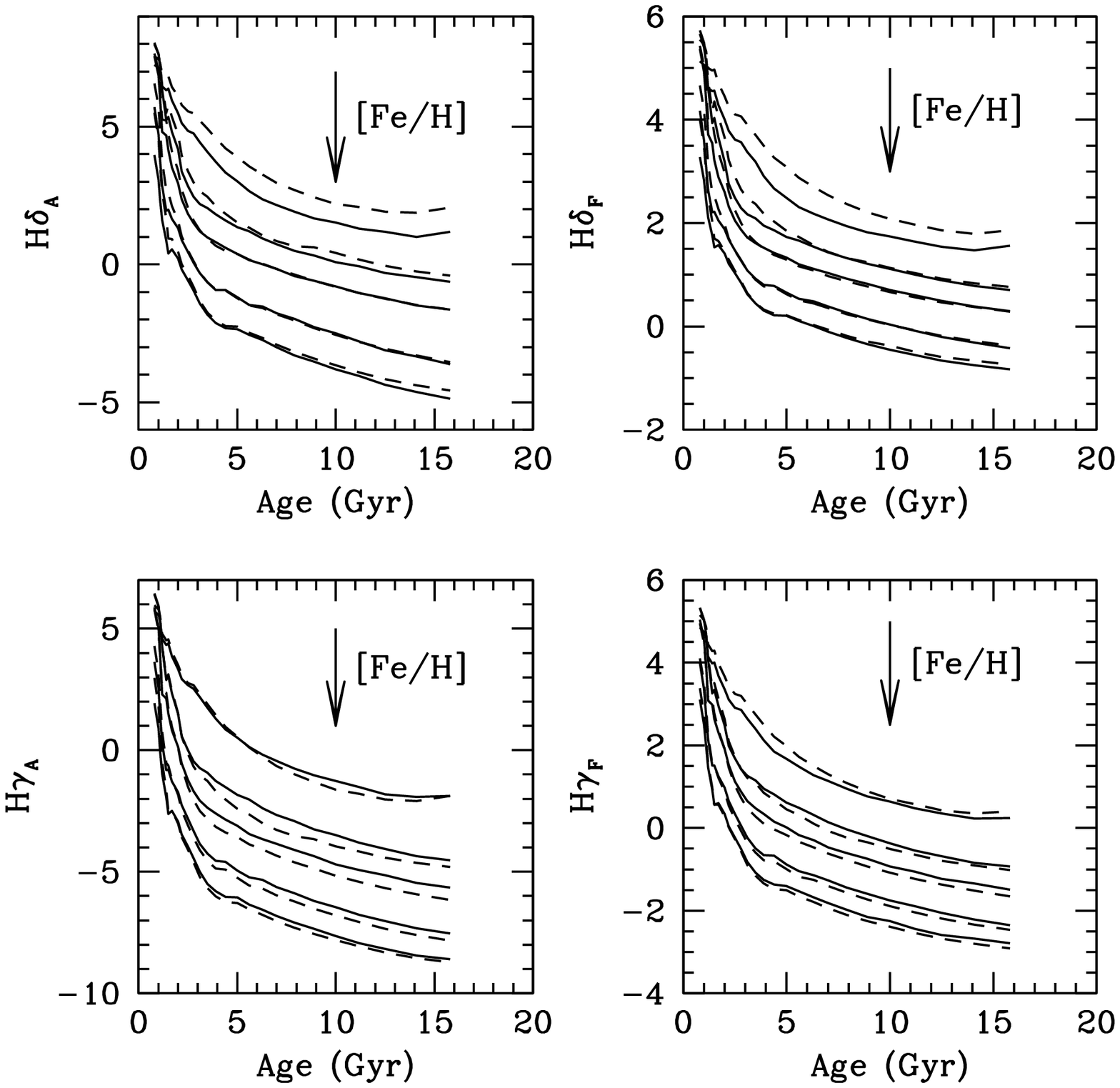}
\caption{a. Comparison of model predictions based on our fitting functions
(solid lines) and on those of Worthey \etal (1994, dashed lines),
computed with the same set of isochrones. Metallicities are [Fe/H] =
--1.3, --0.7, --0.4, 0.0, and +0.2. The arrows indicate the direction
of increasing [Fe/H]. For the lowest metallicities, the Worthey \etal
(1994) fitting functions are not defined for ages lower than 5 Gyr, but
we decided to keep the comparisons for completeness.}
\label{comp_w94a}
\end{figure*}

\setcounter{figure}{6}
\begin{figure*}
\plotone{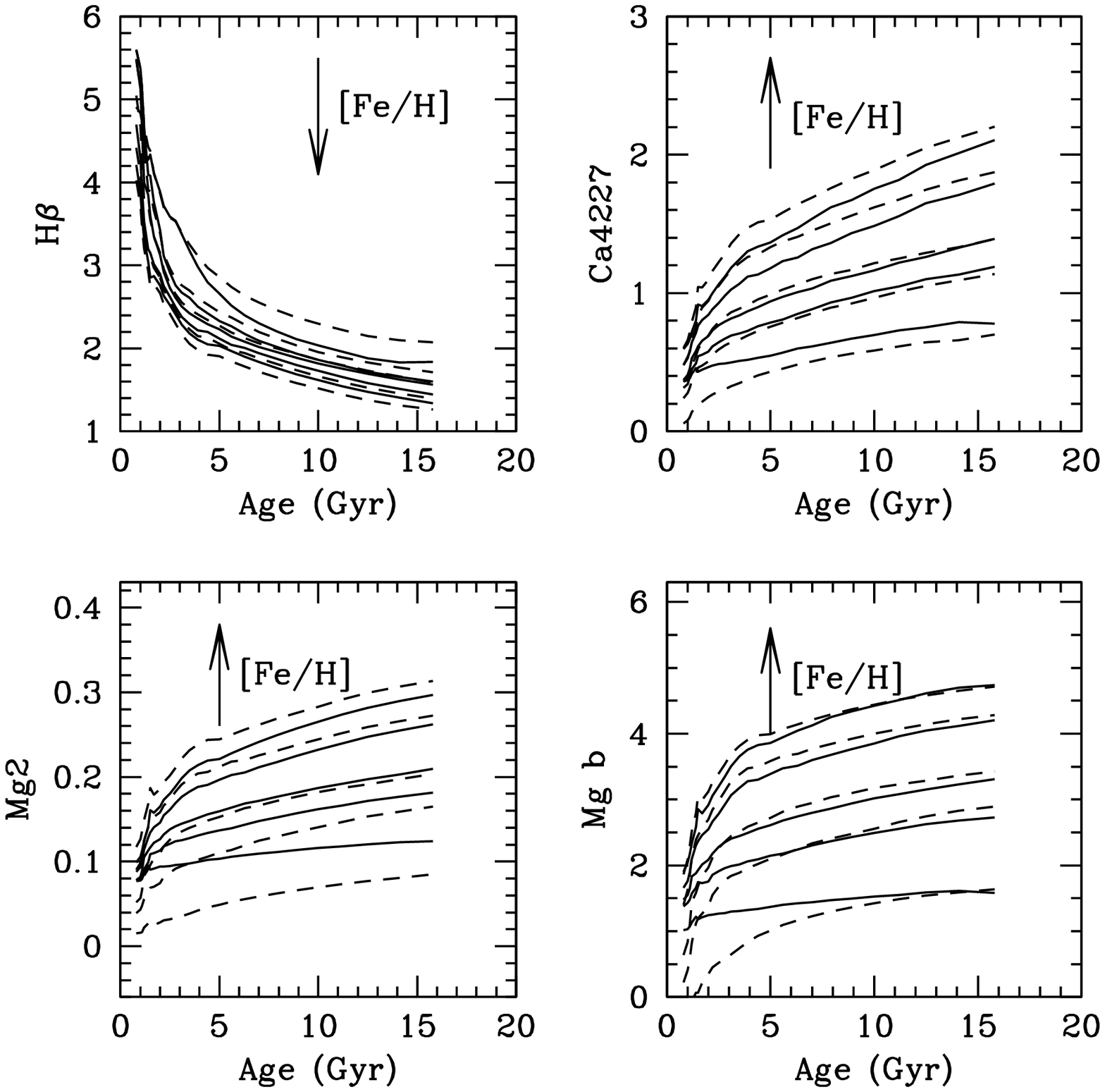}
\caption{b.}
\label{comp_w94b}
\end{figure*}

\setcounter{figure}{6}
\begin{figure*}
\plotone{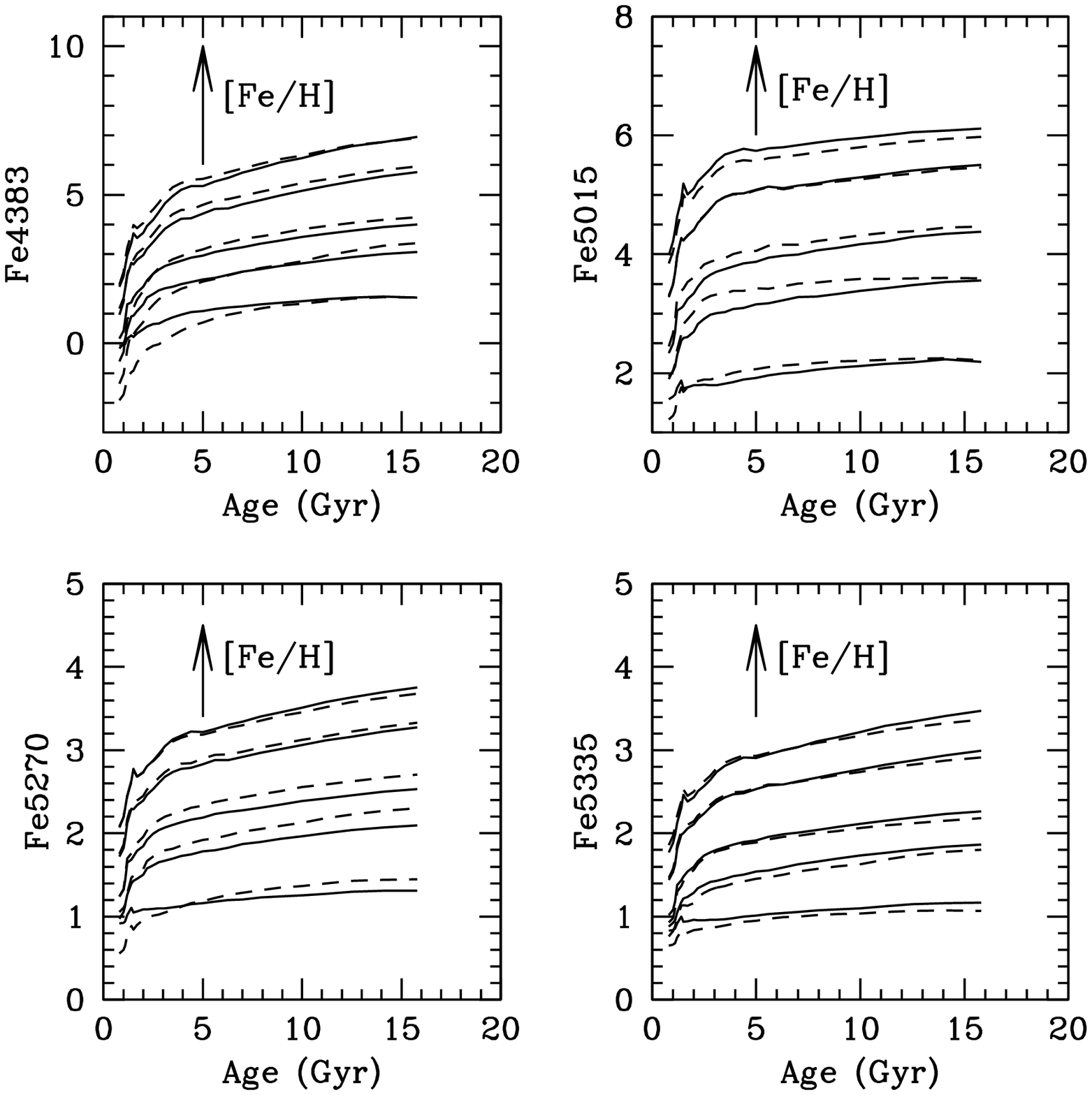}
\caption{c.}
\label{comp_w94c}
\end{figure*}

\setcounter{figure}{6}
\begin{figure*}
\plotone{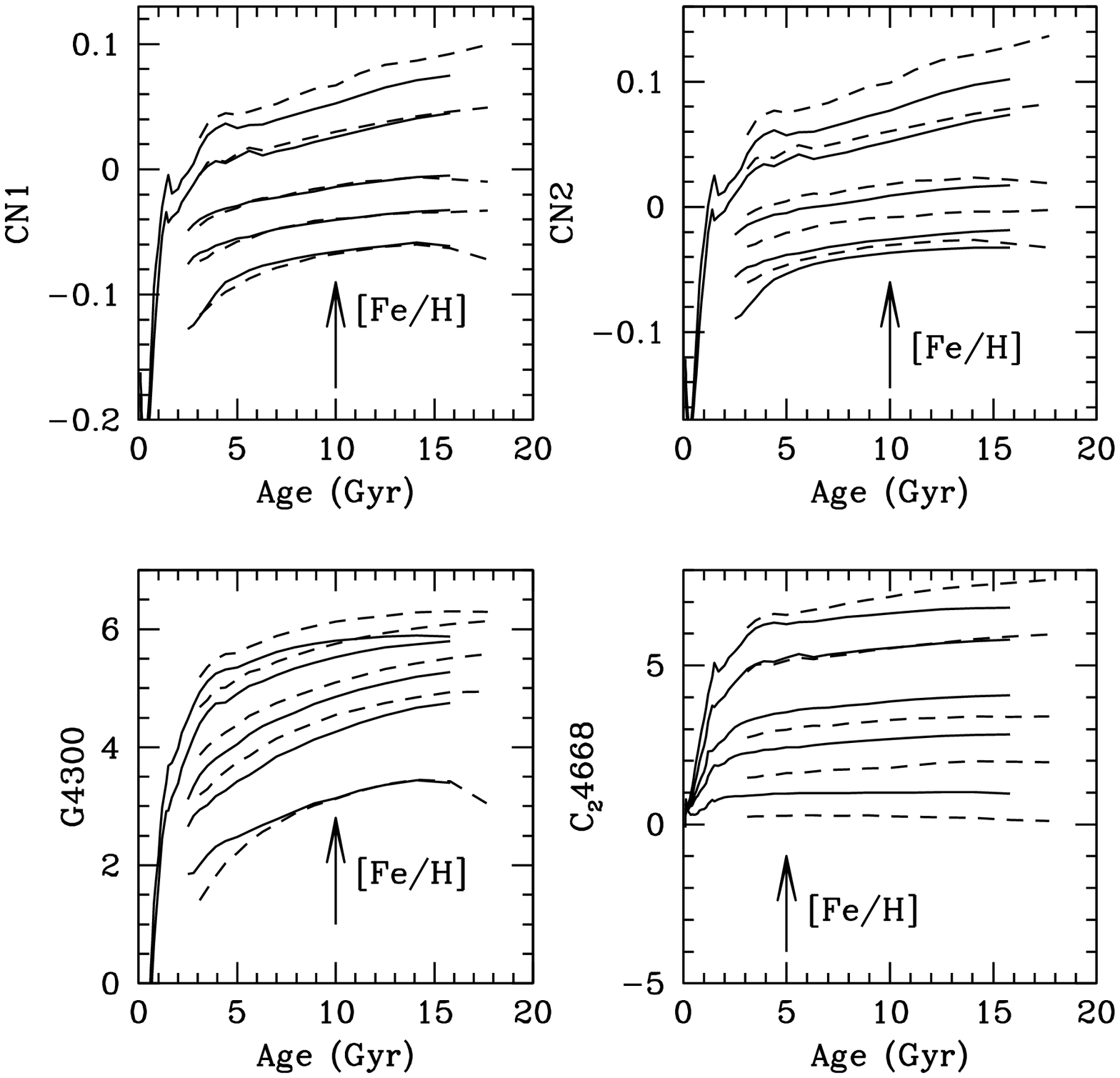}
\caption{d.}
\label{comp_w94d}
\end{figure*}

We restrict our comparison to previous work on Lick/IDS fitting functions
to those computed by G. Worthey and collaborators, because they are
available for all the indices studied here and are based on a very
comprehensive spectral library. Moreover, they are the most widely
used fitting functions in stellar population synthesis work. 

In order to assess the impact of our new fitting functions on model
predictions we proceeded as follows. We computed integrated line indices
for models 1 through 5 in Table~\ref{ssppar} adopting our own fitting
functions and those of Worthey \etal (1994) and Worthey \& Ottaviani
(1997) (henceforth simply Worthey et al.). In this way we isolate the
effect on model predictions due only to the adoption of our new fitting
functions.

The two sets of model predictions are compared in Figures~\ref{comp_w94a}a
through \ref{comp_w94d}d for all indices. The indices computed adopting
the Worthey \etal fitting functions were brought into our system of EWs
using the zero-points listed in Table~\ref{zeropoints}. In each panel
arrows indicate in which sense model metallicity varies, to help the reader 
identify the models with different [Fe/H].

\begin{figure}
\plotone{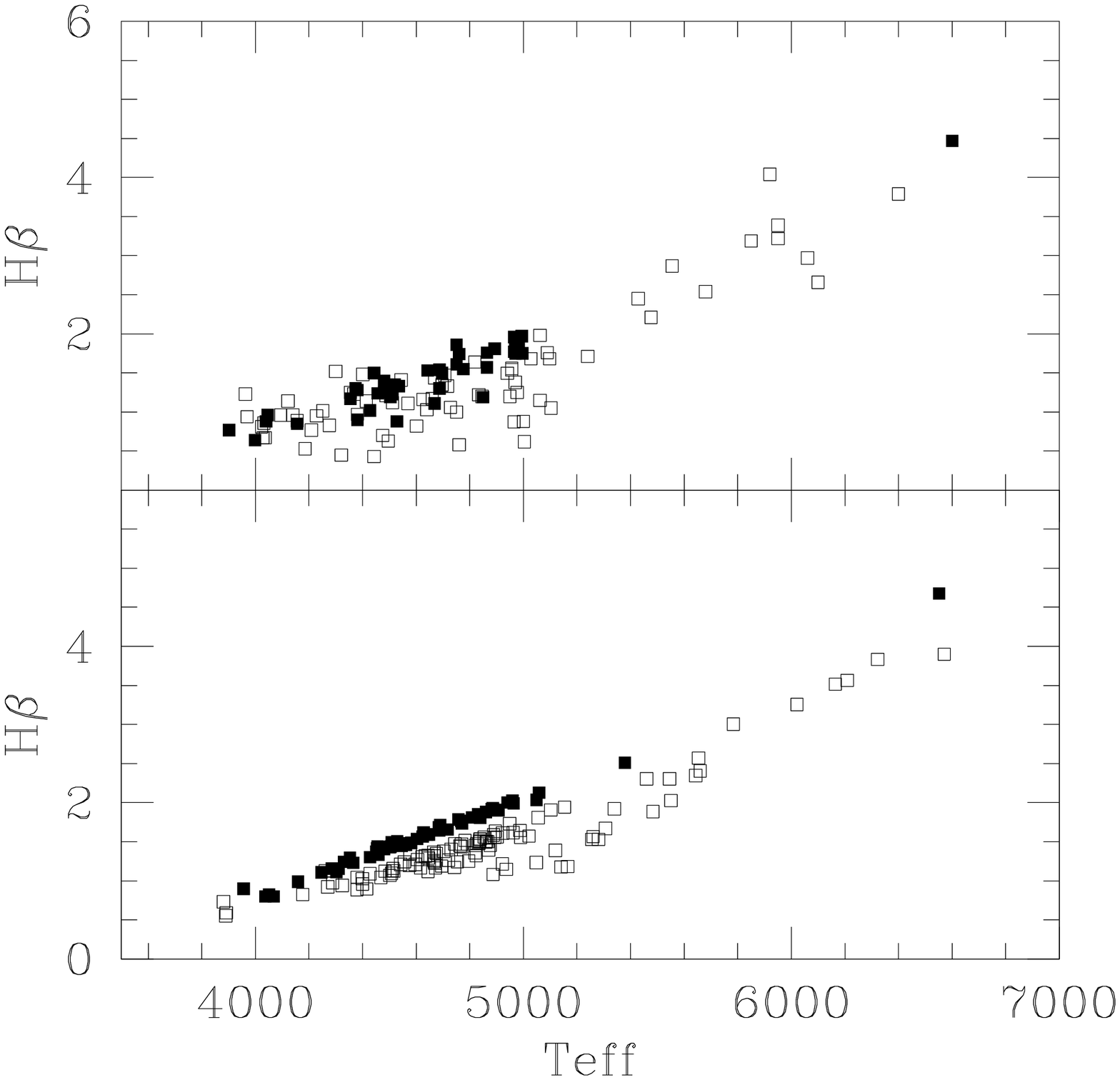}
\caption{{\it Top panel}: $H\beta$ against \Te\ for giant stars in the
Worthey \etal (1994) database. {\it Bottom panel:} Same for our data. {\it
Open squares}: stars with [Fe/H] $<$ --0.3. {\it Filled squares}: stars
with [Fe/H] $>$ 0. The improvement on both index measurements and stellar
parameters allows us to estimate the dependence of the index on [Fe/H]
more accurately.
}
\label{Hbgiants}
\end{figure}

The overall agreement between the two sets of computations is good. Not
unexpectedly, most of the differences are found at low metallicity,
where both sets of fitting functions are more uncertain. Amongst
the Balmer lines, the most important differences are found for
$H\beta$. This index is more metallicity-dependent when the Worthey
\etal fitting functions are adopted. This is a very interesting result
that serves to illustrate how improvements in the accuracy of stellar data 
(both stellar parameters and spectra) can cause a noticeable improvement in
model predictions.  In Figure~\ref{Hbgiants} we compare the input data
used in the computation of both sets of fitting functions. In the upper
panel, $H\beta$ and \Te\ from the Worthey \etal (1994) are plotted against
each other for G and K giants and the same plot is repeated in the lower
panel using our data. Stars in two ranges of metallicity are plotted in
order to highlight the dependence of $H\beta$ on this parameter. Stars
with [Fe/H] $>$ 0 are plotted with solid squares and stars with [Fe/H]
$<$ --0.3 with open squares. A dependence of $H\beta$ on metallicity,
whereby at fixed \Te\ the index becomes {\it stronger} for {\it higher}
[Fe/H], can be seen in both data-sets, but is far more clear-cut in our
data than in those of Worthey \etal (1994). As a result, we can estimate
the dependence of $H\beta$ on metallicity {\it in stellar spectra} more
accurately. We find that $H\beta$ in the spectra of GK giants responds to
variations in [Fe/H] roughly twice as strongly than predicted by Worthey
\etal (1994) in the sense that, we repeat, $H\beta$ becomes stronger 
for higher metallicity. On the other hand, we know that higher metallicity
systems tend to have {\it cooler} turn-offs, which tends to produce {\it weaker} 
$H\beta$. Therefore, the two above effects tend to cancel out, with the
net result that the index in integrated spectra of stellar populations becomes 
{\it less sensitive} to [Fe/H] than predicted by former models. As a result, the 
new fitting functions show that $H\beta$ is a better age indicator (i.e., less
sensitive to [Fe/H]) than previously thought.

\begin{figure}
\plotone{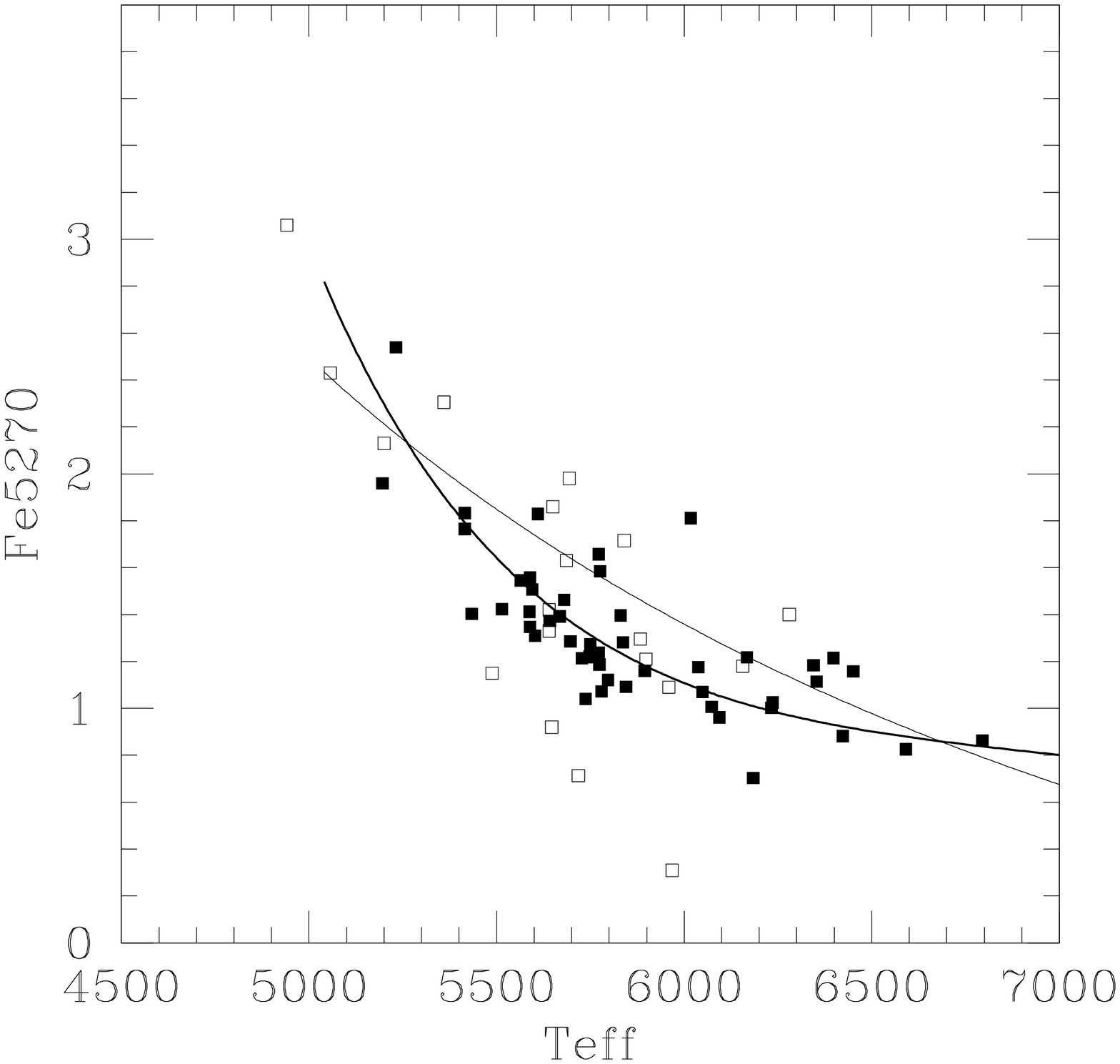}
\caption{\Te\ against Fe5270 for K-F dwarfs. Our data are shown as filled
squares, and Worthey \etal data as open squares. Our fitting function is
shown as a thick line and that of Worthey \etal as a thin line. Only stars
with [Fe/H] = --0.4 $\pm$ 0.15 are displayed and the fitting functions were
computed for [Fe/H] = --0.4.}
\label{Fe52dwarfs}
\end{figure}

The Fe indices are extremely important because they are mostly sensitive
to the abundance of iron (Tripicco \& Bell 1995), thus providing a
close estimate of the mean [Fe/H] of an integrated stellar population.
In Figure~\ref{comp_w94c} we compare our model predictions to those based
on the Worthey \etal fitting functions for all the Fe indices modelled
here. Agreement between the two sets of fitting functions is good for
Fe4383 and Fe5335. The most important differences are found for Fe5270
at metallicities below solar. In Figure~\ref{Fe52dwarfs} we compare the
two sets of fitting functions for dwarfs with [Fe/H]=--0.4. Our data
and fitting functions are represented respectively by the solid squares
and thick solid line. The open squares and thin line indicate Worthey
\etal data and fitting functions. Only dwarfs with [Fe/H] = --0.4 $\pm$
0.15 are plotted. As in the case of Figure~\ref{Hbgiants} the quality
of the new stellar data is quite superior, as can be seen by the lower
scatter in the solid squares. That of course makes it far easier to
compute an accurate fitting function for the index. It can be seen that
our new set of fitting functions provides a better description of the
data for mildly metal-poor dwarfs.  The latter accounts for roughly 2/3
of the mismatch seen in Figure~\ref{comp_w94c}. The rest of the mismatch
is due to smaller differences in the fitting functions for giant stars.

\begin{figure}
\plotone{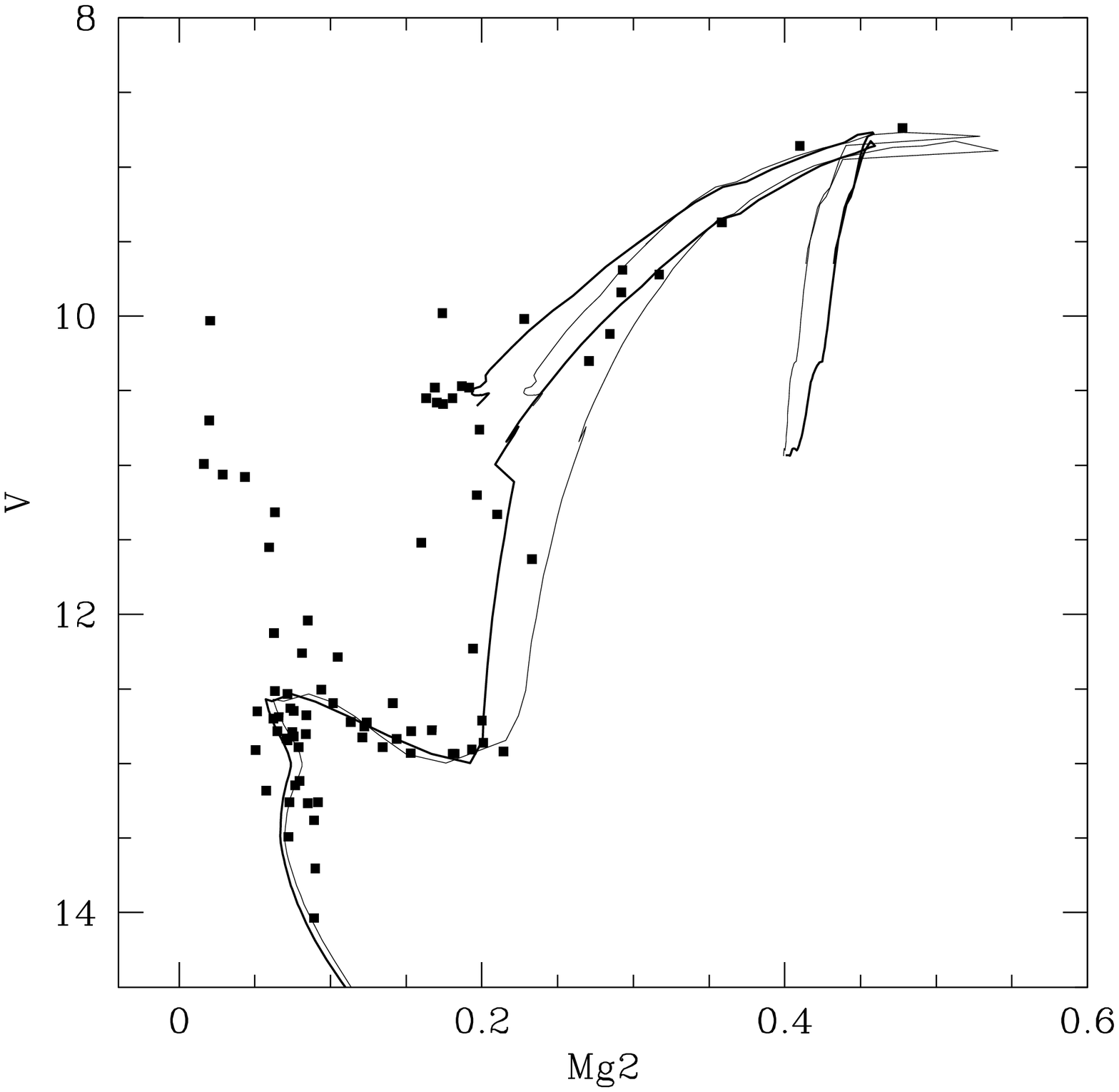}
\caption{An Mg$_2$-magnitude diagram for stars from M~67. The thin line
shows computations adopting the Girardi \etal (2000) isochrone for 3.5
Gyr and solar metallicity and the Worthey \etal fitting functions. The
thick lines were obtained using our fitting functions.}
\label{mg2m67}
\end{figure}

Another interesting case is that of indices that are strongly sensitive to
surface gravity, such as Mg$_2$, Mg $b$, and Ca4227, for which the Worthey
\etal al fitting functions yield higher values for solar metallicity at
all ages. This is because the line strengths in the spectra of giants
are stronger according to the Worthey et al. fitting functions. This
point is illustrated in Figure~\ref{mg2m67} where the two sets of fitting
functions are compared with Mg$_2$ data for M~67 stars in an 
Mg$_2$-magnitude diagram. The data come from Paper III, whereas the 
isochrones were computed by combining the two sets of fitting functions with 
the Girardi \etal (2000) isochrone for an age of 3.5 Gyr and solar metallicity. The
latter was shown to provide an excellent match to the color-magnitude diagram of
the cluster (see Paper III for details). It can be seen from this Figure
that, when the Worthey \etal fitting functions are adopted the index
is over-predicted by $\sim$ 0.05 mag throughout most of the red-giant
sequence and also at the horizontal branch (thick lines). A similar
behavior is seen for Mg $b$ and Ca4227.

\begin{figure}
\plotone{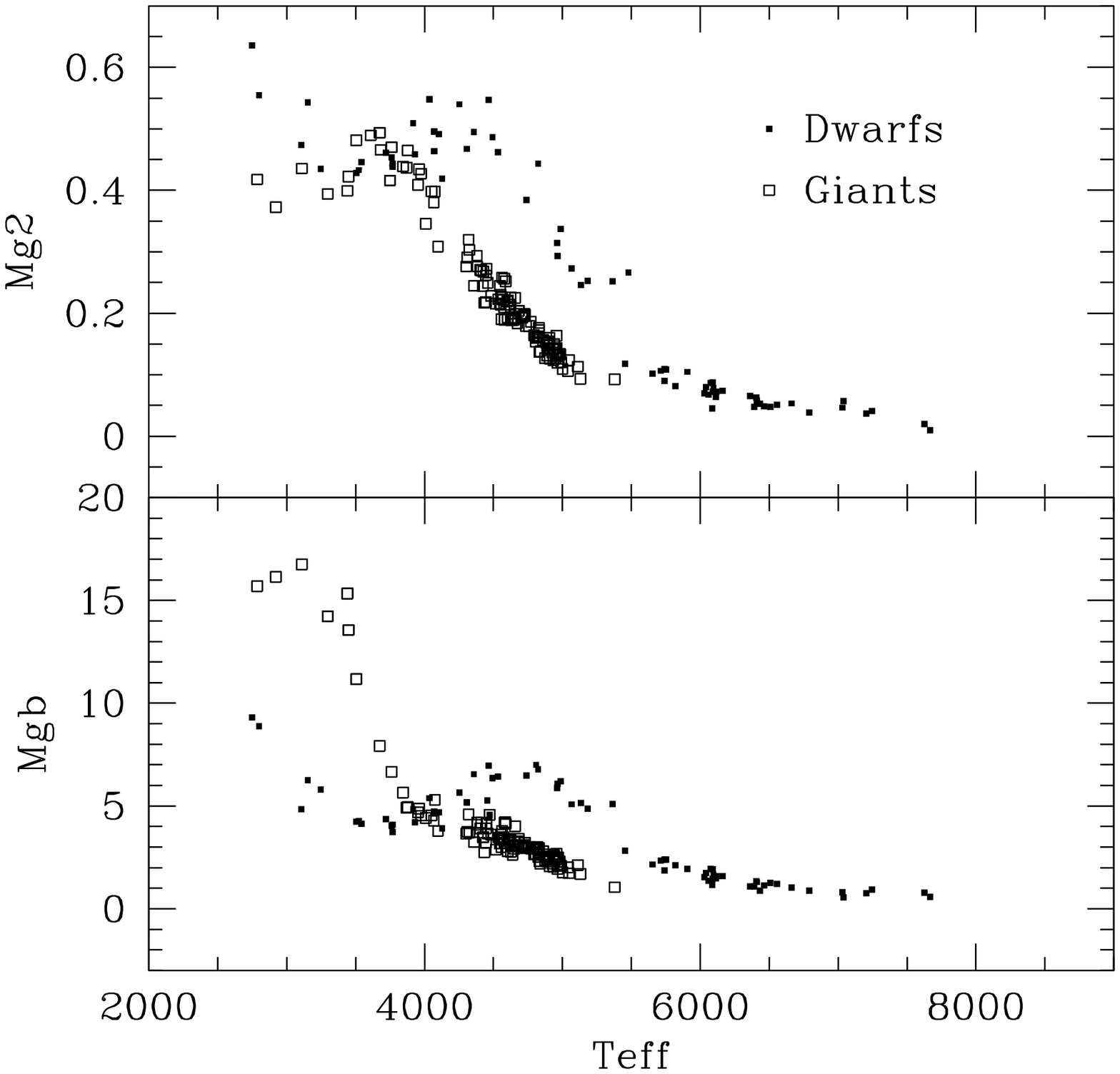}
\caption{Comparison between the behavior of Mg $b$ (lower panel) and
Mg$_2$ as a function of \Te\ and \gl. In both panels, measurements taken
in the spectra of the library stars are plotted as a function of \Te\ for
dwarf and giant stars, as indicated in the upper panel. For K stars (4000
$\simless$ \Te\ $\simless$ 5500 K), the behavior of the two indices as a
function of both stellar parameters is essentially the same. Both indices
are strongly sensitive to temperature and tend to be stronger in dwarf
stars. For cooler, M stars, Mg $b$ tends to be much stronger in giants,
unlike Mg$_2$, which is less dependent on surface gravity. This is
due to the effect of TiO bands on Mg $b$.  See text for details.}
\label{mg2mgb}
\end{figure}

It is also important to point out that the two Mg indices have a
markedly different sensitivity to IMF variations. While Mg$_2$ is
strongly sensitive to the contribution of K dwarfs, Mg $b$ is nearly
insensitive. This can be understood by looking at Figure~\ref{mg2mgb},
where we plot measurements of the two indices in our library star
spectra as a function of \Te\, for dwarf and giant stars. For K stars
(5500 $\simgreat$ \Te\ $\simgreat$ 4000 K), both indices respond to \Te\
and \gl\ in essentially the same way. In particular, they tend to be
stronger in K dwarfs, because both the Mg II lines and the MgH band-head
included in the Mg$_2$ passband are stronger for higher surface gravities
(Barbuy, Erdelyi-Mendes \& Milone 1992). At lower \Te , presumably 
because the Mg II lines saturate, the indices cease to increase for
lower temperatures and its dependence on \gl\ also becomes weaker. In
the M-star regime (\Te\ $\simless$ 4000 K) the two indices behave in
drastically different ways. While Mg $b$ becomes much stronger in giants
than in dwarfs, Mg$_2$ is very little dependent on surface gravity.
The reason for this behavior is that, as pointed out in Paper III,
Mg $b$ is severely affected by a TiO band, which is so strong in the
spectra of M giants that Mg $b$ becomes essentially a TiO indicator
(see Figure 3 in Paper III, for details). Because TiO bands are very
strongly sensitive to \gl\, being stronger in giants than in dwarfs
of the same \Te\ (Schiavon \& Barbuy 1999, Schiavon 1998), the Mg $b$
index becomes much stronger in the former than in the latter. The
Mg$_2$ index, on the other hand, is far less influenced by TiO lines,
because they affect both the pseudo-continuum and index passband in
similar ways. This result has an interesting ramification, namely, that
Mg$_2$ is an IMF-sensitive index, and Mg $b$ is nearly unaffected by IMF
variations. This can be understood by looking at Figure~\ref{mg2mgb}. The
Mg$_2$ index is IMF-sensitive because it is much stronger in dwarf stars,
so that it tends to be stronger for dwarf-enriched IMFs. The same is not
true for Mg $b$, because the index is so strong in cool giants that its
sensitivity to the contribution by K-dwarfs is washed away. As a result,
when used in combination, the Mg$_2$ and Mg $b$ indices can be used to
constrain both the magnesium abundance and the shape of the IMF in the
low-mass regime. We return to this topic in Section~\ref{47tuclight}.

There is a caveat here that needs to be highlighted. When we first
computed the model predictions with our fitting functions we obtained
too weak Mg$_2$ values for stars in the lower giant branch ($12.5 \simless$
V $\simless 11.5$ in M67, cf. Figure~\ref{mg2m67}). That region of the
diagram is inhabited by K stars with intermediate surface gravities
($3.0 \simless \log g \simless 3.6$), which are scarce in our spectral
library. Therefore, our fitting functions are poorly constrained in
this region of stellar parameter space. For that reason, we decided
to interpolate our predictions for gravity-sensitive indices, using
index-magnitude diagrams such as the one shown in Figure~\ref{mg2m67}
to check the quality of the interpolations.

In Summary, we conclude that our fitting functions are generally in good
agreement with those of Worthey \etal (1994). The differences found are
mostly due to the better quality of our data and the higher accuracy of
our stellar parameters. The latter validates our efforts to refine the
stellar parameter determinations, as described in Section~\ref{stelpar}. 

\subsection{Abundance-Ratio Effects} \label{aenhance}

Absorption-line strengths in the integrated spectrum of a single stellar
population are affected by its abundance pattern for two main reasons:
{\it i)} the effective temperatures and, potentially, the luminosities of two stars 
with same mass, age, helium abundance, and metallicity,
but different abundance patterns, are different, and as a result the
spectra of these two stars are different;
{\it ii)} the spectra of two stars that occupy the same position on
the HR diagram and have the same metallicity are different if they
have different abundance patterns.  The effect of the abundance mix
on the effective temperature of a given star is due to the relative
contribution of different elements to the overall opacity of the stellar
interior.  Oxygen is the most important metallic source of opacity at the high
temperatures prevalent in the stellar interiors (Vandenberg \& Bell 2001).
In the outer layers, iron is the predominant metal source of opacity
for FGK stars, which dominate the light of single stellar populations in
the spectral region of relevance for this study.  It is therefore fair
to say that the effect of abundance ratios on the positions of stars in
the HR diagram is dictated by the relative abundances of oxygen and 
iron\footnote{Carbon and nitrogen, to a lesser extent, also contribute to
determining the temperatures of the outer layers of K stars, via the
back-warming effects due to CN opacity (see Gustafsson \etal 1975 for
a discussion).}.  Line strengths, on the other hand, can be strongly
affected by the individual abundances of elements which are not optically
active enough to produce a substantial change in the star's structure.
Two stars with the same mass, age, metallicity, and helium abundance,
and whose abundance patterns are the same except, for instance, for
their calcium abundances, have virtually the same temperature and luminosity,
and their spectra will be essentially the same, except for differences
in the strengths of lines due to atomic calcium or due to molecules
involving calcium, such as calcium hydride.

Since the realization that the chemical composition of stars in giant
early-type galaxies is enhanced in light elements (e.g., Peterson 1976,
O'Connell 1980, Peletier 1989, Worthey \etal 1992), a great deal
of effort has been invested into producing realistic models with an
$\alpha$-enhanced abundance pattern.  These efforts branch out in two
major directions, aimed at accounting for the two major effects listed
above: {\it i)} computation of stellar evolutionary tracks for metal-rich
stars incorporating an $\alpha$-enhanced mixture (e.g., Weiss \etal
1995, Salasnich \etal 2000, Kim \etal 2002) in order to assess
the impact of $\alpha$-enhancement on model predictions, and {\it ii)}
estimating the effect of the abundance pattern onto stellar spectra and
line indices (e.g., Barbuy 1994, Tripicco \& Bell 1995, Paper I, Barbuy
\etal 2003, Coelho 2004, Mendes de Oliveira \etal 2005, Korn, Maraston \&
Thomas 2005), using spectrum synthesis from model stellar atmospheres.

Before comparing model predictions with data in Section~\ref{clust},
it is interesting to discuss the impact of abundance ratios on
model predictions.  For simplicity, throughout this paper we refer to
{\it i)} above as {\it evolutionary} abundance-ratio effects and
to {\it ii)} as {\it spectroscopic} abundance-ratio effects. In
Section~\ref{aeffect}, we examine evolutionary abundance ratio 
effects, which are those stemming from the influence of the abundance
mixture on the luminosity and effective temperature of a star of
given mass, metallicity and evolutionary stage.  Spectroscopic
abundance-ratio effects are considered in Section~\ref{aratios}.

\subsubsection{Evolutionary Abundance-Ratio Effects} \label{aeffect}

\begin{figure}
\plotone{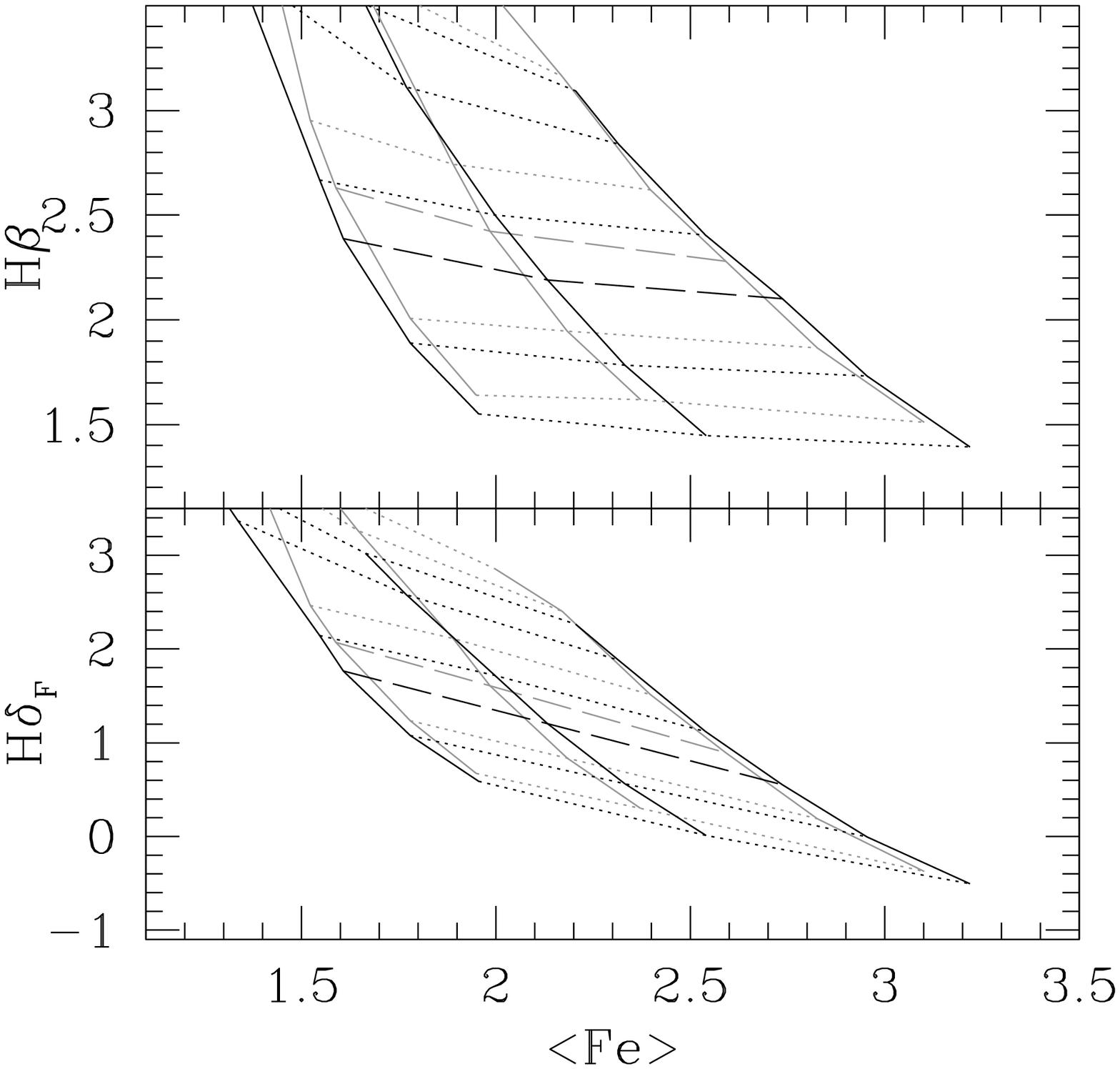}
\caption{Effect of adopting $\alpha$-enhanced isochrone (dark lines),
as opposed to solar-scaled ones (gray lines) on predictions for $<Fe>$
and Balmer lines for single stellar populations. The ages shown are,
from top to bottom, 1.2, 1.5, 2.5, 3.5, 7.9, and 14.1 Gyr. The values for
[Fe/H] are --0.8 (--0.7 for solid lines), --0.4, and 0.0. The $\alpha$
enhanced isochrones yield weaker Balmer lines and slightly weaker $<Fe>$
for the same age and metallicity.  Note that the effect is stronger on
$H\beta$ than on $H\delta_F$.
}
\label{alphaiso}
\end{figure}

In Figure~\ref{alphaiso} our models for single stellar populations
are displayed in the $H\beta$ vs. $<Fe>$ (upper panel) and $H\delta_F$
vs. $<Fe>$ planes. These indices illustrate very well the general effect
of $\alpha$-enhanced isochrones on Balmer and metal lines.  In both
panels, black lines indicate models computed with the $\alpha$-enhanced
Padova isochrones (Salasnich \etal 2000) and gray lines those with the
solar-scaled isochrones (Girardi \etal 2000).  The ages of the models
displayed are, from top to bottom, 1.2, 1.5, 2.5, 3.5, 7.9, and 14.1 Gyr.
For clarity, we restrict the comparison to those amongst the two sets of
models computed with similar [Fe/H] values (2 through 4 and 6 through
8 in Table~\ref{ssppar}).  Solid lines connect same-[Fe/H] models, and
dotted lines connect same-age models.  The models for 3.5 Gyr are
plotted with a long-dashed line, for clarity.

The main difference between models computed with solar-scaled and
$\alpha$-enhanced isochrones is that the latter tend to predict
weaker Balmer lines and slightly stronger metal lines, for the same
age and [Fe/H]. This is because $\alpha$-enhanced turn-off stars
are cooler and fainter than their solar-scaled counterparts {\it at
fixed [Fe/H]}, due to increased opacity, especially due to oxygen,
in the stellar interior (e.g., Vandenberg \& Bell 2001)\footnote{
The effect due to oxygen enhancement is somewhat diminished by the
fact that the $\alpha$-enhanced Padova isochrones are computed with
a higher helium abundance for fixed-[Fe/H] (see Table~\ref{ssppar}).
Higher helium abundances leads to slightly warmer and brighter turnoffs,
which partly offsets the cooling of the turnoff due to increased oxygen
abundances}. In particular, the mixture adopted by Salasnich \etal in
their $\alpha$-enhanced tracks is enhanced in [O/Fe] by +0.5 dex relative
to solar. As a combination of the temperature effects on Balmer and
metal lines, the models based on $\alpha$-enhanced isochrones appear
to ``slide'' relative to the solar-scaled models along same-[Fe/H]
lines, towards weaker $H\beta$.  The final effect is that, for a given
data point, $\alpha$-enhanced models predict {\it younger} ages but,
interestingly, essentially the same [Fe/H]. For the mixture adopted by
Salasnich \etal (Table~\ref{ssppar}) the age effect is of the order of
$\sim$ 1 Gyr at intermediate ages ($\sim$ 4 Gyr), and as large as $\sim$
3 Gyr for the ages of the oldest globular clusters ($\sim$ 14 Gyr). This
effect, together with our improvement to the fitting function of the
$H\beta$ index, leads to significantly younger ages for old stellar
populations, thus ameliorating a long-standing problem, namely, that
stellar population synthesis models tend to predict too old ages for the
oldest stellar systems (e.g., Cohen, Blakeslee \& Rhyzov 1998, Gibson
\etal 1999, Vazdekis \etal 2001, Papers I and II, Proctor, Forbes \&
Beasley 2004, Lee \& Worthey 2005).  This issue is addressed further
in Section~\ref{clust}.

We also note that $H\delta_F$ is substantially less affected than $H\beta$
especially for the oldest models. Comparing the two models for 14 Gyr
in the bottom panel of Figure~\ref{alphaiso} we see that the variation
in $H\delta_F$ corresponds to less than $\simless$ 1 Gyr, compared to
$H\beta$, whose variation amounts to $\simless$ 3 Gyr. 

We would like to call the reader's attention to an important new
development. After submission of the first version of this paper, Weiss
\etal (2006) showed that there was an error in the opacity tables adopted
in the calculation of the Padova $\alpha$-enhanced evolutionary tracks,
by (Salasnich \etal 2000). This error is such that the temperatures
at the red giant branch and turnoff were overestimated by 200 and
100 K, respectively, for solar metallicity\footnote{These numbers were
obtained by comparing tracks A and R for 1 M$_\odot$, from Weiss \etal
2006 (see their Table 2), following a suggestion by A. Weiss}. Even
though new isochrones with the mixture adopted by Salasnich \etal are
not available, we simulated the effect of the corrected opacity tables
by artificially changing the temperatures of giant and turn-off stars
uniformly by 200 and 100 K in Salasnich \etal isochrones with Z=0.04
(nearly solar [Fe/H]). As a result, Balmer lines get weaker and metal
lines get stronger. The change in $H\beta$ ($H\delta_F$) is of the
order of $\sim$ --0.15 (--0.1) ${\rm\AA}$. Because Balmer lines would
tend to get weaker in the models, ages according to these ``corrected''
$\alpha$-enhanced isochrones would get {\it younger}, thus accentuating
the differences seen in Figure~\ref{alphaiso}.  The effect would be of
the order of $\sim$ 2.5 (1) Gyr for an 11 (3) Gyr-old stellar population,
in the case of $H\beta$. Ages according to $H\delta_F$ would get younger
by $\sim$ 1.2 (0.5) Gyr for 11 (3) Gyr-old stellar populations. The change
in \fem\ is of the order of $\sim$ +0.14 ${\rm\AA}$ and it is such that
models change along a line of constant [Fe/H] in the \fem-$H\beta$ plane.
More definitive numbers have to await publication of $\alpha$-enhanced
theoretical isochrones computed on the basis of updated opacities.

According to Weiss (2006, private communication), adoption of new opacities
has a less important impact for lower metallicities. For instance, in the
case of the metallicity of 47~Tuc, changes in age would be of the order of
$\sim$ 1.5 Gyr and essentially zero in [Fe/H], so that our discussion in
Paper II remains entirely valid. 

\subsubsection{Spectroscopic Abundance-Ratio Effects} \label{aratios}

In the previous section we studied how the adoption of $\alpha$-enhanced
isochrones affects our model predictions. Here we show how the
sensitivity of line indices to the various elemental
abundances alters the model predictions.  The method, first proposed by
Trager \etal (2000) and further developed by Thomas \etal (2003a) and Korn
\etal (2005), is based on estimates of how line indices in the spectra of
three types of stars with relevant atmospheric parameters (a turnoff, a
giant, and a lower main sequence star) change as a function of variations
of the abundances of individual elements. Briefly, the method goes as
follows. Initially, synthetic spectra are computed on the basis of model
photospheres, for the three stellar types above, assuming a solar-scaled
abundance pattern. For each stellar type, a new spectrum is then computed
by varying the abundance of a single element.  The impact of this sole
elemental abundance on all spectral indices is assessed by comparing
line indices measured in both the solar-scaled synthetic spectrum and
that computed with the altered abundance pattern. The procedure is
repeated for all the chemical abundances of relevance in the spectral
region of interest and the final results are summarized in the form of
sensitivity tables, such as those provided by Tripicco \& Bell (1995)
and more recently updated by Korn \etal (2005) and Serven, Worthey \&
Briley (2005).  The sensitivity tables are used to estimate incremental
changes in absorption line indices as a function of stellar parameters,
which are integrated along the theoretical isochrone in order to produce
integrated line strength predictions for any desired abundance pattern.

\begin{figure}
\plotone{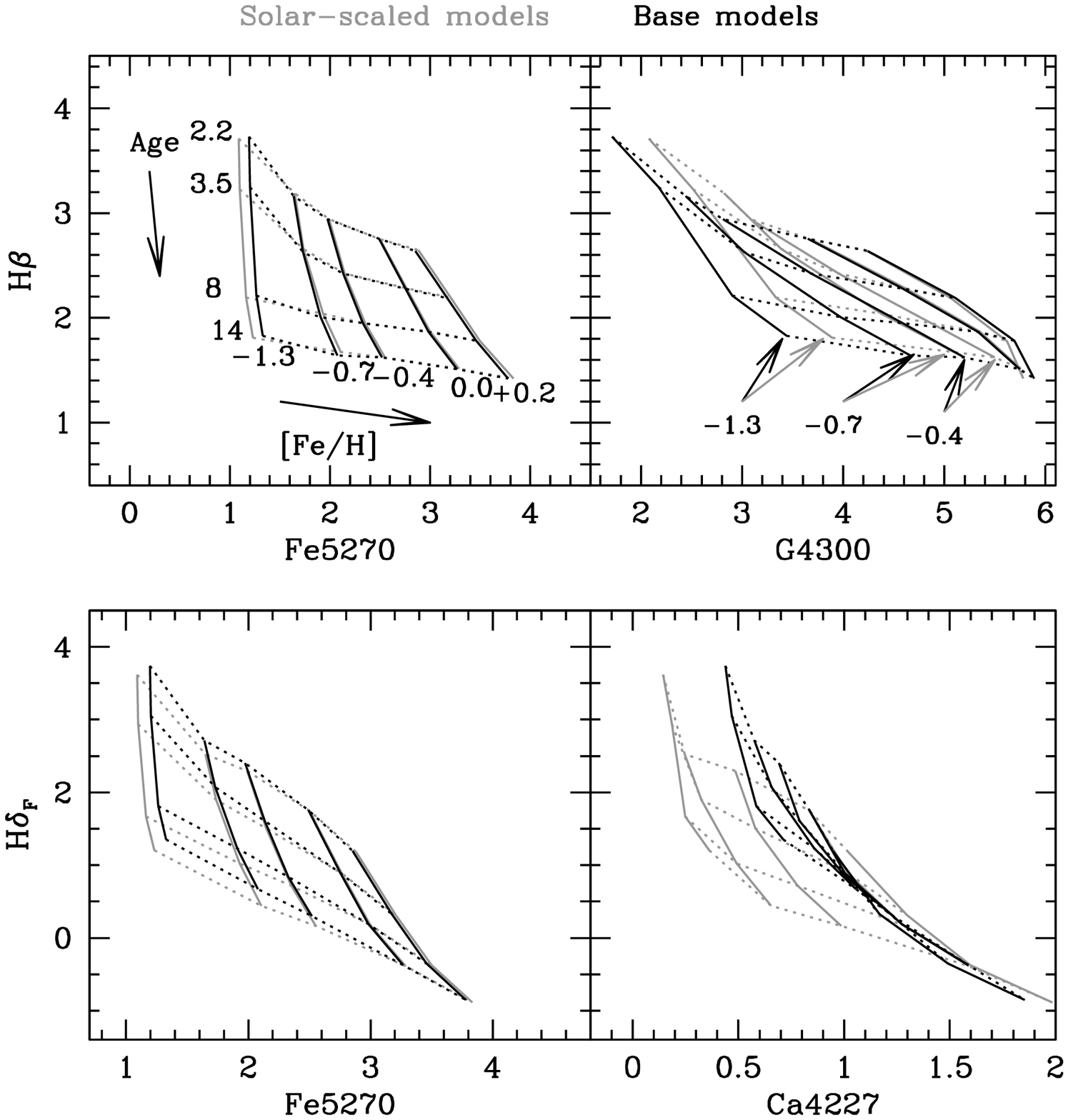}
\caption{Comparison between solar-scaled (gray) and base models (dark)
on a few representative index-index plots. {\it Upper Left:} Both $H\beta$
and Fe5270 are essentially insensitive to abundance ratio variations. {\it
Upper Right:} G4300 is stronger in solar-scaled than in the base models
for [Fe/H] $\leq$ --0.4, because the latter have higher oxygen abundances
(see text). {\it Lower Left:} $H\delta_F$ is weaker in solar-scaled than in
the base models for [Fe/H] $\simless$--0.4, because at these metallicities
the base models have higher magnesium and calcium abundances, indicating
that there are lines due to these two elements in the $H\delta_F$
index passband. {\it Lower Right:} The Ca4227 index is much weaker in
solar-scaled than in the base models for [Fe/H]$\simless$--0.4, having
similar strength for [Fe/H]=0 and +0.2.  This reflects the fact that
the base models have higher calcium abundance than solar-scaled models
for metallicities below solar (see Table~\ref{tabxfe}).
}
\label{araeff_illust}
\end{figure}

In this section we discuss the behavior of all the line indices studied
in this paper as a function of the most important elements affecting
their strengths in the range of stellar population parameters considered.
In order to study this effect in isolation from that discussed in the
previous Section, we perform calculations based on the {\it same} set of
isochrones (Padova, solar-scaled), but varying the line indices according
to their sensitivity to abundance ratio variations.  The sensitivity
tables used are those by Korn \etal (2005).

\bigskip
\centerline{\it Solar-scaled vs.\ Base Models}
\smallskip

The starting point for this discussion are the base models, which are
summarized in Table~\ref{ssppar}. Models for any abundance pattern
can be calculated relative to the base models in a differential
fashion.  As a first step, we generate solar-scaled models, for
which [X$_i$/Fe] = 0 for all elements X$_i$, and at all values of
[Fe/H]. These models are generated by correcting the base models from
the abundance pattern listed in Table~\ref{tabxfe} to a solar-scaled
abundance pattern. Because metallicity in our models is cast in terms
of [Fe/H], this parameter is kept fixed whenever we calculate a model
with a new abundance pattern.  For instance, according to 
Table~\ref{tabxfe}, if one wants to compute
solar-scaled models with [Fe/H]=--0.4, one needs to correct the
indices in the [Fe/H]=--0.4 base models for abundance ratio variations of
$\Delta$~[O/Fe]=--0.2, $\Delta$~[Mg/Fe]=--0.13, $\Delta$~[Ca/Fe]=--0.06,
and $\Delta$~[Ti/Fe]=--0.11. Those are the elemental abundance variations
needed to bring the abundance ratios in Table~\ref{tabxfe} to [X$_i$/Fe]=0
(no need to perform any correction for variations of [C/Fe] and [N/Fe],
as these ratios are solar in the spectral library for all values of
[Fe/H]).  As a result, two models with same [Fe/H] and different abundance
patterns have different total abundances [Z/H]. This is different from
the procedure followed by Trager \etal (2000) and Thomas \etal (2003a),
who cast their models in terms of [Z/H], so that their $\alpha$-enhanced
models have lower [Fe/H] for fixed [Z/H].

\begin{figure}
\plotone{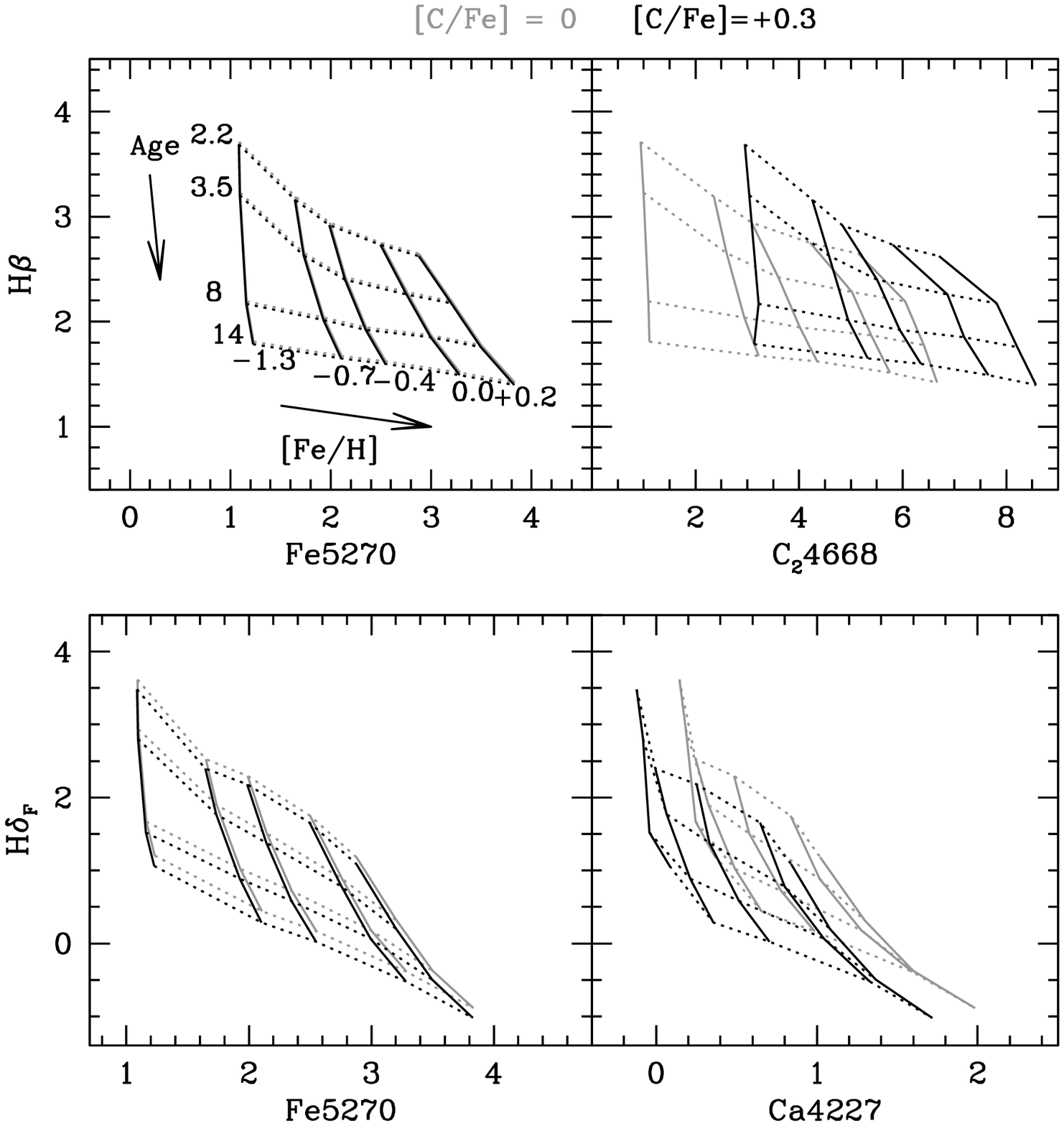}
\caption{Comparison between solar-scaled (gray) and carbon-enhanced
(dark) models. {\it Upper Left:} This panel shows that $H\beta$ and
Fe5270 are virtually insensitive to carbon abundance variations. {\it
Upper Right:} As expected, C$_2$4668 is strongly sensitive to carbon
abundance variations. It is the best carbon-abundance indicator modeled
in this paper. {\it Lower Left:} $H\delta_F$ is very mildly sensitive
to carbon abundance variations, in spite of its being surrounded by CN
lines (see text). {\it Lower Right:} The Ca4227 index is very sensitive
to carbon, being stronger in lower carbon-abundance models. This is due
to the presence of a CN band-head on the index blue pseudo-continuum,
as pointed out by Prochaska \etal (2005, see discussion in the text).
}
\label{ceff14}
\end{figure}

The result is illustrated in Figure~\ref{araeff_illust}, where base and
solar-scaled models (dark and gray lines, respectively) are compared in
four representative index-index planes. Because the abundance pattern
characteristic of the base models only differs importantly from
solar-scaled for [Fe/H]$\leq$--0.4, we only expect to find differences in
this iron abundance interval. In the upper left panel, the two sets of
models are compared in the Fe5270-$H\beta$ plane, where there is hardly
any difference between solar-scaled and base models.  The reason is that
both indices are very little affected by variations of the abundances of
any elements other than Fe.  The only exception happens at [Fe/H]=--1.3,
where the base models have slightly stronger Fe5270, due to the presence
of titanium and calcium lines in the index passband.  Because [Fe/H]
is kept fixed when switching from base to solar-scaled models, the
two model sets are virtually identical in the Fe5270-$H\beta$ plane.
The same is not the case in the upper-right panel, where models are
compared in the G4300-$H\beta$ plane, where the G4300 index appears to
display a complex behavior as a function of [Fe/H]. As expected, the
index is unchanged for models with [Fe/H]=0 and +0.2, but it becomes {\it
stronger} in solar-scaled models for lower [Fe/H] values (same-[Fe/H]
models are indicated with arrows, for clarity). The G4300 index is
mostly a carbon abundance indicator, because of the presence of a strong
vibrational band of the CH molecule in the index passband. However,
Table~\ref{tabxfe} tells us that [C/Fe] is solar in the base models,
so that it is the same for all [Fe/H] in both sets of models. However,
the G4300 index is also affected, in an indirect way, by oxygen abundance
variations, in the sense that the index tends to be weaker for higher
oxygen abundances\footnote{This is because the abundance of oxygen
affects the concentration of free carbon atoms in the stellar plasma
via the dissociation equilibrium of the CO molecule. Of the molecules
involving carbon and oxygen, carbon monoxide is the one with the highest
dissociation potential, so that once the temperature is below a certain
threshold, the free carbon and oxygen atoms in the plasma are
preferentially consumed by CO formation (e.g., Tsuji 1973).  Therefore,
the higher the oxygen abundance, the lower the concentration of free
carbon in the plasma, the fewer carbon atoms are available to form CH,
the weaker the G4300 index.}. Because the abundance of oxygen is lower in
the solar-scaled than in the base models, the former have stronger G4300
for fixed [Fe/H].  One can also conclude from this plot that the G4300 index is 
very strongly sensitive to age, which can be seen by the fact that the model
grids are very far from orthogonal. On the other hand its sensitivity to metallicity
and/or carbon abundance is relatively weak, as can be seen by the fact that 
model lines for different values of [Fe/H] are packed very close together, 
especially for [Fe/H] $\geq$ --0.7. As a result, the G4300 index is a less than 
ideal carbon abundance indicator, which is the reason why we decided to
include the C$_2$4668 index in our models\footnote{The author thanks 
Jenny Graves for insisting on that point!}. A more detailed discussion
of this issue will be presented in Graves \& Schiavon (2006, in preparation).

\begin{figure}
\plotone{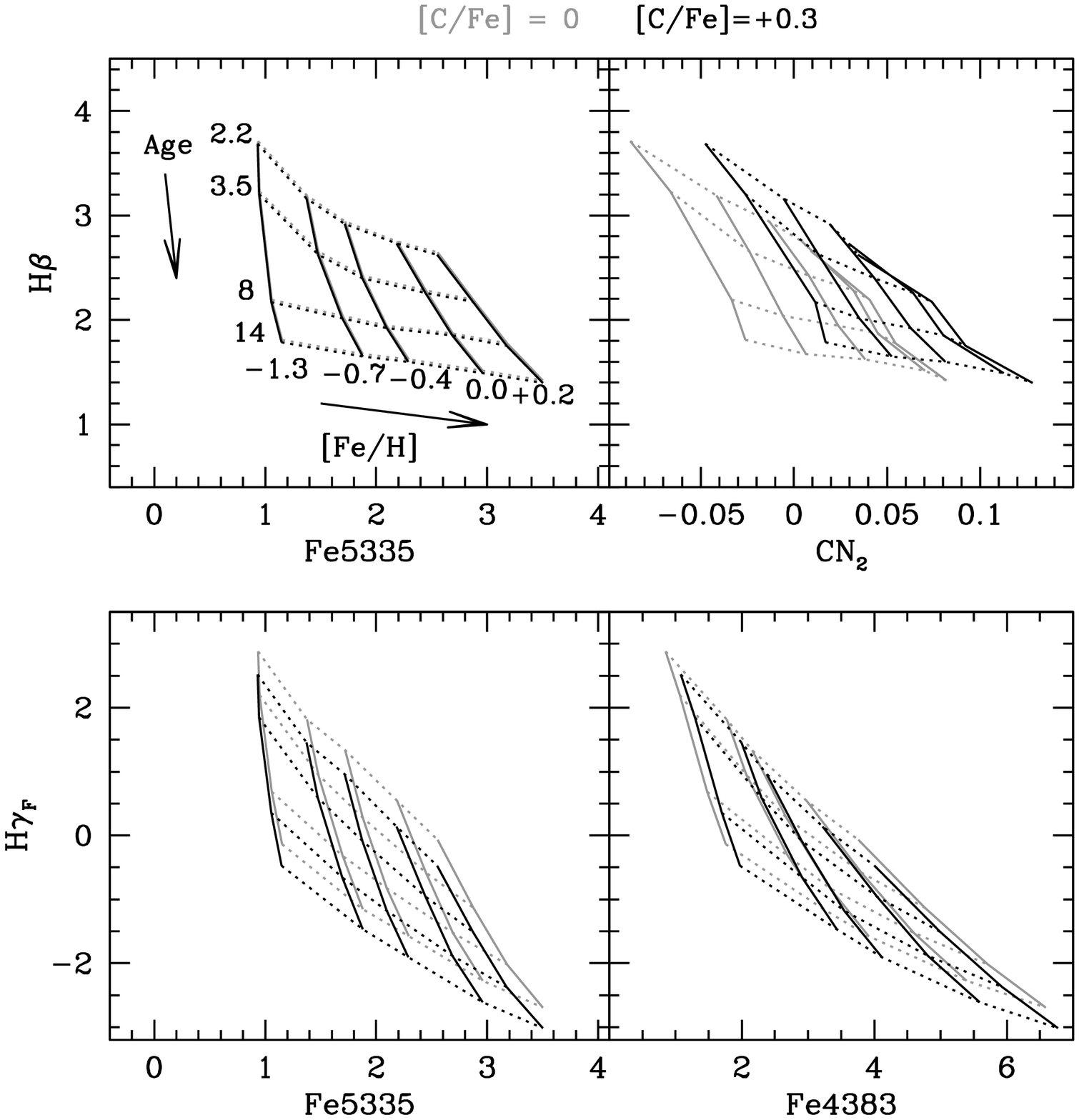}
\caption{Figure~\ref{ceff14} continued. {\it Upper Left:} This
panel shows that the Fe5335 index is also essentially insensitive to
carbon abundance variations. {\it Upper Right:} As expected, CN$_2$
is very sensitive to carbon, being substantially stronger for higher
carbon abundances. This index is also sensitive to nitrogen, but not as
strongly as to carbon. {\it Lower Left:} This plot shows how $H\gamma_F$
is affected by carbon abundance variations. It gets weaker for higher
carbon, due to contamination of the index pseudo-continuum by CH lines.
{\it Lower Right:} The Fe4383 index is almost insensitive to carbon.
The index becomes only slightly stronger for higher carbon abundances.
This effect is also due to the presence of CH lines in the index passband.
}
\label{ceff15}
\end{figure}

In the lower left panel, models are compared in the
Fe5270-$H\delta_F$ plane. It can be seen that $H\delta_F$ is slightly
weaker in solar-scaled models for [Fe/H]$\leq$--0.4. This is because
the passband of the index includes lines due to calcium and magnesium,
whose abundances are higher in the base models at these lower values of
[Fe/H].  Finally, in the lower-right panel, models are compared in the
Ca4227-$H\delta_F$ plane, which allows us to investigate the behavior
of the Ca4227 index as a function of abundance ratios. As expected,
base models have stronger Ca4227 at [Fe/H]$\leq$--0.4 than solar-scaled
models, given that the latter have lower calcium abundances. The effect
is in fact enhanced by the strengthening of CN bands in the solar-scaled
models, due to their lower oxygen abundances, for reasons that will be
discussed in the next section.

\bigskip
\centerline{\it Effects of Carbon Enhancement}
\label{carbon}
\smallskip

The blue spectra of G and K-type stars is pervaded by weak to moderately
strong absorption lines due to the CN and CH molecules.  That is in
fact one of the most important challenges for the reliable modeling
of the spectra of old/intermediate-age stellar populations in the
blue (e.g., Vazdekis 1999, Paper I, Prochaska, Rose \& Schiavon 2005,
Prochaska \etal 2006). For that reason, it is important to investigate
the effects of variations of the abundances of carbon and nitrogen on
blue spectral indices.

In Figures~\ref{ceff14} and \ref{ceff15} we
compare solar-scaled models with models where only the abundance
of carbon is enhanced by +0.3 dex.  In the upper left panel
of Figure~\ref{ceff14}, the models are compared in the
Fe5270-$H\beta$ plane, where it can be seen that the two indices
are essentially insensitive to carbon abundance variations. In the
upper right panel, on the other hand, one can see that the C$_2$4668
index is extremely sensitive to carbon abundance variations, indeed,
far more sensitive than the G4300 index. This is not unexpected, given 
that the concentration of the C$_2$ molecule in stellar atmospheres 
depends quadratically on the abundance of carbon, while that of CH 
depends only linearly on that parameter. In the lower
left panel, one can see that $H\delta_F$ is very little affected by
carbon abundance variations, in spite of the fact that the index is
immersed in a thick forest of CN lines. This has been discussed in
Paper I, where it was shown that the small sensitivity of $H\delta_F$
to carbon abundance variations was due to the the fact that the effect
of CN lines in both the index passband and pseudo-continuum regions is
partially cancelled\footnote{Essentially the same conclusion is reached
for the $H\delta_A$ index}. For a detailed discussion of the impact of
CN lines on $H\delta$ measurements, see Prochaska \etal (2006). In the
lower right panel of Figure~\ref{ceff14} the models are compared
in the Ca4227-$H\delta_F$ plane, where it can be seen that the Ca4227
index is strongly sensitive to carbon abundance variations, in the sense
that the index becomes substantially {\it weaker} for increasing carbon
abundances. This effect has been studied in detail by Prochaska \etal
(2005) and it is due to the contamination of the blue pseudo-continuum of
the index by a CN band-head. An increase in carbon abundance leads to a
depression of the blue continuum and, consequently, to an artificially
lower Ca4227 index. Prochaska et al. (2005) defined a new index,
Ca4227$_r$, which is far less affected by CN contamination. Analyzing
a large sample of nearby early-type galaxies, they showed that this new
index presents a correlation with velocity dispersion ($\sigma$) which
is much stronger than that of the Lick Ca4227 index, resembling the
behavior of other $\alpha$-elements, such as magnesium. In subsequent
sections, we will show that when the effect of CN lines is accounted
for, one can extract reliable calcium abundances from measurements of
the Lick Ca4227 index.

Figure~\ref{ceff15} illustrates the effect of carbon abundances
on another set of relevant Lick indices. In the upper left panel, one
can see that, like Fe5270, the Fe5335 index is unchanged
when carbon is varied. The same is true of the Fe5015 index
(not shown). In the upper right panel, on the other hand, one can
see that the CN$_2$ index is very sensitive to carbon abundance,
as expected.  It is also sensitive to nitrogen abundance, to a lesser
extent\footnote{Essentially the same conclusions is reached for the
CN$_1$ index}. In the lower left panel, the models are compared in the
Fe5335-$H\gamma_F$ plane, where it can be seen that this Balmer line
is somewhat affected by carbon, in the sense that $H\gamma_F$ is weaker
for higher carbon abundances. This is due to contamination of the index
pseudo-continuum by CH lines\footnote{Essentially the same conclusion is
reached for the $H\gamma_A$ index}. Finally, in the lower right panel,
one can see that the Fe4383 index is only very mildly affected by carbon
abundances, in the sense that the index becomes stronger for higher
carbon, due to contamination of the index passband by CH lines.

Nitrogen abundance variations affect CN lines and therefore indirectly
affect a number of indices, most notably $H\delta_F$, $H\delta_A$, and 
Ca4227.  Since the variation of CN lines as a function
of carbon abundances and their impact on line indices has been discussed
here and since CN lines respond similarly to carbon and nitrogen (see
Korn \etal 2005), there is no need to show model variations as
a function of nitrogen here.

In summary, we find that, among the Balmer line indices, the only ones
that are affected by the abundance of carbon are the $H\gamma_F$ and
$H\gamma_A$ indices. Application of a solar-scaled model to $H\gamma$
measurements taken in the spectrum of a carbon-enhanced stellar population
would lead to an age {\it overestimate}.  For a stellar population with
[C/Fe]=+0.3 the effect would be of the order of $\sim$ 5 (1) Gyr for ages of
$\sim$ 10 (2) Gyr. Amongst the metal line indices, the ones that are the
most sensitive to carbon are C$_2$4668, G4300, CN$_1$, CN$_2$, and Ca4227 (the latter
due to a spurious contamination of the index blue pseudo-continuum). The
cleanest carbon abundance indicator studied in this work is C$_2$4668, as 
it is solely dependent on the abundances of carbon and
iron, and (very weakly) on age. The CN indices
are also very strongly dependent on carbon, but are also sensitive to
nitrogen, as expected.  The iron indices Fe5270, Fe5335, and Fe5015 are
free of the influence of carbon abundance variations, while the Fe4383
index is slightly affected by them.

\bigskip
\centerline{\it Effects of $\alpha$-Element Enhancement}
\smallskip

\begin{figure}
\plotone{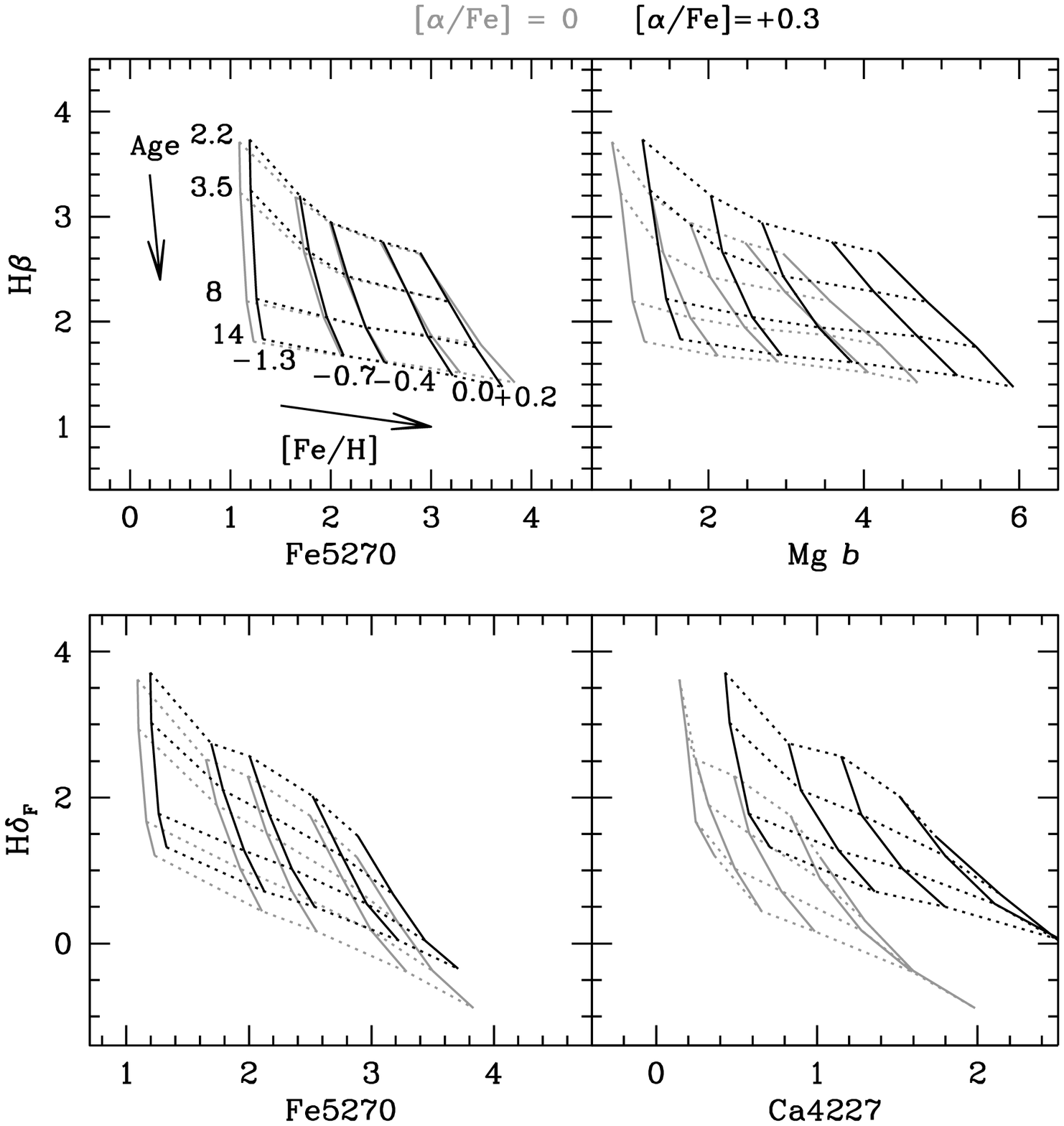}
\caption{Comparison between solar-scaled and $\alpha$-enhanced models.
{\it Upper Left:} This panel shows that $H\beta$ is insensitive to
spectroscopic $\alpha$-enhancement, while Fe5270 is only very mildly
sensitive, and only at very low [Fe/H]. {\it Upper Right:} As expected,
the Mg $b$ index is very sensitive to variations of [Mg/Fe]. It is the
chief magnesium abundance indicator in our models. {\it Lower Left:}
The $H\delta_F$ index is sensitive to $\alpha$-enhancement, being
slightly stronger for higher $\alpha$-element abundances. This is due
to contamination by magnesium and silicon lines in the index passband.
{\it Lower Right:} The Ca4227 index is extremely sensitive to [Ca/Fe]. It
is the only calcium abundance indicator in our models.
}
\label{aeff16}
\end{figure}

\begin{figure}
\plotone{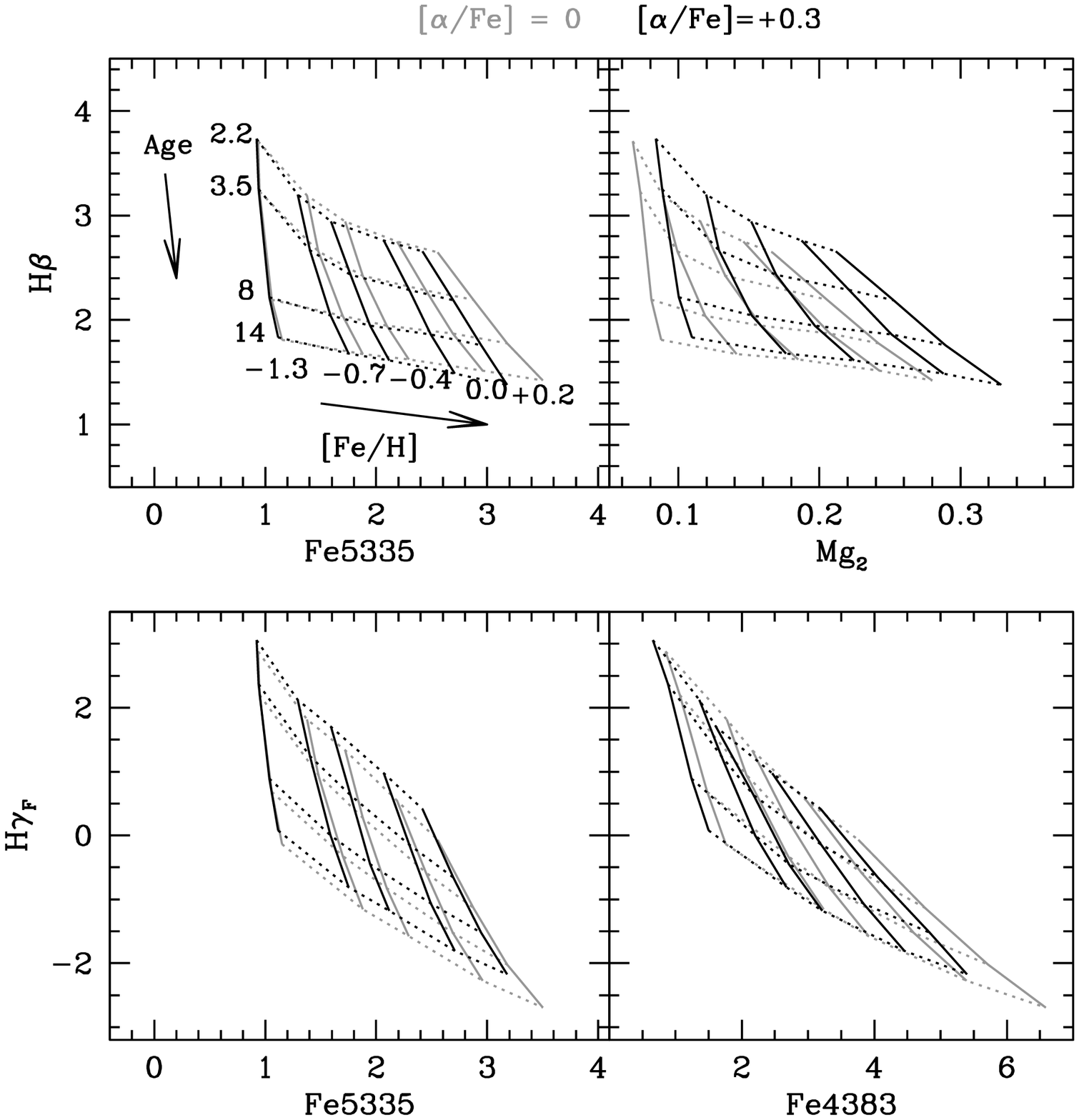}
\caption{Figure~\ref{aeff16} continued. {\it
Upper Left:} The Fe5335 index is shown to be only very mildly sensitive
to $\alpha$-enhancement, only for high [Fe/H]. {\it Upper Right:} As
expected, Mg$_2$ is very sensitive to [Mg/Fe], but not as strongly as
Mg $b$. {\it Lower Left:} The $H\gamma_F$ index shows some sensitivity
to $\alpha$-enhancement, mostly because of the decreased strength of
CN lines, due to enhanced oxygen abundances. {\it Lower Right:} The
Fe4383 index is mildly sensitive to $\alpha$-enhancement, being weaker
for higher $\alpha$-element abundance, mostly due to the presence of
magnesium and calcium lines in the index pseudo-continua.
}
\label{aeff17}
\end{figure}

The abundance of $\alpha$ elements relative to that of iron provides some
of the most fundamental clues available on the history of star formation
and chemical enrichment of galaxies (e.g., Matteucci \& Tornamb\'e 1987,
Wheeler \etal 1989, Peletier 1989, Worthey \etal 1992, Edvardsson \etal
1993, McWilliam 1997, Worthey 1998, Trager \etal 2000). Therefore, it
is crucially important to understand how indices respond to variations of
$\alpha$-element abundances, so that the latter can be estimated from index 
measurements taken in the integrated spectra of galaxies. In Figures~\ref{aeff16} 
and \ref{aeff17} we contrast solar-scaled and $\alpha$-enhanced models in a 
number of index-index diagrams. The $\alpha$-enhanced models are computed 
by increasing by +0.3 dex the abundances of oxygen, magnesium, calcium,
sodium, silicon, and titanium. We emphasize again that these computations
do not take into account the effect of oxygen abundances in the stellar
interiors (see Section~\ref{aeffect}), but only their 
spectroscopic effect, which is mostly due to
changes in the strengths of carbon-based molecules, due to the impact
of oxygen abundances on the concentration of free carbon atoms via the
dissociation equilibrium of the CO molecule. As another caveat, we note
that indices capable of constraining the abundances of sodium, silicon,
or titanium are not modelled in this work.  However, these abundances only
have a small impact on the strengths of some of the indices modelled here
(see Korn \etal 2005 for details), so we choose to force these abundances
to track those of the other $\alpha$-elements.

In the upper left panels of Figures~\ref{aeff16} and
\ref{aeff17}, the models are compared in the Fe5270-$H\beta$
and Fe5335-$H\beta$ planes, respectively. In these plots it can be
seen that $H\beta$ is free of any influence due to the enhancement
of $\alpha$-element abundances. This result, combined with that of
Figure~\ref{ceff14} confirms the finding by other authors (e.g.,
Korn \etal 2005) that $H\beta$ is the cleanest age indicator within the
Lick index family, given its very low dependence on metallicity, and
its virtual independence on any abundance-ratio effects. These figures
also show that Fe5270, Fe5335, and, to a lesser extent, Fe4383 (lower
right panel of Figure~\ref{aeff17}) are only very mildly
sensitive to $\alpha$-enhancement.  This result, combined with those of
Figures~\ref{ceff14} and \ref{ceff15}, implies that
the Fe5270 and Fe5335 indices are essentially only dependent on iron
abundance and (mildly) on age.  Therefore, they are all very reliable
[Fe/H] indicators.  The case of Fe4383 is interesting. According
to the Korn \etal (2005) tables, this index is mostly affected by
the abundance of iron, followed by that of magnesium, in the sense
that the index becomes weaker when [Mg/Fe] increases. Supposedly,
this is due to the presence of magnesium lines in one of the index
pseudo-continuum windows. However, inspection of the Moore, Minnaert \& 
Houtgast (1966) table of absorption lines identified in the solar spectrum 
reveals no such lines.  Also according to Korn \etal (2005), the Fe5015 index
(not shown) is very strongly sensitive to titanium and magnesium, being
stronger (weaker) for higher (lower) [Ti/Fe] ([Mg/Fe]). Presumably this
is due to the presence of a large number of MgH lines in the index red
pseudo-continuum, and strong TiI lines in the index passband (Moore \etal
1966).  Therefore, we caution against using this index in a situation
where the abundances of magnesium and titanium are unconstrained.

In the upper right panels of Figures~\ref{aeff16} and \ref{aeff17}, the
models are compared in the Mg $b$-$H\beta$ and Mg$_2$-$H\beta$ planes,
respectively. One can see that the Mg $b$ and Mg$_2$ indices are very
strongly sensitive to [Mg/Fe], as expected, with Mg $b$ being a little
more sensitive than Mg$_2$. For this reason, these indices have been
used in the literature as the chief $\alpha$-enhancement indicators. For
reasons that will be discussed in Section~\ref{47tuclight}, we will
adopt Mg $b$ as our main indicator of magnesium abundance.

The lower left panels of Figures~\ref{aeff16} and  \ref{aeff17}
illustrate how the Balmer line indices $H\delta_F$ and $H\gamma_F$,
respectively, respond to variations of $\alpha$-element abundances.
In both cases, a mild response to $\alpha$-enhancement is seen,
in the sense that the indices tend to be stronger for higher
$\alpha$-element abundances\footnote{Essentially the same result was
found in the cases of the $H\delta_A$ and $H\gamma_A$ indices.}, so that
non-consideration of $\alpha$-enhancement effects could lead to age
{\it underestimates}. The effect would be of the order of 4-5 Gyr for
old ages and solar metallicity, and about 1 Gyr for ages around 2 Gyr,
for a stellar population with [$\alpha$/Fe] $\sim$ +0.3. While in the
case of the $H\delta$ indices this is, according to Korn \etal (2005), 
due to contamination of the index
passbands by magnesium and silicon lines (though we failed to find any of
the former in the Moore \etal 1966 catalogue), in the case of $H\gamma$
it is because of the weakening of CH lines in the index pseudo-continuum,
due to enhanced oxygen abundances.

The lower right panel of Figure~\ref{aeff16} shows
a comparison of solar-scaled and $\alpha$-enhanced models in the
Ca4227-$H\delta_F$ plane. This plot shows that the Ca4227 is very
strongly sensitive to [Ca/Fe], so that it will be used here as our
chief indicator of calcium abundances. However, as we pointed out in
the previous subsection, this index is also very heavily influenced by
carbon abundance, which needs to be accounted for if one wants to use the
Ca4227 index for calcium abundance determinations. Finally, the lower
right panel of Figure~\ref{aeff17} compares solar-scaled
and $\alpha$-enhanced models in the Fe4383-$H\gamma_F$ plane. This plot
suggests that Fe4383 is mildly sensitive to $\alpha$-enhancement, in the
sense that it is weaker in $\alpha$-enhanced models. This results from
contamination of one of the index's pseudo-continua by Mg lines.

The main conclusions of our investigation of the effects of
$\alpha$-enhancement on the Lick indices studied in this work can
be summarized as follows. Amongst the Balmer line indices, $H\beta$
is the only one that is not affected by $\alpha$-enhancement, or any
other abundance-ratio effects. $H\gamma_F$, $H\gamma_A$, $H\delta_F$,
and $H\delta_A$ are similarly affected in the sense that they are {\it
stronger} in spectra of $\alpha$-enhanced stellar populations, for
fixed age and [Fe/H]. Amongst the metal lines, the Fe5270, and Fe5335
indices are essentially {\it insensitive} to $\alpha$-enhancement,
Fe4383 is only mildly affected by it, and Fe5015 is strongly affected by
it. This is good news, telling us that a safe estimate of [Fe/H]
is warranted.  Our main indicators of $\alpha$-element abundances are
Mg $b$, Mg$_2$, and Ca4227. While Mg $b$ is our cleanest indicator of
an $\alpha$-element abundance (magnesium, in this case), Ca4227 can be
used to estimate calcium, provided that carbon and nitrogen abundances
are known, so that the CN effect on Ca4227 can be accounted for. The
abundances of carbon and nitrogen can be inferred from the combined use
of either G4300 or C$_2$4668 (both sensitive to carbon only, but C$_2$4668
is preferable, see Graves \& Schiavon 2006, in preparation, for a discussion) 
and CN$_1$ and CN$_2$ indices (sensitive to carbon and nitrogen).

\subsection{A Method for Determining Mean Ages and Metal Abundances
of Stellar Populations} \label{amethod}

The understanding acquired in the last Section of the response of line
indices to variations of age and elemental abundances can be used to
establish a method to estimate both the age and abundance pattern of
stars in clusters and galaxies, from the interpretation of Lick indices
measured in their integrated spectra.

Initially let us suppose that measurements for all the line indices
modelled in this work are available for a given stellar system. The
method consists in constraining first the most influential parameters
(i.e., those which affect the largest number of observables) and
then descend hierarchically towards constraining less influential
parameters.  Inspection of Figures~\ref{ceff14} through \ref{aeff17} 
tells us that the parameters that affect the largest number of Lick indices 
are age and metallicity, which in our models is cast in terms of [Fe/H]. 
Virtually all the Lick indices are affected by age and [Fe/H]
variations, to various degrees. Therefore, the starting point of the method is
the determination of age and [Fe/H]. According to our conclusions
from Section~\ref{aratios}, the best way of estimating age and [Fe/H]
is by comparing data with models in a diagram involving an iron index
(preferably Fe5270 or Fe5335, or some combination of these) and
$H\beta$.  Therefore, we assume that the Fe and $H\beta$ indices are
only sensitive to [Fe/H] and age and estimate those parameters on the
basis of solar-scaled models (the choice of models here is unimportant,
provided our assumption that these indices are unaffected by abundance
ratios is approximately correct). The second most influential parameter
is the abundance of carbon, which affects a large number of line indices,
though not all of them (for instance, $H\beta$, Fe5270, and Fe5335 are
essentially not affected by carbon abundances), via the contamination
of index pseudo-continua and passbands by lines due to CN, CH, or 
C$_2$, which pervade the spectral region under study. Of all the indices 
modelled in this paper,
the C$_2$4668 index is best suited for carbon abundance determinations,
so the next step in our method consists of searching the [C/Fe] value
that best matches the $C_2$4668 index for the same age and [Fe/H] as
estimated from the analysis of $H\beta$ and Fe indices.  Once [C/Fe]
is estimated, the next step consists of determining [N/Fe], as the
abundance of nitrogen affects a large number of indices, because of its
influence on the strength of CN lines.  The best indicator of nitrogen
abundances are the CN bands themselves, so the next step in the procedure
consists of searching the [N/Fe] value for which the CN$_1$ and/or CN$_2$
indices are matched for the same age, [Fe/H], and [C/Fe] that match
the measurements of $H\beta$, C$_2$4668 and Fe indices. The remaining
parameters in the hierarchical sequence would be [Mg/Fe] and [Ca/Fe], as
they influence only a very few line indices, such as Mg $b$, Mg$_2$, and
Ca4227. Therefore, the final step of our procedure consists of estimating
[Mg/Fe] and [Ca/Fe] by searching the values that match Mg $b$/Mg$_2$
and Ca4227, respectively, for the same age, [Fe/H], [C/Fe], and [N/Fe]
as estimated from the match to all the previous indices. Once the latter
is achieved, a first estimate of age, [Fe/H], [C/Fe], [N/Fe], [Mg/Fe],
and [Ca/Fe] has been reached. The process now needs to be iterated, given
that we initially supposed that $H\beta$ and the Fe index/indices of
choice were essentially independent of any parameters other than age and
[Fe/H], which is not entirely correct. Experience shows that, for most
applications and depending on the degree of internal consistency aimed,
one iteration is good enough.

\begin{deluxetable*}{ccccccccccc} 
\tabletypesize{\scriptsize}
\tablecaption{Cluster Parameters}
\tablewidth{0pt}
\tablehead{\colhead{Cluster ID} & \colhead{Age (Gyr)} & \colhead{[Fe/H]} &
           \colhead{[O/Fe]} & \colhead{[N/Fe]$^a$} & \colhead{[C/Fe]$^a$} & 
           \colhead{[Mg/Fe]} & \colhead{[Ca/Fe]} & \colhead{[Ti/Fe]} &
           \colhead{[Na/Fe]} & \colhead{[Si/Fe]}     }
\startdata
\object[M5]{M~5}       & 11 & --1.3$\pm$0.1 & +0.3$\pm$0.2 & +1.2/0.0   & --1.0/--0.3 & 
                  +0.3$\pm$0.1 & +0.3$\pm$0.1 & +0.2$\pm$0.1  \\
\object[47 Tuc]{47~Tuc}  & 12 & --0.7$\pm$0.05 & +0.5$\pm$0.1 &  +1.1/+0.3 & --0.2/0.0 & 
                  +0.4$\pm$0.1 & +0.2$\pm$0.1 & +0.3$\pm$0.1 &
                  +0.2$\pm$0.1 & +0.3$\pm$0.1 \\
\object[NGC6528]{NGC~6528}$^b$  & 11 &  --0.15/+0.1 & +0.15/+0.07 &   &   & 
                  +0.07/+0.14 & --0.40/+0.23 & --0.10/+0.23 &
                  +0.43/+0.40 & +0.08/+0.36 \\
\object[M67]{M~67}      & 3.5&   0.0$\pm$0.1 &  0.0$\pm$0.1 &  0.0$\pm$0.1 &  0.0$\pm$0.1 &  
                   0.0$\pm$0.1 &  0.0$\pm$0.1 &   0.0$\pm$0.1 &
                  +0.2$\pm$0.1 & +0.1$\pm$0.1\\
\enddata
\tablenotetext{a}{CN-strong/CN-weak}
\tablenotetext{b}{Abundances sources: Zoccali \etal (2004) and Barbuy \etal (2004)/
Carretta \etal (2001)}
\label{clusterdata}
\end{deluxetable*}
\subsubsection{Caveats} \label{caveats}

Two important caveats are worthy of mention.  First of all, one
of the single most important parameters influencing the integrated
properties of stellar populations has been left out of the procedure
outlined above: the abundance of oxygen.  Oxygen abundances affect the 
stars'  interior opacities, thus having a strong impact on their structure
and evolution.  In our case, the abundance of oxygen therefore affects the 
choice of  theoretical isochrones adopted in the model synthesis.  Unfortunately, 
oxygen abundances for non-resolved, old and intermediate-age, stellar
populations are unknown and very difficult to constrain. It has become
standard practice in the literature to address this problem by adopting
not only the assumption that oxygen tracks magnesium, but also the more
far-reaching corollary that [Z/H] can be determined once Fe and Mg indices
(or a combination therefrom) are matched by the models. The former, more
fundamental, assumption, while grounded on the theoretical expectation
that magnesium and oxygen have a similar nucleosynthetic origin (e.g.,
Matteucci \& Tornamb\'e 1987, Wheeler \etal 1989, Woosley \& Weaver 1995)
has recently been challenged by the finding that metal-rich Galactic
bulge stars, which have strongly super-solar [Mg/Fe], appear to have
[O/Fe] below or around solar (Fulbright, Rich \& McWilliam 2005, Cunha
\& Smith 2006). Therefore, we emphasize that, while the assumption of
oxygen tracking magnesium might be, if not entirely reasonable, the only
possible way out of this quandary, it is no more than an assumption,
which still awaits confirmation from a compelling observational result.
For the time being, any results coming from the method proposed here
should be taken with caution. A good way of dealing with this uncertainty
would be to adopt two widely different values of [O/Fe] and carrying
out the procedure to the end, so that the final impact of an [O/Fe]
assumption on the final results can be assessed.

Another important caveat regards the uncertainties in the outputs
of this method. Except for age and [Fe/H], error propagation should
lead to very large error bars in the abundances of the elements at the
bottom of the hierarchy devised above. Therefore, the uncertainty in the
abundance of calcium, for instance, should be the highest, as the Ca4227
index is strongly influenced by the abundances of carbon and nitrogen,
whose uncertainties in turn depend on the uncertainties in [Fe/H]
and age. The case of magnesium is less serious, as the Mg $b$ index is
strongly affected only by age, [Fe/H], besides the output parameter,
[Mg/Fe]. Nitrogen stands in between those two cases, as the CN band is
affected by age, [Fe/H], and [C/Fe], besides [N/Fe]. In view of these
difficulties, we strongly recommend the reader to restrict application
of this method to only very high S/N data, so as to prevent the very
large uncertainties inherent to the method from rendering the results
meaningless.

This method has been implemented by G. Graves, from Lick Observatory, as
an IDL routine which will soon be made available publicly. The routine,
called ``EZ\_Ages'', and the mathematical algorithm used to search the
best solutions for a given set of index measurements are described in
detail in Graves \& Schiavon (2006, in preparation).

\section{Comparison with Cluster Data} \label{clust}

A fundamental test to which every stellar population synthesis model
must be submitted is the comparison to Galactic clusters. If the models
cannot reproduce the data for these well-known resolved systems, using
the right set of input parameters, their application to galaxy evolution
can be rightly called into question. It is not surprising, therefore,
that there is a vast literature dedicated to such comparisons (e.g.,
Rose 1994, Bruzual \etal 1997, Schiavon \& Barbuy 1999, Gibson \etal
1999, Vazdekis \etal 2001, Schiavon \etal (2002a,b), Maraston \etal 2003,
Schiavon \etal 2004b, Proctor \etal 2004, Schiavon \etal (2004a,b), Lee
\& Worthey 2005, Lilly \& Fritze-v.\ Alvensleben 2006, and references
therein). In this section, we show that our models match the data of
four well-known Galactic clusters to high accuracy. Our discussion is
focussed on a few well known representative clusters for which very
high quality CMD data and abundance analyses of cluster members are
available in the literature.  The absorption line indices for our globular
clusters of choice were measured in the very high S/N integrated spectra
collected by Schiavon \etal (2005) and, in the case of M~67, they were
taken from Paper III and converted to our system of equivalent widths
(but see Section~\ref{m67}). While essentially all results presented
here are valid for the entire sample of 40 globular clusters observed
by Schiavon \etal (2005), we defer an in-depth discussion of the whole
sample to a forthcoming paper.

The clusters examined here are M~5 (=NGC~5904), 47~Tuc (=NGC~104),
NGC~6528, and M~67 (=NGC~2682). Relevant data for these clusters are
summarized in Table~\ref{clusterdata}.  Ages come from analyses of CMD
data by Salaris \& Weiss (2002, M~5), Paper II (47~Tuc), Ortolani \etal
(2001, NGC~6528), and Paper III (M~67). The iron abundance for M~67 comes
from the compilation in Paper III.  Abundances for M~5 come from Cohen,
Briley \& Stetson (2002, [C/Fe] and [N/Fe]) and Ram\'\i rez \& Cohen
(2003, other ratios).  Carbon and nitrogen abundances for 47~Tuc stars
come from Briley \etal (2004), while for other elements abundances come from
Carretta \etal (2004) and Alves-Brito \etal (2005). Oxygen abundances 
in 47~Tuc dwarfs come from Carretta \etal (2005). Abundances
for NGC~6528 are averages of the determinations by Carretta \etal
(2001), Zoccali \etal (2004), and Origlia, Valenti \& Rich (2005) (but see
discussion below and in Section~\ref{n6528}).  Regarding the abundances
of carbon and nitrogen, we note that stars in M~5 and 47~Tuc (and perhaps
in all globular clusters) are known to present a bimodal distribution of
the abundances of these elements (e.g., Dickens, Bell \& Gustafsson 1979,
Norris \& Freeman 1979, Smith, Bell \& Hesser 1989, Cannon \etal 1998,
Cohen \etal 2002, Briley \etal 2004, Carretta \etal 2005, Lee 2005,
Smith \& Briley 2006). For these elements, we list the extremes of the
range of values spanned by cluster main sequence stars. Unfortunately,
no such study is available for the C and N abundances of main sequence
stars in NGC~6528. Origlia, Valenti \& Rich (2005) determined carbon
abundances in four bright NGC~6528 giants, reporting [C/Fe]=--0.4. This
value probably reflects the workings of internal mixing, and we choose
not to consider it. For M~67, the abundance ratios were taken from the
compilation in Paper III, except for oxygen, calcium and titanium, which
were taken from Tautvaisiene \etal (2000) and Shetrone \& Sandquist
(2000). Finally, because different groups disagree as to the metal
abundances of NGC~6528 (Carretta \etal 2001, Barbuy \etal 2004, Zoccali
\etal 2004), abundances by these two different groups are listed for
this cluster. This issue is further discussed in Section~\ref{n6528}.

\begin{figure}
\plotone{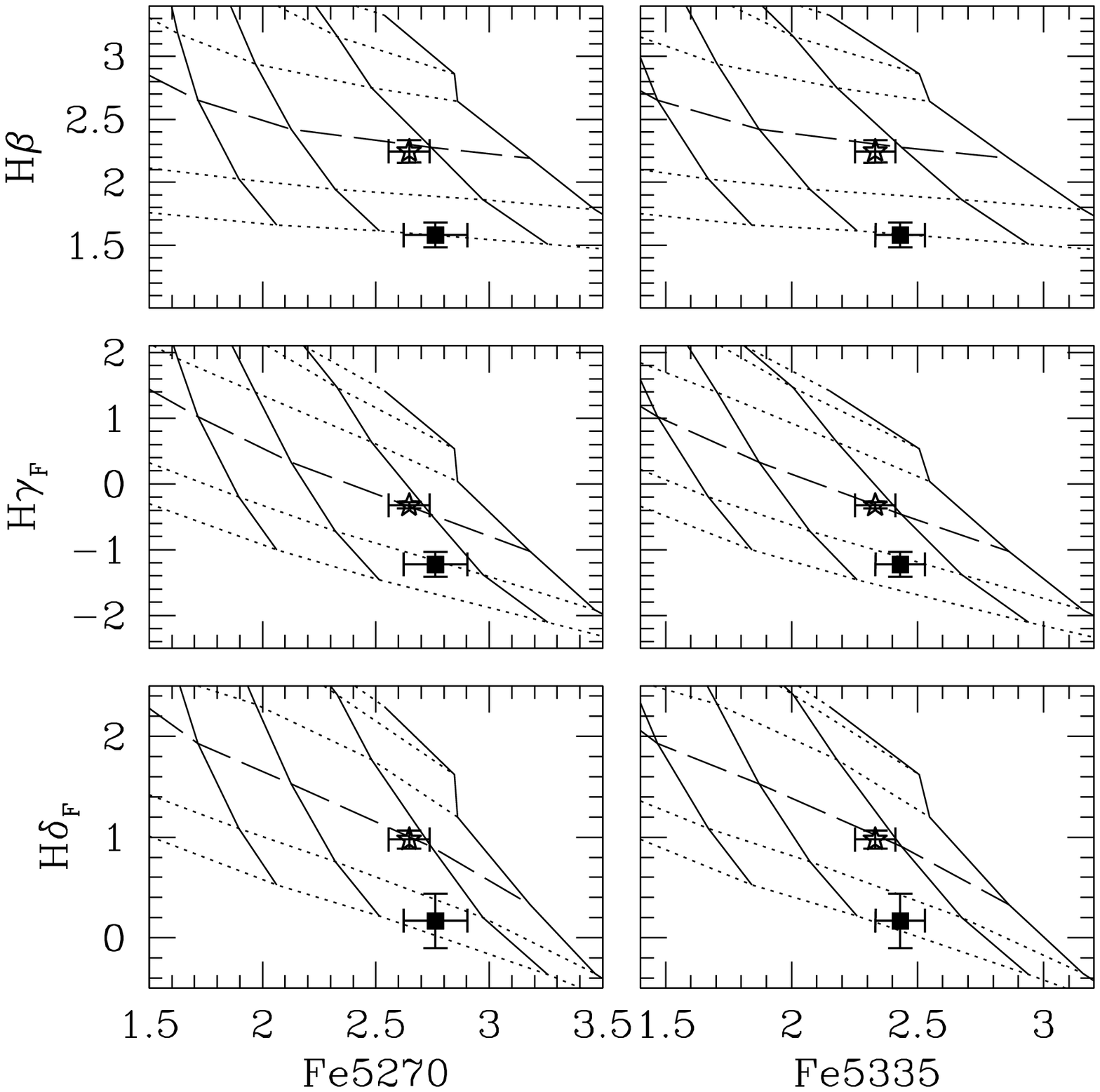}
\caption{Data on M~67 (star) and NGC~6528 (filled square) compared
to solar-scaled model predictions.  Same-age models are connected by
dotted lines, except for the 3.5 Gyr models, which are connected by
dashed lines, for clarity. Same-[Fe/H] lines are solid. The ages of
the models displayed are, from top to bottom, 1.2 (barely visible), 1.5,
2.5, 3.5, 7.9, and 14.1 Gyr. The values for [Fe/H] are, from left to
right, --0.7, --0.4, 0.0, and +0.2. The best-fitting model for M~67 has
an age of $\sim$ 3.8 Gyr and [Fe/H]=--0.08. Note the consistency with 
which all indices are matched by the models at essentially the same
position on the grid.
} 
\label{m67_65281} 
\end{figure}

Comparison of the data on elemental abundances shown in
Tables~\ref{tabxfe} and \ref{clusterdata} suggests that the overall
abundance pattern of the cluster stars is mostly quite similar (to within
0.1 dex) to that of our library stars with the same [Fe/H]. In fact,
in the case of M~67 there is an almost perfect match between cluster
and field star abundance patterns. For the globular clusters, though,
there are a few important exceptions that need to be kept in mind. The
relative abundances of carbon and nitrogen in globular-cluster stars
deviate strongly from those of field stars. The main consequence is that,
in integrated light, globular clusters tend to show strongly enhanced
CN lines and slightly weaker CH lines when compared to models based on
spectra of field stars (see Paper I for a detailed discussion). This
should mainly affect the comparison of our model predictions to data
on the CN$_1$, CN$_2$, G4300, and C$_2$4668 indices but can also
disturb other indices whose passband and/or pseudo-continua contain
lines due to these molecules (for instance, Ca4227, see below, or the
$H\gamma$ indices). Another important difference is the one between the
abundance pattern of NGC~6528 and that of library stars of near-solar
metallicity. This bulge cluster has higher [O/Fe] by $\sim$ 0.1 dex and
[Ca/Fe] (potentially) lower by $\sim$ 0.4 than field stars of the same
metallicity. Its relative magnesium abundance seems to be slightly
higher as well. These few exceptions aside, it is fair to say that the
abundance patterns of the clusters and the field stars employed in our
model construction are a relatively good match. Therefore, we first
compare our base-model predictions with cluster data without performing
any correction to bring the models into the cluster abundance pattern
(except in the case of M~67, for which such comparisons were discussed in
Paper III and will not be repeated here). Confrontation between cluster
data and models tuned to the clusters abundance patterns are discussed
in Sections~\ref{47tucratios} and \ref{n6528}.

\subsection{M~67} \label{m67}

\begin{figure}
\plotone{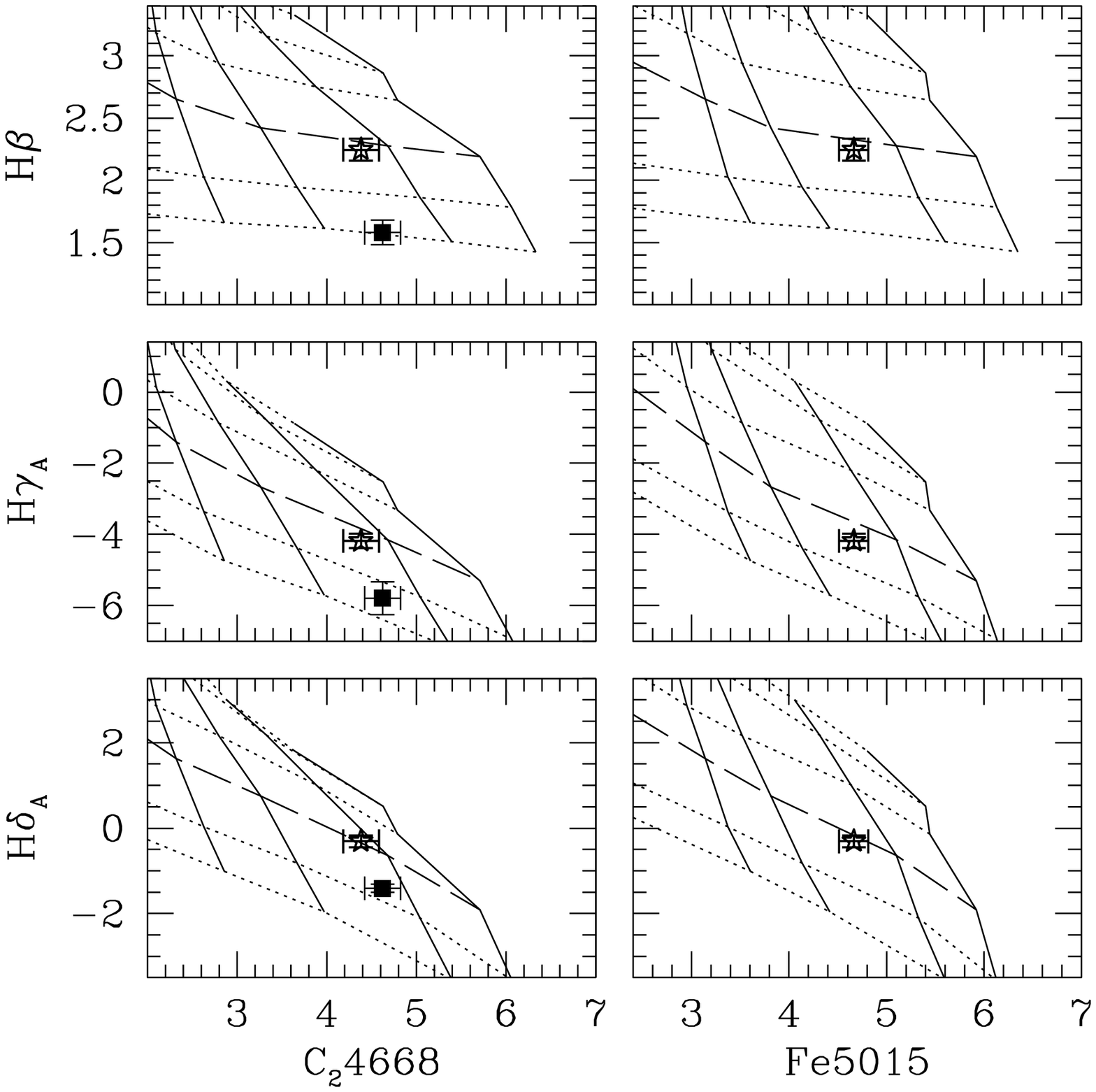}
\caption{Figure~\ref{m67_65281} continued. Model age and [Fe/H] are the
same as in Figure~\ref{m67_65281}.  Note the outstanding
consistency with which all the indices from M~67 are matched by the models
(i.e., same age and [Fe/H] everywhere).  Due to a spectral blemish in
the Schiavon \etal (2005) data, the Fe5015 is not available for NGC~6528.
}
\label{m67_65282}
\end{figure}

The open cluster M~67 has an age of $\sim$ 3.5$^{+1.0}_{-0.5}$ Gyr
and nearly solar metallicity ([Fe/H] = 0.0 $\pm$ 0.1). Therefore, it
inhabits a region of parameter space that is very important for studies
of the integrated light of galaxies, both nearby and at large distances,
where early-type galaxies are often found to have similar mean ages and
metallicities (e.g., Trager \etal 2000, Kelson \etal 2001, Caldwell \etal
2003, Denicol\'o \etal 2005a,b, Gallazzi \etal 2005, Schiavon \etal 2006).

\subsubsection{The Integrated Spectrum of M~67} 

In Paper III an integrated spectrum of M~67 was obtained from coaddition
of individual spectra of cluster members, weighted according to their
luminosities and relative numbers, assuming a Salpeter IMF. Since publication 
of that work a few revisions have been made to the cluster integrated spectrum 
and Lick indices measured in it, so that an update on these observables is
made necessary here. The first important change relative to the values
published in Paper III refers to the EW measurements, which had to be
retaken, for the reasons exposed in Section~\ref{abslines}.  The second
important revision relates to star \#6472 (ID from Montgomery, Marschall
\& Janes 1993), which has been erroneously included in the coaddition as
a first-ascent giant star. Inspection of Figure 1 in Paper III suggests
that this star is too blue (by $\simgreat$ 0.2 mag in {\it B--V}) to be on
the red giant branch of M~67, which is confirmed by its warm spectrum. It
is also too bright and too blue to be an early-AGB star. Because of its
uncertain evolutionary stage, we decided to remove this star from the
coaddition to produce the cluster integrated spectrum. In the third
relevant change, we performed a test which showed that main sequence
M~67 stars fainter than V $\sim$ 15 (the magnitude limit in Paper III
sample) contribute significantly to the integrated Lick indices of
the cluster. In order to correct for this effect, we used the Padova
isochrones that best matched the cluster color-magnitude diagram (Girardi
\etal 2000, solar-scaled, solar metallicity, 3.5 Gyr-old) to compute model
predictions including and excluding stars less massive than $\sim$ 0.85
$M_\odot$. The difference between these two model predictions was used to
correct the Lick indices measured in the cluster integrated spectrum for
the contribution of low-mass stars. Both corrected and uncorrected Lick
indices for M~67 are listed in Table~\ref{m67ind}. Comparison between the
two sets of indices shows the corrections are small, but not negligible
for some indices. These numbers supersede the values provided in Paper
III and will be used throughout this paper in our comparison with model
predictions.

\begin{deluxetable*}{ccccccccccccccccc}
\tabletypesize{\scriptsize}
\tablecaption{Revised Lick Indices for M~67}
\tablewidth{0pt}
\tablehead{
\colhead {}& $H\delta_F$ &  $H\delta_A$ &  CN$_1$   &    CN$_2$    &  
            Ca4227 & G4300 & 
           $H\gamma_F$ &  $H\gamma_A$ & Fe4383 &  C$_2$4668 & $H\beta$ &  
	   Fe5015 & Mg$_2$ & 
             Mgb     &    Fe5270    &  Fe5335 } 
\startdata
M67 & 1.0 & --0.1 & --0.005 & 0.019 & 0.82 & 4.7 & --0.21 &  --4.0  & 
3.8 & 4.5 & 2.4 & 4.7 & 0.161 & 2.8 & 2.6 & 2.2 \\
M67--Corr$^1$ & 1.0 & --0.3 & --0.007 & 0.018 & 0.93 & 4.7 & --0.32 &  --4.2  & 
3.8 & 4.4 & 2.2 & 4.7 & 0.175 & 2.9 & 2.6 & 2.3 \\
Error & 0.1 & 0.2 & 0.006 & 0.007 & 0.08 & 0.1 & 0.07 & 0.2 & 
0.2 & 0.1 & 0.1 & 0.4 & 0.006 & 0.2 & 0.1 & 0.1 \\
\enddata
\tablenotetext{1}{Indices corrected to include contribution by low-mass
stars (see text).}
\label{m67ind}
\end{deluxetable*}

\subsubsection{Confronting Models with M~67 Data} \label{m67spec}

\begin{figure}
\plotone{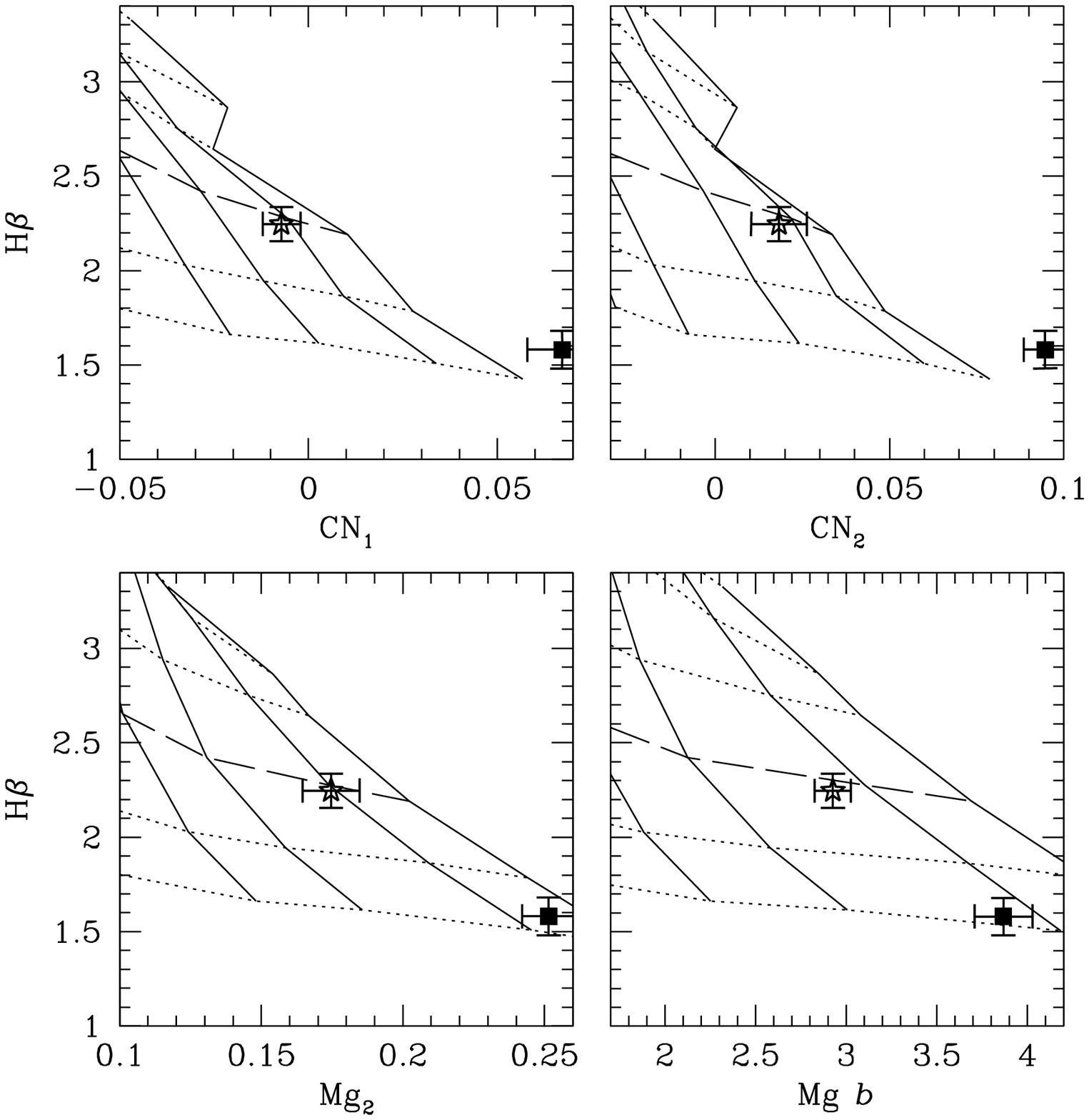}
\caption{Figure~\ref{m67_65281} continued, now showing comparisons 
between models and data for indicators of carbon, nitrogen, and magnesium
abundances. Again, data for M~67 are matched consistently for the same
age and [Fe/H] as for indices in the previous figures.}
\label{m67_65283}
\end{figure}

\begin{figure}
\plotone{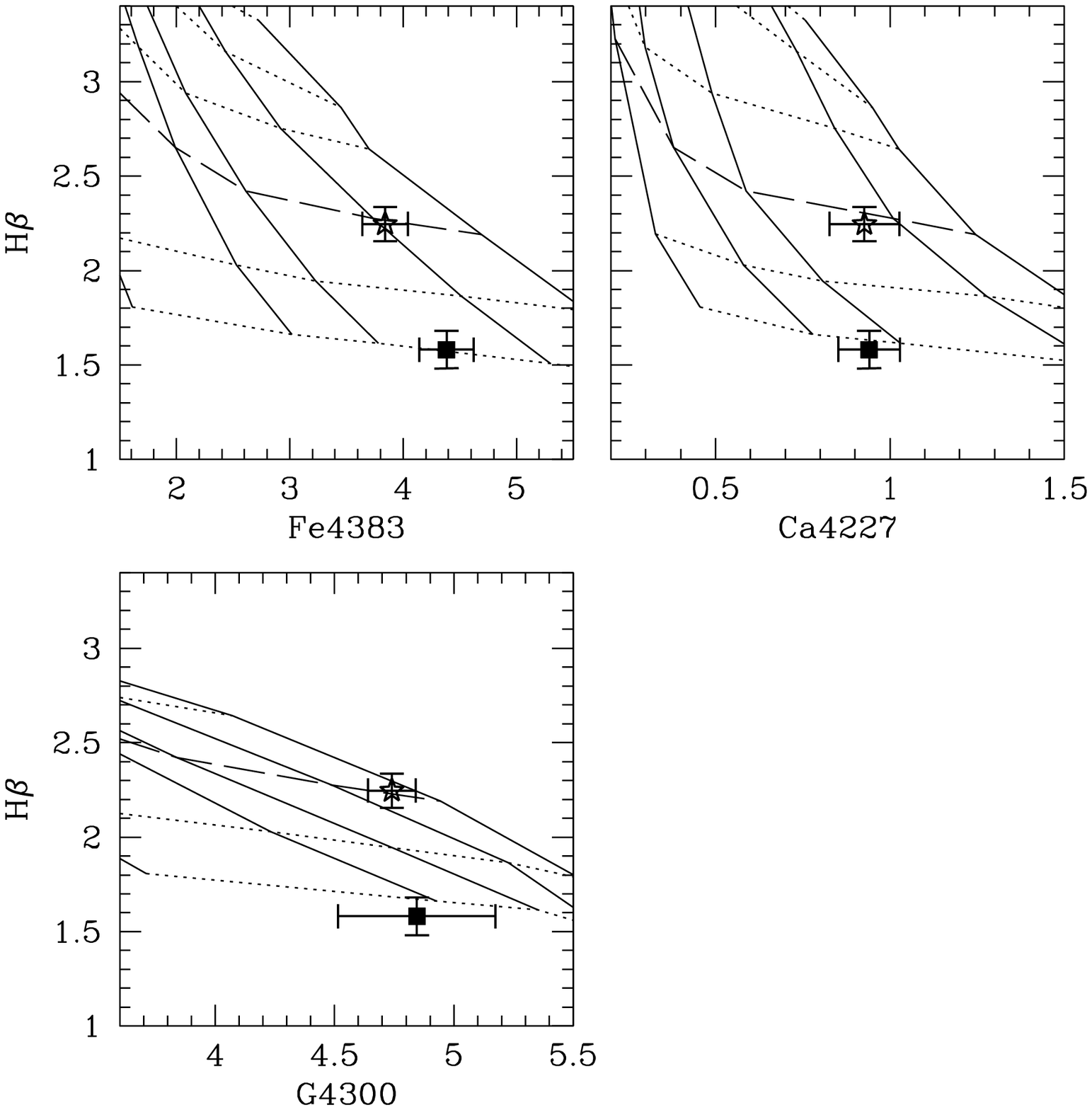}
\caption{Figure~\ref{m67_65281} continued, now showing comparisons
between models and data for indices sensitive to iron, calcium and carbon.
The G4300 index is the one for which the models are the poorest match to
the data (almost 0.2 dex too metal-rich, according to the models).}
\label{m67_65284}
\end{figure}

In Paper III we presented a detailed comparison between our solar-scaled
models and Lick indices for M~67. We showed that our models match the
Balmer line indices $H\beta$, $H\gamma_F$, and $H\delta_F$ for the same
age that is inferred from the fit of the cluster's color-magnitude
diagram, to within $\sim$ 0.5 Gyr.  Moreover, we showed that the
indices \fem\ (the average of indices Fe5270 and Fe5335), Mg $b$,
CN$_1$, and CN$_2$ were matched for the known cluster elemental
abundances to within $\pm$ 0.1 dex. Here we extend these comparisons to
include the remaining indices modelled in this paper. In order to find
the model that best matches the data for M~67, we applied the method
described in Section~\ref{amethod} and found the following best-fitting
parameters, using the Padova solar-scaled isochrones and assuming solar
[O/Fe]: age $\sim$ 3.8 Gyr, [Fe/H]=--0.08, [Mg/Fe]=0.01, [C/Fe]=--0.03,
[N/Fe]=+0.02, and [Ca/Fe]=--0.03. The models computed for this abundance
pattern are provided in Table A6 in the Appendix. Inspection
of Table~\ref{clusterdata} shows that these results agree with the known
cluster values to within 0.3 Gyr and 0.08 dex in age and metal abundances,
respectively. In Figures~\ref{m67_65281} through \ref{m67_65284} these
best-fitting models are compared with Lick indices for M~67 in a number of
relevant index-index diagrams.  The model ages and metallicities plotted
are as follows: 1.2, 1.5, 2.5, 3.5, 7.9, and 14.1 Gyr, and [Fe/H] = --0.7,
--0.4, 0.0, and +0.2. The filled squares represent data for NGC~6528,
which will be discussed in Section~\ref{n6528}.

Figures~\ref{m67_65281} and \ref{m67_65282} compare data to models
in metal vs. age-sensitive-index plots. For almost all indices,
the data for M~67 fall within 1-$\sigma$ of our model prediction
for a solar-metallicity, 3.5 Gyr-old, single stellar population.
In Figures~\ref{m67_65283} and \ref{m67_65284} the data for indices
sensitive to carbon, nitrogen, magnesium, and calcium are compared
with the models, using $H\beta$ as age indicator. Again, the models
are a satisfactory match to the data for the correct age and metal
abundances. We call attention for the very good match
obtained for Ca4227, which was not discussed in Paper III.  We note that
in that paper we could not match the M~67 data for G4300. The match here
is slightly improved, though still not entirely satisfactory. The G4300
index is matched for a model with [Fe/H]$\sim$+0.1, which is almost 0.2
dex too metal-rich compared with the value obtained from the match to the 
C$_2$4668 index, which is in much better agreement with the cluster
known carbon abundance. This result should serve as a warning against
use of the G4300 index for carbon abundance determinations in metal-rich
stellar populations (see Graves \& Schiavon 2006, in preparation).

In summary, the models match the age of M~67 and its abundances of iron,
carbon, nitrogen, magnesium and calcium to within 1 Gyr and 0.1
dex. Most importantly, the ages according to all Balmer line indices
are consistent to within 0.5 Gyr. This is a very encouraging result that
ratifies the application of our models for estimation of mean ages and
metal abundances of stellar systems in a very important age/metallicity
regime. To some extent, this is not surprising, given that the integrated
light of the cluster over most of the spectral interval involved in
this study is dominated by turnoff stars which, in the case of M~67,
are characterized by mid-F spectral types of solar metallicity, for
which most of the physical input and calibrations underlying the models
are very well established. The fact that the abundance pattern of our
models for solar metallicity are a close match to that of the cluster is
also very helpful. In the next sections we focus on the more challenging
task of matching the data for Galactic globular clusters, which depart
considerably from the regime where such favorable conditions are enjoyed.

\subsection{47~Tuc and M~5} \label{47tucm5}

\begin{figure}
\plotone{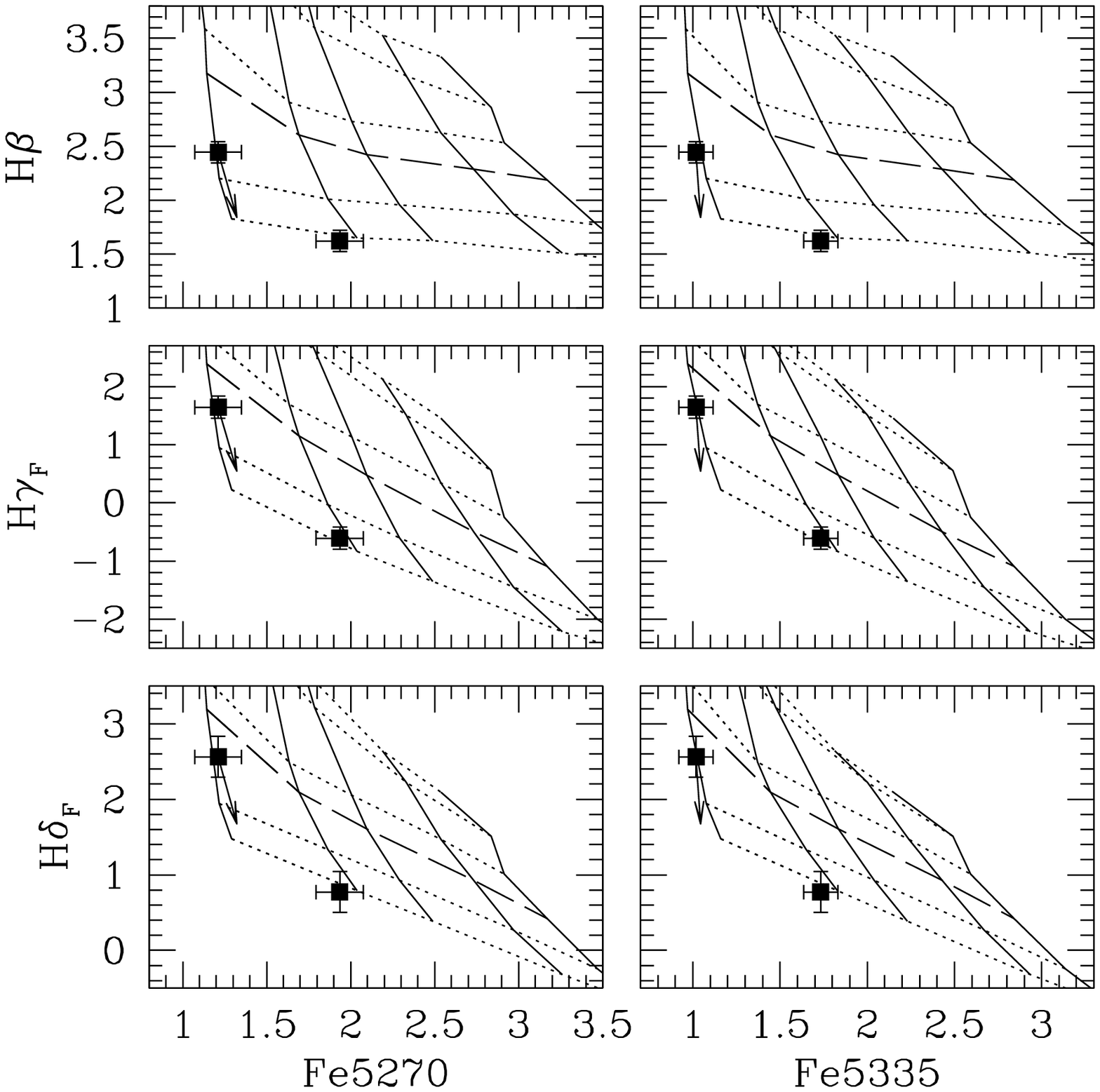}
\caption{a. Data for M~5 (left) and 47~Tuc. The arrow indicates how the
predictions for M~5 change when the contribution due to blue HB stars is
``removed'' from the cluster data. Ages are 1.2, 1.5, 2.5, 3.5, 7.9, and
14.1 Gyr, and metallicities are [Fe/H]=--1.3, --0.7, --0.4,
0.0, and +0.2. The models for 3.5 Gyr are connected by a 
dashed line, to guide the eye. The models for [Fe/H]=--1.3 are not shown in 
Figure~\ref{47tuc_m51}e. See detailed discussion in the text.
}
\label{47tuc_m51}
\end{figure}

In this Section, we compare our model predictions to data for two of
the best studied Galactic globular clusters, 47~Tuc and M~5. It is
very interesting to test our models in the region of stellar population
parameter space occupied by metal-poor to mildly metal-rich old globular
clusters, in view of the on-going efforts to constrain the history of
merging/star-formation of external galaxies through the determination of
the spectroscopic ages and metallicities of their globular clusters
(e.g., Cohen \etal 1998, Cohen, Blakeslee \& C\^ot\'e 2003, Larsen \etal
2003, Burstein \etal 2004, Beasley \etal 2005, Puzia \etal 2005, Brodie
\& Strader 2006, and references therein).

The data for these two clusters are compared to the models in
Figures~\ref{47tuc_m51}a through \ref{47tuc_m52}f. The abundance patterns
of 47~Tuc and M~5 are $\alpha$-enhanced (Table~\ref{clusterdata}), in
particular with [O/Fe] = +0.5 and +0.3, respectively.  We recall from
Section~\ref{aeffect} that the $\alpha$-enhanced Padova isochrones
were computed assuming [O/Fe]=+0.5. However, they do not extend to
low enough metallicities for a comparison with M~5 data.  Moreover,
Weiss \etal (2006) showed that the Padova $\alpha$-enhanced isochrones
need to be revised (see discussion in Section~\ref{aeffect}) so that we
choose to compare the data with models generated using the solar-scaled
Padova isochrones (models 1--5 in Table~\ref{ssppar}). This set of
theoretical isochrones does not match the oxygen abundances of the two
clusters.  However, based on our discussion in Section~\ref{aeffect},
the evolutionary $\alpha$-enhancement effects are well-understood, so
that we can use models based on solar-scaled isochrones, while keeping
in mind the effect of that inconsistency. In particular, we expect to
predict slightly too old ages, regardless of the Balmer line used.

\subsubsection{Fe vs Balmer Lines}

\setcounter{figure}{21}
\begin{figure}
\plotone{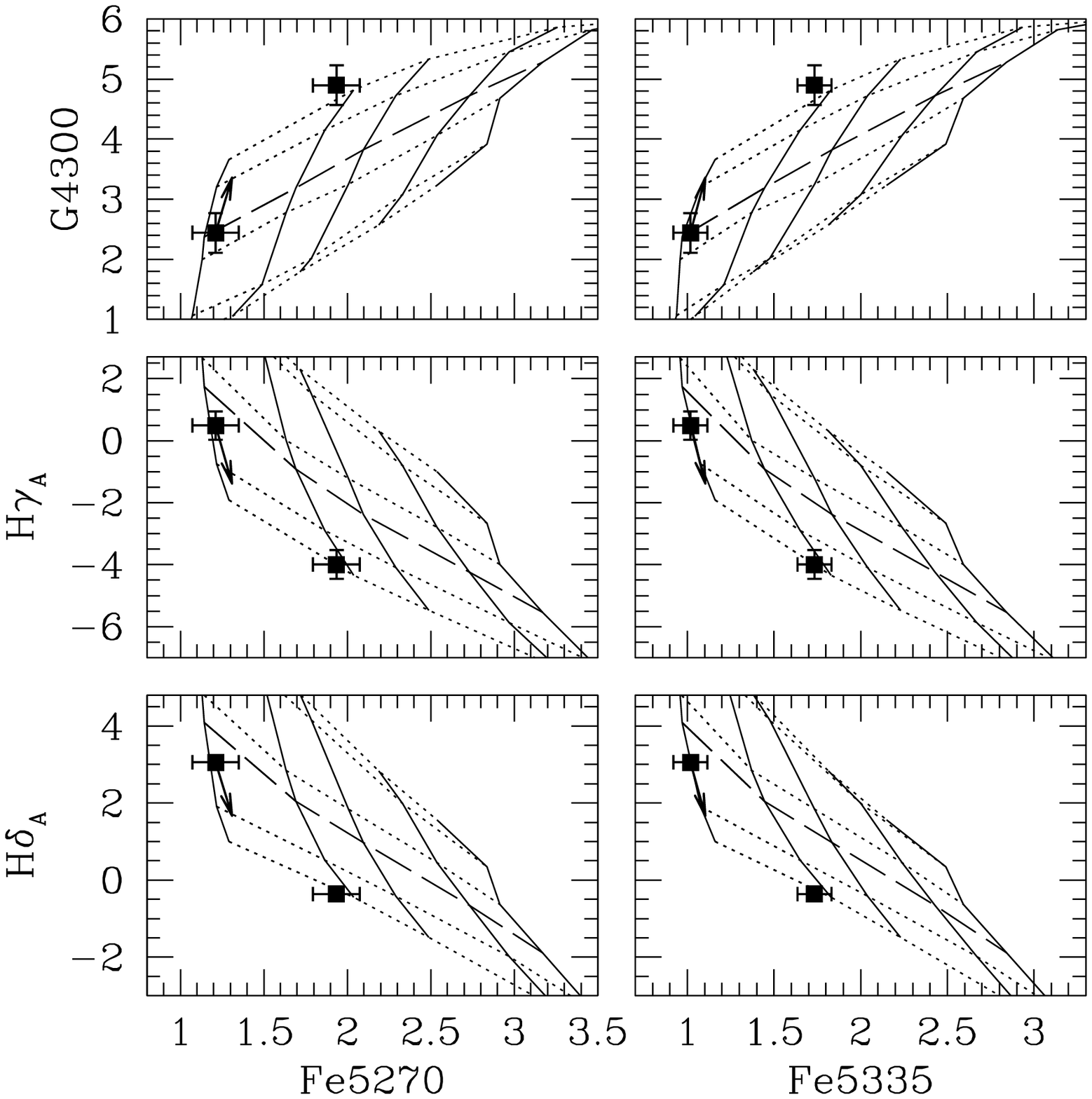}
\caption{b.}
\label{47tuc_m52}
\end{figure}

As in the previous Section, we first focus on plots between Fe and
Balmer lines, because these indices are mostly sensitive to age and
[Fe/H] (see discussion in Section~\ref{aratios}). We first consider the
case of 47~Tuc. In all the panels in Figures~\ref{47tuc_m51}a through
\ref{47tuc_m53}c the models predict the correct [Fe/H] of the cluster to
within $\pm$ 0.05 dex. The spectroscopic age for this cluster according
to the models is $\sim$ 14 Gyr, which is older by $\sim$ 2--3 Gyr than
the age based on analysis of the cluster CMD using the $\alpha$-enhanced
Padova isochrones (Paper II). Roughly half of this mismatch is due
to our adoption of solar-scaled isochrones in the current calculation.
The rest of the discrepancy is due to an effect pointed out in Paper II,
where it was shown that it is motivated by a mismatch between the observed
luminosity function of the cluster and theoretical predictions, which
underestimate the number of giant stars brighter than the horizontal
branch (HB), relative to main sequence stars.  While it is not clear
whether this mismatch between data and theory in the luminosity function
space is restricted to 47~Tuc and a few other clusters (e.g., Langer,
Bolte \& Sandquist 2000) Zoccali \& Piotto (2000) found an apparent
trend according to which models seem to under-predict the relative
number of giants in more metal-rich clusters. Zoccali \& Piotto point
out that uncertainties in the bolometric corrections for metal-rich
cool giants might be responsible for the mismatch, but to our knowledge
this hypothesis has not yet been tested. Clearly, more work is needed
to clarify this matter.

The case of M~5 is very interesting. The Balmer lines in this cluster's
spectrum are too strong for its age. In Figures~\ref{47tuc_m51}a through
\ref{47tuc_m51}c, the spectroscopic age of M~5 according to the models
is somewhere between 4 and 6 Gyr (probably even a little younger, if the
[Fe/H]=--1.3 models were based on $\alpha$-enhanced isochrones). This
is in stark contrast with the known CMD-based age of the cluster
($\sim$ 11 Gyr). This effect has been pointed out before (Freitas
Pacheco \& Barbuy 1995, Lee, Yoon \& Lee 2000, Maraston \& Thomas 2000,
Schiavon \etal 2004b) and is due to the influence of blue HB stars
which are not accounted for by the theoretical isochrones adopted in
our models. These old and bright A-F-type stars have very strong Balmer
lines which can mimic a young turnoff if not properly accounted for by
the models. Schiavon \etal (2004b) studied this problem and devised a
method to disentangle this degeneracy between age and HB-morphology,
which explores the differential sensitivity of $H\delta$ and $H\beta$
to the influence of blue HB stars.

\setcounter{figure}{21}
\begin{figure}
\plotone{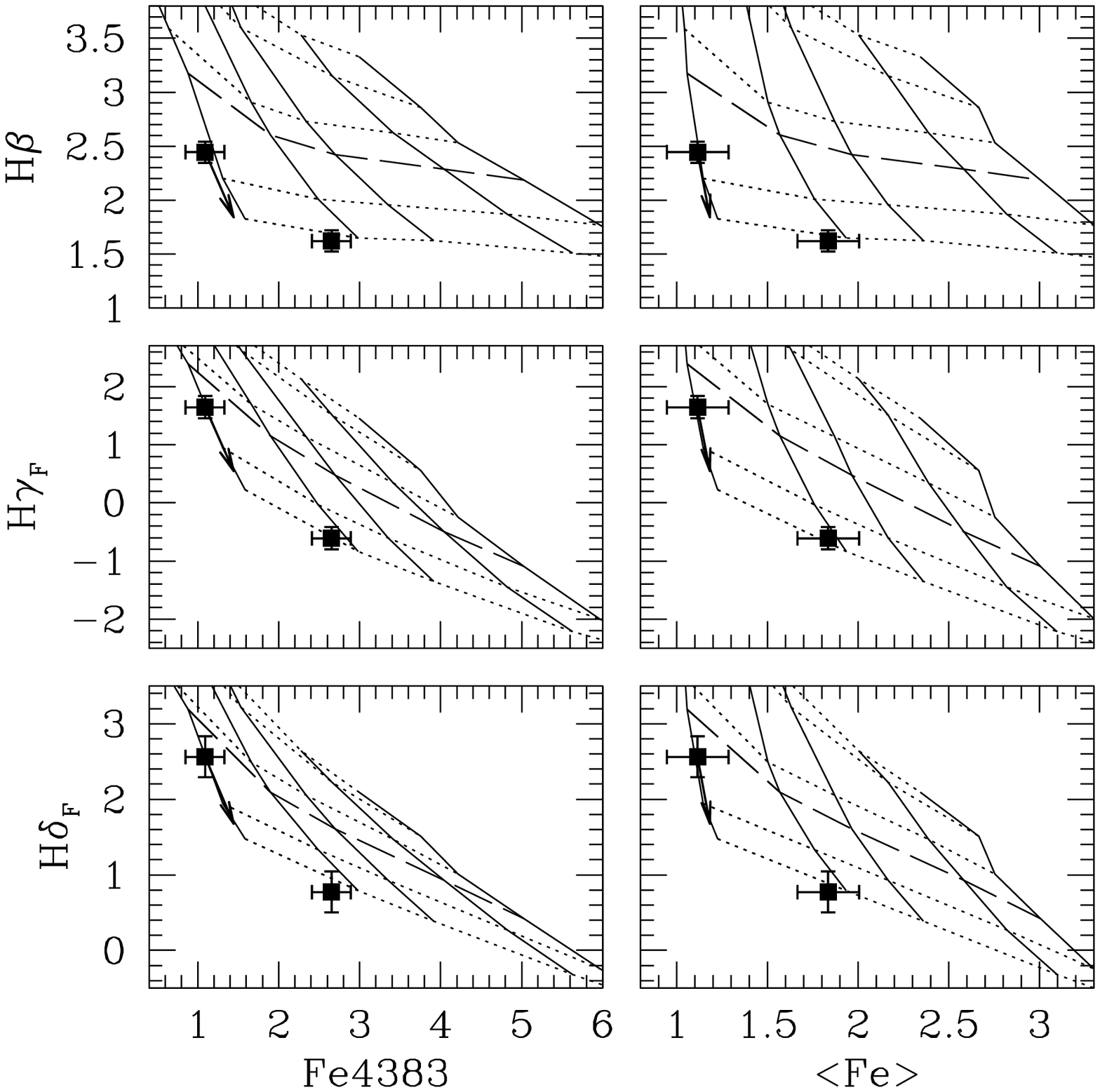}
\caption{c.}
\label{47tuc_m53}
\end{figure}

The theoretical isochrones adopted in our models do not produce blue
HB stars in the metallicity range considered here.  The morphology of
the HB is chiefly dictated by mass loss along the giant branch phase,
a phenomenon for which a deterministic theory is still lacking. As a
result, models for the mass loss along the giant branch rely on empirical
calibrations of mass-loss rates as a function of stellar parameters, thus
having limited predictive power. Therefore, we adopt a more conservative
path and just correct the observations of M~5 for the effect of blue HB
stars, on the basis of a high-quality CMD for the cluster (see details
in Schiavon \etal 2004b). In this way we can at least check whether
our models predict correctly the cluster properties in the absence of
blue HB stars. The arrows attached to the data for M~5 indicate how
the line indices change when the contribution from blue HB stars is
removed. These arrows were computed by Schiavon \etal (2004b), from a
combination of the color-magnitude diagram of M~5 (from Piotto \etal 2002)
and the fitting functions presented here. For details, see Schiavon \etal
(2004b).  In all panels of Figures~\ref{47tuc_m51}a--c we can see that
the age predicted by the models for the HB-free version of M~5 is $\sim$
10 Gyr and [Fe/H] $\sim$ --1.3, in outstanding agreement with the values
listed in Table~\ref{clusterdata}. Given the uncertainties involved in the
model calibrations at the low metallicity end and in the correction for
the effects of blue HB stars, we consider this a very satisfactory result.

\setcounter{figure}{21}
\begin{figure}
\plotone{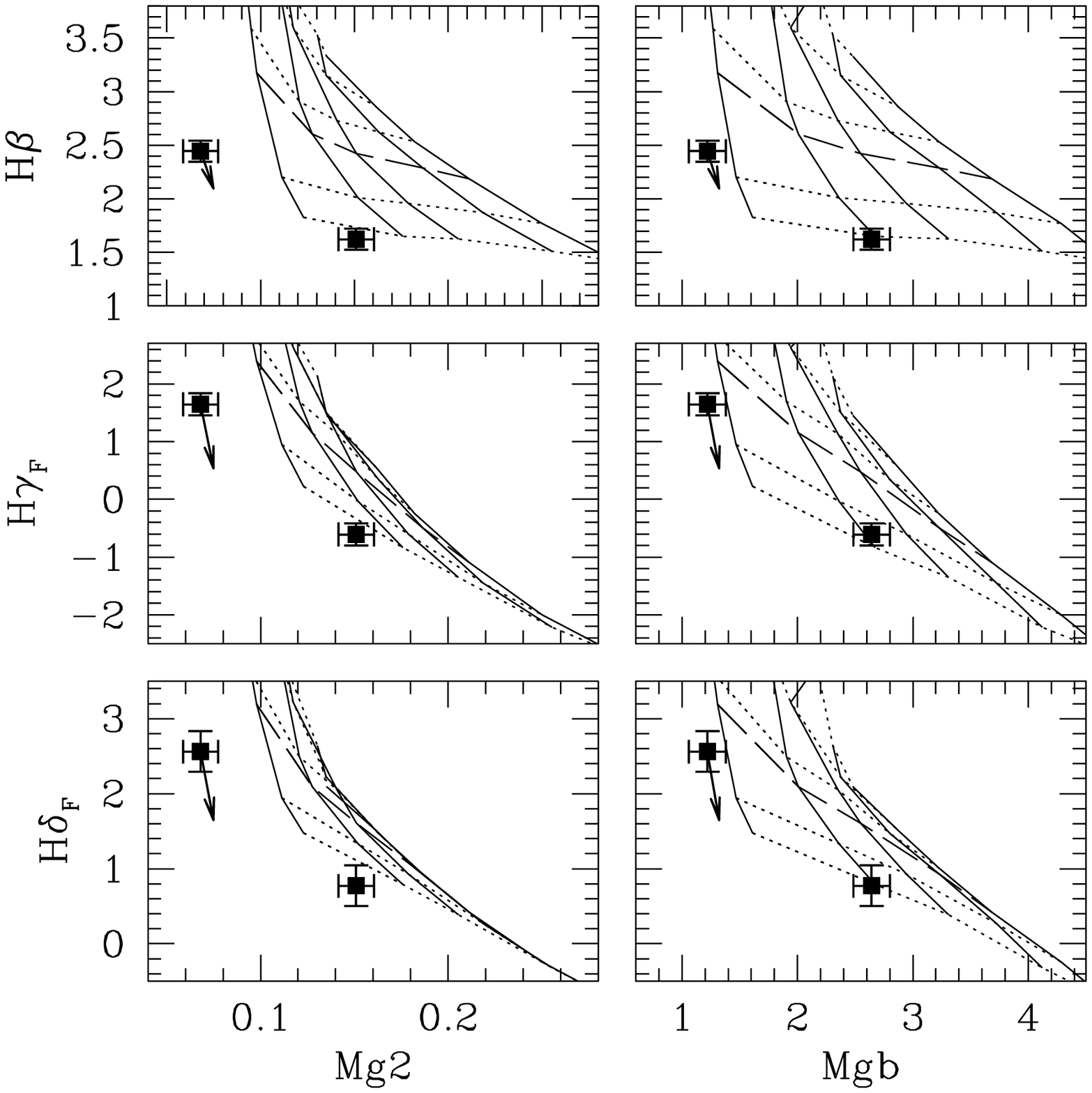}
\caption{d.}
\label{47tuc_m54}
\end{figure}

\subsubsection{Light-element Indices vs Balmer Lines} \label{47tuclight}

\setcounter{figure}{21}
\begin{figure}
\plotone{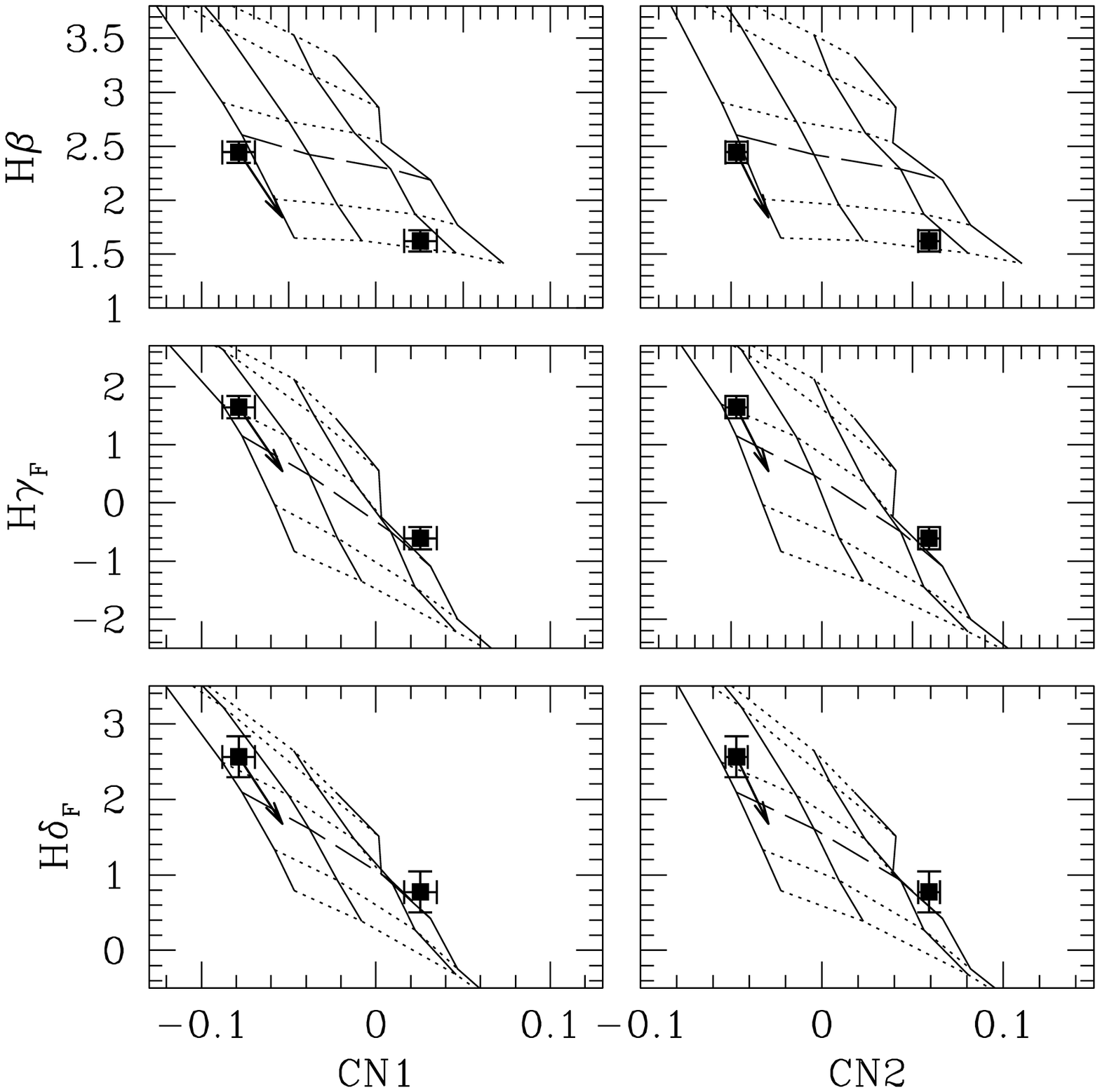}
\caption{e.}
\label{47tuc_m55}
\end{figure}

In Figures~\ref{47tuc_m54}d--f we show the cluster data compared with the
models in light-element-index vs Balmer-line planes. We first focus on the
case of the the indices Mg$_2$ and Mg~$b$, which are mostly sensitive to
the abundance of magnesium. In the case of Mg $b$, the agreement between
models and data is very good, especially for 47~Tuc, for which the 14
Gyr-old model for [Fe/H]=--0.7 falls right on top of the data points in
all three panels. This is not surprising, given that [Mg/Fe] for both the
models and the cluster (Tables~\ref{tabxfe} and ~\ref{clusterdata}) differ
by only $\sim$ 0.1 dex. The same is not true for Mg$_2$, for which the
model for the same age and metallicity is too strong by 0.03 mag, which
would lead to an underestimate of roughly 0.3 dex in [Mg/H]. We suggest
that this mismatch is due to the extreme mass segregation in the cluster
cores.  Because the core of 47~Tuc is strongly depleted of low-mass stars
(e.g., De Marchi \& Paresce 1995, Howell \etal 2001, Monkman \etal 2006),
Mg$_2$ tends to be weaker than the value predicted for a Salpeter IMF,
while Mg $b$, which is not so sensitive to the contribution by low mass
stars, is less affected (see discussion in Section~\ref{comparisons}).
The most recent determination of the mass function in the core of
47~Tuc was performed by Guhathakurta \etal (2006, in preparation, but
see Monkman \etal 2006), who found that, within the cluster core, the
mass function below the turn-off is well matched by a power law with $x
\sim -4.0$.  In Figure~\ref{ieffect} we illustrate the effect of mass
segregation by comparing calculations performed with a Salpeter IMF ($x
= 1.35$, bottom panels) and a dwarf-depleted mass function ($x=-4.0$).
It can be seen that dwarf-depleted match Mg$_2$ considerably better than
those based on a Salpeter IMF. The predictions for Mg $b$ change very
little in comparison, with the models agreeing with the data to within
0.1 ${\rm\AA}$.  Because both indices are subject to the influence of
elemental abundances that may be somewhat uncertain, we only suggest that
the initial inconsistency between the magnesium abundances based on Mg
$b$ and Mg$_2$ might be due to mass segregation effects. In any case,
the conclusion that a combination of these two indices can be used to
constrain the low-mass end of the mass function is robust, provided
other variables such as abundance ratios are tightly constrained.

\setcounter{figure}{21}
\begin{figure}
\plotone{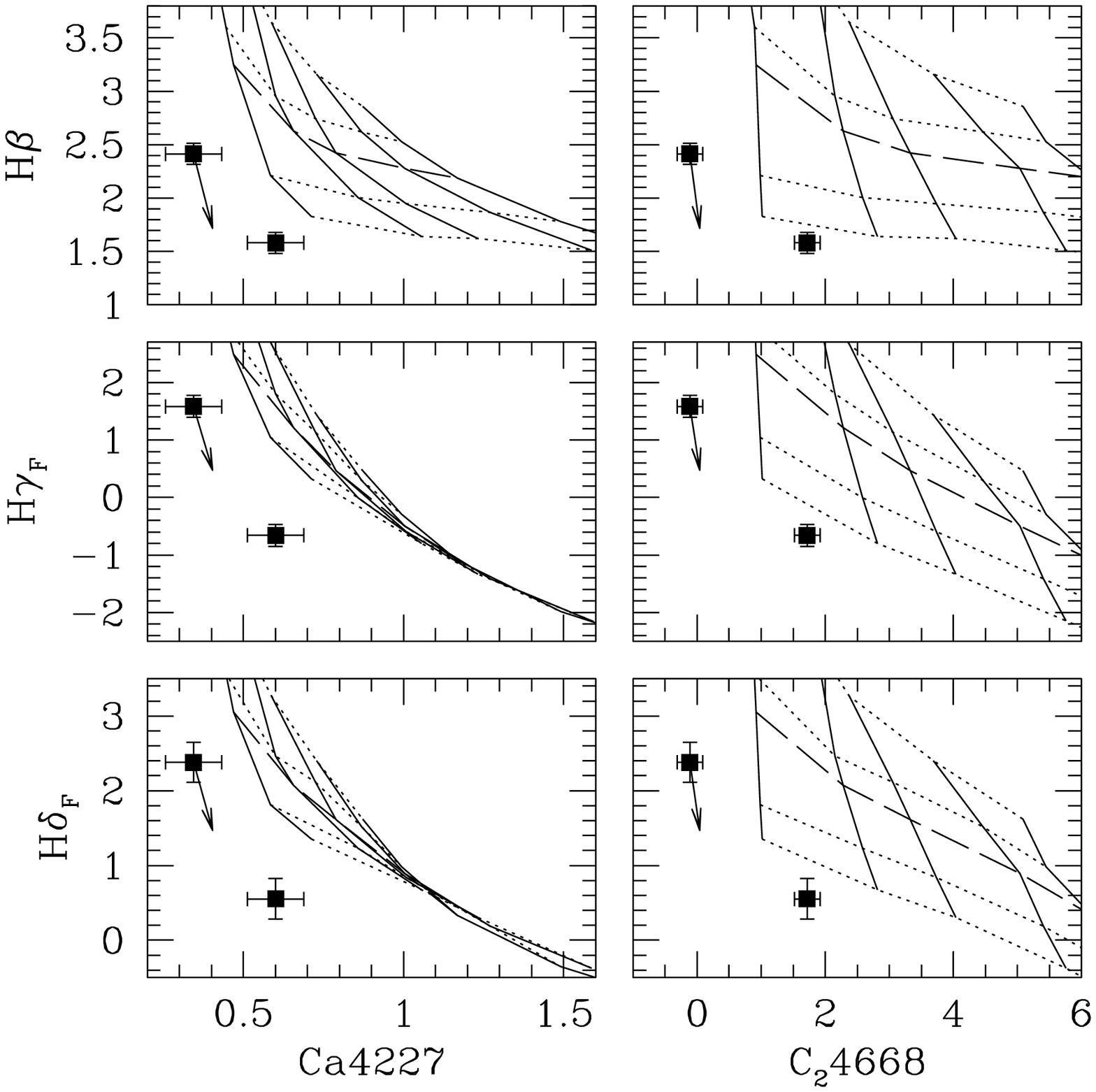}
\caption{f.}
\label{47tuc_m56}
\end{figure}

We note that in Figure~\ref{ieffect} the dwarf-depleted models
have slightly stronger $H\beta$ than those computed with a Salpeter
IMF. This can be understood in terms of the relative contribution to
the integrated light at $\sim$ 4861 ${\rm\AA}$ by giant, turnoff, and
lower main sequence stars.  Comparing the numbers for dwarf-depleted
and Salpeter-IMF models one finds that, because the initial masses of
turnoff and giant stars are essentially the same, the (giant-to-dwarf)
ratio between the contribution by these two types of stars to the
integrated light varies only by a factor of 1.5 between the two
models (being of course higher in dwarf-depleted models). On the
other hand, the ratio between giants and lower main-sequence stars
varies by roughly a factor of 6. Because giants have stronger $H\beta$
than lower main-sequence stars, the net result is that dwarf-depleted
models have stronger $H\beta$.  Of more interest to us is the fact that
the dwarf-depleted models over-predict the value of $H\beta$ in 47~Tuc
by $\sim$ 0.2 ${\rm\AA}$, so that the spectroscopic age of the cluster
according to those models is a bit too old.  This is not unexpected. As
we mentioned in Section~\ref{47tucm5}, the oxygen abundance of 47~Tuc
is much higher than that adopted in the models of Figure~\ref{ieffect}.
The effect of adopting the right oxygen abundance for 47~Tuc can be
gauged by comparing the thick lines in Figure~\ref{ieffect}, which
correspond to computations performed adopting the $\alpha$-enhanced
Padova isochrones, for which [O/Fe]=+0.5. While agreement with $H\beta$
in 47~Tuc is improved, the discrepancy is not entirely removed. This is
because, as discussed in Section~\ref{aeffect}, the $\alpha$-enhanced
Padova isochrones overpredict the temperatures of turn-off and giant
stars, as pointed out by Weiss \etal (2006). In Section~\ref{aeffect},
we estimated an approximate correction to our model predictions and
found that $H\beta$ in old $\alpha$-enhanced models should be weaker
by roughly 0.15 ${\rm\AA}$, which would bring our models into very good
agreement with 47~Tuc data.

\begin{figure}
\plotone{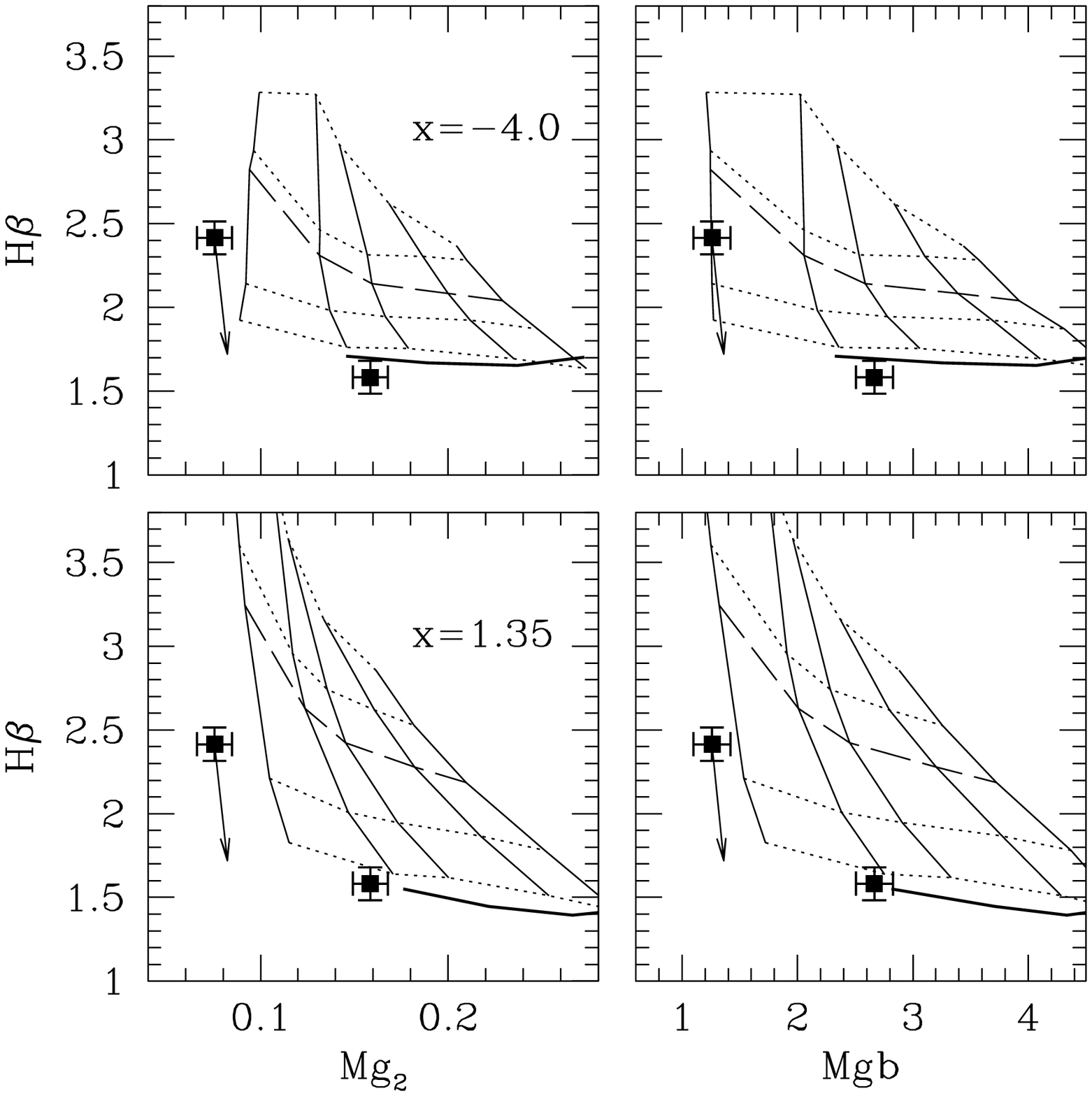}
\caption{Effects of mass function on the predictions for the Mg
indices. {\it Upper panels}: a dwarf-depleted IMF is adopted, consistent with
the clusters' heavy mass-segregation. {\it Lower panels}: Same as the top
panels of Figure~\ref{47tuc_m54}d, where a Salpeter IMF is adopted in the
computations.  Note the improved agreement between model and data for
Mg$_2$ in the upper panels.}
\label{ieffect}
\end{figure}

In Figure~\ref{47tuc_m55}e the data are compared in CN-index vs
Balmer-line planes. It can be seen that the clusters have much stronger
CN indices than predicted by the models. In the case of 47~Tuc, CN$_1$
(CN$_2$) is stronger by 0.06 (0.08) mag than the model for [Fe/H]=--0.7
and 14 Gyr, so that in all panels the cluster falls near the solar
metallicity locus. This is not a new result (e.g.,  Vazdekis 1999, Paper
I, Puzia \etal 2002) neither is it unexpected. As anticipated in the
beginning of Section~\ref{clust}, the relative abundances of carbon and
nitrogen of models and cluster are largely discrepant (Tables~\ref{tabxfe}
and \ref{clusterdata}).  In Figure~\ref{47tuc_m56}f, model predictions
for C$_2$4668 and Ca4227 are compared with the data. Interestingly,
models overpredict the values of both indices. In the case of C$_2$4668,
this probably indicates that the mean luminosity-weighted [C/Fe] is
below solar, which is not surprising, given the abundances measured
in individual stars (Table~\ref{clusterdata}).  The case of Ca4227 is
interesting. Despite the fact that the relative abundance of calcium in
the models is a very close match to that of the clusters, there is a very
large mismatch with the data. In the case of 47~Tuc, the 14 Gyr-old,
[Fe/H]=--0.7 model over-predicts Ca4227 by $\sim$ 0.6 ${\rm\AA}$,
leading to an underestimate in [Ca/H] by more than 0.7 dex. We suspect
that this might in part be due to the influence of CN lines on the Ca4227
index. These results clearly indicate that indices sensitive to carbon and
nitrogen abundances need to be corrected for abundance-ratio effects. This
is the topic of Section~\ref{47tucratios}.

\subsubsection{CN-Strong Models for Globular Clusters}
\label{47tucratios}

We have seen in the previous section that the base models do an excellent
job of reproducing the Balmer, Fe, and Mg indices for one metal-poor
and one mildly metal-rich Galactic globular cluster. The same level
of agreement was not reached, however, for the carbon, nitrogen and
calcium-sensitive indices. In order to address this problem we computed
a new set of models adopting the known abundance pattern of 47~Tuc,
following the procedure outlined in Section~\ref{aratios}, so as to
generate CN-strong models for this cluster. Such models may be very
useful given the fact that the CN-strong phenomenon in integrated spectra
of globular clusters seems to be ubiquitous, being probably caused by
very high nitrogen abundances (e.g., Burstein \etal 1984, Brodie \&
Huchra 1991, Papers I and II, Li \& Burstein 2003, Beasley \etal 2004,
Burstein \etal 2004).

The elemental abundances adopted for 47~Tuc are those listed in
Table~\ref{clusterdata}. Because carbon and nitrogen have a bimodal
distribution, we computed two sets of models, corresponding to the
CN-strong and CN-weak abundance patterns. These models were then
combined, with weights determined by the relative numbers of CN-strong
and CN-weak stars in the core of 47~Tuc, from Briley (1997), who found
CN-strong/CN-weak $\sim$ 2.  In view of the mass-segregation effects
discussed in the previous section, we adopted a dwarf-poor IMF in these
computations, with x=--4 (equation~\ref{eq3}).

\begin{figure}
\plotone{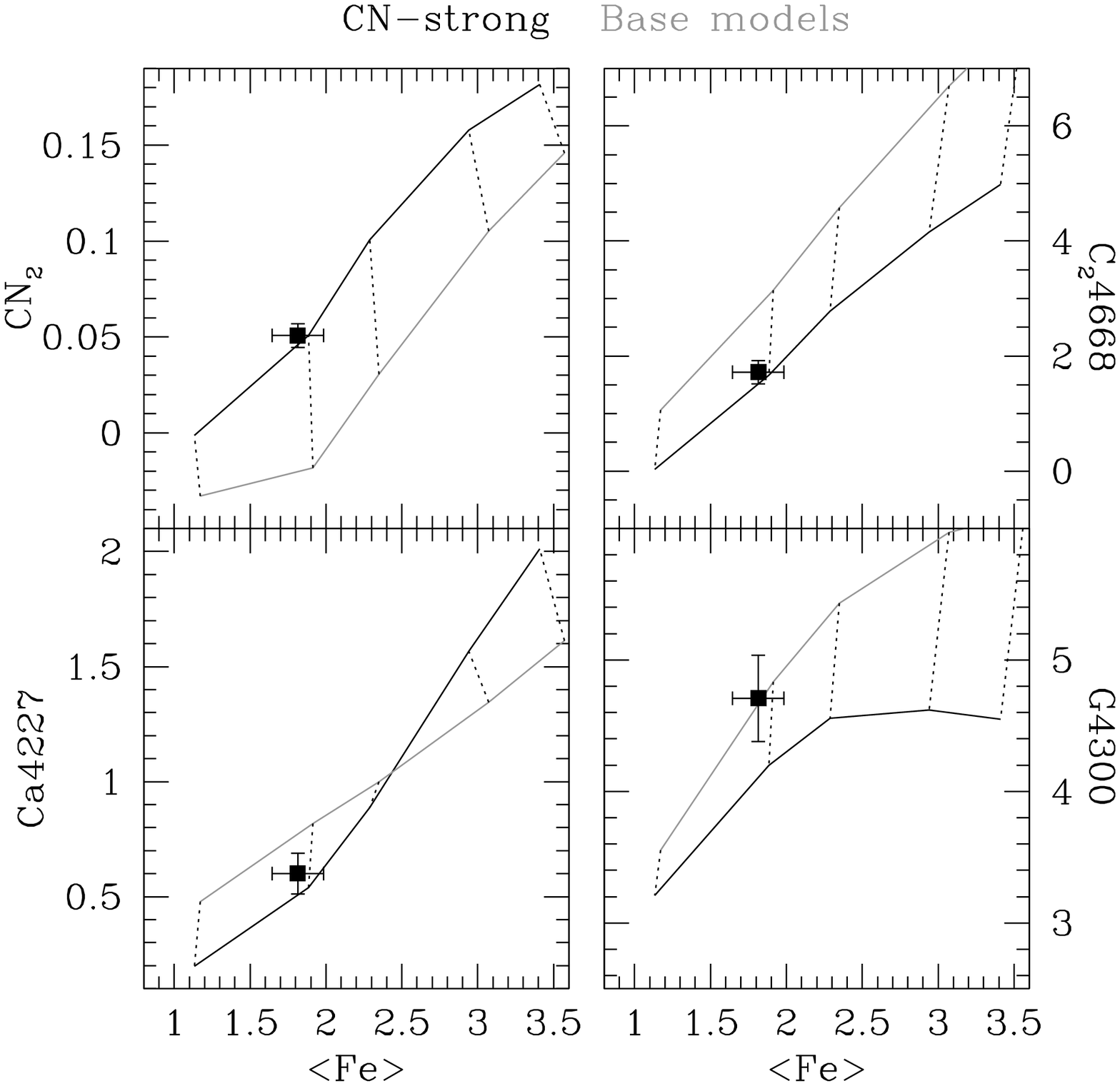}
\caption{Comparison, between data for 47~Tuc and models in iron vs.
CN-sensitive data plots. Models shown are all for an age of 14.1 
Gyr.  Gray lines: base models from Table~\ref{ssppar}.  Dark lines: 
models computed for the abundance pattern of 47~Tuc. Dotted lines 
connect same-[Fe/H] models (from left to right, [Fe/H]=--1.3, --0.7,
--0.4, 0.0, +0.2).  The models with [Fe/H]=--0.7 and the cluster
abundance pattern reproduce very well all indices except for G4300.
See text.
}
\label{47tucabund}
\end{figure}

In Figure~\ref{47tucabund} these models are compared with our base models
(1-5 in Table~\ref{ssppar}, in a few interesting index-index plots. Gray
lines represent the base models, while dark lines represent the CN-strong
models described above. Same-[Fe/H] models are connected by dotted lines.
As expected, CN-strong models are characterized by stronger CN$_2$, and
weaker carbon indices (G4300 and C$_2$4668). The behavior of Ca4227 is
more complex. It is weaker in metal-poor models, where [Ca/Fe] and [O/Fe]
in the CN-strong and base models are similar, so that the differences
arise purely due to the stronger CN strengths in the former. At higher
metallicities, Ca4227 becomes stronger in the CN-strong models because
those have higher [Ca/Fe] and [O/Fe] than the base models, by 0.1 and 0.6
dex, respectively.

Also shown in Figure~\ref{47tucabund} are line indices in the spectrum
of 47~Tuc.  It can be seen that in all plots 47~Tuc is well matched by
the 14 Gyr-old, CN-strong, models with [Fe/H]=--0.7 (the second dotted
line from left to right), except for the case of G4300.  For Ca4227,
C$_2$4668, and CN$_2$ the CN-strong models are a vast improvement over
CN-normal models (the same is true of CN$_1$, not shown).  In the case
of Ca4227, this result highlights an important fact, seldom appreciated
in the literature: the Ca4227 index is {\it very} sensitive to the
abundances of carbon and nitrogen, due to a contamination of its blue
pseudo-continuum window by a CN band-head. In order to circumvent this
problem, Prochaska \etal (2005) propose the definition of a new index,
Ca4227r, which is far less sensitive to the CN contamination, thus being
a more reliable indicator of the calcium abundances. See Prochaska \etal
(2005) for details.  Finally, we note that adoption of these CN-strong
models only affects predictions of these carbon/nitrogen-sensitive
indices, so that the quality of the match to all other indices is the
same as in the previous plots.  Models computed for both the CN-strong
and CN-normal abundance patterns of 47~Tuc stars (Table~\ref{clusterdata})
and a dwarf-depleted IMF ($x=-4$) are provided in Tables A7 and A8
in the Appendix.

\subsection{NGC~6528} \label{n6528}

\begin{figure}
\plotone{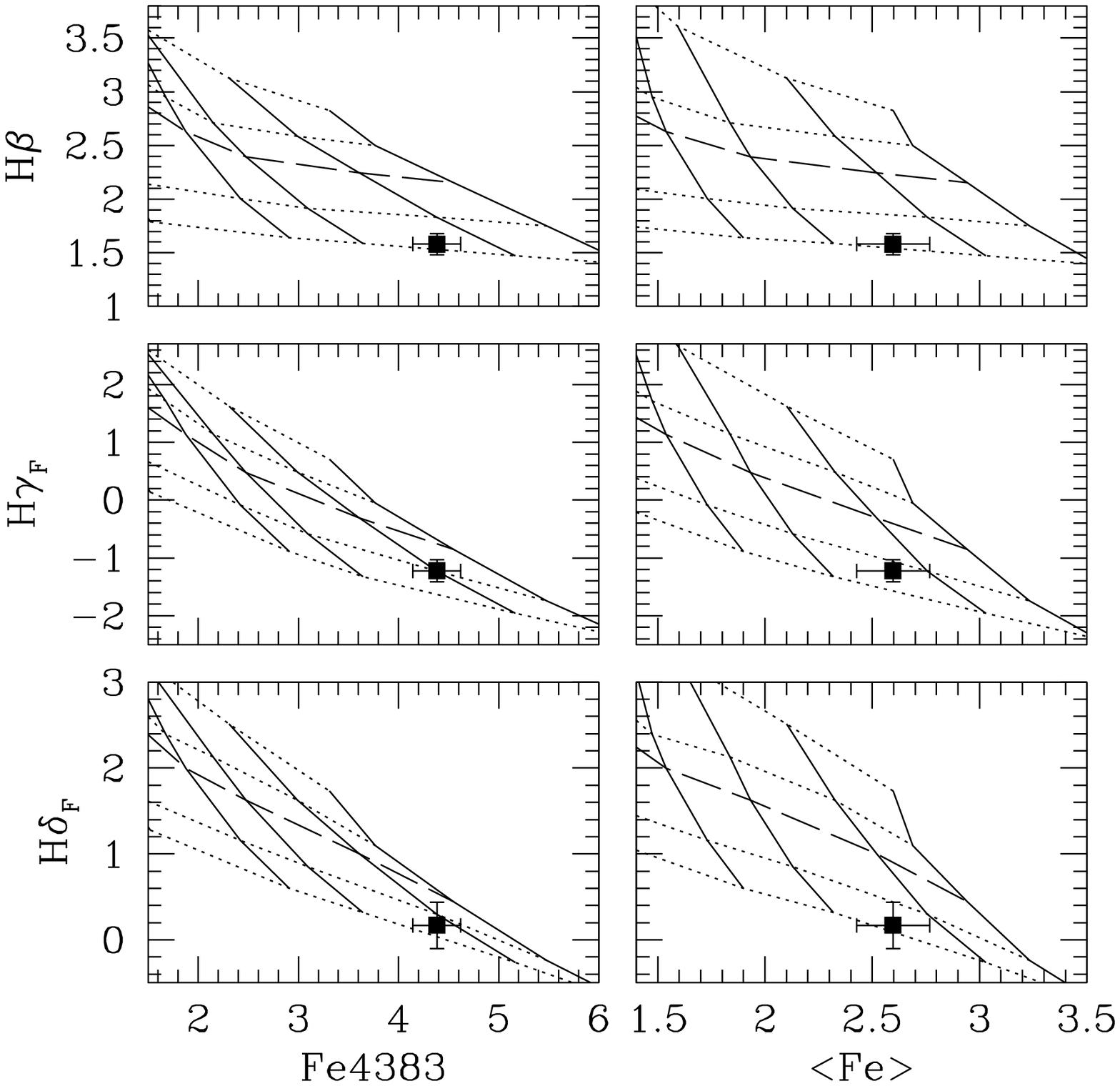}
\caption{Comparison of data for NGC~6528 with model predictions assuming
the input parameters listed in Table~\ref{mod6528}. Note the consistency
with which models match all indices for the same age and [Fe/H]. Ages
plotted are 1.5, 2.5, 3.5 (dashed), 7.9, and 14.1 Gyr, and metallicities
are [Fe/H] = --0.7, --0.4, 0.0, and +0.2.
}
\label{6528_mod25}
\end{figure}

The bulge cluster NGC~6528 is one of the most metal-rich Galactic
globular clusters, with [Fe/H] determinations ranging between --0.15
and +0.1 (Carretta \etal 2001, Zoccali \etal 2004, Origlia \etal
2005). Besides, it is known to be old ($\sim$ 11 Gyr, Ortolani \etal
1995, Feltzing \& Johnson 2002). The abundance pattern of NGC~6528 is
still a subject of debate, as the three recent studies mentioned above
quote significantly different abundances for some very important elements
(Table~\ref{clusterdata}). For instance, abundance determinations in these
studies differ by as much as 0.2 dex in the case of iron, 0.3 dex in the
case of oxygen, magnesium, and silicon, 0.4 dex for titanium, and 0.8 dex
in the case of calcium. Moreover, the carbon and nitrogen abundances of
main sequence stars are unknown.  These uncertainties are probably due to
difficulties associated to the cluster's distance and severe reddening,
which has made it so far impossible to obtain high-resolution spectra
of unevolved stars, for which both the uncertainties involved in the
abundance determinations and star-to-star variations are less important.

As a prelude to our effort towards matching the data on NGC~6528
with models based on the cluster abundance pattern, we show in
Figures~\ref{m67_65281} through \ref{m67_65284}, the indices for
NGC~6528 over-plotted on models with a nearly solar abundance
pattern. The abundance pattern of these models (models 1--5 in
Table~\ref{ssppar}, and Table~\ref{tabxfe}) differs from that of the
cluster (Table~\ref{clusterdata}), so we do not expect a perfect match
to the data, but the comparison might be nonetheless instructive. From
these figures, it can be seen that: 1) the age of the cluster according
to $H\beta$ is $\sim$ 14 Gyr, which is slightly too old, while it is a
bit younger according to the $H\gamma$ and $H\delta$ indices (roughly
10 and 12 Gyr, respectively). This is not unexpected given that,
while NGC~6528 stars seem to be at least slightly oxygen-enhanced, the
isochrones adopted in the model computations are solar-scaled.  Recall
that different Balmer lines respond differently to $\alpha$-enhancement
(Section~\ref{aenhance}) and therefore should change in different ways
if we switched to $\alpha$-enhanced models.  2) The iron abundances,
according to Fe4383, Fe5270, and Fe5335 range between --0.3 and --0.2 dex,
in rough agreement with the lower value in Table~\ref{clusterdata}; 3)
relative to the models, the cluster looks too strong in the CN indices,
which suggests the existence of CN-strong stars, just as in the case of
47~Tuc and M~5; 4) NGC~6528 looks mildly too strong in the Mg indices,
which is consistent with its measured [Mg/Fe]; 5) as in the case of
47~Tuc and M~5, NGC~6528 is very weak in Ca4227 which, given its very
strong CN indices, is hinting that the effect of CN lines on Ca4227
discussed in the case of 47~Tuc might be in operation also for NGC~6528.

\begin{deluxetable*}{ccccccccccc} 
\tablecaption{Best Fitting Model for NGC~6528}
\tablewidth{0pt}
\tablehead{\colhead{Isochrone} & Age & [Fe/H] &
           \colhead{[O/Fe]} & \colhead{[N/Fe]} & \colhead{[C/Fe]} & 
           \colhead{[Mg/Fe]} & \colhead{[Ca/Fe]} & \colhead{[Ti/Fe]} &
           \colhead{[Na/Fe]} & \colhead{[Si/Fe]}     }
\startdata
Girardi \etal & 13 & --0.2 & +0.15 & +0.5 & --0.1 & +0.1 & --0.1 & --0.1 & +0.4 & +0.1 \\
\enddata
\label{mod6528}
\end{deluxetable*}

Now we turn to the task of producing models that mirror the abundance
pattern of NGC~6528. Ideally, we would perform an exercise similar
to that of Section~\ref{47tucratios}, but given the above mentioned
uncertainties in the cluster elemental abundances, we have to
proceed differently. Instead, we adopt the method discussed in
Section~\ref{amethod} in order to search the age, metallicity, and
abundance pattern that are a best match to the cluster data.  We adopt
solar-scaled Padova isochrones in this exercise, because they match more
closely the abundance pattern of the cluster, particularly the abundance
of oxygen.  The $\alpha$-enhanced Padova isochrones were computed
assuming [O/Fe] = +0.5, whereas the cluster, according to analyses of
individual stars by Zoccali \etal and Carretta \etal has at most [O/Fe]
$\sim$ +0.15.\footnote{The author thanks Paula Coelho for pointing
this out.} Other $\alpha$-elements, like magnesium and titanium, are
roughly +0.4 dex more enhanced relative to iron in the $\alpha$-enhanced
Padova isochrones, which is roughly +0.3 dex higher than measured in
cluster stars.  Moreover, as discussed in Section~\ref{aeffect}, Weiss
\etal (2006) have shown that there is a problem with the metal-rich
$\alpha$-enhanced Padova isochrones, so we refrain from adopting them in
the analysis of NGC~6528.  The input abundances of titanium, sodium, and
silicon were taken from Zoccali \etal (2004, see Table~\ref{clusterdata}).

\begin{figure}
\plotone{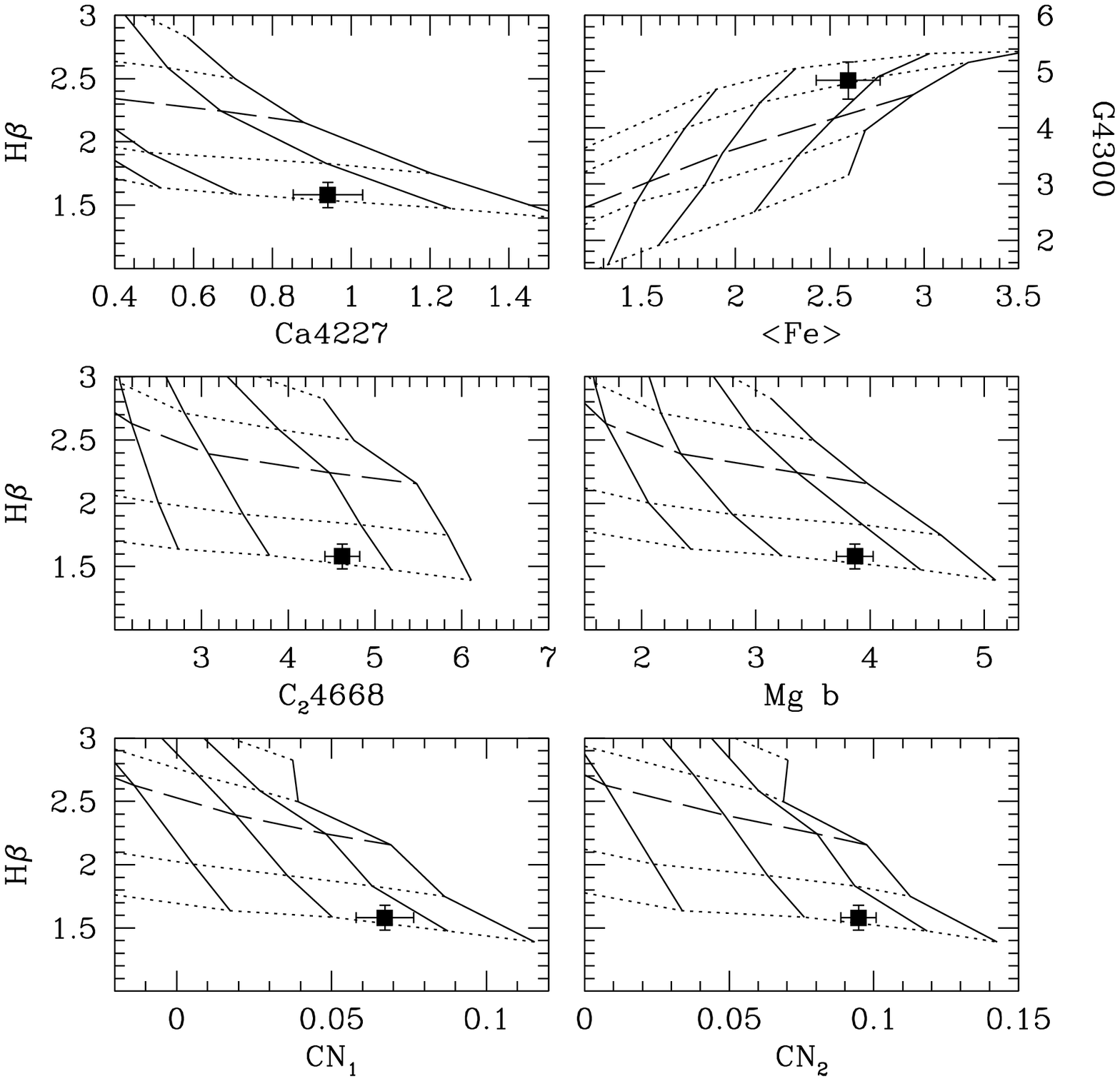}
\caption{Figure~\ref{6528_mod25} continued. Here the models are compared
to data for indices sensitive to the abundances of calcium, carbon, 
nitrogen, and magnesium. Note that the best matching models have the
same age and [Fe/H] as those in Figure~\ref{6528_mod25} (except for 
G4300, which is matched for a slightly too young age---see text).
}
\label{6528_mod26}
\end{figure}

The parameters of the best-fitting model are listed in
Table~\ref{mod6528}, and it is compared with cluster data in
representative index-index diagrams in Figures~\ref{6528_mod25} and
\ref{6528_mod26} Comparing the numbers in Tables~\ref{clusterdata} and
\ref{mod6528}, one can see that the best-fitting spectroscopic age (based
on $H\beta$, according to the method described in Section~\ref{amethod})
is 2-Gyr older than the CMD-based age from Feltzing \& Johnson
(2002). While on one hand this difference is comfortably within the
errors, given the error bars in both studies ($\pm$ 2 Gyr), on the other it
can be traced to Feltzing \& Johnson's adoption of the $\alpha$-enhanced
Padova isochrones for Z=0.04.  According to Table~\ref{ssppar}, this
model has [Fe/H] = 0.01 and [O/Fe] = +0.5, which are respectively $\sim$
0.2 and 0.4 dex higher than found in this study and in spectroscopic
abundance determinations of cluster stars (which mostly preceded Feltzing
\& Johnson's study).  Accounting for both [Fe/H] and [O/Fe] differences
would bring the ages in both studies into agreement.

Consistency between age estimates based on the various Balmer line
indices has also been largely achieved. Ages according to $H\gamma_F$,
(10 Gyr), $H\delta_F$ (12 Gyr), and $H\gamma_A$ (13 Gyr, not shown)
agree very well with the $H\beta$-based age. The only exception is
that of $H\delta_A$ (not shown), according to which the spectroscopic
age of the cluster is $\sim$ 8 Gyr. We recall that no such effect was
seen for more metal-poor and younger clusters in Sections~\ref{47tucm5}
and \ref{m67spec}.  Inspection of the Korn \etal (2005) tables reveals
that the elemental abundance that affects $H\delta_A$ the most strongly
(after iron) is that of silicon. If we adopt the [Si/Fe] determination
by Carretta \etal (2001), which is higher than that of Zoccali \etal
(2004) by $\sim$ 0.3 dex, the $H\delta_A$-based age becomes 10 Gyr,
which is in much better agreement with the ages based on the other
Balmer lines.  That might also explain why no such discrepancy was found
for the other clusters, for which the abundance of silicon used as input
is well constrained.  While this result could be construed as favoring a
[Si/Fe] value at the higher end of the wide range allowed by abundance
determinations from the literature, we prefer to wait for the matter to
be settled by further detailed abundance studies. We therefore conclude
that the Balmer line indices are indicating consistent ages for NGC~6528,
the only exception being $H\delta_A$, which indicates mildly too young
ages, possibly because the index is affected by the (poorly constrained)
abundance of silicon. Finally, we note that, within the uncertainties,
there is a trend of slightly younger ages towards higher-order
Balmer lines.  We speculate that this mismatch is partially due to the
adoption of theoretical isochrones whose [O/Fe] is too low. While the
Girardi \etal (2000) isochrones adopted in the mild-$\alpha$ models
have [O/Fe]=0, spectroscopic determinations tell us that the cluster
has at least [O/Fe] $\sim$ +0.1 and might be slightly higher. As
discussed in Section~\ref{aeffect}, model predictions for $H\beta$
are substantially more affected by oxygen abundances than the higher
order Balmer lines. In fact, it can be seen in Figure~\ref{alphaiso}
that in the old, metal-rich, regime ($\sim$ 14 Gyr, [Fe/H] $\simgreat$
0) $H\delta_F$ is essentially unaffected by the oxygen abundance of
the theoretical isochrones adopted. Therefore, adoption of theoretical
isochrones with slightly higher [O/Fe] would decrease the $H\beta$-based
ages and bring it into better agreement with those based on the higher
order Balmer lines and analysis of the cluster CMD.

The best-fitting abundances of iron and magnesium are also in very
good agreement with values from the literature, though in both cases
our estimates fall at the low end of the range allowed by abundance
determinations from the literature. We call attention for the remarkable
consistency of the [Fe/H] estimates coming from Fe4383, Fe5270, and Fe5335
(Fe5015 is not available for NGC~6528), which agree with each other within
0.05 dex. Magnesium abundances according to Mg $b$ and Mg$_2$ differ by
$\sim$ 0.2 dex, in the sense that the latter are higher. It is hard to
understand the reason for this discrepancy.  One possible explanation
may be the existence of line opacity sources that
are unaccounted in the Korn \etal (2005) sensitivity tables. A strong
candidate would be the TiO molecule, which is very strong in cool giants
which must be present in metal-rich systems such as NGC~6528. However,
from the discussion in Section~\ref{comparisons}, we would expect Mg $b$
to indicate higher magnesium abundances than Mg$_2$, which is the opposite
of what we are observing. While this issue certainly deserves further
scrutiny, we believe that the Mg $b$-based abundance is more reliable,
since this index is not affected by flux-calibration or IMF uncertainties
(Sections~\ref{abslines} and \ref{47tuclight}, respectively).

The abundances of carbon and nitrogen in NGC~6528 stars were not
determined in the studies summarized in Table~\ref{clusterdata}, so that
our estimates listed in Table~\ref{mod6528} are a first attempt in that
direction. We find that NGC~6528 follows a pattern that is similar to
that 47~Tuc (Section~\ref{47tucm5}), being slightly carbon-depleted and
very strongly nitrogen-enhanced. It is reasonable to suppose that the
same type of dichotomy in the abundances of carbon and nitrogen that
is found in M~5, 47~Tuc, and many other clusters (e.g., Dickens \etal
1979 Norris \& Freeman 1979, Smith \etal 1989, Cannon \etal 1998, Cohen
\etal 2002, Briley \etal 2004, Carretta \etal 2005, Lee 2005, Smith \&
Briley 2006, and references therein) may also be present in NGC~6528,
though this needs to be confirmed by spectroscopy of individual stars.
We note that there is a slight disagreement between the carbon abundances
obtained from matching the G4300 and C$_2$4668 indices, in that the latter
are higher by $\sim$ 0.1 dex. While this discrepancy is minor it should be 
subject to further investigation in the future (Graves \& Schiavon 2006, in 
preparation). Finally, we point out that our
value for [Ca/Fe] falls within the range of abundance determinations from
the literature, which is no great accomplishment, given the sizable
disagreement between the estimates by the different groups. Clearly,
more work is needed in this front.

We conclude that the data for NGC~6528 are very well matched for a
mildly $\alpha$-enhanced abundance pattern and for an age that is in
very good agreement with determinations based on analysis of the cluster
color-magnitude diagram.  Furthermore, we find that the cluster data are
well matched for a [C/Fe]= $\sim$ --0.1 and [N/Fe] $\sim$ +0.5, which
mirrors the abundance pattern of other Galactic clusters, indicating that
NGC~6528 stars are liable to present similar bimodal distributions in
their carbon and nitrogen abundances.  We found outstanding consistency
between the ages and iron abundances determined from different indices,
but small discrepancies in the cases of carbon and magnesium.

The models computed for the abundance pattern given in Table~\ref{mod6528}
are provided in Table A9 in the Appendix.

\subsection{Summary}

We have performed a detailed comparison of our models with high quality
data from a representative set of Galactic clusters. For the clusters with
very well known ages, metallicities, and abundance ratios (M~67, 47~Tuc,
and M~5), our models matched essentially all the line indices with a
very high degree of consistency, for the right set of input parameters (even
though for M~5 that was only possible once the contribution by blue horizontal
branch stars to the integrated light of the cluster was removed).
For NGC~6528, where the metal abundances are more uncertain and, in
the case of carbon and nitrogen, unknown, we followed the procedure
described in Section~\ref{amethod} in order to estimate the cluster
age and metal abundances. The input parameters of the best-fitting
model are well within the range allowed from previous work on stellar
abundances and ages.  In the process, we learned a few important lessons:
1) Outstanding consistency was reached for ages and iron abundances
estimated on the basis of blue and red indices. We recall that one of
the main goals of this modeling effort is to extend the accuracy and
reliability of Lick index modeling into the blue, with an eye towards
their application into distant galaxy work. The results of this Section
positively qualify our models for such applications (for an initial
effort, see Schiavon \etal 2006); 2) Use of C$_2$4668 in
conjunction with the CN indices allows us to estimate carbon and nitrogen
abundances reliably. While this is not surprising, previous attempts were
hampered by difficulties in the modeling of these indices. Combining those 
indices with Ca4227 which is strongly influenced by contamination of the
blue pseudo-continuum by
CN lines, allows determination of calcium abundances; 3) Because Mg$_2$
can be strongly influenced by the contribution by stars in the lower main sequence, 
agreement between the magnesium abundances that are obtained from that index 
and Mg $b$ can only be achieved if the models include the correct input IMF;
4) Finally, the oxygen abundances are very important to help deciding
what are the adequate theoretical isochrones used in the models. While
this was known before, we showed that a substantial mismatch between the
oxygen abundances of the input isochrones and that of the target stellar
population can generate a small, but detectable, systematic effect in
the ages that are inferred from the different Balmer lines.

We carry this newly acquired knowledge on to the next section, where we
take a brief look at some galaxy data from the literature.

\section{Comparison with Galaxy Data} \label{galaxies}

The chief motivation behind our modeling efforts is to develop
tools to constrain the history of star formation and chemical enrichment
of galaxies from analysis of their integrated spectra. In this Section we
briefly discuss the comparison between our models and a few high quality
data sets of line indices measured in integrated spectra of galaxies.
We are aware of the fact that matters become far more complicated when
one makes the transition from clusters to galaxies, so we do not intend
to discuss these comparisons exhaustively, to avoid lengthening what
is already a very long paper. So we just discuss some general trends
coming from a comparison of our models to data from the literature,
as a prelude to a more in-depth analysis of very-high S/N data for 
nearby galaxies, which will be presented in a forthcoming paper.

Galaxies are complex systems that have undergone a more prolonged 
history of star formation than that of globular clusters. Therefore, they are
liable to host a mixture of stellar populations of different ages and
metallicities. Yet we are comparing them with models for single-burst
stellar populations. Therefore, any time we refer to ages and metal
abundances in this Section what we really mean is the luminosity-weighted
mean ages and metal abundances.

\subsection{The Trager \etal Sample: Intermediate ages and
$\alpha$-Enhancement in Giant Early-Type Galaxies} \label{tragersect}

\begin{figure}
\plotone{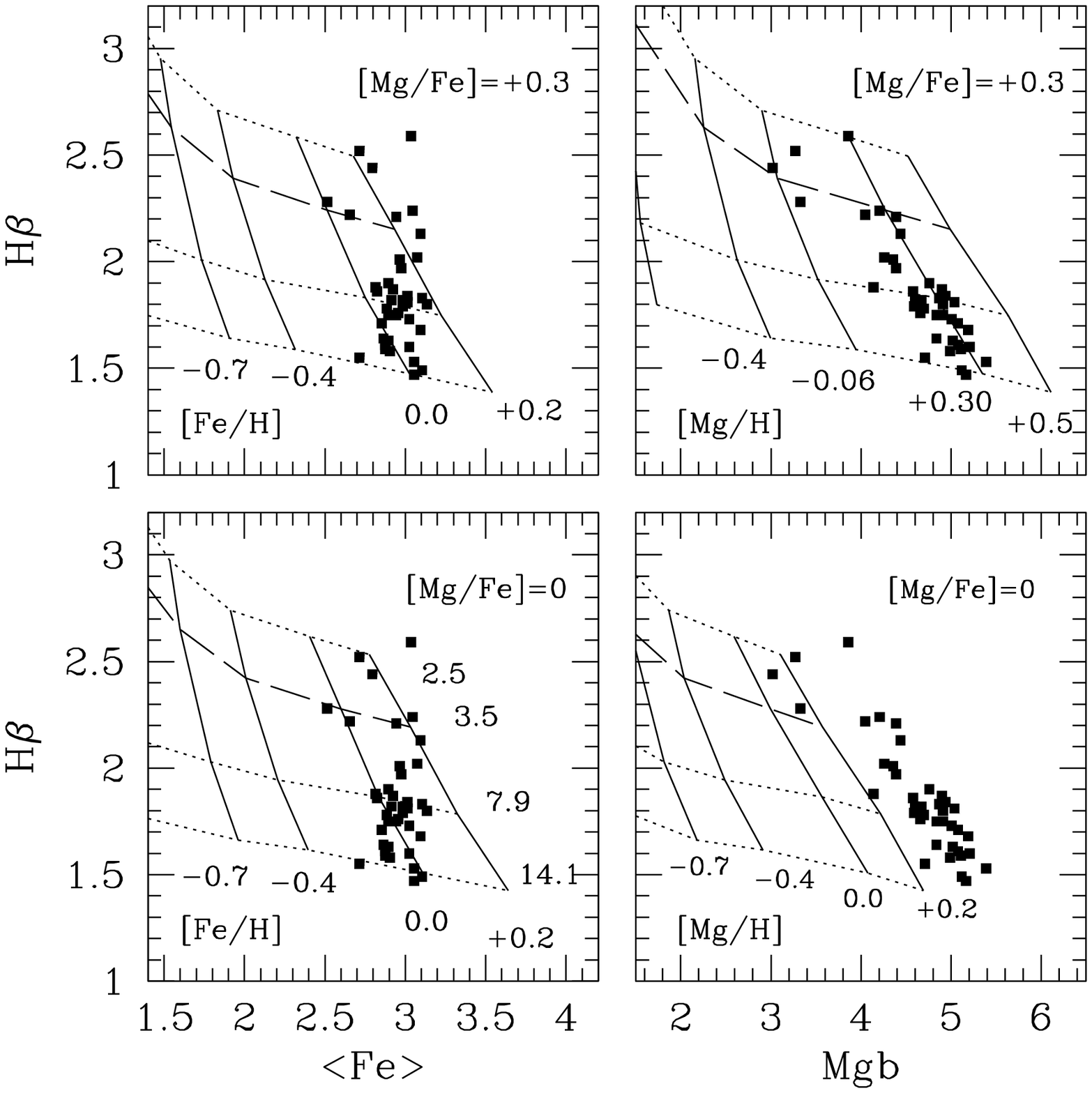}
\caption{Comparison of our models to data by Trager \etal (2000). {\it
Bottom panels:} solar-scaled models; {\it Upper panels:} models with
[Mg/Fe]=+0.3.  The labels on the left (right) panels show [Fe/H]
([Mg/H]). The bulk (3/4) of the sample galaxies have mean ages between 7
and 14 Gyr (lower left panel). Solar-scaled models cannot match \fem\ and
Mg $b$ for the same age and metallicity (lower panels). The upper panels
show that Mg-enhanced models are a much better match to the data, as they
match both \fem\ and Mg $b$ for the same age and [Fe/H] (roughly solar).}
\label{tragerdata}
\end{figure}

We start by displaying in Figure~\ref{tragerdata} the data from Trager
\etal (2000) for nearby early-type galaxies on top of our model grids
in the $<Fe>$ vs. $H\beta$ and Mg $b$ vs $H\beta$ planes. In the lower
panels we compare the data with solar-scaled models, computed adopting the
solar-scaled Padova isochrones.  Same-[Fe/H] lines are labeled according
to the abundances of iron, [Fe/H], (left panel) and magnesium, [Mg/H],
(right panel), which of course are the same for these [Mg/Fe]=0 models.
Model ages are indicated in the lower left panel and are the same on
all the other panels. As discussed in previous sections, both \fem\ and
$H\beta$ are very insensitive to abundance ratios, so they can be used to
estimate the mean [Fe/H] and age of the stellar populations in galaxies.
Therefore, according to the lower left panel of Figure~\ref{tragerdata},
the bulk ($\sim$ 3/4) of the Trager \etal sample has roughly solar [Fe/H]
and ages between $\sim$ 7 and 14 Gyr. The remaining 1/4 of the sample has
mean ages lower than $\sim$ 7 Gyr, with some galaxies reaching ages of the
order of 2.5 Gyr. We focus here on the 3/4 of the sample with ages older
than $\sim$ 7 Gyr. In the lower right panel, the same models are compared
with data on the Mg $b$-$H\beta$ plane. In this panel, we can see that
the same models that match \fem\ data under-predict Mg $b$ by $\sim$
1 ${\rm\AA}$.  This is a well-known result, which is telling us that
stars in giant early-type galaxies are magnesium-enhanced ([Mg/Fe] $>$ 0).
In the top panels we compare the same data with models computed adopting
[Mg/Fe]=+0.3 and keeping all other abundance ratios solar.  These models
are computed adopting the {\it solar-scaled} Padova isochrones (see
Section~\ref{anote}).  Note that the models plotted in the upper panels
have the same [Fe/H] values as in the lower panels, but the values for
[Mg/H] are +0.3 dex higher in the upper panels.  The plots in the upper
panels show that the [Mg/Fe]=+0.3 models match the \fem\ and $H\beta$
data for the same ages and [Fe/H] as the solar-scaled models, with the
oldest 3/4 of the sample having roughly solar [Fe/H]. On the other hand,
the Mg-enhanced models with [Fe/H]=0 are a much better match to the Mg
$b$ data, indicating that the old galaxies in the Trager \etal sample
have mean [Mg/Fe] ([Mg/H]) of the order of +0.3 (+0.3).  Therefore,
we reproduce the results by Worthey \etal (1992), who found that giant
early-type galaxies have [Mg/Fe] higher than solar. The models employed
in Figure~\ref{tragerdata} are presented in Tables A10 through A13 in
the Appendix.

\begin{figure}
\plotone{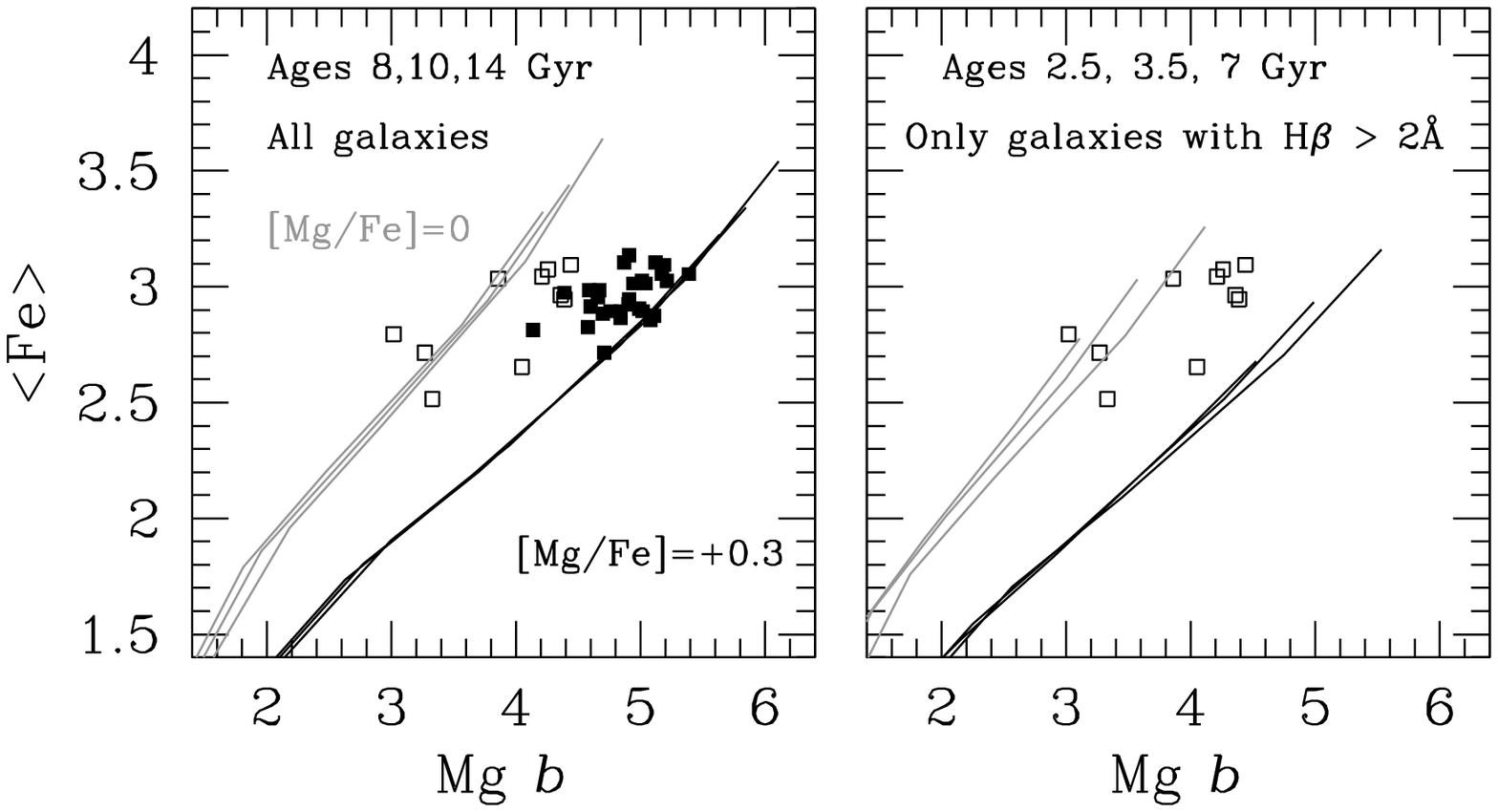}
\caption{Trager \etal data compared with our models in \fem--Mg $b$
diagrams. {\it Left Panel:} All galaxies are compared with solar-scaled
(gray) and Mg-enhanced models. Galaxies with mean stellar ages younger
than $\sim$ 7 Gyr ($H\beta > 2 {\rm \AA}$) are plotted with open
symbols. Older galaxies (solid squares) tend to have higher [Mg/Fe] than
their younger counterparts.  {\it Right panel:} Same plot, comparing only
galaxies with ages lower than $\sim$ 7 Gyr with models for younger single
stellar populations with solar-scaled (gray) and Mg-enhanced abundance
patterns. Galaxies with younger mean stellar ages in the Trager \etal
sample tend to have lower [Mg/Fe].
}
\label{trag_mgfe}
\end{figure}

We also confirm previous results indicating that early-type
galaxies have a large spread in mean ages, hinting at the presence
of an intermediate age component in their stellar populations (e.g.,
Trager \etal 2000, Kuntschner 2000, Caldwell, Rose \& Concannon 2003,
Denicol\'o \etal 2005, Thomas \etal 2005, Mendes de Oliveira \etal
2005). We note that a pattern of younger galaxies having higher
[Fe/H] and lower [Mg/Fe] is also found, which is also in agreement with
the findings by previous authors. The latter result is more clearly seen
in Figure~\ref{trag_mgfe} where models and data are compared in \fem-Mg
$b$ space. In the left panel, all galaxies are over-plotted on solar-scaled
and [Mg/Fe]=+0.3 models for 8 Gyr and older models. Galaxies with ages 
younger than $\sim$ 7 Gyr ($H\beta > 2$ in Figure~\ref{tragerdata}) are
plotted with open squares, while older galaxies are represented by solid
symbols. There is a clear trend in the sense that younger galaxies tend
to be closer to the solar-scaled models, whereas older galaxies lie
closer to the Mg-enhanced lines. In the right panel, only younger galaxies
are compared with model predictions for single stellar populations with
comparable ages, showing that the trend is confirmed.

Finally, we point out that there are no galaxies in
Figure~\ref{tragerdata} with mean stellar ages older than $\sim$ 14 Gyr,
which means no galaxies older than the universe (Spergel \etal 2003),
demonstrating that stellar population synthesis models are reaching a
state of maturity whereby mean stellar ages obtained from comparison
with high quality data are meaningful in an absolute sense.

\subsubsection{A Note on the Theoretical Isochrones Adopted}
\label{anote}

As discussed in Section~\ref{caveats}, it has become standard in
the literature to assume that not only magnesium, but all other
$\alpha$-elements, including oxygen, are equally enhanced in massive
early-type galaxies.  However, there is mounting evidence that this
might not be the case.  For instance, calcium seems at least not to
be as enhanced as magnesium (e.g., Vazdekis \etal 1997, Worthey 1998,
Trager \etal 1998, Henry \& Worthey 1999, Thomas \etal 2003, Prochaska
\etal 2005, Section~\ref{chemevol}).  Likewise, oxygen has been found
not to track magnesium in Galactic bulge metal-rich stars (Fulbright
\etal 2005, Cunha \& Smith 2006).  To our knowledge, there has been no
compelling determination of oxygen abundances in the stellar populations
of early-type galaxies to this date, so that it is fair to say that the
abundance of that element in early-type galaxies is unknown.
This is unfortunate because, as discussed in Section~\ref{aeffect},
of all $\alpha$ elements, oxygen is the most relevant for the interior
structure and evolution of stars.  In view of this uncertainty and the
problems with the $\alpha$-enhanced theoretical isochrones adopted in this
paper (Section~\ref{aeffect}), we feel justified in adopting solar-scaled
theoretical isochrones in Figure~\ref{tragerdata} and in the remaining
of this paper.  We refer the reader interested in assessing the effect
on our final results of adopting $\alpha$-enhanced isochrones to the
discussion in Section~\ref{aeffect}, where it was shown that this effect
is minor.

\subsection{Stellar Populations in the Blue} \label{spblue}

\subsubsection{The Eisenstein \etal (2003) Sample}

Finally, we get to the point of comparing our model predictions to galaxy
data covering the full range of line indices considered in this paper. We 
are especially interested in contrasting the results based on blue and red
indices. Until recently, high quality data for blue Lick indices were
rare, if not entirely absent. The situation is changing quickly, as
surveys of galaxies in the local and distant universe call for the need
of a better understanding of the blue spectral region. A number of recent
studies provided blue Lick index measurements based on moderate-to-high
S/N spectra (e.g., Denicol\'o \etal 2005a,b, Rampazzo \etal 2005, Nelan
\etal 2005, S\'anchez-Bl\'azquez \etal 2006a). We choose to analyze the
stacked spectra from the Sloan Digital Sky Survey (SDSS) by Eisenstein
\etal (2003). Our choice is chiefly motivated by the the fact that the
stacked spectra are available publicly, so that we can perform our own
index measurements and, most importantly, apply our own corrections
for the effects of $\sigma$-broadening and Balmer line emission in-fill.
Moreover, the stacked spectra have very high S/N ($>$ 400/${\rm\AA}$).
Our procedure was described in Schiavon \etal (2006), but we briefly
discuss the main aspects here.

\begin{figure}
\plotone{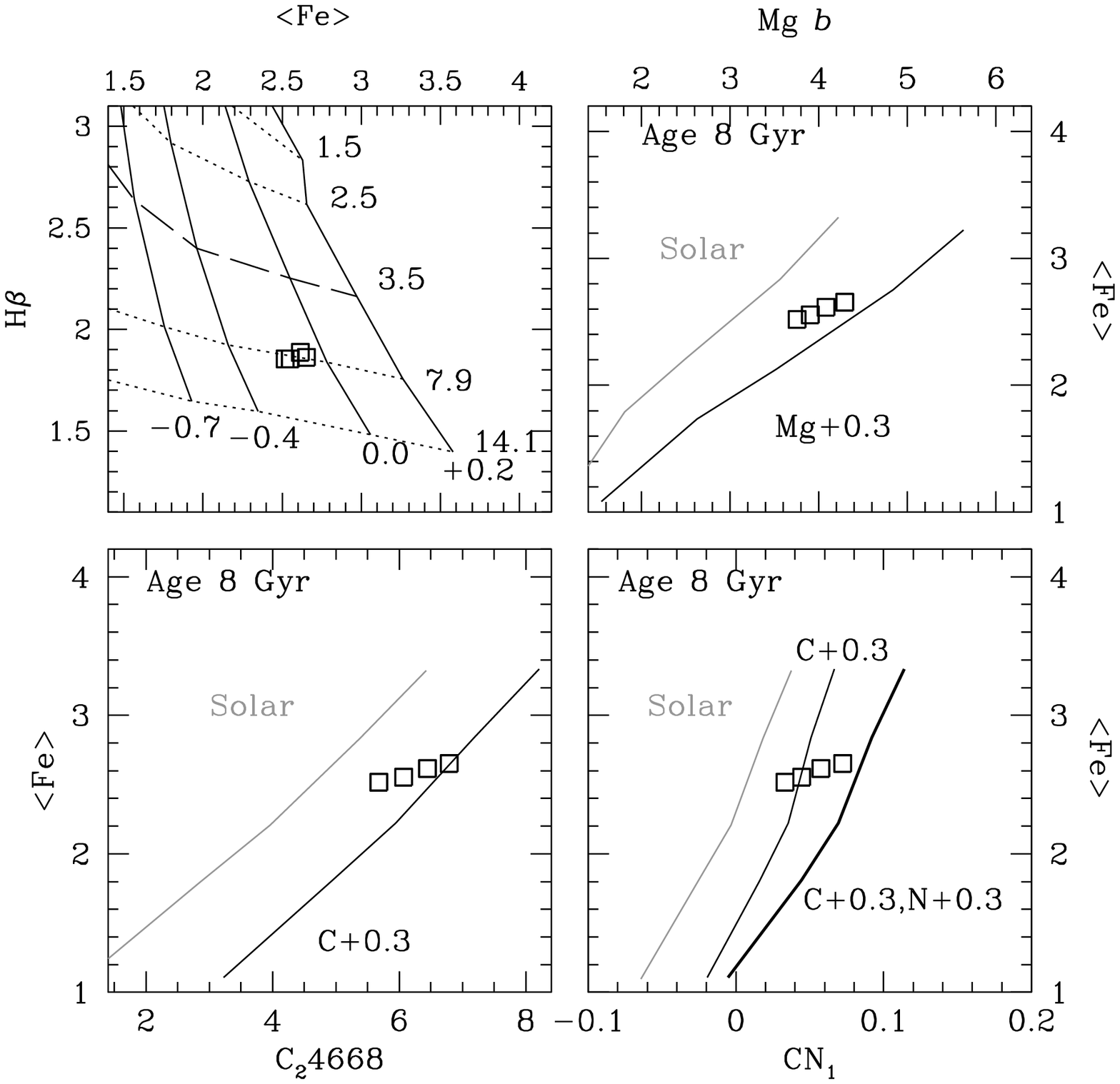}
\caption{Comparison of Lick indices measured in SDSS stacked spectra
(``All'' from Eisenstein \etal 2003) and our predictions for single
stellar populations. Galaxy absolute luminosity increases from left to
right (see $M_r$ values in Table~\ref{sdssdata}). Errorbars are always
smaller than symbol size. {\it Upper left:} The SDSS sample under study
has a mean age of $\sim$ 8 Gyr (regardless of luminosity) and [Fe/H]
just below solar (slightly dependent on luminosity) {\it Upper right:}
In this and the remaining panels, galaxies are compared with solar-scaled
(gray lines) and enhanced models (enhanced elements indicated in the
labels).  SDSS galaxies are Mg-enhanced, with more luminous galaxies
being slightly more enhanced. {\it Lower left:} Galaxies are C-enhanced,
with more luminous galaxies being more enhanced. {\it Lower right:}
Three model predictions are shown: solar scaled (gray), C-enhanced
(thin), and C,N-enhanced (thick).  C-enhancement alone is not enough
to match CN data, which require N-enhancement as well. More luminous
galaxies are clearly more enhanced here.
} 
\label{sdssinds} 
\end{figure}

The Eisenstein \etal (2003) sample consists of spectra of thousands
of early-type galaxies from SDSS selected in terms of color (red) and
morphology (bulge-dominated), with redshifts between $0.1 < z < 0.2$ (MAIN
sample).  The individual galaxy spectra sample a circular 3\arcsec\ region
centered on each galaxy and the spectral resolution is $\sim$ 170 km
s$^{-1}$ (FWHM).  Galaxies were assigned to bins according to luminosity
and environment, and all individual spectra within each bin were coadded
so as to generate very high S/N spectra as a function of luminosity and
environmental density. The main properties of the Eisenstein \etal sample
are summarized in their Table 1. We choose to analyze their ``All''
sample, which refers to galaxies binned only by luminosity, regardless
of environmental density.  Because each luminosity bin includes galaxies
of all environments, the number of coadded spectra per bin is always in
excess of 2,500, so that the S/N of each stacked spectrum is extremely
high. Our sample therefore consists of 4 spectra, one for each of the 4
luminosity bins in Eisenstein \etal (2003) sample (Table~\ref{sdssdata}
and Table 1 in Eisenstein \etal). We note that $M_r^\star \approx -20.8$
for this sample (Blanton \etal 2001), so that our results refer to bright
galaxies only.  The spectra were downloaded from D. Eisenstein's website
({\tt http://cmb.as.arizona.edu/$^\sim$eisenste/}).

\subsubsection{Index Measurements}

\begin{figure}
\plotone{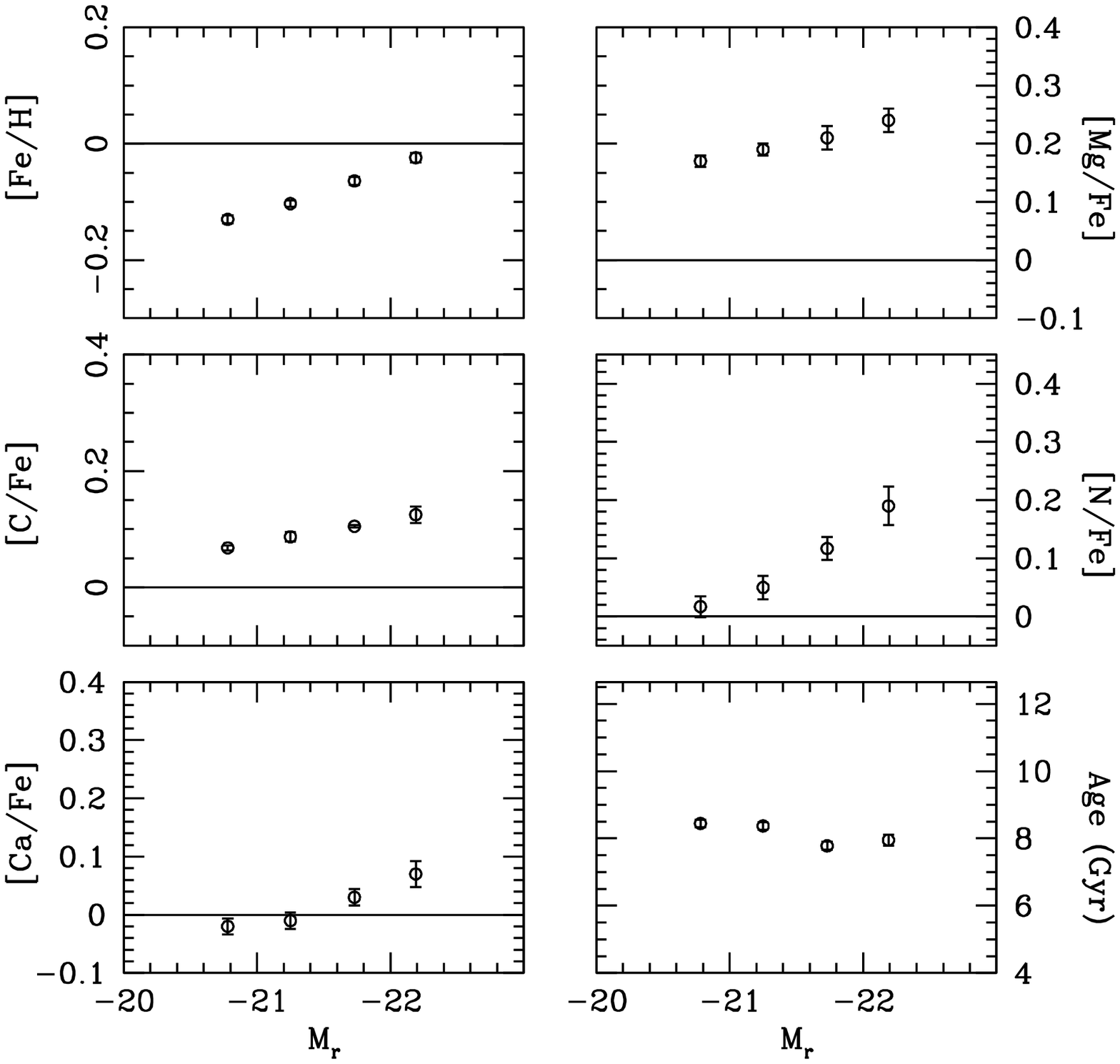}
\caption{Mean luminosity-weighted metal abundances and ages of the
stellar populations in the SDSS galaxies from the Eisenstein \etal (2003)
``All'' sample, as a function of absolute magnitude in the SDSS $r$
band. Note that $M^\star_r \approx -20.8$ at the involved redshifts. All
panels have a vertical scale of 0.5 dex. Iron abundances are slightly
below solar for all luminosities, and abundance ratios are above solar
for all elements in almost all luminosity bins. All abundances and all
abundance ratios appear to be correlated with luminosity at different
levels. Nitrogen is the most strongly enhanced element and also the
one whose enhancement appears to be the most strongly correlated with
galaxy luminosity. We find no correlation between mean age and luminosity
(but see Figure~\ref{sdssages}).
}
\label{sdssabunds}
\end{figure}

\begin{deluxetable*}{lcccccccccccccccccc}
\tabletypesize{\scriptsize}
\tablecaption{Data for stacked SDSS spectra of early-type galaxies from
Eisenstein \etal (2003)\label{sdssdata}}
\tablewidth{0pt}
\tablehead{
\colhead{$<M_r>$} & \colhead{[OII] $\lambda$3727} &
\colhead{$H\delta_A$} & \colhead{$H\delta_F$} & \colhead{CN$_1$} & \colhead{CN$_2$}&
\colhead{Ca4227} & \colhead{G4300} & \colhead{$H\gamma_A$} & \colhead{$H\gamma_F$}&
\colhead{Fe4383} & \colhead{C$_2$4668} & \colhead{$H\beta$} & \colhead{Fe5015} &
\colhead{Mg$_2$} & \colhead{Mg $b$} & \colhead{Fe5270} & \colhead{Fe5335} &
\colhead{$\sigma$ (km/s)} }
\startdata
--20.78 & 1.73 & --0.56 & 0.85 & 0.0332 & 0.0607 & 0.94 & 4.82 & --4.58 & --0.85 & 4.08 & 5.68 & 1.85 & 4.10 & 0.2288 & 3.76 & 2.63 & 2.41  & 190 \\
--21.25 & 1.49 & --0.79 & 0.76 & 0.0447 & 0.0725 & 0.94 & 4.89 & --4.77 & --0.95 & 4.20 & 6.07 & 1.85 & 4.32 & 0.2367 & 3.90 & 2.66 & 2.44 & 210 \\
--21.73 & 0.82 & --1.13 & 0.61 & 0.0575 & 0.0870 & 0.98 & 5.00 & --5.10 & --1.14 & 4.32 & 6.45 & 1.89 & 4.53 & 0.2464 & 4.08 & 2.72 & 2.51 & 240 \\
--22.19 & 0.06 & --1.45 & 0.47 & 0.0724 & 0.1046 & 1.04 & 5.17 & --5.41 & --1.33 & 4.57 & 6.79 & 1.86 & 4.85 & 0.2601 & 4.30 & 2.75 & 2.56 & 270 \\
error & 0.02 & 0.02 & 0.01 & 0.0005 & 0.0005 & 0.01 & 0.02 & 0.02 & 0.01 & 0.03 & 0.03 & 
0.01 & 0.04 & 0.0002 & 0.01 & 0.02 & 0.02 & 20 \\
\enddata
\end{deluxetable*}

\begin{deluxetable*}{lccccccc}
\tabletypesize{\scriptsize}
\tablecaption{Mean Ages (Gyr) and Metal Abundances from the Eisenstein \etal (2003) SDSS
Stacked Spectra \label{sdssresults}\label{sdssaa}} 
\tablewidth{0pt}
\tablehead{
\colhead{$<M_r>$} & 
\colhead{Age($H\beta$)} & 
\colhead{Age($H\delta_F$)} & 
\colhead{[Fe/H]} & 
\colhead{[C/Fe]} & 
\colhead{[N/Fe]} &
\colhead{[Mg/Fe]} & 
\colhead{[Ca/Fe]} 
}
\startdata
--20.78 & 8.2 & 3.8  & --0.17  &   0.08  &   0.12  &  0.19  &   0.02 \\
--21.25 & 8.1 & 4.6  & --0.15  &   0.12  &   0.17  &  0.21  &   0.06 \\
--21.73 & 7.4 & 5.2  & --0.10  &   0.13  &   0.28  &  0.22  &   0.10 \\
--22.19 & 7.6 & 6.1  & --0.08  &   0.15  &   0.38  &  0.26  &   0.17 \\
  error & 0.3 & 0.2  &   0.01  &   0.03  &   0.04  &  0.02  &   0.05 \\
\enddata
\end{deluxetable*}

All Lick indices of interest were measured in the stacked spectra using
{\tt lick\_ew}\footnote{see http://www.ucolick.org/$^\sim$graves}, an idl 
routine written by G. Graves (for a description, see Graves \& Schiavon
2006, in preparation). Before we can compare these measurements with
our models, we need to correct the line indices for the effect of
$\sigma$-broadening. That of course requires measuring first the velocity
dispersions directly in the stacked spectra. We proceeded as follows.
Velocity dispersions were measured through Fourier cross correlation
using the IRAF {\tt rv.fxcor} routine. The templates adopted were
model single stellar population spectra calculated for a range of ages
and metallicities, as described in Paper I. The choice of template is
very important and can introduce important systematic effects if not
carefully done. For each spectrum, a first guess of $\sigma$ is
made, the indices are corrected and initial values of age and [Fe/H] are
determined, if these values do not agree with those of the initial single
stellar population template, a new template is adopted with the estimated
age and [Fe/H] and the process is iterated until convergence is attained.

After velocity dispersions are determined, the indices can be
corrected to their standard $\sigma = 0$ values. These corrections were
estimated again using model spectra of single stellar populations with
appropriate ages and metallicities. Listings of such corrections for
a set of representative ages and velocity dispersions are provided in
Tables A14 through A17 in the Appendix.

The last step before we can compare models and data is the correction of
Balmer line indices for the effect of emission line in-fill. Balmer line
emission was estimated from the equivalent width of the [OII] $\lambda$
3727 ${\rm\AA}$ line adopting a ratio between that line and $H\alpha$
given by EW[OII]/EW($H\alpha$) = 6. This value was taken from Yan \etal
(2006), who studied the emission line properties of a large sample of
SDSS galaxies. They found that most line-emitting red galaxies in SDSS
tend to present LINER-like line ratios (see also, e.g., Phillips \etal
1986 and Rampazzo \etal 2005). Emission EWs for higher-order Balmer
lines are obtained from EW($H\alpha$) by assuming standard values
for the Balmer decrement (in the absence of reddening) and continuum
fluxes measured in the stacked spectra. In this way, one obtains
EW($H\beta$)/EW($H\alpha$)=0.36, EW($H\gamma$)/EW($H\alpha$)=0.19,
and EW($H\delta$)/EW($H\alpha$)=0.13. Corrected indices and the
velocity dispersions measured in the stacked spectra are listed in
Table~\ref{sdssdata}. The equivalent widths of the [OII] line, measured
according to the definition of Fisher \etal (1998), are also provided in
that Table\footnote{We note that, before estimating $H\alpha$ EWs from
measurements of the [OII] $\lambda$ 3727 line, one has to determine the
zero point of that line, i.e., what is the value measured in the spectrum
in the total absence of line emission. That value is not zero because
of the presence of absorption lines blended with [OII] emission and also
contaminating the index pseudo-continua. A
zero-point of $\sim$ 3.7 ${\rm\AA}$ was estimated by Konidaris \etal
(2006, in preparation) on the basis of stellar population synthesis
models.  That value therefore needs to be added to the EWs listed in
Table~\ref{sdssdata} before those EWs can be used to infer in-fill
corrections.}

Before attempting quantitative estimates of mean ages and metal abundances
of SDSS early-type galaxies, we compare the indices, measured as described
above, with our model predictions in Figure~\ref{sdssinds}. In all plots,
error bars are smaller than symbol sizes and galaxy luminosity increases
from left to right. In the upper left panel, data are compared with
solar-scaled models in the \fem-$H\beta$ plane. Because these indices are
essentially insensitive to abundance ratio effects, this plot allows us
to obtain a first estimate of mean age and [Fe/H]. The result is that the
stacked spectra indicate approximately the same mean age ($\sim$ 8 Gyr),
regardless of luminosity. On the other hand, [Fe/H] is just below solar,
and seems to be weakly correlated with luminosity.

In the remaining panels, SDSS early-type galaxies are compared with models
in index-index planes that are sensitive to the abundances of magnesium,
carbon, and nitrogen. In all these diagrams, solar-scaled (gray lines)
models are plotted along with models computed with the abundances of
a few key elements enhanced by +0.3 dex.  In the upper right panel,
the data are compared with solar-scaled and magnesium-enhanced models
for an age of 8 Gyr, in the \fem-Mg $b$ plane. This plot suggests that
early-type galaxies are enhanced in magnesium, with [Mg/Fe] slightly
below +0.3. A slight correlation between Mg-enhancement and luminosity
is apparent. In the lower left panel, data are compared with solar-scaled
and carbon-enhanced models in the \fem-C$_2$4668 plane. Again in this case
there is a clear indication of carbon-enhancement in early-type galaxies,
with a more clear trend of carbon-enhancement as a function of luminosity
than in the case of magnesium enhancement. Finally, in the lower right
plot, data and models are compared in the \fem-CN$_1$ diagram.  Because
the CN$_1$ index is sensitive to both carbon and nitrogen enhancements,
three models are displayed: solar scaled (gray) and carbon-enhanced (thin)
models, plus models where both carbon and nitrogen are enhanced (thick).
One can see that, while carbon-enhanced models do a good job of matching
the C$_2$4668 index, the same is not true for CN$_1$ data, which are
stronger than predicted by [C/Fe]=+0.3 models. In fact, matching CN$_1$
data requires that the abundance of nitrogen be also enhanced---thick
lines do reach the high CN$_1$ values observed.  One can also note that
the correlation between enhancement and luminosity here is even stronger
than in the case of the other plots.

\begin{figure}
\plotone{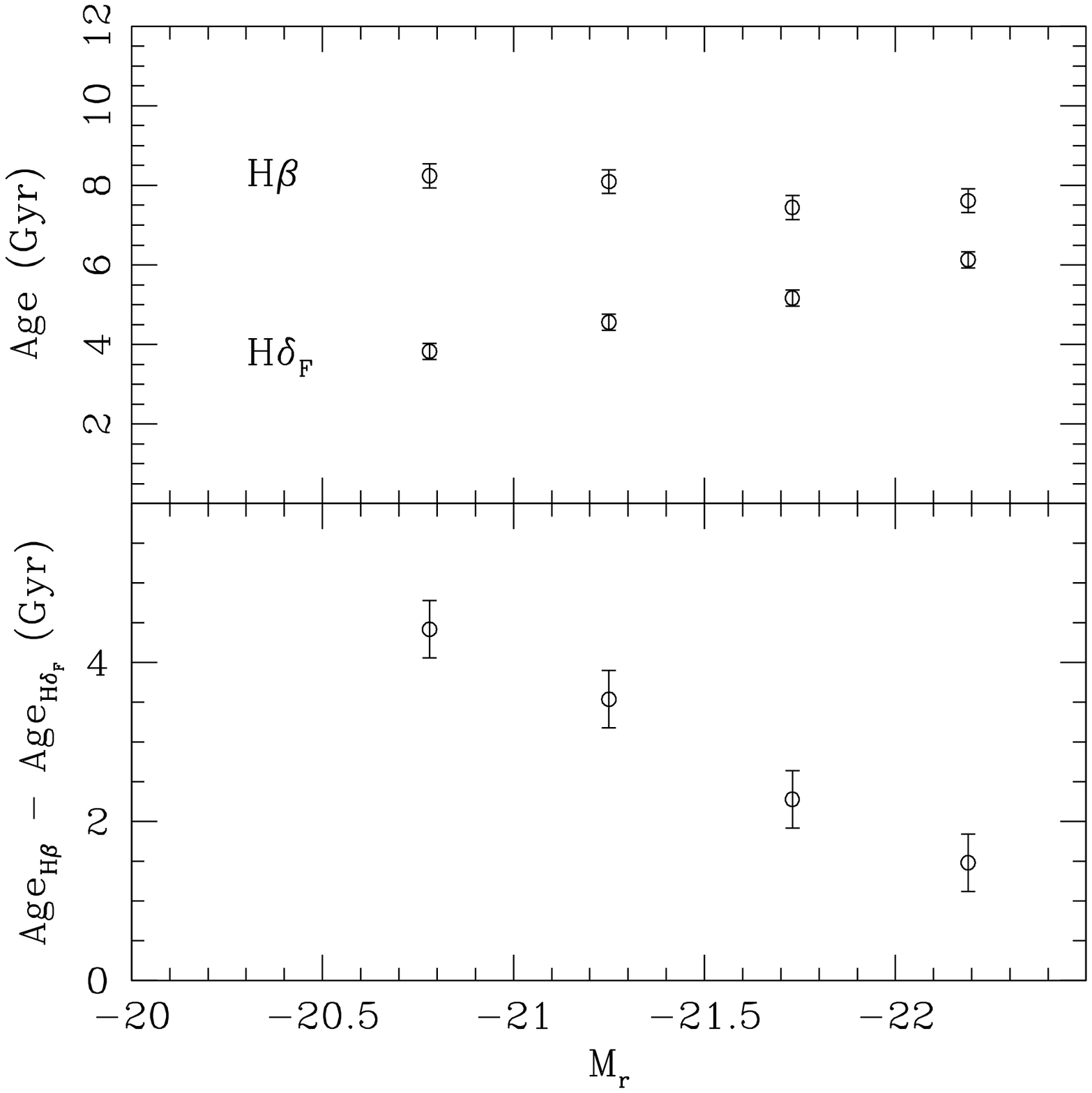}
\caption{Mean ages of SDSS galaxies in the Eisenstein \etal (2003)
sample, estimated from models including the abundance ratios
from Figure~\ref{sdssabunds} and Table~\ref{sdssaa}.  {\it Upper
panel:} Ages according to $H\beta$ and $H\delta_F$. As expected from
Figure~\ref{sdssinds}, mean ages according to $H\beta$ are $\sim$ 8 Gyr,
regardless of luminosity. On the other hand, ages according to $H\delta_F$
are substantially (2-4 Gyr) younger, with ages according to $H\gamma_F$
(not shown) lying somewhere inbetween. {\it Lower panel:} Differences
between $H\delta_F$-based and $H\beta$-based ages, as a function of
luminosity.  The mismatch between $H\beta$ and $H\delta_F$-based ages
is very luminosity dependent, being larger for lower luminosity galaxies.
} 
\label{sdssages} 
\end{figure}

\begin{figure}
\plotone{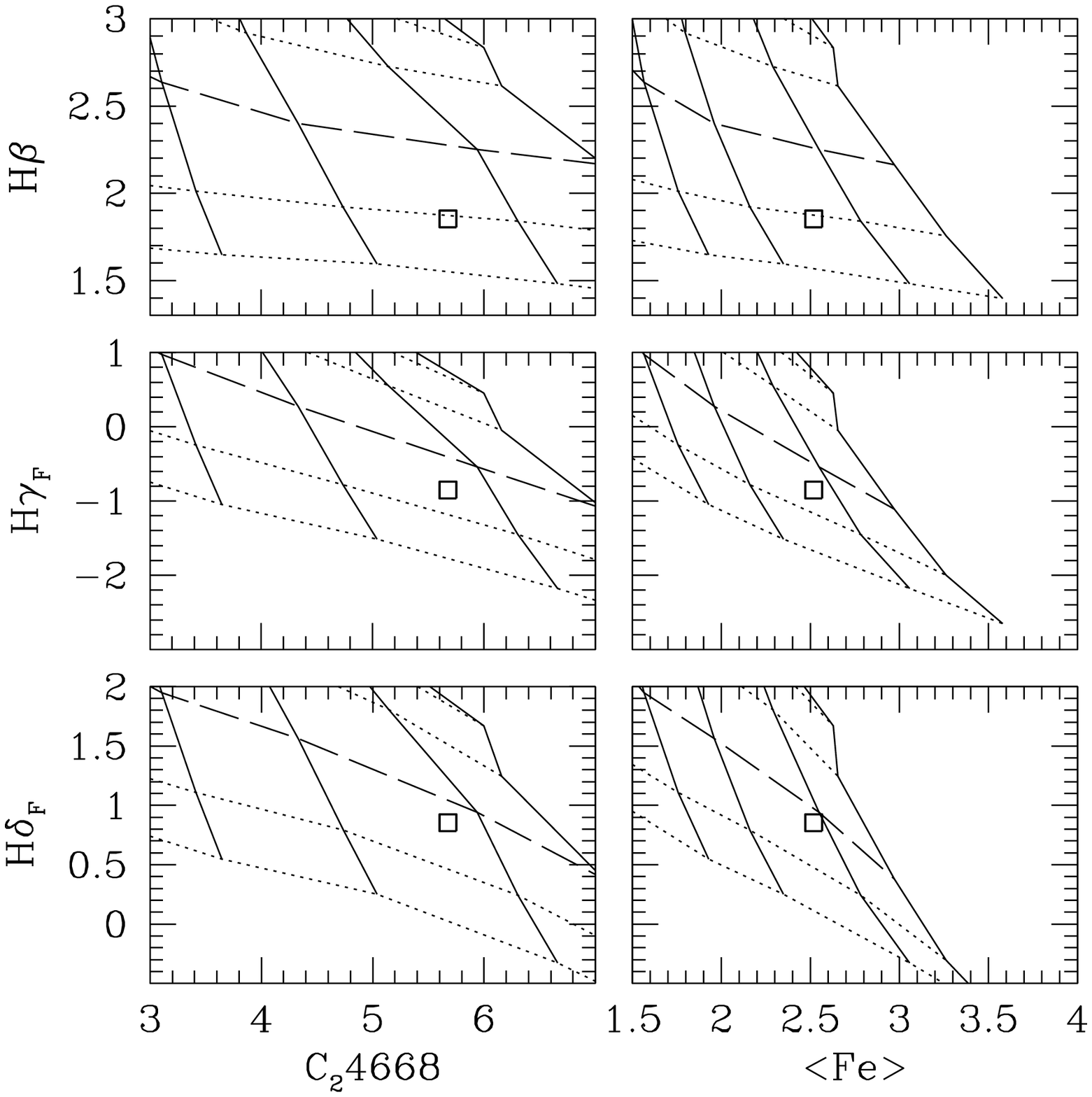}
\caption{Data for the lowest luminosity bin ($<M_r>$ = --20.8) compared
with single stellar population models in Balmer-metal line planes. Model
ages are the same as upper left panel of Figure~\ref{sdssinds}. To
help guiding the eye, the 3.5 Gyr models are connected by dashed
lines. The models were computed for the best-matching abundance pattern
(Table~\ref{sdssaa}), as can be seen by the good match to C$_2$4668
(see Section~\ref{chemevol} and Figure~\ref{sdssabunds}). Therefore,
abundance-ratio effects on $H\gamma_F$ and $H\delta_F$ are accounted
for. The bluer the Balmer line, the younger the mean age that is
estimated. The effect is independent of the metal index adopted.
}
\label{hbhdfe}
\end{figure}

With these qualitative results in mind, we apply our method described in
Section~\ref{amethod} in order to obtain quantitative estimates of mean
ages and metal abundances of the stellar populations of galaxies in the
Eisenstein \etal (2003) sample, on the basis of the line indices measured
in their spectra.  The results are listed in Table~\ref{sdssresults}
and shown in Figure~\ref{sdssabunds}, where resulting abundance ratios
and mean ages are plotted as a function of mean $r$-band absolute
magnitude. As expected from the discussion above, abundance ratios vary
strongly as a function of mean luminosity.  We will return to this issue
in Section~\ref{chemevol}, but first discuss the mean ages of early-type
galaxies in the next section.

\subsubsection{Mean Ages and the History of Star Formation of Early-type
Galaxies} \label{ages}

We start by looking at the mean ages estimated from the different
Balmer lines.  According to Figures~\ref{sdssinds} and \ref{sdssabunds},
early-type galaxies have luminosity-weighted mean ages of 8 Gyr, and
no correlation is found between mean age and luminosity.  The method
described in Section~\ref{amethod} uses $H\beta$ as the only age
indicator, due to its outstanding insensitivity to spectroscopic
abundance ratio effects (Section~\ref{aratios}).  It is reasonable to
expect, however, that other line indices should yield consistent ages
when data are compared with models computed for the right abundance
pattern. The latter is an important precaution, as the $H\delta$
and $H\gamma$ indices studied here are susceptible to spectroscopic
abundance ratio effects, so that these indices can only be used
for age estimates using models computed for the abundance pattern
estimated above (Figure~\ref{sdssabunds}).  
In Figure~\ref{sdssages}, ages estimated from $H\beta$ and
$H\delta_F$ are plotted against $<M_r>$. Ages according to $H\delta_F$
are consistently {\it lower}, with the difference ranging between 2
(25\%) and 4 Gyr (50\%). The differences are several times larger than
the internal error bars (which are very small, thanks to the exceedingly
high S/N of the stacked spectra).

\begin{figure}
\plotone{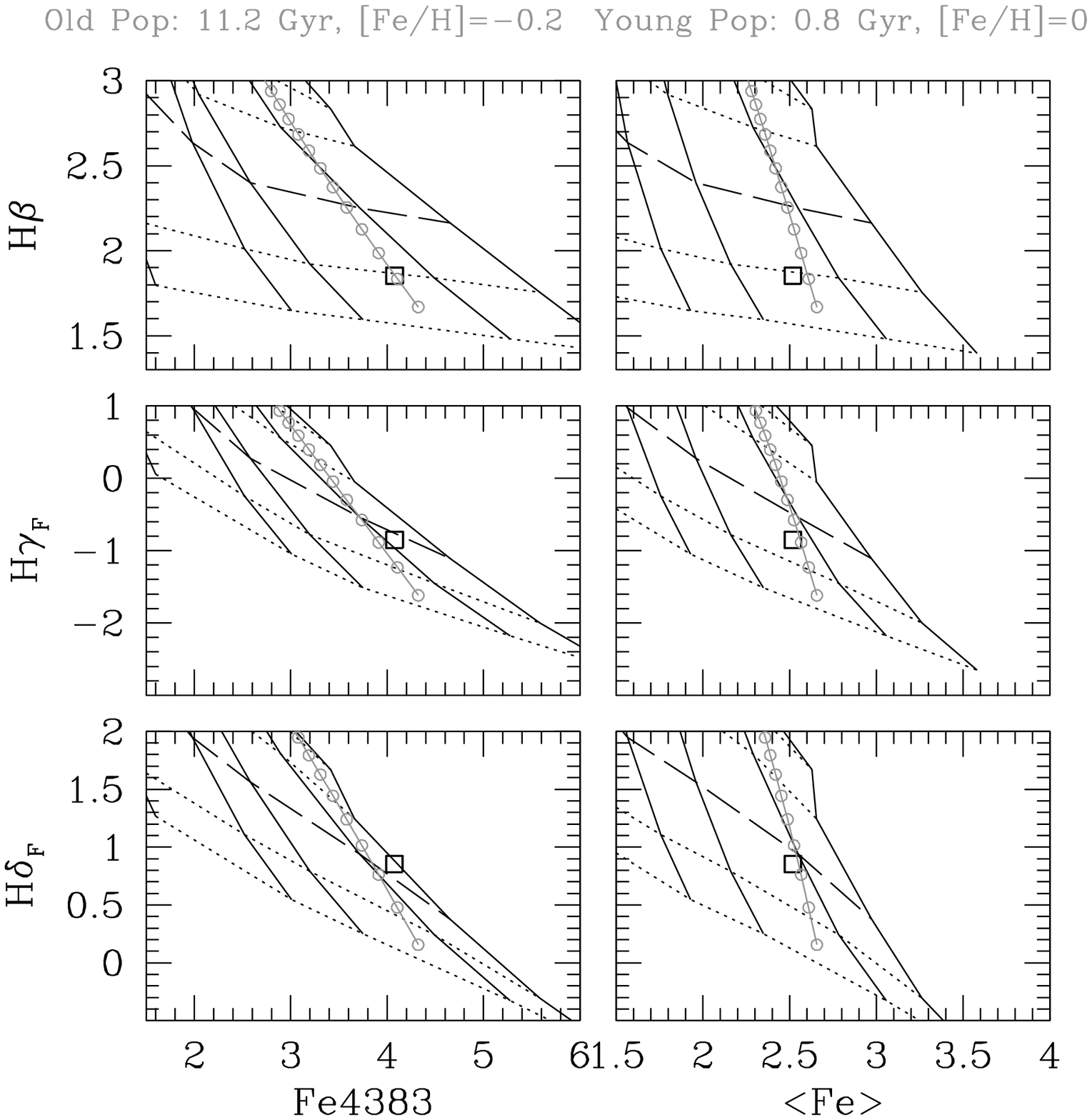}
\caption{Comparison between data for the lowest luminosity bin and a
family of composite two-population models, consisting of an old single
stellar population with an age of 11.2 Gyr and [Fe/H]=--0.2 combined
with a young population with age 0.8 Gyr and solar metallicity. The
contribution of the young population to the total mass budget is varied
in steps of 0.5\%, indicated by the gray open symbols. The largest the
contribution of the young population to the mass budget, the stronger
(weaker) the Balmer (metal) index for the composite population. A best
match is obtained when the mass allocated in the young population is
somewhere between $\sim$ 0.5 and 1\% of the total mass. A similarly
good match is obtained when the young component is $\sim$ 1 Gyr old,
but in that case the mass fraction is an order of magnitude higher. See
discussion in text.
}
\label{young}
\end{figure}

It is interesting to verify how this effect manifests itself in
index-index plots where Balmer and metal line indices are compared
with models. Unfortunately, we cannot compare models and data for the 
different luminosity bins all in the same set of plots, as $H\delta$ and
$H\gamma$ indices are strongly sensitive to abundance-ratio effects
requiring different sets of single-stellar population models to be computed
for each luminosity bin, 
so that plotting various sets of models in the same graph would make
things look rather confusing. Therefore, we henceforth focus on the
lowest luminosity bin, for which the age differences are the highest.
In Figure~\ref{hbhdfe}, data for the $<M_r> = -20.8$ bin are compared
to best-fitting models in \fem-Balmer and C$_2$4668 planes.  From
this figure, we learn that the age according to $H\delta_F$ is
$\sim$ 4 Gyr (the data are just below the 3.5 Gyr dashed line),
while that according to $H\beta$ is twice as old. The age according
to $H\gamma_F$ lies somewhere in-between those values. In order for
the $H\delta_F$-based age to agree with that based on $H\beta$,
$H\delta_F$ would have to be $\sim$ 0.45 ${\rm\AA}$ weaker than the
measured value which is entirely ruled out by the very small error bars
of the measurements (Table~\ref{sdssdata}). We note that the models
are a very good match to the C$_2$4668 and CN (not shown) indices,
so that in principle abundance-ratio effects on $H\delta_F$, which
are mostly due to carbon and nitrogen abundances, are accounted
for.  While the focus of this discussion is on $H\delta_F$, we
emphasize that this trend is quite general, in that {\it bluer}
Balmer lines tend to indicate {\it younger} ages.

\begin{figure}
\plotone{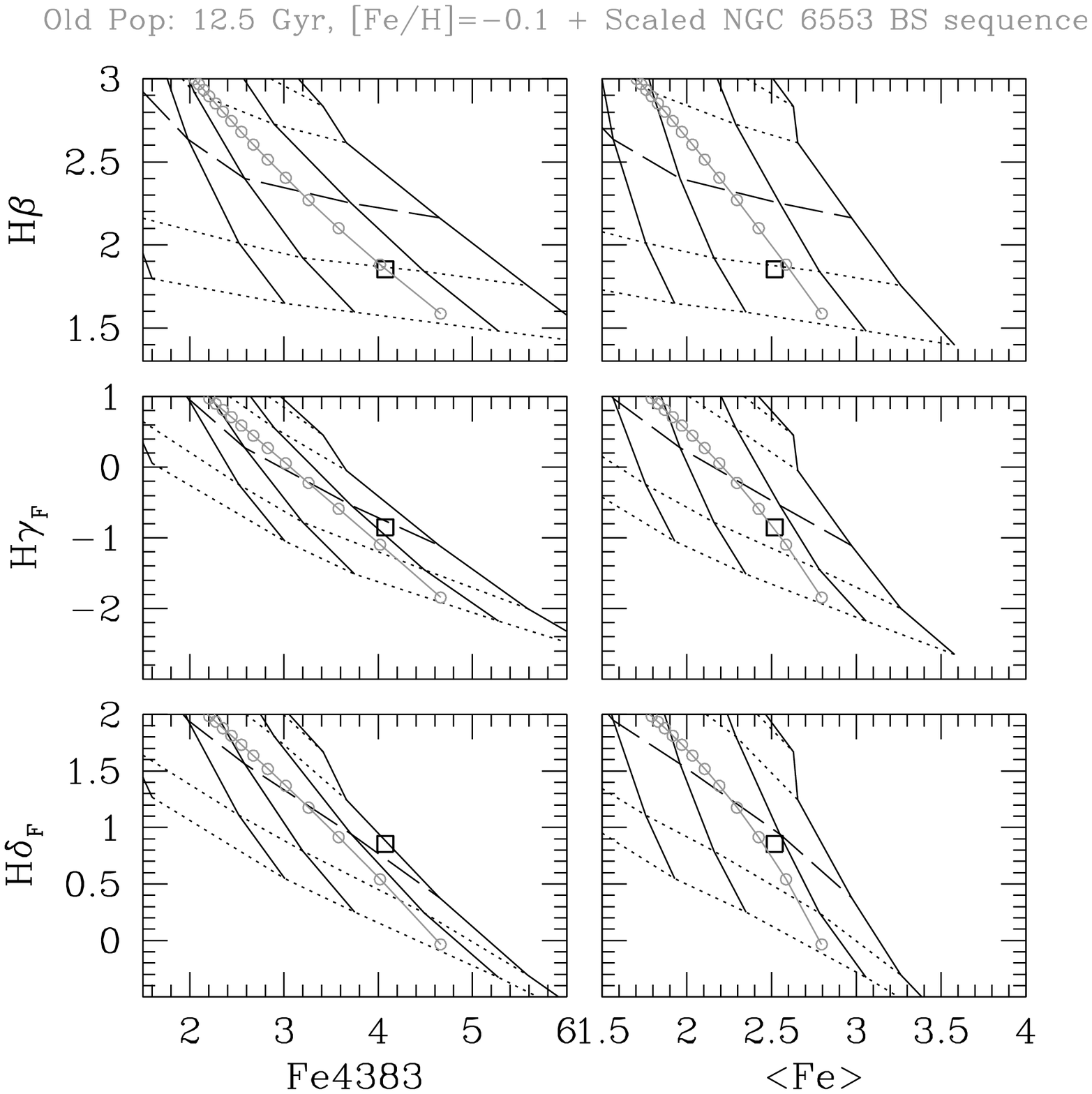}
\caption{Comparison between data for the lowest luminosity bin and a
family of composite two-population models, consisting of an old single
stellar population with an age of 11.2 Gyr and [Fe/H]=--0.2 combined
with blue stragglers from the metal-rich Galactic globular cluster
NGC~6553. The specific frequency of blue stragglers is varied in steps
of 100 stars per 10$^4$ $L_\odot$.  Adding blue stragglers to an old
population, one can match the data reasonably well, but only for extremely
high blue-straggler specific frequencies. See details in text.
}
\label{bstrag}
\end{figure}

No such inconsistency was found when models were confronted with
cluster data (Section~\ref{clust}), especially in the case of M~67,
whose age and metallicity are comparable with those of SDSS early-type
galaxies. Therefore, model inconsistencies cannot be blamed for the age
differences apparent in Figures~\ref{sdssabunds} through ~\ref{hbhdfe}.
We must seek other explanations. It is conceivable that the corrections
for emission line in-fill are introducing a systematic effect in our
Balmer line indices, since the relative corrections depend on assumptions
on the size of the Balmer decrement and on the absence of reddening.
However, it is very hard to attribute this discrepancy to errors in the
Balmer in-fill corrections, for the following reason. The corrections for
$H\beta$ and $H\delta_F$ were respectively $\sim$ 0.3 and 0.1 ${\rm\AA}$. 
Supposing we have overestimated the correction, the discrepancy between the 
two ages would be enhanced. In the extreme case, of no correction at all,
the age according to $H\beta$ would be $\sim$ 14 Gyr and that according to
$H\delta_F$ would be about 5 Gyr, which would increase the age discrepancy
from a factor of two to a factor of three. Admitting that our in-fill corrections
were instead underestimated, we can compute the size of the correction
needed to bring the $H\beta$ and $H\delta_F$-based ages into agreement.
The result is that the correction to $H\beta$ would have to be $\sim$
0.7 ${\rm\AA}$, or more than twice the correction adopted. That is
clearly ruled out by the small errors in the [OII] equivalent widths,
and continuum measurements, as well as reasonable assumptions for the
uncertainties in the Balmer decrement. Moreover, it would drive all the ages
to much lower values, below 3.5 Gyr for ages according to $H\beta$, which 
would be in stark disagreement with all previous work on stellar age estimates
in early-type galaxies. Therefore we conclude that systematic
errors in our emission line in-fill corrections cannot account for the
age discrepancies found.

\begin{figure}
\plotone{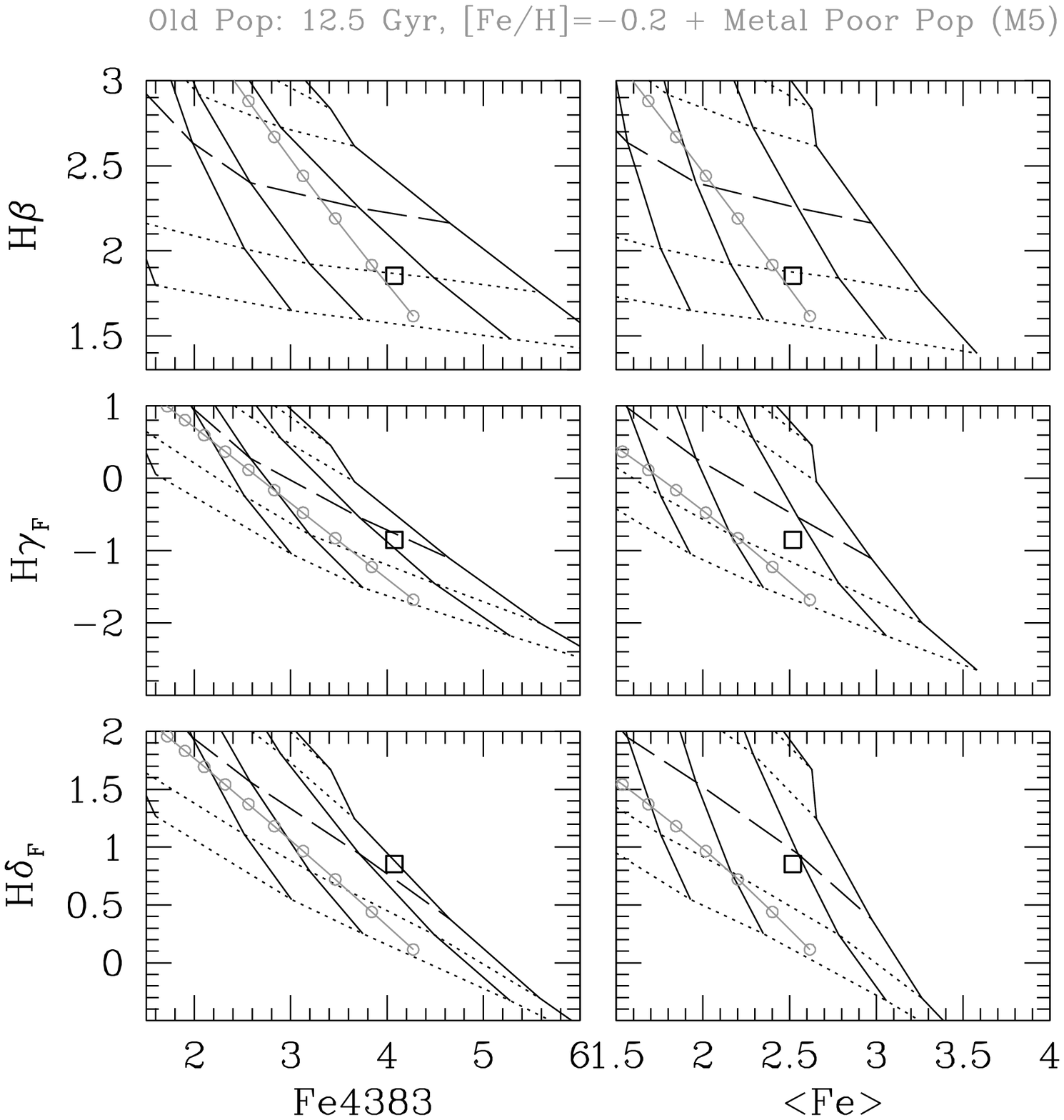}
\caption{Comparison between data for the lowest luminosity bin and a
family of composite two-population models, consisting of an old single
stellar population with an age of 11.2 Gyr and [Fe/H]=--0.2 combined
with an old metal-poor population, represented by data from a Galactic
globular cluster (M~5/NGC~5904). The contribution of the metal-poor
population to the total mass budget is varied in steps of 5\%, indicated
by the gray open symbols.  The match to the data is very poor, as these
models cannot reproduce the strengths of all Balmer lines.
}
\label{mpoor}
\end{figure}

In the absence of systematic effects introduced by our in-fill
corrections, another very likely possibility is that stellar
populations that are unaccounted by the models may be affecting the
line indices differentially.  It is conceivable that the trend seen
in Figure~\ref{hbhdfe} may be caused by the contribution of warm stars
(spectral types A to early-F), which are characterized by strong
Balmer lines. The contribution by these stars to the integrated
light peaks in the far blue, and falls steeply longward, so that
they tend to affect more strongly higher order Balmer lines, located
further into the blue.  A similar effect was found by Schiavon \etal
(2004b) in data for Galactic globular clusters. In that case, the
warm star component was identified with blue HB stars, but
young/intermediate-age stars are likely to have the same effect on
the integrated light of galaxies.  Abundance ratios that are not
accounted for in our models could in principle also generate this
type of systematic effect.  In what follows we consider four possible
explanations for this trend of mean ages with Balmer line: 1)
Contamination of the integrated light by a small fraction, by mass,
of a young/intermediate-age stellar population; 2) Contamination
by blue stragglers; 3) Metal-poor stellar populations with a blue
horizontal branch; and 4) Abundance-ratio effects. In what follows
we examine briefly each one of these scenarios.

{\it Young/Intermediate-age Stars:} In order to test this possibility,
we extended our model calculations to ages as low as 0.1 Gyr, for
[Fe/H]=0 and +0.2. These computations should be taken with caution,
because the fitting functions underlying our models did not take into
account the effect of metallicity for stars hotter than $\sim$ 7500
K. The latter effect, however, is likely to be small, so we can adopt
these calculations at least for a first examination of our working
hypothesis. We generate families of two-component models, whose input
parameters are the age and metallicity of the old component, the age of
the young component and the fractional contribution of the latter to
the total mass.  The abundance pattern assumed for the old component
is that listed in Table~\ref{sdssresults}, and both the metallicity
and abundance pattern of the young component are assumed to be solar.
The input ages considered for the old component varied between 10 and 14
Gyr and [Fe/H] varied between --0.4 and 0.  The age of the young component
was varied between 0.1 and 1.2 Gyr. The fractional contribution of the
young component to the total mass of the system was varied between 0
and 20\%. A $\chi$-square minimization procedure was adopted to find
the model that best matches the data for the Balmer and Fe indices.
For simplicity, indices which are prime indicators of other abundances
than iron were not included in the minimization, in order to avoid
having to include the abundance ratios of the young component as another
parameter in the two-component models.  In Figure~\ref{young} the gray
lines show the predictions of the family of two-component models,
containing the best fitting model.  The gray open circles indicate
increments of the mass fraction allocated to the young component in
steps of 0.5\%. One can see that, as that fraction increases, Balmer
lines get stronger and metal lines get weaker, due to the increasing
contribution to the integrated light by hot stars. It is also apparent
that the bluer indices are more strongly affected. For instance, when
the young stars make up 2\% of the total mass (fourth open circle from
the bottom up) $H\delta_F$ indicates a single stellar population age
lower than 3.5 Gyr, while the $H\beta$-based age would be $\sim$ 5 Gyr.
The old component of the best fitting model has an age of 11.2 Gyr and
[Fe/H]=--0.15, whereas the young component is 0.8 Gyr-old, contributing
$\sim$ 0.5--1\% of the total mass of the system.  We note that this
result is somewhat degenerate, as other families of models provide
an equally good fit. If one increases the age of the young component,
a good match can still be found to the data if its contribution to the
total mass budget is increased by an adequate factor. This degeneracy is
well known from studies of post-starbust galaxies and it can be broken
by introducing age indicators further to the blue (Leonardi \& Rose
1996). Most importantly, we note that the two-component model is a better
match to the data than single stellar population models. It is also a
better match to the data than the alternative models considered below.
We note that a similar result was obtained by S\'anchez-Bl\'azquez
\etal (2006a), who estimated younger ages from r.m.s. fitting of the
blue part of their galaxy data. They interpreted their result as being
due to a spread of stellar population ages in their sample galaxies.
Interestingly, we find that the age discrepancy is larger for lower
luminosity galaxies. If our interpretation is correct, this result tells
us that star formation was more extended in lower mass galaxies, which is
in agreement with findings by other authors (e.g., Caldwell \etal 2003,
Bernardi \etal 2005, Gallazzi \etal 2006) and seems to lend support to
the downsizing scenario (Cowie \etal 1996).

{\it Blue Stragglers:} Another family of warm stars which can
potentially enhance Balmer lines in integrated spectra of galaxies
are the blue stragglers. In paper III we showed that blue stragglers
have a strong impact on the integrated spectrum of M~67, by markedly
increasing the Balmer lines strengths in the cluster spectrum, which
indicates substantially younger ages ($\sim$ 1.5 Gyr) than the known 3.5
Gyr age of the cluster. We also found that, as in the case of the galaxy
spectra discussed here, the age of the cluster seems younger according
to bluer Balmer lines. In order to test the blue straggler hypothesis,
we perform the following test. We extracted colors and magnitudes of blue
stragglers from the VI color-magnitude diagram of the metal-rich globular
cluster NGC~6553 by Zoccali \etal (2001). Blue stragglers are known to
be very abundant in this cluster (e.g., Beaulieu \etal 2001). We choose
to use the Zoccali \etal data because these authors used observations at
different epochs in order to measure the proper motion of the cluster,
so that we can minimize foreground contamination of our blue straggler
selection.  Stars were considered as blue stragglers in NGC~6553 if they
met the following criteria: $V > 19.4$, $(V-I) < 1.68$, and relative
proper motion smaller than 0.1 arcsec (see Zoccali \etal for details).
The resulting blue straggler sample consisted of 50 stars, whose colors
and magnitudes were de-reddened and converted into absolute magnitudes
adopting E(V--I) = 0.90 (Barbuy \etal 1998) and (m-M)$_0$=13.64 (Zoccali
\etal 2001). The latter were used to compute effective temperatures and
surface gravities using the Lejeune \etal (1998) calibrations and assuming
[Fe/H]=--0.2 (Cohen \etal 1999, Barbuy \etal 2004) and 1$M_\odot$. The
latter stellar parameters were used to generate line indices using the
fitting functions presented in Section~\ref{ffs} for each blue-straggler
star. The latter were integrated using the stars' absolute magnitudes
and colors in order to generate integrated indices of the blue straggler
stars.  The blue straggler indices were then used to generate a family
of two-component models, by ``contaminating'' a reference old stellar
population model with blue straggler light, assuming a range of blue
straggler specific frequencies. The best match to the data was obtained
when the old component was assumed to be 12.5 Gyr old, with [Fe/H]=--0.1.
These models are compared with the observations in Figure~\ref{bstrag},
where the blue straggler specific frequency was varied from 0 to 1000 blue
stragglers per 10$^4$ $L_\odot$. The gray open symbols indicate steps
of 100/10$^4$ $L_\odot$.  Figure~\ref{bstrag} shows that adding blue
stragglers to an old stellar population can also account for the trends
observed in the data.  However, the specific frequency needed to match
the data is very high, ranging between 100 and 200 blue stragglers per
10$^4$ $L_\odot$.  As noted by Trager \etal (2000), typical specific
frequencies found in Galactic globular clusters, whose characteristic
core densities are much higher than those of early-type galaxies,
range from a few to a few tens of stars per 10$^4$$L_\odot$ (Ferraro,
Fusi Pecci \& Belazzini 1995).  The case of M~67 is also definitely very
extreme. In Paper III we saw that the spectrum of M~67 that includes blue
stragglers has much stronger Balmer lines than the BS-free spectrum. But
that spectrum includes star \# 6481 (ID from Montgomery \etal 1993), with
$T_{\rm eff} \simgreat 12,000K$ (Landsman \etal 1998) and several stars
hotter than $\sim$ 8500 K. Such high temperatures are not expected to be
found among lower mass blue stragglers characteristic of old, metal-rich,
stellar populations.  Moreover, M~67 has a specific blue straggler
frequency that far exceeds that of other open clusters with the same
central density (Landsman \etal 1998, Ahumada \& Lapasset 1995), which is
possibly the result of severe mass-segregation followed by evaporation
of low mass stars (Hurley \etal 2001). In summary, while we cannot rule
out the blue stragglers as the responsible for the apparent younger ages
we are getting from higher-order Balmer lines, we deem this a less likely
scenario, given the extreme conditions required to satisfy the data.

{\it Old, Metal-Poor Populations:} This scenario has been examined
recently by Trager \etal (2005), so we will address it very briefly.
It has been proposed by Maraston \& Thomas (2000) that an old metal-poor
stellar population component with a blue horizontal branch can account
for the strong Balmer lines observed in integrated spectra of early-type
galaxies. Moreover, their signature would be very similar to that
found here, where higher order Balmer lines are more strongly affected
(Schiavon \etal 2004b). In order to test this scenario, we adopt the
line indices measured in the integrated spectrum of M~5, which were
discussed in Section~\ref{clust}, and the integrated UBV colors of
the cluster, taken from Van den Bergh (1967). The old, metal-poor,
single stellar population thus produced is added to the old, metal-rich
fiducial adopted in the previous exercises, to produce the two-component
models shown in Figure~\ref{mpoor}. The open symbols indicate increments of
5\% of the mass fraction of the metal-poor component, which is varied
from 0 to 50\%. While this model matches reasonably well the data for
$H\beta$, it fails to satisfy the observations for $H\gamma_F$ and in
particular those for $H\delta_F$. (see Trager \etal 2005 for a more detailed
discussion). Varying the age of the prevalent old metal-rich population
allows accommodating the data for one Balmer lines, at the expense of
deteriorating the match to the others. We also tried using data for
other globular clusters in the Schiavon \etal (2005) spectral library
(NGC~2298 and 5986), but the quality of the match did not improve. The
reason, we speculate, is that the temperature distribution of the warm
stars in these old metal-poor stellar populations is not the one needed
to match the data.  We conclude that an old metal-poor component can
not explain the behavior of Balmer lines in the data.

{\it Abundance Ratios:} Our models are a good match to the galaxy data for
a number of spectral indices, which poses constraints on the abundances
of iron, magnesium, carbon, nitrogen and calcium. Those are the elemental
abundances that influence the most strongly the line indices studied
in this paper. However, abundances of other elements such as oxygen,
titanium, silicon, and sodium are largely unconstrained. Titanium,
silicon, and sodium are spectroscopically active in the atmospheres of
the stars that dominate the light in the systems under study, and the
sensitivities of line indices as a function of these elements (Tripicco \&
Bell 1995, Korn \etal 2005) can be used as a guide to gauge the possible
effects on the Balmer lines studied. These studies tell us that $H\beta$
is largely unaffected by abundance variations of elements other than iron,
so we turn to $H\delta_F$. In order to bring the $H\delta_F$-based age
into agreement with that based on $H\beta$, $H\delta_F$ would have to
be decreased by roughly 0.45 ${\rm\AA}$. Of the three spectroscopically
active elements, titanium is the one which affects $H\delta_F$ the most
strongly. Inspection of the Korn \etal (2005) tables shows that a +0.3 dex
variation in [Ti/Fe] causes $H\delta_F$ to drop by 0.31 ${\rm\AA}$ in the
spectrum of a super-metal-rich giant star, 0.09 ${\rm\AA}$ for a turn-off
star, and 0.22 ${\rm\AA}$ for a cool main sequence star. Therefore, in
order to increase the model predictions to match $H\delta_F$
through a change in [Ti/Fe], the latter would have to be {\it decreased} by
more than --0.7 dex (if we can trust linear extrapolations of the Korn
\etal sensitivities). This sounds extreme. Similar reasoning poses even
stronger lower limits on variations of silicon and sodium. Oxygen is
more complicated, because it indirectly affects the line strengths via
the dissociation equilibrium of CO and its impact on the strengths of
more spectroscopically active carbon molecules, like CN and CH. This
is accounted for in Korn et al.'s calculations, and consulting their
tables we verify that only extreme variations of [O/Fe] can explain the
$H\delta_F$ ages. However, oxygen can also play a role through its impact
on stellar evolution. We saw in Section~\ref{aeffect} that this effect
is stronger on $H\beta$ than on $H\delta_F$ and found a slight, similar
age trend on our comparisons with NGC~6528 data in Section~\ref{n6528},
which we attribute to a slight mismatch between the oxygen abundances of
the cluster and that of the isochrones. The trend is such that, in order
to match $H\beta$ without affecting $H\delta_F$ substantially, [O/Fe]
would have to be increased. Figure~\ref{alphaiso} shows how $H\beta$
changes when [O/Fe] varies from 0 to +0.5. Such a variation would account
for about half of the effect seen in Figure~\ref{hbhdfe}, so that
in order to account for the $H\delta_F$/$H\beta$ age mismatch [O/Fe] would
probably need to be raised to $\sim$ +1.0 (again if linear extrapolations
are to be trusted). While [O/Fe]=+1.0 may sound contriving, it cannot
be ruled out. However, abundance determinations of stars in our closest
proxy to the cores of early-type galaxies, the Galactic bulge field,
seem to indicate much lower values for [O/Fe] (Fulbright \etal 2005).
Therefore we conclude that, unless there is an important opacity source
missing in the Korn \etal (2005) tables, and/or the effect of oxygen abundances
on the evolutionary tracks of low-mass stars is quite substantially underestimated
in the Padova isochrones, abundance ratio effects are a unlikely explanation
for the differences we are finding between the ages determined from different
Balmer lines.

In summary, while neither the blue straggler nor the abundance ratio
scenarios can be completely ruled out, they seem to require extreme conditions in
order to satisfy the observations. We conclude that contamination of
the integrated light of the cores of galaxies by small mass fractions
of young/intermediate-age stellar populations is the {\it most likely}
scenario to account for the trends found. If this result is confirmed,
the inference is that stellar population synthesis models are now able to
constrain not only the mean ages of the stellar populations of galaxies
from their integrated light, but also their distribution. This has been
possible because the models adopted are extended to a wider baseline
than previously considered and also because they match the data for known
systems spanning the relevant range of stellar population parameters in
an accurate and consistent fashion. This result also implies that the
early-type galaxies studied have undergone a prolonged history of star 
formation, possibly with a small fraction of their stars being formed in the 
very recent past, as proposed in a number of previous works (e.g., O'Connell
1976, O'Connell 1980, Trager \etal 2000, to name a few). The ideal way
of testing this scenario involves extending model and data accuracy
towards an even wider baseline, preferably including the far blue and
the ultraviolet.


\subsubsection{Metal Abundances and the History of Chemical Enrichment
of Early-type Galaxies} \label{chemevol}

The abundance ratios obtained from application of our method described
in Section~\ref{amethod} to estimate stellar population parameters from
Lick indices are displayed in Figure~\ref{sdssabunds}. As expected from
Figure~\ref{sdssinds}, we find all galaxies to have iron abundances
slightly below solar. There is a correlation between [Fe/H] and $M_r$,
where [Fe/H] ranges between $\sim$ --0.15 for the faintest and just
below solar for the brightest bin. Abundance ratios of all the elements
studied relative to iron are solar or above solar, and are all correlated
with luminosity to different degrees.  That all elemental abundances
are correlated with luminosity is an expected result, which derives
from the more fundamental relation between mass and metallicity (e.g.,
Tinsley 1978). The interpretation of our results for abundance ratios
is more subtle.

First and foremost, the most striking result in Figure~\ref{sdssabunds}
is the behavior of nitrogen abundances, both in absolute terms and as
a function of luminosity. We caution that this result is sensitive to
the abundance of carbon, which might be subject to systematics due to
unknown oxygen abundances and/or theoretical uncertainties in the sensitivity of CN formation to carbon
abundance variations (see Korn \etal 2005 for a discussion).  Taking our
results at face value, we find nitrogen to be enhanced in this SDSS
sample, with [N/Fe] varying from just above solar, in the low luminosity
end, to $\sim$ +0.2 for the highest luminosity bin.  In the Galaxy,
[N/Fe] is essentially solar for stars in a wide range of iron abundance
(c.f. Figure~\ref{abrat}).  The only stellar systems known where [N/Fe]
departs strongly from solar are globular clusters, where its {\it mean}
value can be as high as $\sim$ +0.8.  (e.g., Cannon \etal 1998, Cohen
\etal 2002, Briley \etal 2004, Carretta \etal 2005, Lee 2005, Smith \&
Briley 2006).  In fact, stars in globular clusters present a wide range
of nitrogen abundances, and the distribution of this parameter seems to be
bimodal. Globular clusters in M~31 seem to be even more nitrogen-rich than
those in the Galaxy (Burstein \etal 1984, Li \& Burstein 2003). However,
the leading scenarios attempting to explain those nitrogen abundances tend
to invoke conditions that are only met in globular clusters (e.g. Cannon
\etal 1998, Beasley \etal 2004, Carretta \etal 2005).

Nitrogen is one of the elements whose history of enrichment is the most
uncertain.  The main source of nitrogen enrichment seems to be mass loss
by intermediate and low mass AGB stars (e.g., Timmes \etal 1995, Henry \&
Worthey 1999, Chiappini, Romano \& Matteucci 2003, Gavil\'an, Moll\'a \&
Buell 2006), but explosive nucleosynthesis in high-mass stars can also
contribute nitrogen, especially at early times (e.g., Chiappini, Matteucci
\& Ballero 2005). The strong correlation of [N/Fe] with luminosity and
(presumably) metallicity, seems to be indicating a strong secondary
contribution to the enrichment of nitrogen in the sample studied (e.g.,
Tinsley 1979). If this interpretation is correct, nitrogen abundances may
pose a novel constraint on the timescale for star formation in early-type
galaxies. Secondary contribution to nitrogen enrichment is predominantly
due to stellar winds from AGB stars with zero-age-main-sequence masses
in the 4-8 $M_\odot$ range (Chiappini \etal 2003), whose lifetimes,
according to the Geneva evolutionary tracks (Lejeune \& Schaerer 2001),
are of the order of 40-200 Myr. If the strong dependence of [N/Fe] on
galaxy luminosity (and, presumably, mass) is a signature of secondary
nitrogen enrichment, star formation in early-type galaxies must have
lasted for at least 40-200 Myr in order for nitrogen contributed by
these intermediate mass stars to be incorporated into new generations
of stars. Therefore, our result for the run of nitrogen abundances as
a function of galaxy luminosity may be setting a lower limit for the
duration of star formation in early-type galaxies.  This is a {\it
new} constraint on the timescale of star formation in these systems.
It is clearly possible to obtain tighter constraints on the basis of
calculations from chemical evolution models, taking into consideration
up-to-date stellar yields as a function of mass and a realistic IMF.


We find that all galaxies in the sample under study are magnesium-enhanced
([Mg/Fe]$>$0), and that more luminous galaxies are more enhanced than
their fainter counterparts. As discussed in Section~\ref{tragersect},
this is a well known result, commonly interpreted as being due to the
fact that the bulk of the stars in these galaxies were formed in a major
event which lasted no longer than $\sim$ 1 Gyr, so that supernova type
Ia could not contribute significantly to chemical enrichment (e.g.,
Wheeler \etal 1989). The correlation between [Mg/Fe] and luminosity has
also been found by other authors (e.g., Trager \etal 2000, Denicol\'o
\etal 2005, Thomas \etal 2005, Mendes de Oliveira \etal 2005) and is
usually interpreted as being due to shorter star formation time-scales
in more massive galaxies. This result is in sync with our finding that
star formation in lower luminosity galaxies seems to have lasted longer
than in their more luminous counterparts (Section~\ref{ages}), based
on the $H\delta_F$-based ages. IMF variations as a function of galaxy
mass could also account for these trends, but this hypothesis is more
difficult to test.

As in previous studies (e.g., Vazdekis \etal 1997, Henry \& Worthey
1999, Saglia \etal 2002, Thomas \etal 2003b, Prochaska \etal 2005)
we find that calcium is not as enhanced relative to iron as magnesium.
However, unlike most previous studies, we find that [Ca/Fe] seems to
be well correlated with galaxy luminosity. As shown by Prochaska \etal
(2005), and discussed in Section~\ref{carbon} the Ca4227 index is very
strongly affected by CN. Once this effect is accounted for, either by
redefining the index in order to minimize CN contamination (Prochaska
\etal 2005) or by estimating the impact of CN lines on the index on the
basis of spectrum synthesis calculations (this work), calcium is seen
to be as correlated with galaxy luminosity as magnesium, which is the
other $\alpha$-element in our analysis.



\section{Conclusions}

We presented a new set of models for the integrated absorption line
indices and UBV magnitudes of single stellar populations. The models are
based on the Jones (1999) spectral library, for which we determined a
new set of homogeneous and accurate stellar parameters and a new set of
fitting functions for Lick indices. Adopting theoretical isochrones from
the Padova group, a set of reliable calibrations relating fundamental
stellar parameters and colors, stellar abundances from the literature,
which characterize the abundance pattern of the library stars, and
sensitivity tables describing how line indices vary as a function
of elemental abundance variations, we produced a new set of model
predictions for Lick indices in single stellar populations. The models
were extensively compared with superb data for Galactic clusters and
nearby galaxies.  Our main results can be summarized as follows:

1) Stellar parameters (effective temperature, surface gravity,
and iron abundance) were determined for the 624 stars from the Jones
(1999) library, based on semi-empirical calibrations of photometric and
spectroscopic indices and fundamental stellar parameters. These parameters
were contrasted with previous determinations from the literature and were
found to be devoid of significant systematic effects. All stars for which
substantial deviations from determinations by other groups were found were
carefully examined and the best set of parameters was chosen. We show that
accurate stellar parameters lie in the core of our ability to predict
accurately the integrated indices of single stellar populations. The
abundance pattern of the stellar library was characterized by surveying
the literature for determinations of abundances of such key elements as
iron, oxygen, carbon, nitrogen, magnesium, titanium, silicon, and calcium.

2) The above stellar parameters were combined with equivalent widths of
line indices measured in the spectra of Jones library stars. This new
equivalent width system is based on flux-calibrated spectra smoothed to
the same resolution as that of the Lick/IDS system. Index measurements in
the new system are substantially more accurate than those upon which the
old Lick/IDS system is based. The two systems differ by small zero-point
shifts. Polynomial fitting functions describing the behavior of 16 line
indices as a function of effective temperature, surface gravity and iron
abundance were computed. We compare our new fitting functions to those
by Worthey \etal (1994) and found small but important improvements for
some indices, which are mostly due to the best quality of the spectra
and of the stellar parameters adopted.

3) The fitting functions are combined with theoretical isochrones from the
Padova group in order to produce predictions of integrated line indices
for single stellar populations. These are compared with high S/N data
from four Galactic clusters spanning a representative range of ages,
metallicities, and abundance patterns. We successfully match essentially
all 16 line indices for the known input parameters (age, metallicity,
abundance pattern, mass function) for each of the clusters considered.
Our predictions for Fe indices, the Mg indices, C$_2$4668, the CN indices,
and Ca4227 match the cluster data for [Fe/H], [Mg/H], [C/H], [N/H], and
[Ca/H] within 0.1 dex of the known cluster values. Spectroscopic ages
based on all Balmer line indices agree with those based on CMD analyses
to within 1--2 Gyr for all clusters. We showed that the Ca4227 index
is strongly influenced by CN lines. It is also shown that consistent
results can only be obtained for Mg$_2$ and Mg $b$ if the appropriate
(dwarf-depleted) mass function for the clusters is adopted, which means
that a combination of these indices can potentially be used to set
constraints on the IMF in the low mass regime.

4) Combining our single stellar population models with the abundance
ratios for the library stars and the sensitivities of line indices to
elemental abundances from Korn \etal (2005), we computed models for
single stellar populations with several different abundance patterns.
We devised a method that employs these models in order to estimate mean
stellar ages and abundances of iron, magnesium, calcium, nitrogen and
carbon.  These models and method were tested against the observations of
Galactic clusters with known abundance patterns with very satisfactory
results. The models predict the cluster elemental abundances in agreement
with the known values to within $\sim$ 0.1 dex. Spectroscopic ages
agree with those based on analysis of cluster CMD data to within 1--2
Gyr, for {\it all} Balmer line indices considered. Using our model
predictions for variable abundance ratios, we found that models with
[Fe/H] = --0.1 $\pm$ 0.1 and a mild $\alpha$-enhancement, [$\alpha$/Fe]
$\sim$ +0.1, are a better match to the data on NGC~6528 than those with
higher enhancement. Matching the G band and CN indices, we found [C/Fe]
$\sim$ -0.1 and [N/Fe] $\sim$ +0.5 as mean values for the stars in the
core of NGC~6528. This mean abundance pattern resembles that of other
well-known Galactic globular clusters such as 47~Tuc, M~71, and M~5,
among others. This might suggest the carbon-nitrogen abundance dichotomy
that characterizes these clusters is also present in NGC~6528.

5) The very good match obtained to data on Galactic clusters encouraged
us to apply our models to observations of nearby galaxies. Initially,
models were compared with the high quality measurements of \fem,
Mg $b$, and $H\beta$ by Trager \etal (2000) for a sample of nearby
galaxies. We reproduce their results, finding mean [Mg/Fe] $\sim$ +0.3,
a spread in mean ages between 2.5 and 14 Gyr. We also found that [Fe/H]
seems to decrease and [Mg/Fe] to increase when one goes from strong to
low-$H\beta$ galaxies, in agreement with previous findings.

6) Our models were compared to high S/N measurements taken on stacked SDSS
spectra of early-type galaxies from Eisenstein \etal (2003). Applying
our method to determine mean ages and abundance ratios from Lick
index measurements, we were able to estimate the abundances of iron,
carbon, nitrogen, magnesium and calcium in these galaxies. We found
that, while iron abundances are slightly below solar, the galaxies
are enhanced relative to iron in all other elements. Iron abundances
and all abundance ratios are shown to be positively correlated with
luminosity. Unlike previous studies, we find that [Ca/Fe] is slightly
enhanced in the sample studied.  Among the elements studied, nitrogen
is the one displaying the most conspicuous behavior, as it is the one
whose enhancement is most strongly correlated with luminosity. This result
might be subject to systematics due to uncertainties in the response of CN
formation to carbon abundance variations, but we nevertheless speculate
that it might be indicating the presence of a strong contribution by
a secondary production mode. If this interpretation is correct, our
result poses a constraint on the lower limit for the timescale for star
formation in early-type galaxies (40--200 Myr). Magnesium is also seen
to be enhanced and its enhancement is also correlated with luminosity,
which is consistent with our finding that lower luminosity galaxies form
stars for longer time periods (see below). More work is clearly needed
to interpret these abundance ratio results.

7) Comparing the spectroscopic ages inferred from $H\beta$ and
higher-order Balmer lines we found a systematic trend whereby the
latter, especially $H\delta$, yield systematically younger ages than
the former. Moreover, this discrepancy is stronger for lower luminosity
galaxies. This is in strong contrast with the results from our analysis
of cluster data, where we found that spectroscopic ages from all Balmer
lines were remarkably consistent. We examined four possible scenarios
to account for the observations: abundance-ratio effects, contamination
by a small fraction of young/intermediate-age stars, by blue stragglers,
and by metal-poor stellar populations with a blue horizontal branch. We
argue that the metal-poor scenario cannot match the data and that the
blue-straggler and abundance-ratio scenarios require extreme conditions
to do so.  Therefore, we conclude that the most likely explanation is a
spread in the ages of the stellar populations in early-type galaxies. The
implications are two-fold. On one hand, if this scenario is confirmed,
stellar population synthesis will have evolved to a stage where it is now
able to constrain not only the mean ages of stars in galaxies, but also
their distribution. On the other, it implies that small amounts of star
formation have occurred in the recent past in these nearby early-type
galaxies. Our results suggest that on average lower luminosity galaxies
formed stars until more recently than their more luminous counterparts,
which lends further support to the ``downsizing'' scenario.  Extending the
accuracy of models and data blue-ward is the best way of further testing
these results, posing stronger and more refined constraints on the
history of star formation in early-type galaxies.

\acknowledgments

This work was initiated when I was a postdoc at Lick Observatory, as
a member of the DEEP group, first as a Gemini Fellow, then as a CNPq
fellow. I would like to thank my parents, Ennio and Norma, without whose
relentless support this work would never have become a reality. Ingrid
Gnerlich is thanked for her permanent encouragement. I would also like
to thank Sandy Faber, Jim Rose, Bob O'Connell, Beatriz Barbuy, David
Koo, Bob Rood, and Ruth Peterson for inspiration, encouragement, and
continuous support. Nico Cardiel is thanked for initial suggestions on
the calculation of fitting functions, Daniel Thomas for making available
the Korn \etal sensitivities in advance of publication, and Achim Weiss
for expert advice on evolutionary tracks. Many thanks go also to Jenny
Graves, who provided the $C_2$4668 measurements in Indo-US spectra
and implemented the models presented here in a slick IDL code. An
anonymous referee is thanked for a thorough and very helpful reading of
the first version of this paper.  This work has made extensive use of
the Simbad and ADS databases. The author acknowledges financial support
from the National Science Foundation through grant GF-1002-99, from
the Association of Universities for Research in Astronomy, Inc., under
NSF cooperative agreement AST 96-13615, from the NSF, through grant AST
00-71198 to the University of California, Santa Cruz, from CNPq/Brazil,
under grant 200510/99-1, and finally from HST Treasury Program grant
GO-09455.05-A to the University of Virginia.

\appendix

\section{Tables}

\begin{deluxetable*}{lrrrrrrrrrrr} 
\label{stars}
\tabletypesize{\scriptsize}
\tablecaption{The stellar library: fundamental parameters and line index
measurements. 
}
\tablewidth{0pt}
\tablehead{
\colhead{ID} & 
\colhead{$T_{\rm eff}$} & 
\colhead{$\log g$} &
\colhead{[Fe/H]}   &
\colhead{$M_{\rm bol}$} &
\colhead{$M_{\rm V}$}  &
\colhead{$(B-V)_0$}   &
\colhead{$H\delta_F$} &
\colhead{$H\delta_A$} &
\colhead{CN$_1$}   &
\colhead{CN$_2$}   &
\colhead{Ca4227}  \\
                &
\colhead{G4300}  &
\colhead{$H\gamma_F$} &
\colhead{$H\gamma_A$} &
\colhead{Fe4383}  &
\colhead{C$_2$4668}  &
\colhead{$H\beta$} &
\colhead{Fe5015}  &
\colhead{Mg $b$}  &
\colhead{Mg$_2$}  &
\colhead{Fe5270}  &
\colhead{Fe5335}  }
\startdata
G12-24     &     5551 &   4.37 &  -0.11 &   4.82 &   4.87 &   0.67 &  -0.588 &   0.794 &  -0.052 &  -0.036 &   0.764 \\
 &   5.340 &  -4.333 &  -0.615 &   3.139 &   3.139 &   2.500 &   3.547 &   3.333 &   0.130 &   2.058 &   1.698 \\
G82-12     &     5610 &   3.93 &  -0.32 &   3.68 &   3.70 &   0.65 &   0.442 &   1.224 &  -0.071 &  -0.055 &   0.740 \\
 &   5.066 &  -3.539 &  -0.177 &   2.634 &   2.934 &   2.508 &   3.249 &   2.522 &   0.125 &   1.846 &   1.539 \\
G43-33     &     5774 &   3.98 &  -0.50 &   3.69 &   3.62 &   0.54 &   2.025 &   1.852 &  -0.082 &  -0.063 &   0.404 \\
 &   4.088 &  -1.088 &   1.124 &   1.347 &   1.498 &   3.019 &   2.352 &   1.434 &   0.075 &   1.195 &   0.909 \\
G44-6      &     5434 &   4.27 &  -0.46 &   4.79 &   4.90 &   0.60 &   0.633 &   1.275 &  -0.076 &  -0.061 &   0.635 \\
 &   4.912 &  -3.210 &  -0.273 &   2.022 &   2.402 &   2.362 &   2.425 &   2.762 &   0.119 &   1.415 &   1.103 \\
G74-5      &     5463 &   4.29 &  -0.91 &   4.92 &   4.99 &   0.57 &   1.155 &   1.494 &  -0.075 &  -0.059 &   0.520 \\
 &   4.280 &  -2.231 &   0.229 &   1.553 &   0.783 &   2.303 &   1.674 &   2.313 &   0.092 &   1.135 &   0.856 \\
\enddata
\tablenotetext{1}{The complete version of this table is in the electronic edition of
the Journal.  The printed edition contains only a sample.}
\end{deluxetable*}

\begin{deluxetable*}{rrrrrrrrrrrrrrrrrr} 
\label{basemod1}
\tabletypesize{\scriptsize}
\tablecaption{Lick indices for single stellar populations: the base
models computed with solar-scaled isochrones.
}
\tablewidth{0pt}
\tablehead{
\colhead{Age} & 
\colhead{[Fe/H]} & 
\colhead{$H\delta_A$} &
\colhead{$H\delta_F$} &
\colhead{CN$_1$}   &
\colhead{CN$_2$}   &
\colhead{Ca4227}  &
\colhead{G4300}  &
\colhead{$H\gamma_A$} &
\colhead{$H\gamma_F$} &
\colhead{Fe4383}  &
\colhead{C$_2$4668}  &
\colhead{$H\beta$} &
\colhead{Fe5015}  &
\colhead{Mg $b$}  &
\colhead{Mg$_2$}  &
\colhead{Fe5270}  &
\colhead{Fe5335}  }
\startdata
  0.10 & 0.0 & 6.9379 &  5.8308 & -0.1619 & -0.1213 & -0.0086 & -1.1189 &  6.8475  & 5.8282 &  0.1518 &  0.6257 &  5.6665 &  1.4216 &  0.6712 &  0.0559 &  0.8318 &  0.7518 \\
  0.10 & 0.2 & 7.9228 &  6.4156 & -0.1788 & -0.1360 &  0.0330 & -1.7570 &  7.9978  & 6.4615 & -0.1183 &  0.4930 &  6.3683 &  1.2715 &  0.7950 &  0.0534 &  0.8059 &  0.7555 \\
  0.11 & 0.0 & 7.3434 &  6.0405 & -0.1689 & -0.1272 &  0.0049 & -1.3225 &  7.1792  & 6.0209 &  0.0856 &  0.6180 &  5.9275 &  1.4137 &  0.7218 &  0.0548 &  0.8486 &  0.7458 \\
  0.11 & 0.2 & 8.3321 &  6.6142 & -0.1873 & -0.1437 &  0.0380 & -1.9326 &  8.3291  & 6.6514 & -0.1928 &  0.4743 &  6.5600 &  1.2435 &  0.7854 &  0.0516 &  0.7924 &  0.7217 \\
  0.13 & 0.0 & 7.8073 &  6.2780 & -0.1760 & -0.1329 &  0.0394 & -1.5656 &  7.6137  & 6.2436 & -0.0435 &  0.6610 &  6.2368 &  1.4856 &  0.8159 &  0.0505 &  0.8423 &  0.7302 \\
\enddata
\tablenotetext{1}{The complete version of this table is in the electronic edition of
the Journal.  The printed edition contains only a sample.}
\end{deluxetable*}

\begin{deluxetable*}{rrrrrrrrrrrrrrrrrr} 
\label{basemod2}
\tabletypesize{\scriptsize}
\tablecaption{Lick indices for single stellar populations: the base
models computed with $\alpha$-enhanced isochrones.
}
\tablewidth{0pt}
\tablehead{
\colhead{Age} & 
\colhead{[Fe/H]} & 
\colhead{$H\delta_A$} &
\colhead{$H\delta_F$} &
\colhead{CN$_1$}   &
\colhead{CN$_2$}   &
\colhead{Ca4227}  &
\colhead{G4300}  &
\colhead{$H\gamma_A$} &
\colhead{$H\gamma_F$} &
\colhead{Fe4383}  &
\colhead{C$_2$4668}  &
\colhead{$H\beta$} &
\colhead{Fe5015}  &
\colhead{Mg $b$}  &
\colhead{Mg$_2$}  &
\colhead{Fe5270}  &
\colhead{Fe5335}  }
\startdata
  0.10 & 0.0 & 8.1686 &  6.4731 & -0.1824 & -0.1385 &  0.0612 & -1.7169 &  7.8668  & 6.3867 & -0.0488  & 0.7528 &  6.3299 &  1.5202 &  0.9085  & 0.0582  & 0.9395  & 0.8202 \\
  0.10 & 0.3 & 8.9895 &  6.8871 & -0.1970 & -0.1517 &  0.1137 & -2.0943 &  8.5524  & 6.7879 & -0.0745  & 1.0056 &  6.6498 &  1.8321 &  1.0910  & 0.0707  & 1.1891  & 1.0672 \\
  0.11 & 0.0 & 8.5311 &  6.6415 & -0.1899 & -0.1453 &  0.0681 & -1.8564 &  8.1572  & 6.5464 & -0.1168  & 0.7293 &  6.4850 &  1.4952 &  0.9100  & 0.0572  & 0.9374  & 0.8045 \\
  0.11 & 0.3 & 9.3018 &  7.0218 & -0.2029 & -0.1570 &  0.1237 & -2.1802 &  8.7989  & 6.9133 & -0.1290  & 1.0121 &  6.7670 &  1.8428 &  1.0977  & 0.0700  & 1.1991  & 1.0660 \\
  0.13 & 0.0 & 8.8966 &  6.8068 & -0.1974 & -0.1521 &  0.0774 & -1.9834 &  8.4562  & 6.7060 & -0.1872  & 0.7037 &  6.6352 &  1.4751 &  0.9131  & 0.0564  & 0.9353  & 0.7914 \\
\enddata
\tablenotetext{1}{The complete version of this table is in the electronic edition of
the Journal.  The printed edition contains only a sample.}
\end{deluxetable*}

\begin{deluxetable*}{cccccccc} 
\label{phot1}
\tablecaption{Magnitudes computed using the solar-scaled isochrones.}
\tablewidth{0pt}
\tablehead{
\colhead{Age} & 
\colhead{[Fe/H]} & 
\colhead{$M_V^1$} &
\colhead{$(B-V)^1$} &
\colhead{$(U-V)^1$} &
\colhead{$M_V^2$} &
\colhead{$(B-V)^2$} &
\colhead{$(U-V)^2$} \\
}
\startdata
  0.10 & 0.00 & 2.751 & 0.105 &-0.375 & 2.661 & 0.121& -0.307  \\ 
  0.10 & 0.22 & 2.792 & 0.067 &-0.247 & 2.763 & 0.088& -0.296  \\ 
  0.11 & 0.00 & 2.809 & 0.110 &-0.321 & 2.736 & 0.126& -0.278  \\ 
  0.11 & 0.22 & 2.835 & 0.098 &-0.247 & 2.834 & 0.091& -0.264  \\ 
  0.13 & 0.00 & 2.877 & 0.115 &-0.272 & 2.805 & 0.137& -0.248  \\ 
  0.13 & 0.22 & 2.941 & 0.093 &-0.213 & 2.896 & 0.102& -0.227  \\ 
\enddata
\tablenotetext{1}{Obtained using the calibrations presented in Paper I}
\tablenotetext{2}{Obtained integrating the colors provided by Girardi \etal
(2000)}
\tablenotetext{3}{The complete version of this table is in the electronic edition of
the Journal.  The printed edition contains only a sample.}
\end{deluxetable*}

\begin{deluxetable*}{cccccccc} 
\label{phot2}
\tablecaption{Magnitudes computed using the $\alpha$-enhanced isochrones.}
\tablewidth{0pt}
\tablehead{
\colhead{Age} & 
\colhead{[Fe/H]} & 
\colhead{$M_V^1$} &
\colhead{$(B-V)^1$} &
\colhead{$(U-V)^1$} &
\colhead{$M_V^2$} &
\colhead{$(B-V)^2$} &
\colhead{$(U-V)^2$} \\
}
\startdata
  0.10 &  0.01& 2.999 & 0.138 &-0.187 & 2.924 & 0.150 &-0.224 \\
  0.10 & 0.33 & 3.042 & 0.180 &-0.146 & 2.992 & 0.173 &-0.150 \\
  0.11 & 0.01 & 3.066 & 0.145 &-0.184 & 2.995 & 0.154 &-0.191 \\
  0.11 & 0.33 & 3.112 & 0.187 &-0.110 & 3.063 & 0.181 &-0.113 \\
  0.13 & 0.01 & 3.135 & 0.151 &-0.149 & 3.070 & 0.158 &-0.157 \\
  0.13 & 0.33 & 3.183 & 0.196 &-0.074 & 3.136 & 0.189 &-0.076 \\
\enddata
\tablenotetext{1}{Obtained using the calibrations presented in Paper I}
\tablenotetext{2}{Obtained integrating the colors provided by Salasnich \etal
(2000)}
\tablenotetext{3}{The complete version of this table is in the electronic edition of
the Journal.  The printed edition contains only a sample.}
\end{deluxetable*}

\begin{deluxetable*}{cccccccccccccccccc} 
\label{m67mod}
\tabletypesize{\scriptsize}
\tablecaption{Lick indices computed for the abundance pattern of M~67
(Table~\ref{clust}).
}
\tablewidth{0pt}
\tablehead{
\colhead{Age} & 
\colhead{[Fe/H]} & 
\colhead{$H\delta_A$} &
\colhead{$H\delta_F$} &
\colhead{CN$_1$}   &
\colhead{CN$_2$}   &
\colhead{Ca4227}  &
\colhead{G4300}  &
\colhead{$H\gamma_A$} &
\colhead{$H\gamma_F$} &
\colhead{Fe4383}  &
\colhead{C$_2$4668}  &
\colhead{$H\beta$} &
\colhead{Fe5015}  &
\colhead{Mg $b$}  &
\colhead{Mg$_2$}  &
\colhead{Fe5270}  &
\colhead{Fe5335}  }
\startdata
  1.20 & -1.3 & 6.0966 &  4.5507 & -0.1510  &-0.1167 &  0.1300 &  0.8990 &  4.2798 &  3.9359 &  0.4553  & 0.6471 &  4.3633 &  1.5833 &  0.6659 &  0.0613  & 0.9834 &  0.8992  \\
  1.20 & -0.7 & 5.9204 &  4.4215 & -0.1427  &-0.1078 &  0.1403 &  1.2615 &  3.9890 &  3.9140 &  0.7978  & 1.5563 &  4.6787 &  2.3875 &  1.0419 &  0.0647  & 1.3361 &  1.0691  \\
  1.20 & -0.4 & 4.5117 &  3.6227 & -0.0982  &-0.0676 &  0.2991 &  1.9786 &  2.0874 &  2.9905 &  1.3462  & 2.2351 &  3.9965 &  3.1119 &  1.5158 &  0.0879  & 1.7320 &  1.4159  \\
  1.20 &  0.0 & 3.0096 &  2.8627 & -0.0612  &-0.0296 &  0.6182 &  2.3479 &  0.3124 &  2.1006 &  2.0694  & 2.8798 &  3.5384 &  4.0534 &  2.0061 &  0.1050  & 2.1810 &  1.8123  \\
  1.20 &  0.2 & 1.8031 &  2.2922 & -0.0469  &-0.0175 &  0.7568 &  2.8072 & -0.8854 &  1.4197 &  2.7212  & 3.6400 &  3.3265 &  4.8045 &  2.3176 &  0.1170  & 2.5374 &  2.1519  \\
\enddata
\tablenotetext{1}{The complete version of this table is in the electronic edition of
the Journal.  The printed edition contains only a sample.}
\end{deluxetable*}

\begin{deluxetable*}{cccccccccccccccccc} 
\label{47tucstrong}
\tabletypesize{\scriptsize}
\tablecaption{ Lick indices computed for the abundance pattern of
CN-strong stars in 47~Tuc, adopting the solar-scaled Padova isochrones
by Girardi \etal (2000), and a dwarf-poor IMF ($x=-4$).
}
\tablewidth{0pt}
\tablehead{
\colhead{Age} & 
\colhead{[Fe/H]} & 
\colhead{$H\delta_A$} &
\colhead{$H\delta_F$} &
\colhead{CN$_1$}   &
\colhead{CN$_2$}   &
\colhead{Ca4227}  &
\colhead{G4300}  &
\colhead{$H\gamma_A$} &
\colhead{$H\gamma_F$} &
\colhead{Fe4383}  &
\colhead{C$_2$4668}  &
\colhead{$H\beta$} &
\colhead{Fe5015}  &
\colhead{Mg $b$}  &
\colhead{Mg$_2$}  &
\colhead{Fe5270}  &
\colhead{Fe5335}  }
\startdata
  1.20 & -1.3 &  4.0428 &  3.2702 & -0.0698 & -0.0372 &  0.1607 &  1.6833 &  1.7633 &  2.5923 &  0.8294 & -0.7163 &  3.1227 &  2.5457 &  1.2904 &  0.0877 &  1.4054 &  1.1899  \\
  1.20 & -0.7 &  4.5019 &  3.1747 & -0.0260 &  0.0089 &  0.1922 &  1.7658 &  1.5785 &  2.6268 &  1.1471 &  0.4849 &  3.5153 &  3.6216 &  2.4293 &  0.1171 &  1.7448 &  1.3788  \\
  1.20 & -0.4 &  3.3331 &  2.4454 &  0.0463 &  0.0790 &  0.4001 &  2.2906 & -0.2106 &  1.7504 &  1.7669 &  1.1358 &  3.0013 &  4.3459 &  3.2696 &  0.1546 &  2.1175 &  1.7232  \\
  1.20 &  0.0 &  2.3064 &  1.9848 &  0.1121 &  0.1565 &  0.7977 &  2.1264 & -1.3286 &  1.2597 &  2.5900 &  1.9413 &  2.7866 &  5.1696 &  4.1125 &  0.1899 &  2.6231 &  2.1925  \\
  1.20 &  0.2 &  1.3420 &  1.5803 &  0.1470 &  0.1896 &  0.9784 &  2.2197 & -2.0882 &  0.8663 &  3.2612 &  2.7738 &  2.7021 &  5.8576 &  4.5784 &  0.2115 &  2.9997 &  2.5611  \\
\enddata
\tablenotetext{1}{The complete version of this table is in the electronic edition of
the Journal.  The printed edition contains only a sample.}
\end{deluxetable*}

\begin{deluxetable*}{cccccccccccccccccc} 
\label{47tucweak}
\tabletypesize{\scriptsize}
\tablecaption{ Lick indices computed for the abundance pattern of
CN-weak stars in 47~Tuc, adopting the solar-scaled Padova isochrones
by Girardi \etal (2000), and a dwarf-poor IMF ($x=-4$).
}
\tablewidth{0pt}
\tablehead{
\colhead{Age} & 
\colhead{[Fe/H]} & 
\colhead{$H\delta_A$} &
\colhead{$H\delta_F$} &
\colhead{CN$_1$}   &
\colhead{CN$_2$}   &
\colhead{Ca4227}  &
\colhead{G4300}  &
\colhead{$H\gamma_A$} &
\colhead{$H\gamma_F$} &
\colhead{Fe4383}  &
\colhead{C$_2$4668}  &
\colhead{$H\beta$} &
\colhead{Fe5015}  &
\colhead{Mg $b$}  &
\colhead{Mg$_2$}  &
\colhead{Fe5270}  &
\colhead{Fe5335}  }
\startdata
  1.20 & -1.3 -4.00 &  4.0421 &  3.2061 & -0.0893 & -0.0571 &  0.3187 &  2.1494 &  1.4408 &  2.4116 &  0.9833 &  1.1672 &  3.1022 &  2.4769 &  1.2449 &  0.0999 &  1.4003 &  1.1890  \\
  1.20 & -0.7 -4.00 &  4.1488 &  3.1160 & -0.0944 & -0.0646 &  0.4962 &  2.2658 &  1.2158 &  2.4324 &  1.3044 &  2.2431 &  3.4959 &  3.5577 &  2.3856 &  0.1286 &  1.7403 &  1.3775  \\
  1.20 & -0.4 -4.00 &  2.8222 &  2.3877 & -0.0483 & -0.0229 &  0.7826 &  2.7939 & -0.5636 &  1.5598 &  1.9242 &  2.9232 &  2.9816 &  4.2806 &  3.2263 &  0.1659 &  2.1128 &  1.7219  \\
  1.20 &  0.0 -4.00 &  1.7065 &  1.9373 & -0.0086 &  0.0260 &  1.2329 &  2.6752 & -1.7236 &  1.0545 &  2.7519 &  3.5962 &  2.7680 &  5.1090 &  4.0713 &  0.2006 &  2.6189 &  2.1907  \\
  1.20 &  0.2 -4.00 &  0.7096 &  1.5373 &  0.0127 &  0.0436 &  1.4399 &  2.7860 & -2.4991 &  0.6556 &  3.4245 &  4.3793 &  2.6839 &  5.7985 &  4.5380 &  0.2219 &  2.9956 &  2.5591  \\
\enddata
\tablenotetext{1}{The complete version of this table is in the electronic edition of
the Journal.  The printed edition contains only a sample.}
\end{deluxetable*}

\begin{deluxetable*}{cccccccccccccccccc} 
\label{n6528mod}
\tabletypesize{\scriptsize}
\tablecaption{Lick indices computed for the (CN-strong) abundance pattern
of NGC~6528, adopting the solar-scaled Padova isochrones by Girardi
\etal (2000) and a Salpeter IMF.
}
\tablewidth{0pt}
\tablehead{
\colhead{Age} & 
\colhead{[Fe/H]} & 
\colhead{$H\delta_A$} &
\colhead{$H\delta_F$} &
\colhead{CN$_1$}   &
\colhead{CN$_2$}   &
\colhead{Ca4227}  &
\colhead{G4300}  &
\colhead{$H\gamma_A$} &
\colhead{$H\gamma_F$} &
\colhead{Fe4383}  &
\colhead{C$_2$4668}  &
\colhead{$H\beta$} &
\colhead{Fe5015}  &
\colhead{Mg $b$}  &
\colhead{Mg$_2$}  &
\colhead{Fe5270}  &
\colhead{Fe5335}  }
\startdata
  1.20 & -1.3 & 6.2125 &  4.6109 & -0.1263 & -0.0879 & -0.0422 &  0.7027 &  4.5141 &  4.0340 &  0.3101 &  0.5845 &  4.3444 &  1.4873 &  0.7496 &  0.0634 &  0.9345 &  0.8754 \\
  1.20 & -0.7 & 6.3753 &  4.5593 & -0.1031 & -0.0621 & -0.1291 &  0.9388 &  4.4001 &  4.0775 &  0.6039 &  1.3198 &  4.6511 &  2.2329 &  1.1794 &  0.0680 &  1.2514 &  1.0113 \\
  1.20 & -0.4 & 5.1351 &  3.8033 & -0.0488 & -0.0108 & -0.0327 &  1.5774 &  2.6077 &  3.1942 &  1.1094 &  1.8991 &  3.9629 &  2.9192 &  1.6805 &  0.0916 &  1.6247 &  1.3373 \\
  1.20 &  0.0 & 3.6539 &  3.0563 & -0.0061 &  0.0331 &  0.2698 &  1.8513 &  0.9273 &  2.3510 &  1.8210 &  2.4877 &  3.4998 &  3.8611 &  2.2017 &  0.1097 &  2.0674 &  1.7295 \\
  1.20 &  0.2 & 2.5017 &  2.5061 &  0.0132 &  0.0507 &  0.3812 &  2.2481 & -0.1853 &  1.7031 &  2.4469 &  3.2099 &  3.2841 &  4.5913 &  2.5298 &  0.1220 &  2.4105 &  2.0574 \\
\enddata
\tablenotetext{1}{The complete version of this table is in the electronic edition of
the Journal.  The printed edition contains only a sample.}
\end{deluxetable*}

\begin{deluxetable*}{cccccccccccccccccc} 
\label{lum205mod}
\tabletypesize{\scriptsize}
\tablecaption{Model that best matches indices in the stacked SDSS spectrum
for $<M_r> = -20.78$.
}
\tablewidth{0pt}
\tablehead{
\colhead{Age} & 
\colhead{[Fe/H]} & 
\colhead{$H\delta_A$} &
\colhead{$H\delta_F$} &
\colhead{CN$_1$}   &
\colhead{CN$_2$}   &
\colhead{Ca4227}  &
\colhead{G4300}  &
\colhead{$H\gamma_A$} &
\colhead{$H\gamma_F$} &
\colhead{Fe4383}  &
\colhead{C$_2$4668}  &
\colhead{$H\beta$} &
\colhead{Fe5015}  &
\colhead{Mg $b$}  &
\colhead{Mg$_2$}  &
\colhead{Fe5270}  &
\colhead{Fe5335}  }
\startdata
  1.20 & -1.3 & 6.0624 &  4.5588 & -0.1415 & -0.1069 & 0.0754 & 0.9920 & 4.2726 & 3.9342 & 0.4451 & 0.8816 & 4.3535 & 1.5240 & 0.9808 & 0.0775 & 0.9854 & 0.8854 \\
  1.20 & -0.7 & 5.9609 &  4.4427 & -0.1194 & -0.0835 & 0.0352 & 1.4565 & 3.9224 & 3.8695 & 0.7880 & 2.3418 & 4.6637 & 2.2493 & 1.4704 & 0.0916 & 1.3197 & 1.0346 \\
  1.20 & -0.4 & 4.5990 &  3.6567 & -0.0672 & -0.0353 & 0.1684 & 2.2135 & 2.0183 & 2.9363 & 1.3131 & 3.3004 & 3.9759 & 2.9203 & 2.0386 & 0.1220 & 1.7033 & 1.3671 \\
  1.20 &  0.0 & 3.1165 &  2.9026 & -0.0284 &  0.0048 & 0.4895 & 2.6023 & 0.2010 & 2.0263 & 2.0477 & 4.1422 & 3.5140 & 3.8443 & 2.6546 & 0.1452 & 2.1507 & 1.7608 \\
  1.20 &  0.2 & 1.9267 &  2.3403 & -0.0111 &  0.0201 & 0.6218 & 3.0820 &-0.9986 & 1.3393 & 2.6822 & 5.0101 & 3.2985 & 4.5711 & 3.0414 & 0.1615 & 2.5004 & 2.0916 \\
\enddata
\tablenotetext{1}{The complete version of this table is in the electronic edition of
the Journal.  The printed edition contains only a sample.}
\end{deluxetable*}

\begin{deluxetable*}{cccccccccccccccccc} 
\label{lum210mod}
\tabletypesize{\scriptsize}
\tablecaption{Model that best matches indices in the stacked SDSS spectrum
for $<M_r> = -21.25$.
}
\tablewidth{0pt}
\tablehead{
\colhead{Age} & 
\colhead{[Fe/H]} & 
\colhead{$H\delta_A$} &
\colhead{$H\delta_F$} &
\colhead{CN$_1$}   &
\colhead{CN$_2$}   &
\colhead{Ca4227}  &
\colhead{G4300}  &
\colhead{$H\gamma_A$} &
\colhead{$H\gamma_F$} &
\colhead{Fe4383}  &
\colhead{C$_2$4668}  &
\colhead{$H\beta$} &
\colhead{Fe5015}  &
\colhead{Mg $b$}  &
\colhead{Mg$_2$}  &
\colhead{Fe5270}  &
\colhead{Fe5335}  }
\startdata
  1.20 & -1.3 & 6.0144 & 4.5497 &-0.1376 &-0.1029 & 0.0555 & 1.0543 & 4.2310 & 3.9145 & 0.4758 & 0.9799 & 4.3532 & 1.5203 & 0.9923 & 0.0792 & 0.9951 & 0.8851 \\
  1.20 & -0.7 & 5.9490 & 4.4404 &-0.1111 &-0.0748 & 0.0188 & 1.5535 & 3.8486 & 3.8336 & 0.8088 & 2.6239 & 4.6610 & 2.2299 & 1.5033 & 0.0961 & 1.3279 & 1.0304 \\
  1.20 & -0.4 & 4.5994 & 3.6570 &-0.0564 &-0.0239 & 0.1518 & 2.3294 & 1.9304 & 2.8929 & 1.3299 & 3.6811 & 3.9719 & 2.8918 & 2.0813 & 0.1279 & 1.7107 & 1.3605 \\
  1.20 &  0.0 & 3.1213 & 2.9041 &-0.0171 & 0.0168 & 0.4739 & 2.7376 & 0.0842 & 1.9689 & 2.0720 & 4.5838 & 3.5092 & 3.8130 & 2.7098 & 0.1522 & 2.1583 & 1.7540 \\
  1.20 &  0.2 & 1.9318 & 2.3425 & 0.0013 & 0.0333 & 0.6068 & 3.2311 &-1.1282 & 1.2750 & 2.7054 & 5.4880 & 3.2930 & 4.5365 & 3.1032 & 0.1692 & 2.5078 & 2.0833 \\
\enddata
\tablenotetext{1}{The complete version of this table is in the electronic edition of
the Journal.  The printed edition contains only a sample.}
\end{deluxetable*}

\begin{deluxetable*}{cccccccccccccccccc} 
\label{lum215mod}
\tabletypesize{\scriptsize}
\tablecaption{Model that best matches indices in the stacked SDSS spectrum
for $<M_r> = -21.73$.
}
\tablewidth{0pt}
\tablehead{
\colhead{Age} & 
\colhead{[Fe/H]} & 
\colhead{$H\delta_A$} &
\colhead{$H\delta_F$} &
\colhead{CN$_1$}   &
\colhead{CN$_2$}   &
\colhead{Ca4227}  &
\colhead{G4300}  &
\colhead{$H\gamma_A$} &
\colhead{$H\gamma_F$} &
\colhead{Fe4383}  &
\colhead{C$_2$4668}  &
\colhead{$H\beta$} &
\colhead{Fe5015}  &
\colhead{Mg $b$}  &
\colhead{Mg$_2$}  &
\colhead{Fe5270}  &
\colhead{Fe5335}  }
\startdata
  1.20 &-1.3  & 5.9878 & 4.5453 &-0.1322 &-0.0971 & 0.0352 & 1.0749 & 4.2175 & 3.9084 & 0.4772 & 1.0092 & 4.3527 & 1.5206 & 0.9990 & 0.0798 & 1.0050 & 0.8853 \\
  1.20 &-0.7  & 5.9632 & 4.4432 &-0.1015 &-0.0646 & 0.0079 & 1.5825 & 3.8307 & 3.8246 & 0.8006 & 2.6977 & 4.6594 & 2.2251 & 1.5230 & 0.0976 & 1.3378 & 1.0282 \\
  1.20 &-0.4  & 4.6299 & 3.6629 &-0.0442 &-0.0109 & 0.1437 & 2.3632 & 1.9111 & 2.8825 & 1.3160 & 3.7787 & 3.9696 & 2.8840 & 2.1069 & 0.1299 & 1.7207 & 1.3569 \\
  1.20 & 0.0  & 3.1547 & 2.9112 &-0.0037 & 0.0310 & 0.4651 & 2.7741 & 0.0596 & 1.9560 & 2.0592 & 4.6963 & 3.5066 & 3.8051 & 2.7428 & 0.1546 & 2.1687 & 1.7505 \\
  1.20 & 0.2  & 1.9670 & 2.3508 & 0.0157 & 0.0487 & 0.5993 & 3.2710 &-1.1545 & 1.2609 & 2.6893 & 5.6094 & 3.2901 & 4.5275 & 3.1397 & 0.1718 & 2.5184 & 2.0790 \\
\enddata
\tablenotetext{1}{The complete version of this table is in the electronic edition of
the Journal.  The printed edition contains only a sample.}
\end{deluxetable*}

\begin{deluxetable*}{cccccccccccccccccc} 
\label{lum220mod}
\tabletypesize{\scriptsize}
\tablecaption{Model that best matches indices in the stacked SDSS spectrum
for $<M_r> = -22.19$.
}
\tablewidth{0pt}
\tablehead{
\colhead{Age} & 
\colhead{[Fe/H]} & 
\colhead{$H\delta_A$} &
\colhead{$H\delta_F$} &
\colhead{CN$_1$}   &
\colhead{CN$_2$}   &
\colhead{Ca4227}  &
\colhead{G4300}  &
\colhead{$H\gamma_A$} &
\colhead{$H\gamma_F$} &
\colhead{Fe4383}  &
\colhead{C$_2$4668}  &
\colhead{$H\beta$} &
\colhead{Fe5015}  &
\colhead{Mg $b$}  &
\colhead{Mg$_2$}  &
\colhead{Fe5270}  &
\colhead{Fe5335}  }
\startdata
  1.20 &-1.3 &   5.9570 & 4.5428 &-0.1271 &-0.0918 & 0.0285 & 1.1130 & 4.1964 & 3.8988 & 0.4798 & 1.0721 & 4.3520 & 1.5181 & 1.0296 & 0.0815 & 1.0217 & 0.8848 \\
  1.20 &-0.7 &   5.9926 & 4.4585 &-0.0924 &-0.0551 & 0.0309 & 1.6296 & 3.8164 & 3.8148 & 0.7752 & 2.8472 & 4.6552 & 2.2018 & 1.6140 & 0.1028 & 1.3513 & 1.0202 \\
  1.20 &-0.4 &   4.6851 & 3.6868 &-0.0327 & 0.0010 & 0.1802 & 2.4160 & 1.9022 & 2.8729 & 1.2741 & 3.9741 & 3.9635 & 2.8485 & 2.2269 & 0.1368 & 1.7324 & 1.3444 \\
  1.20 & 0.0 &   3.2149 & 2.9379 & 0.0086 & 0.0439 & 0.5040 & 2.8265 & 0.0471 & 1.9449 & 2.0175 & 4.9247 & 3.4996 & 3.7676 & 2.8937 & 0.1629 & 2.1806 & 1.7377 \\
  1.20 & 0.2 &   2.0336 & 2.3811 & 0.0290 & 0.0627 & 0.6430 & 3.3268 &-1.1647 & 1.2495 & 2.6386 & 5.8562 & 3.2820 & 4.4852 & 3.3072 & 0.1809 & 2.5297 & 2.0636 \\
\enddata
\tablenotetext{1}{The complete version of this table is in the electronic edition of
the Journal.  The printed edition contains only a sample.}
\end{deluxetable*}

\begin{deluxetable*}{ccccccccccccccccc} 
\label{sigcorr1}
\tabletypesize{\scriptsize}
\tablecaption{Corrections to $\sigma=0$ km s$^{-1}$,
for a 1.5 Gyr-old stellar population with solar metallicity.
}
\tablewidth{0pt}
\tablehead{
\colhead{$\sigma$ (km s$^{-1}$)} & 
\colhead{$H\delta_A$} &
\colhead{$H\delta_F$} &
\colhead{CN$_1$}   &
\colhead{CN$_2$}   &
\colhead{Ca4227}  &
\colhead{G4300}  &
\colhead{$H\gamma_A$} &
\colhead{$H\gamma_F$} &
\colhead{Fe4383}  &
\colhead{C$_2$4668}  &
\colhead{$H\beta$} &
\colhead{Fe5015}  &
\colhead{Mg $b$}  &
\colhead{Mg$_2$}  &
\colhead{Fe5270}  &
\colhead{Fe5335}  }
\startdata
     25 & 0.9993& 1.0007& 0.0000& 0.0002& 1.0028& 1.0006& 0.9998& 1.0005& 1.0015& 1.0005& 1.0004& 1.0018& 0.0000& 1.0012& 1.0018& 1.0033    \\ 
     50 & 0.9973& 1.0029& 0.0001& 0.0005& 1.0112& 1.0023& 0.9992& 1.0023& 1.0062& 1.0020& 1.0016& 1.0072& 0.0001& 1.0048& 1.0071& 1.0132    \\ 
     75 & 0.9939& 1.0065& 0.0003& 0.0010& 1.0253& 1.0050& 0.9983& 1.0054& 1.0139& 1.0046& 1.0036& 1.0160& 0.0002& 1.0110& 1.0159& 1.0298    \\ 
    100 & 0.9893& 1.0117& 0.0005& 0.0018& 1.0453& 1.0089& 0.9974& 1.0101& 1.0246& 1.0083& 1.0064& 1.0280& 0.0004& 1.0200& 1.0280& 1.0531    \\ 
    125 & 0.9839& 1.0184& 0.0008& 0.0027& 1.0713& 1.0138& 0.9968& 1.0166& 1.0384& 1.0132& 1.0100& 1.0428& 0.0005& 1.0319& 1.0431& 1.0832    \\ 
    150 & 0.9776& 1.0268& 0.0011& 0.0038& 1.1035& 1.0198& 0.9968& 1.0253& 1.0551& 1.0194& 1.0143& 1.0600& 0.0007& 1.0471& 1.0610& 1.1200    \\ 
    175 & 0.9708& 1.0370& 0.0015& 0.0050& 1.1425& 1.0268& 0.9977& 1.0367& 1.0748& 1.0268& 1.0195& 1.0793& 0.0009& 1.0655& 1.0813& 1.1636    \\ 
    200 & 0.9636& 1.0490& 0.0019& 0.0063& 1.1885& 1.0348& 0.9998& 1.0511& 1.0974& 1.0356& 1.0257& 1.1004& 0.0012& 1.0870& 1.1036& 1.2140    \\ 
    225 & 0.9563& 1.0630& 0.0023& 0.0077& 1.2422& 1.0439& 1.0035& 1.0691& 1.1230& 1.0457& 1.0329& 1.1231& 0.0014& 1.1116& 1.1278& 1.2711    \\ 
    250 & 0.9491& 1.0790& 0.0027& 0.0092& 1.3040& 1.0541& 1.0090& 1.0911& 1.1515& 1.0571& 1.0415& 1.1471& 0.0017& 1.1389& 1.1535& 1.3350    \\ 
    275 & 0.9421& 1.0971& 0.0031& 0.0106& 1.3747& 1.0653& 1.0164& 1.1175& 1.1831& 1.0700& 1.0514& 1.1724& 0.0019& 1.1686& 1.1807& 1.4053    \\ 
    300 & 0.9355& 1.1174& 0.0035& 0.0121& 1.4555& 1.0777& 1.0261& 1.1486& 1.2178& 1.0842& 1.0630& 1.1987& 0.0023& 1.2007& 1.2092& 1.4821    \\ 
    325 & 0.9293& 1.1400& 0.0039& 0.0136& 1.5470& 1.0913& 1.0380& 1.1850& 1.2559& 1.0998& 1.0762& 1.2261& 0.0026& 1.2348& 1.2391& 1.5651    \\ 
    350 & 0.9239& 1.1649& 0.0043& 0.0150& 1.6508& 1.1062& 1.0525& 1.2271& 1.2973& 1.1167& 1.0913& 1.2545& 0.0030& 1.2708& 1.2705& 1.6540    \\ 
    375 & 0.9191& 1.1920& 0.0046& 0.0164& 1.7678& 1.1226& 1.0695& 1.2754& 1.3424& 1.1349& 1.1084& 1.2837& 0.0034& 1.3087& 1.3035& 1.7491    \\ 
    400 & 0.9151& 1.2216& 0.0050& 0.0177& 1.8997& 1.1404& 1.0892& 1.3303& 1.3915& 1.1543& 1.1276& 1.3138& 0.0039& 1.3485& 1.3386& 1.8500    \\ 
\enddata
\tablenotetext{1}{Corrections are multiplicative for indices defined as EWs
and additive for those defined in magnitudes. }
\end{deluxetable*}

\begin{deluxetable*}{ccccccccccccccccc} 
\label{sigcorr3}
\tabletypesize{\scriptsize}
\tablecaption{Corrections to $\sigma=0$ km s$^{-1}$,
for a 3.5 Gyr-old stellar population with solar metallicity.
}
\tablewidth{0pt}
\tablehead{
\colhead{$\sigma$ (km s$^{-1}$)} & 
\colhead{$H\delta_A$} &
\colhead{$H\delta_F$} &
\colhead{CN$_1$}   &
\colhead{CN$_2$}   &
\colhead{Ca4227}  &
\colhead{G4300}  &
\colhead{$H\gamma_A$} &
\colhead{$H\gamma_F$} &
\colhead{Fe4383}  &
\colhead{C$_2$4668}  &
\colhead{$H\beta$} &
\colhead{Fe5015}  &
\colhead{Mg $b$}  &
\colhead{Mg$_2$}  &
\colhead{Fe5270}  &
\colhead{Fe5335}  }
\startdata
     25 & 1.0031& 1.0011& 0.0000& 0.0001& 1.0028& 1.0003& 0.9999& 1.0004& 1.0014& 1.0006& 1.0004& 1.0018& 0.0001& 1.0011& 1.0018& 1.0032    \\ 
     50 & 1.0117& 1.0047& 0.0001& 0.0004& 1.0110& 1.0014& 0.9997& 1.0017& 1.0054& 1.0022& 1.0015& 1.0072& 0.0002& 1.0045& 1.0072& 1.0129    \\ 
     75 & 1.0266& 1.0106& 0.0004& 0.0010& 1.0246& 1.0032& 0.9994& 1.0036& 1.0121& 1.0050& 1.0032& 1.0161& 0.0003& 1.0104& 1.0161& 1.0290    \\ 
    100 & 1.0474& 1.0190& 0.0006& 0.0018& 1.0441& 1.0057& 0.9992& 1.0051& 1.0214& 1.0090& 1.0057& 1.0281& 0.0005& 1.0190& 1.0282& 1.0517    \\ 
    125 & 1.0746& 1.0303& 0.0010& 0.0028& 1.0693& 1.0089& 0.9991& 1.0058& 1.0333& 1.0142& 1.0087& 1.0428& 0.0007& 1.0303& 1.0435& 1.0809    \\ 
    150 & 1.1082& 1.0446& 0.0014& 0.0039& 1.1008& 1.0128& 0.9994& 1.0043& 1.0478& 1.0208& 1.0124& 1.0600& 0.0009& 1.0447& 1.0615& 1.1168    \\ 
    175 & 1.1488& 1.0622& 0.0018& 0.0052& 1.1387& 1.0175& 1.0002& 1.0000& 1.0647& 1.0286& 1.0167& 1.0792& 0.0011& 1.0621& 1.0820& 1.1595    \\ 
    200 & 1.1965& 1.0832& 0.0023& 0.0066& 1.1834& 1.0229& 1.0017& 0.9920& 1.0840& 1.0379& 1.0219& 1.1001& 0.0014& 1.0824& 1.1046& 1.2087    \\ 
    225 & 1.2520& 1.1084& 0.0028& 0.0081& 1.2356& 1.0292& 1.0040& 0.9795& 1.1056& 1.0485& 1.0279& 1.1225& 0.0017& 1.1056& 1.1291& 1.2648    \\ 
    250 & 1.3157& 1.1379& 0.0034& 0.0097& 1.2958& 1.0362& 1.0072& 0.9630& 1.1296& 1.0605& 1.0350& 1.1461& 0.0020& 1.1314& 1.1550& 1.3274    \\ 
    275 & 1.3884& 1.1722& 0.0039& 0.0113& 1.3647& 1.0441& 1.0115& 0.9425& 1.1559& 1.0739& 1.0434& 1.1709& 0.0023& 1.1595& 1.1825& 1.3966    \\ 
    300 & 1.4706& 1.2116& 0.0045& 0.0129& 1.4431& 1.0530& 1.0168& 0.9185& 1.1844& 1.0887& 1.0533& 1.1966& 0.0027& 1.1899& 1.2114& 1.4722    \\ 
    325 & 1.5632& 1.2569& 0.0051& 0.0145& 1.5322& 1.0630& 1.0234& 0.8918& 1.2153& 1.1048& 1.0648& 1.2233& 0.0031& 1.2224& 1.2417& 1.5541    \\ 
    350 & 1.6674& 1.3085& 0.0056& 0.0161& 1.6327& 1.0740& 1.0312& 0.8637& 1.2485& 1.1223& 1.0780& 1.2508& 0.0036& 1.2568& 1.2735& 1.6421    \\ 
    375 & 1.7838& 1.3668& 0.0061& 0.0177& 1.7462& 1.0863& 1.0403& 0.8348& 1.2841& 1.1411& 1.0931& 1.2791& 0.0041& 1.2931& 1.3070& 1.7362    \\ 
    400 & 1.9137& 1.4330& 0.0067& 0.0192& 1.8742& 1.0999& 1.0507& 0.8059& 1.3220& 1.1611& 1.1102& 1.3081& 0.0047& 1.3315& 1.3425& 1.8362    \\ 
\enddata
\tablenotetext{1}{Corrections are multiplicative for indices defined as EWs
and additive for those defined in magnitudes. }
\end{deluxetable*}

\begin{deluxetable*}{ccccccccccccccccc} 
\label{sigcorr8}
\tabletypesize{\scriptsize}
\tablecaption{Corrections to $\sigma=0$ km s$^{-1}$,
for a 7.9 Gyr-old stellar population with solar metallicity.
}
\tablewidth{0pt}
\tablehead{
\colhead{$\sigma$ (km s$^{-1}$)} & 
\colhead{$H\delta_A$} &
\colhead{$H\delta_F$} &
\colhead{CN$_1$}   &
\colhead{CN$_2$}   &
\colhead{Ca4227}  &
\colhead{G4300}  &
\colhead{$H\gamma_A$} &
\colhead{$H\gamma_F$} &
\colhead{Fe4383}  &
\colhead{C$_2$4668}  &
\colhead{$H\beta$} &
\colhead{Fe5015}  &
\colhead{Mg $b$}  &
\colhead{Mg$_2$}  &
\colhead{Fe5270}  &
\colhead{Fe5335}  }
\startdata
     25 & 1.0012& 1.0034& 0.0001& 0.0001& 1.0027& 1.0003& 0.9999& 1.0005& 1.0013& 1.0006& 1.0004& 1.0019& 0.0001& 1.0011& 1.0018& 1.0032    \\ 
     50 & 1.0047& 1.0131& 0.0002& 0.0005& 1.0106& 1.0011& 0.9997& 1.0021& 1.0051& 1.0022& 1.0014& 1.0075& 0.0002& 1.0045& 1.0071& 1.0127    \\ 
     75 & 1.0104& 1.0302& 0.0004& 0.0011& 1.0239& 1.0025& 0.9994& 1.0047& 1.0114& 1.0051& 1.0029& 1.0165& 0.0003& 1.0104& 1.0158& 1.0285    \\ 
    100 & 1.0184& 1.0552& 0.0007& 0.0019& 1.0427& 1.0044& 0.9992& 1.0079& 1.0202& 1.0091& 1.0051& 1.0288& 0.0005& 1.0189& 1.0279& 1.0509    \\ 
    125 & 1.0286& 1.0898& 0.0011& 0.0029& 1.0670& 1.0070& 0.9991& 1.0114& 1.0314& 1.0145& 1.0078& 1.0440& 0.0007& 1.0302& 1.0428& 1.0797    \\ 
    150 & 1.0408& 1.1359& 0.0015& 0.0040& 1.0975& 1.0101& 0.9993& 1.0149& 1.0450& 1.0211& 1.0109& 1.0617& 0.0010& 1.0444& 1.0606& 1.1152    \\ 
    175 & 1.0550& 1.1967& 0.0020& 0.0054& 1.1341& 1.0139& 1.0000& 1.0180& 1.0610& 1.0291& 1.0146& 1.0815& 0.0013& 1.0616& 1.0808& 1.1575    \\ 
    200 & 1.0711& 1.2766& 0.0025& 0.0068& 1.1773& 1.0182& 1.0012& 1.0202& 1.0791& 1.0385& 1.0190& 1.1030& 0.0016& 1.0816& 1.1029& 1.2063    \\ 
    225 & 1.0889& 1.3813& 0.0031& 0.0083& 1.2276& 1.0233& 1.0031& 1.0214& 1.0994& 1.0492& 1.0241& 1.1261& 0.0019& 1.1043& 1.1269& 1.2621    \\ 
    250 & 1.1083& 1.5210& 0.0037& 0.0099& 1.2855& 1.0291& 1.0059& 1.0211& 1.1219& 1.0614& 1.0302& 1.1505& 0.0022& 1.1297& 1.1524& 1.3246    \\ 
    275 & 1.1293& 1.7115& 0.0043& 0.0116& 1.3519& 1.0358& 1.0096& 1.0195& 1.1464& 1.0749& 1.0374& 1.1761& 0.0026& 1.1573& 1.1793& 1.3937    \\ 
    300 & 1.1516& 1.9809& 0.0049& 0.0132& 1.4271& 1.0433& 1.0143& 1.0163& 1.1729& 1.0898& 1.0459& 1.2027& 0.0030& 1.1871& 1.2076& 1.4693    \\ 
    325 & 1.1753& 2.3791& 0.0055& 0.0149& 1.5125& 1.0518& 1.0201& 1.0117& 1.2015& 1.1062& 1.0560& 1.2302& 0.0035& 1.2191& 1.2373& 1.5516    \\ 
    350 & 1.2001& 3.0189& 0.0061& 0.0166& 1.6087& 1.0613& 1.0270& 1.0060& 1.2321& 1.1238& 1.0677& 1.2586& 0.0041& 1.2528& 1.2684& 1.6400    \\ 
    375 & 1.2261& 4.1801& 0.0067& 0.0182& 1.7171& 1.0721& 1.0350& 0.9994& 1.2646& 1.1428& 1.0811& 1.2877& 0.0047& 1.2885& 1.3013& 1.7347    \\ 
    400 & 1.2529& 6.9104& 0.0073& 0.0198& 1.8389& 1.0840& 1.0442& 0.9921& 1.2991& 1.1631& 1.0965& 1.3176& 0.0053& 1.3262& 1.3363& 1.8356    \\ 
\enddata
\tablenotetext{1}{Corrections are multiplicative for indices defined as EWs
and additive for those defined in magnitudes. }
\end{deluxetable*}

\begin{deluxetable*}{ccccccccccccccccc} 
\label{sigcorr14}
\tabletypesize{\scriptsize}
\tablecaption{Corrections to $\sigma=0$ km s$^{-1}$,
for a 14.1 Gyr-old stellar population with solar metallicity.
}
\tablewidth{0pt}
\tablehead{
\colhead{$\sigma$ (km s$^{-1}$)} & 
\colhead{$H\delta_A$} &
\colhead{$H\delta_F$} &
\colhead{CN$_1$}   &
\colhead{CN$_2$}   &
\colhead{Ca4227}  &
\colhead{G4300}  &
\colhead{$H\gamma_A$} &
\colhead{$H\gamma_F$} &
\colhead{Fe4383}  &
\colhead{C$_2$4668}  &
\colhead{$H\beta$} &
\colhead{Fe5015}  &
\colhead{Mg $b$}  &
\colhead{Mg$_2$}  &
\colhead{Fe5270}  &
\colhead{Fe5335}  }
\startdata
     25 & 1.0009& 0.9972& 0.0000& 0.0001& 1.0026& 1.0002& 0.9999& 1.0006& 1.0012& 1.0006& 1.0003& 1.0019& 0.0000& 1.0011& 1.0018& 1.0031    \\ 
     50 & 1.0034& 0.9900& 0.0002& 0.0005& 1.0104& 1.0009& 0.9997& 1.0023& 1.0049& 1.0023& 1.0012& 1.0076& 0.0001& 1.0045& 1.0071& 1.0125    \\ 
     75 & 1.0077& 0.9774& 0.0004& 0.0011& 1.0236& 1.0020& 0.9994& 1.0051& 1.0110& 1.0052& 1.0026& 1.0169& 0.0003& 1.0103& 1.0158& 1.0281    \\ 
    100 & 1.0136& 0.9602& 0.0007& 0.0019& 1.0422& 1.0036& 0.9991& 1.0087& 1.0195& 1.0094& 1.0045& 1.0295& 0.0005& 1.0187& 1.0278& 1.0502    \\ 
    125 & 1.0210& 0.9387& 0.0011& 0.0030& 1.0664& 1.0056& 0.9990& 1.0128& 1.0303& 1.0148& 1.0068& 1.0450& 0.0008& 1.0298& 1.0428& 1.0786    \\ 
    150 & 1.0299& 0.9133& 0.0016& 0.0042& 1.0966& 1.0082& 0.9991& 1.0174& 1.0434& 1.0216& 1.0094& 1.0630& 0.0010& 1.0437& 1.0605& 1.1137    \\ 
    175 & 1.0402& 0.8844& 0.0021& 0.0056& 1.1328& 1.0112& 0.9996& 1.0219& 1.0587& 1.0298& 1.0123& 1.0833& 0.0013& 1.0605& 1.0806& 1.1554    \\ 
    200 & 1.0519& 0.8529& 0.0027& 0.0070& 1.1755& 1.0149& 1.0007& 1.0263& 1.0762& 1.0394& 1.0157& 1.1053& 0.0016& 1.0801& 1.1028& 1.2037    \\ 
    225 & 1.0647& 0.8191& 0.0033& 0.0086& 1.2254& 1.0191& 1.0024& 1.0301& 1.0957& 1.0504& 1.0196& 1.1289& 0.0020& 1.1023& 1.1267& 1.2589    \\ 
    250 & 1.0787& 0.7847& 0.0040& 0.0103& 1.2828& 1.0240& 1.0048& 1.0333& 1.1173& 1.0628& 1.0244& 1.1538& 0.0023& 1.1270& 1.1522& 1.3208    \\ 
    275 & 1.0937& 0.7496& 0.0047& 0.0120& 1.3482& 1.0297& 1.0082& 1.0356& 1.1408& 1.0767& 1.0301& 1.1798& 0.0027& 1.1540& 1.1791& 1.3895    \\ 
    300 & 1.1096& 0.7152& 0.0053& 0.0138& 1.4226& 1.0362& 1.0124& 1.0371& 1.1661& 1.0920& 1.0369& 1.2069& 0.0032& 1.1831& 1.2072& 1.4647    \\ 
    325 & 1.1265& 0.6820& 0.0060& 0.0156& 1.5067& 1.0437& 1.0176& 1.0377& 1.1934& 1.1087& 1.0451& 1.2349& 0.0037& 1.2142& 1.2369& 1.5464    \\ 
    350 & 1.1441& 0.6501& 0.0067& 0.0173& 1.6017& 1.0522& 1.0239& 1.0376& 1.2224& 1.1268& 1.0548& 1.2637& 0.0043& 1.2472& 1.2681& 1.6345    \\ 
    375 & 1.1625& 0.6203& 0.0074& 0.0191& 1.7082& 1.0618& 1.0312& 1.0369& 1.2532& 1.1463& 1.0660& 1.2932& 0.0049& 1.2820& 1.3009& 1.7289    \\ 
    400 & 1.1816& 0.5926& 0.0081& 0.0208& 1.8281& 1.0726& 1.0397& 1.0357& 1.2857& 1.1671& 1.0791& 1.3234& 0.0056& 1.3187& 1.3358& 1.8295    \\ 
\enddata
\tablenotetext{1}{Corrections are multiplicative for indices defined as EWs
and additive for those defined in magnitudes. }
\end{deluxetable*}


\clearpage










\end{document}